\documentclass[onecolumn,showpacs,preprintnumbers,amsmath,amssymb]{revtex4-1}

\usepackage{epsfig}% Include figure files
\usepackage{graphicx}% Include figure files
\usepackage{dcolumn}% Align table columns on decimal point
\usepackage{bm}% bold math
\usepackage{epstopdf}
\usepackage{subfigure}

\newcommand{\EQ}{\begin{equation}}
\newcommand{\EN}{\end{equation}}
\newcommand{\ea}{\end{eqnarray}}
\newcommand{\ba}{\begin{eqnarray}}

\newcommand{\bear}{\begin{eqnarray}}
\newcommand{\ear}{\end{eqnarray}}

\begin{document}

%%%%%%%%%%%%%%%%%%%%%%%%%%%%%%%%%%%%%%%%%%%%%%%%%%%%%%%%%%%%%%%%%%%%%%%%%%
%                               Title                                    
%%%%%%%%%%%%%%%%%%%%%%%%%%%%%%%%%%%%%%%%%%%%%%%%%%%%%%%%%%%%%%%%%%%%%%%%%%

\title{Pseudoparticle approach to 1D integrable quantum models}
\author{J. M. P. Carmelo$^{a,b,c}$ and P. D. Sacramento$^{d,c}$}

\affiliation{$^{a}$Department of Physics, University of Minho, Campus Gualtar, P-4710-057 Braga, Portugal}
\affiliation{$^{b}$Center of Physics of University of Minho and University of Porto, P-4169-007 Oporto, Portugal}
\affiliation{$^{c}$Beijing Computational Science Research Center, Beijing 100193, China}
\affiliation{$^{d}$Departamento de Fisica and CeFEMA, Instituto Superior T\'ecnico, Universidade de Lisboa, 
Av. Rovisco Pais, P-1049-001 Lisboa, Portugal}

\date{23 October 2017}
%\date{\today}

%%%%%%%%%%%%%%%%%%%%%%%%%%%%%%%%%%%%%%%%%%%%%%%%%%%%%%%%%%%%%%%%%%%%%%%%%%
%                              abstract                                  
%%%%%%%%%%%%%%%%%%%%%%%%%%%%%%%%%%%%%%%%%%%%%%%%%%%%%%%%%%%%%%%%%%%%%%%%%%

\begin{abstract}
Over the last three decades a large number of experimental studies on several quasi one-dimensional (1D) metals and quasi 
1D Mott-Hubbard insulators have produced evidence for distinct spectral features identified 
with charge-only and spin-only fractionalized particles. They can be also observed in ultra-cold atomic 1D optical lattices and quantum wires.
1D exactly solvable models provide nontrivial tests of the approaches for these systems relying on field theories.
Different schemes such as the pseudofermion dynamical theory (PDT) and the mobile quantum impurity model (MQIM) have 
revealed that the 1D correlated models high-energy physics is
qualitatively different from that of a low-energy Tomonaga-Luttinger liquid (TLL). This includes the momentum 
dependence of the exponents that control the one- and two-particle dynamical correlation functions near their 
spectra edges and in the vicinity of one-particle singular spectral features. 

On the one hand, the low-energy 
charge-only and spin-only fractionalized particles are usually identified with holons and spinons,
respectively. On the other hand, ``particle-like'' representations in terms of
{\it pseudoparticles}, related PDT {\it pseudofermions}, and MQIM particles are suitable for the description of 
both the low-energy TLL physics and high-energy spectral and dynamical properties of 1D correlated systems. 

The main goal of this review is to revisit the usefulness of pseudoparticle and PDT pseudofermion representations for the study of 
both static and high-energy spectral and dynamical properties of the 1D Lieb-Liniger Bose gas, spin-$1/2$ isotropic Heisenberg chain,
and 1D Hubbard model. Moreover, the relation between the PDT and the MQIM is clarified. The fractionalized particles and related 
composite pseudoparticles/pseudofermions emerging within such non-perturbative 1D correlated systems are qualitatively different 
from the Fermi-liquid quasiparticles. In contrast to the holons and spinons, the relation to the electron creation and 
annihilation operators of the operators associated with the 1D Hubbard model three fractionalized particles is uniquely defined. 
The occupancy configurations of such fractionalized particles generate {\it all} energy and momentum eigenstates 
of that model. Both the static and dynamical properties of the three models under review are shown to be 
controlled at all energy scales by pseudofermion phase shifts associated with only zero-momentum forward scattering.
The corresponding microscopic processes are much simpler than those of the underlying particles non-perturbative interactions. 
\end{abstract}

\pacs{71.10.Pm, 71.10.Fd, 71.10.Hf, 72.15.Nj}

\maketitle
\newpage
\tableofcontents
%%%%%%%%%%%%%%%%%%%%%%%%%%%%%%%%%%%%%%%%%%%%%%%%%%%%%%%%%%%%%%%%%%%%%%%%%%
%                              body of paper                             
%%%%%%%%%%%%%%%%%%%%%%%%%%%%%%%%%%%%%%%%%%%%%%%%%%%%%%%%%%%%%%%%%%%%%%%%%%
\section{Introduction}
\label{elem-obj-intr}

Within quantum physics the isolated electrons are split into smaller components, earning them the designation of a fundamental particle.
However this does not necessarily apply when several electrons are brought together. Also the collective behavior
of interacting bosonic particles may not be described in terms of that of isolated bosons.

On the one hand, most of the current understanding of many-particle quantum systems involves the concept of 
a quasiparticle. In three dimensions, starting from free constituents obeying either fermionic or bosonic statistics, one builds 
a many-particle ground state as either a Fermi sea or a Bose-Einstein condensate. Interactions then Òadiabatically deformÓ the 
ground state into a Fermi liquid with well-defined electron-like excitations, or a condensate state with Bogoliubov-like 
modes, respectively. In both cases, these well-defined excitations are conveniently described as quasiparticles. 
Moreover, they reveal themselves via sharp lines in dynamical correlation functions, indicative of free-particle-like 
coherently propagating modes.

On the other hand, one-dimensional (1D) interacting systems are characterized by a breakdown of the 
basic Fermi liquid quasiparticle picture. Indeed, no quasiparticles occur when the electrons range of motion is 
restricted to a single spatial dimension \cite{Sutherland-04,Voit}. In a 1D 
chain correlated electrons rather split into basic fractionalized charge-only and spin-only particles \cite{Voit,Schulz-90,Lederer-00}.
These fractionalized particles can move with different speeds and even in different directions in the 1D 
many-electron system. Electrons in that system have this ability because they behave like waves. 
When excited, such waves can split into multiple waves, each carrying different characteristics of the electron. 
This occurs because collective modes take over. Indeed, applying perturbations does not create single Fermi-liquid 
quasiparticles. It rather originates an energy continuum of excitations described by exotic fractionalized particles. The latter emerge
within 1D many-particle systems. However, they cannot exist independently outside such systems. Moreover, they are not adiabatically connected 
to free particles. Hence they must be described using a different language. 

In electronic systems and spin chains these characteristic fractionalized-particle continua 
of excitations have been observed \cite{Kim-06,Ralph-12,Schlappa-12,Bisogni-15}. 
Realizations of 1D quantum liquids are numerous. They can take the form of, for example, quasi-1D materials.
Over the last three decades, many angle-resolved photoemission spectroscopy (ARPES) studies on several quasi-1D 
metals and quasi-1D Mott-Hubbard insulators have indeed revealed separate charge and spin spectral features. This includes the ARPES 
spectra of the compound K$_{0.3}$MoO$_3$ and other quasi-1D materials \cite{Allen-95},
quasi-1D organic metals (Bechgaard salts) (TMTSF)$_2$PF$_6$, (TMTSF)$_2$Cl O$_4$
(where TMTSF is tetramethyltetraselenafulvalene), and
(TMTTF)$_2$PF$_6$ (where TMTTF is tetramethyltetratiafulvalene) \cite{Zwick-97},
quasi-1D metal Li$_{0.9}$Mo$_6$O$_{17}$ \cite{Allen-02,Gweon-02},
quasi-1D organic conductor tetrathiafulvalene tetracyanoquinodimethane (TTF-TCNQ) \cite{TTF,Ralph-02,spectral0,spectral},
quasi-1D Mott-Hubbard insulators SrCuO$_2$ and Sr$_2$CuO$_3$	
\cite{Kim-06,Kim-96,Kim-97,Fujisawa-99,Hasan-02},
NaV$_2$O$_5$ \cite{Kobayashi-98}, Na$_{0.96}$V$_2$O$_5$ \cite{Kobayashi-99},
SrCuO$_2$ and V$_6$O$_{13}$ \cite{Benthien-04B},
doped quasi-1D Mott-Hubbard insulator Sr$_2$CuO$_{3+\delta}$ \cite{Kidd-08},
1D metallic surface state on an anisotropic InSb(001) surface covered with Bi \cite{Ohtsubo-15}, and
metallic 1D line defects in transition dichalcogenides such as MoSe$_2$ \cite{MoSe-17}.

Moreover, similar charge-like and spin-like spectral features were seen as well by
electron energy-loss spectroscopic studies on quasi-1D metals and other low-dimensional 
materials \cite{Fink-01}. They were also seen in high-resolution resonant inelastic X-ray scattering on the quasi-1D Mott-Hubbard 
insulator Sr$_2$CuO$_3$ \cite{Ralph-12,Schlappa-12}. The high-resolution resonant inelastic X-ray scattering experiments
on CaCu$_2$O$_3$ reported in Ref. \cite{Bisogni-15} reveal that the orbital hopping 
in that compound can select different degrees of dimensionality. A spin-orbital fractionalization along the leg direction 
$x$ through the $xz$ orbital channel was observed as in a 1D system.
The mode separation in 1D correlated bosonic and fermionic models and corresponding fractionalized particles are
also observed in 1D trapped ultracold atomic gases and ultra-cold atoms on 1D optical lattices
\cite{Guan-13,Guan-15,Cazalilla-11,Petrov-00,Murmann-15,Golovach-09,Fabbri-15,Zoller-05,Massel-05,Huo-12,Cheneau-12}
and quantum wires \cite{Jompol-09}. 

The non-perturbative nature of 1D correlated systems prevents their study by conventional perturbative many-body
techniques. Nonetheless, some of these systems are exactly solvable by the Bethe ansatz (BA). This method was
developed by Bethe in 1931 \cite{Bethe}. He applied it to the spin-$1/2$ isotropic Heisenberg chain \cite{Heisenberg}.
The BA turned out to be not only useful for that model, but also a very powerful method for a wide range of integrable 
models. This applies both within and outside the scope of condensed matter physics. The BA provides the exact energy eigenvalues
and some thermodynamic quantities. Combined with bosonization \cite{Voit,Schulz-90,Lederer-00,Luther-74,Luther-75,Solyom-79} 
or the conformal-invariance associated with the spectra finite-size corrections \cite{Blote-85,Affleck-85},
the BA allows the computation of low-energy physics quantities. This has revealed
that 1D correlated models share common low-energy properties associated with the universal class of the Tomonaga-Luttinger liquid (TLL) \cite{Tomonaga-50,Luttinger-63,Solyom-79}.

One of the main challenges in the study of the 1D correlated systems properties is the 
calculation of dynamical correlation functions. Indeed, it has been difficult to apply the BA to the derivation of high-energy 
dynamical correlation functions. (In this review ``high energy'' means excitation energy values beyond 
those of the low-energy TLL validity.) The high-energy dynamical correlation functions of some integrable models with spectral gap 
\cite{Smirnov} and spin lattice systems \cite{Jimbo} can be studied by the form-factor approach.
However, form factors of more complex integrable systems such as the 1D Hubbard model remains an 
unsolved problem. 

The pseudofermion dynamical theory (PDT) has allowed to access that model high-energy dynamical correlation functions 
beyond the low-energy TLL limit \cite{V-1}. The theory relies on a suitable pseudofermion representation of the 1D Hubbard 
model BA solution. Shortly after the PDT was introduced, novel approaches that rely on a mobile quantum impurity 
model (MQIM) method have been developed to tackle the 
high-energy physics of both non-integrable and integrable 1D correlated quantum problems
\cite{Glazman-09,Glazman-12,Essler-10,Seabra-14,DSF-n1,Glazman-BG-08}. The exponents characterizing the dynamical 
correlation functions singularities have been found by both such schemes to be functions 
of momenta. They differ significantly from the predictions of the linear TLL 
theory \cite{V-1,TTF,spectral-06,VI,Glazman-09,Glazman-12,Essler-10,Seabra-14,DSF-n1,Moreno-13,Glazman-BG-08,DCFBo-16,CPJD-15,CarCadez-16,CarCadez-17}.

There are several methods and representations for the study of some of the quantum problems
reviewed in this paper. The charge-only and spin-only fractionalized particles that emerge in 1D correlated electronic systems
\cite{Sutherland-04,Voit} are within them usually identified with holons and spinons, respectively \cite{Natan-94,Essler-94,Essler-94-B}. 
The conventional holons and spinons have been constructed inherently to be associated with the charge and spin elementary excitations 
of integrable electronic models, respectively. Moreover, spinons are used to describe the elementary excitations of spin chains.
Holons and spinons are defined in terms of the deviation of the charge and spin BA distributions, respectively, from their ground-state value 
\cite{Bethe,Cloizeaux-62,Lieb,Lieb-03,Ovchi-70,Takahashi-71,Takahashi-72,Coll-74,Fowler-78,Natan-79,Natan-80,Faddeev-81,Anderson-87}.
This general definition was implemented for the spin excitations of the spin-$1/2$ $XXX$ chain 
in Ref. \cite{Cloizeaux-62}. For the charge and spin excitations of the 1D Hubbard model it was used in Refs.
\cite{Lieb,Ovchi-70,Coll-74}. A corresponding preliminary example of a spin-$1/2$ spinon is the spin-$1/2$ 
color spinor introduced in Ref. \cite{Natan-79} for the solvable 1D Gross-Neveu model \cite{Natan-80}. 
Its spectrum is associated with one ``hole'' emerging under a transition from a spin-singlet ground state to an 
excited energy eigenstate, in a sequence of BA spin quantum numbers. The spin-$1/2$ spin waves introduced in 
Ref. \cite{Faddeev-81} for the spin-$1/2$ $XXX$ chain have a similar definition. 

The "particle-like" representations in terms of the {\it pseudoparticles} and related {\it pseudofermions} 
discussed in this paper have a uniquely defined yet non-perturbative relation to the models physical particles. 
(In this review ``physical particles'' refer to the bosons, spins $1/2$, and electrons associated with the
operators in the models Hamiltonian usual expressions.)
The term ``pseudoparticle'' appeared early in the literature of the Hubbard model \cite{Mott-68,Bartel-74}. More 
recently it has been used for particles other than those reviewed here \cite{Marbach-12}. The latter are the
pseudoparticles that emerge within 1D integrable models \cite{Korepin-79,Korepin-80}. This includes in models
with Abelian global $U(1)$ symmetry \cite{DCFBo-16,Bosegas-94}, spin-$1/2$ chains with a single non-Abelian global $SU(2)$ symmetry \cite{XXXchain-94,CTD-15,CT-17,CPJD-15}, and more complex electronic models
\cite{Carmelo-91-A,Carmelo-91,Carmelo-92,Carmelo-92-B,Carmelo-92-C,Carmelo-93,Carmelo-93-B,Carmelo-94,Carmelo-94-B,Carmelo-97,Carmelo-97-B,Carmelo-97-C,Carmelo-99,Carmelo-00,Carmelo-00A,Carmelo-03,Carmelo-04}. 
In spite of the non-perturbative one-particle properties of integrable 1D correlated systems, the description of their
two-particle static properties is within the pseudoparticle representation very similar to that of a Fermi
liquid. Indeed, it is controlled by Landau parameters associated with pseudoparticle residual interactions 
$f$ functions \cite{Carmelo-91-A,Carmelo-91,Carmelo-92,Carmelo-92-B,Carmelo-92-C}.

The PDT pseudofermions are generated from the pseudoparticles under a unitary transformation. It is such that the
pseudofermion energy spectrum lacks the pseudoparticle $f$ functions term. As a result, within the thermodynamic 
limit (TL) the pseudofermions spectrum has no energy interaction terms \cite{CarCadez-17}. Under the transitions from a ground state 
to one- or two-particle excited states, the pseudofermions scatter off those created or annihilated under the transition. 
Under such scattering events, the pseudofermions merely acquire a phase shift. Such pseudofermion zero-momentum forward-scattering 
processes control both the low-energy and high-energy dynamical correlation functions of integrable 1D 
correlated models \cite{V-1,CarCadez-16,CarCadez-17}.  This renders the pseudofermion representation particularly suitable to the study 
of high-energy dynamical correlation functions.

The pseudoparticle and related pseudofermion representations are here discussed 
within the constructs of three prominent 1D correlated systems: The 1D Bose gas with two-body repulsive interaction
\cite{Liniger-63,Lieb-63,Berezin-64,McGuire-64,Yang-69,Thacker-81,Korepin-93,Guan-16}, 
the spin-$1/2$ isotropic Heisenberg chain \cite{Bethe,Takahashi-71,Muller-81,Faddeev-81,Takahashi-99}, and the repulsive 1D Hubbard model
\cite{Gutzwiller-63,Hubbard,Lieb,Lieb-03,Ovchi-70,Takahashi-72,Woy-82,Woy,Shastry-86,Shastry-86A,CM,Olmedilla-87,Olmedilla-88,Wadati-87,Shiroishi-95,Martins-97,Martins-98,Yue-97,1D-05}. One of the motivations of this review is the physical interest of the pseudoparticle and
pseudofermion representations of such low-dimensional correlated systems, which is not purely theoretical.
The studies of this review refer to the TL within which the imaginary part of the complex 
rapidities in the BA equations of models with non-Abelian global symmetries simplify \cite{Takahashi-71,Takahashi-72}.
The BA complex rapidities string deviations \cite{Caux-07,Deguchi-00} from such ideal strings 
do not affect in that limit the properties reviewed here.

One of the first applications of the BA to models with Abelian $U(1)$ symmetry was the study of 
a continuum problem of bosons interacting by a two-body $\delta$-function potential with interaction parameter $c$.
Now it is known as 1D Lieb-Liniger Bose gas \cite{Liniger-63,Lieb-63,Berezin-64,McGuire-64,Yang-69,Thacker-81,Korepin-93,Guan-16}. 
The model properties depend on the ratio $c/n_b$. Here $n_b$ is the boson density.
As a field theory, this is the repulsive quantum nonlinear Schr\"odinger model \cite{Thacker-81}. Although the 
original quantum problem is given in terms of bosons, the occupancies of model BA distribution have
a Pauli-like character, being only zero or one. The model can be simulated in systems of ultra-cold bosonic 
atomic 1D optical lattices \cite{Guan-13,Cazalilla-11,Petrov-00}.
Its charge dynamical structure factor can be probed in experimental Bragg spectra of such ultra-cold atoms \cite{Golovach-09,Fabbri-15}.
The static properties of the 1D Lieb-Liniger Bose gas revisited in this review are shown to be naturally described by
fermionic-like pseudoparticles with no internal degrees of freedom \cite{DCFBo-16,Bosegas-94}.
The high-energy one- and two-boson dynamical correlation functions spectral weights are 
derived within the related pseudofermion representation. They are controlled by pseudofermion phase shifts.

The spin-$1/2$ isotropic Heisenberg chain \cite{Heisenberg} is a 1D model of spins $1/2$ with a coupling constant $J$.
It was the first quantum system ever to be solved by the BA in 1931 \cite{Bethe}. This spin-chain model remains of great interest due 
to its underlying richness. In crystals where there is some 1D anisotropy, the model spin chains actually 
appear and describe the dominant physical behavior \cite{SPA-11,Ralph-11}. 
Several crystals are known to realize a 1D spin chain described by the spin-$1/2$ isotropic Heisenberg model. 
Examples are KCuF$_3$, Sr$_2$CuO$_3$ and CuPzN, which have been probed by neutron scattering 
\cite{Nagler-91,Stone-03,Zaliznyak-04,Lake-05,Enderle-05,Nilsen-08,Walters-09}.
High-resolution resonant inelastic X-ray scattering revealed the model spectrum 
in TiOCl \cite{Ralph-11} and La$_2$CuO$_4$ \cite{Braicovich-10}. The spin-$1/2$ $XXX$ chain can
as well be prepared in a 1D ultra-cold atomic trap \cite{Murmann-15}. 
The static properties of the spin-$1/2$ $XXX$ chain 
discussed in this paper are discussed by use of a representation in
terms of spin-neutral composite pseudoparticles \cite{CPJD-15}. Their constituents 
are spin-singlet pairs of the model physical spins $1/2$. The spin currents of the
energy eigenstates are also revisited. It is shown in terms of an exact spin current
expression that the elementary currents associated with the model conventional spinons describe
the translational degrees of freedom of the model physical spins $1/2$ in multiplet configurations.
However, such spinons are shown not contain the spin internal degrees of freedom of
the latter spins. The model spin dynamical correlation functions are also discussed.
They are found to be controlled by the scattering phase shifts of spin-neutral composite pseudofermions. 

The general Hubbard model was originally introduced as a toy model to 
study d-electrons in transition metals \cite{Gutzwiller-63,Hubbard}. It features electrons that can hop between 
nearest-neighbor lattice sites due to the finite hopping integral $t$. Its sites represent atoms, 
that are arranged in an ordered, crystalline pattern of well-defined geometry. When two electrons of 
opposite spin projection are on the same site, they have to pay 
the energy $U$ due to their mutual repulsion. This introduces additional electronic correlations beyond 
those of a statistical nature due to the Pauli principle. The model properties depend on the ratio $U/t$. In this paper the 
parameter $u\equiv U/4t$ is often used. 

One of the few rigorous results for the Hubbard model on any bipartite lattice refers to its global symmetry. 
It is well known that on such lattices the model Hamiltonian has two global $SU(2)$ symmetries 
\cite{HL,Yang,Yang-90,Lieb-89}. Consistently, in the early nineties of the past century it was found that for $u\neq 0$ the Hubbard model on a 
bipartite lattice has at least a $SO(4) = [SU(2)\otimes SU(2)]/Z_2$ symmetry. It contains the 
$\eta$-spin and spin $SU(2)$ symmetries \cite{Yang,Yang-90}. More recently it was found in Ref. \cite{bipartite} 
that for $u\neq 0$ and on any bipartite lattice its global symmetry is actually larger and given by 
$[SO(4)\otimes U(1)]/Z_2=[SU(2)\otimes SU(2)\otimes U(1)]/Z_2^2$. (This is equivalent to $SO(3)\otimes SO(3)\otimes U(1)$.)
The $SU(2)$ and $U(2)=SU(2)\otimes U(1)$ symmetries in the model $[SU(2)\otimes SU(2)\otimes U(1)]/Z_2^2$ 
global symmetry refer to the spin and charge degrees of freedom, respectively. The charge $U(2)=SU(2)\otimes U(1)$
symmetry includes a $SU(2)$ $\eta$-spin symmetry and a $U(1)$ lattice hidden symmetry beyond $SO(4)$. 
The latter symmetry is called {\it $c$-lattice $U(1)$ symmetry} in this review. It is indeed associated with the lattice degrees of freedom. 

The Hubbard chain is the simplest condensed-matter toy model for the description of the role of correlations 
in the exotic properties of quasi-1D materials \cite{TTF,Dionys-87,Ralph-02,spectral0,spectral}.  
It is an important correlated electronic system whose BA solution was first derived by the coordinate BA \cite{Lieb,Lieb-03}.
This has followed a similar solution for a related continuous model with repulsive $\delta$-function interaction \cite{Yang-67}. 
Its BA solution was also reached by the inverse-scattering method \cite{Shastry-86,Shastry-86A,CM,Olmedilla-87,Olmedilla-88,Wadati-87,Shiroishi-95,Martins-97}. 
The non-perturbative relation of the pseudoparticles and related pseudofermions to the electrons
involves in the case of the 1D Hubbard model an electron-rotated-electron unitary transformation.
It is performed by the exact BA solution. The well-known $U/t\rightarrow\infty$ $N_e$-electron wave functions factorization 
\cite{Lederer-00,Woy,Ogata-90,Ogata-91,Parola-90,Parola-92,Weng,Weng-94,Karlo-95,Karlo-96,Karlo-97,Wang-95,Gallagher-97,Gebhard-97,Sidorova-13,Poilblanc-13} gives rise under such a unitary transformation to a finite-$U/t$ factorization of the $N_e$-rotated-electron wave functions. The corresponding
electron-rotated-electron unitary operator is uniquely defined in terms of the matrix elements 
between the model energy and momentum eigenstates.

The static properties of the 1D Hubbard model discussed in this review are shown to be described by
a quantum liquid of several pseudoparticle branches. This includes $c$ pseudoparticles without internal degrees of freedom, 
spin-neutral composite pseudoparticles, and a third type of $\eta$-spin-neutral composite pseudoparticles.
The constituents of the latter are $\eta$-spin-singlet pairs of 
$\eta$-spin-$1/2$ fractionalized particles \cite{Carmelo-04,V-1,CarCadez-16,CarCadez-17}.
The model one- and two-electron dynamical correlation functions are in this paper reviewed 
in the suitable pseudofermion representation. Within it such functions
spectral weights are controlled by pseudofermion scattering phase shifts 
\cite{V-1,TTF,spectral-06,VI,CarCadez-16,CarCadez-17}. 

Concerning new developments, the fact that fractionalized particles are observed in quantum wires \cite{Jompol-09} 
renders them candidates for technological applications. Quantum wires are widely used to connect quantum ``dots". 
They may in the future form the basis of quantum computers. Thus the further understanding of the properties of these 
fractionalized particles may be important for such quantum technologies. It could as well help to develop more complete 
theories of superconductivity and conduction in low-dimensional condensed-matter systems. 

Another new development is the study of the effects of electron finite-range interactions. In the case
of 1D, the most physically interesting problems with such interactions are described by non-integrable models. 
This would be an important new development, since such effects occur in actual low-dimensional systems. 
For instance, the exponent that controls the suppression of the density of states 
of most quasi-1D metals \cite{Ralph-02,spectral0} and metallic 1D line defects in transition 
dichalcogenides \cite{MoSe-17} is larger than $1/8$. This is a unmistakable signature of electron 
finite-range interactions \cite{Schulz-90}.

The MQIM was developed to tackle the high-energy physics of both non-integrable and integrable 1D correlated 
quantum problems \cite{Glazman-09,Glazman-12}. In this review the relation between the PDT pseudofermions and the MQIM particles
is clarified. The extension of the MQIM to non-integrable 1D correlated electronic systems 
with long-range interactions is a complex problem. The universality in the vicinity of high-energy one-electron spectral functions singular lines found within the 
MQIM in Ref. \cite{Glazman-09} has been preliminarily used to construct a finite-range renormalized model generated 
from the 1D Hubbard model by a transformation, upon gently turning on suitable potentials \cite{MoSe-17}.

In general, units of both Planck constant $\hbar$ and lattice constant $a$ are used in this paper.
In the figures $u$ and $U$ stand for $U/4t$ and $U/t$, respectively. If not stated otherwise, the word {\it state} 
refers in this review to an energy and momentum
eigenstate. Concerning the general layout of what will be discussed in the following sections, the paper is
organized as follows. 

In Section \ref{Bg} the use of a pseudoparticle representation for the 1D Lieb-Liniger Bose gas is addressed. 
This includes the study of the model static and low-temperature quantities within that representation.
A related pseudofermion representation is used to study the high-energy 
behavior of the one-boson spectral function and charge dynamical structure factor 
near their spectra edges. 

The representation of the spin-$1/2$ isotropic Heisenberg chain in terms of
$n$-pseudoparticles is the topic addressed in Section \ref{Heichain}. Here $n=1,...,\infty$
refers to the number of singlet pairs of the model physical spins $1/2$
that are bound within a composite $n$-pseudoparticle. The use of such a representation simplifies 
the study and derivation of the model static and low-temperature quantities.
Moreover, a related pseudofermion representation is used in that section to compute the spin
longitudinal and transverse dynamical structure factors in the vicinity of their spectra lower thresholds. 
The relation of the dynamical structure factors peaks to the inelastic neutron scattering experiments 
on actual spin-chain compounds is discussed. The spinon representation of the model is also addressed.
A related extended BA $n$-bands hole representation valid for the model in its full Hilbert
space is discussed. Combining the $n$-pseudoparticle and $n$-bands hole representations
provides valuable physical information on the processes that control the model spin currents.
The relation of both the $n$-pseudoparticles and $n$-band holes to the model
physical spins $1/2$ is also clarified.

In Section \ref{rot-symm-1} the 1D Hubbard model is introduced. Out of the infinite choices of rotated electrons
that follow from its global $[SU(2)\otimes SU(2)\otimes U(1)]/Z_2^2$ symmetry, those that emerge
from the specific electron-rotated-electron unitary transformation performed by
the BA solution are considered. The latter rotated electrons degrees of freedom
separation is shown to lead to three types of fractionalized particles: The $c$ pseudoparticles without internal degrees of freedom, 
the rotated spins $1/2$, and the rotated $\eta$-spins $1/2$. 

As in the case of the spin-$1/2$ $XXX$ chain,
there emerge within the many-particle system composite $s n$ pseudoparticles that have
$n=1,...,\infty$ neutral pairs of rotated spins $1/2$ bound within them. 
This issue is revisited in Section \ref{rot-symm-2}. Additional
composite $\eta n$ pseudoparticles that have $n=1,...,\infty$ neutral pairs of rotated $\eta$-spins $1/2$ 
bound within them emerge within the system as well. The relation of the different pseudoparticle types 
and corresponding $n$-band holes to the rotated electrons and electrons is discussed and clarified.

In Section \ref{exc-spectra} the 1D Hubbard model $c$ and $\alpha n$ pseudoparticle quantum liquid is 
the issue under review. Several physical quantities are derived within the framework of such a quantum liquid by methods that resemble 
those used in Fermi-liquid theory. The internal configurations of the spin-singlet pairs and $\eta$-spin-singlet pairs 
are found to have a binding and anti-binding character, respectively.

The dynamical correlation functions within the pseudofermion representation 
is the general issue addressed in Section \ref{PRPST}. The PDT version 
suitable to the 1D Hubbard model is shortly reviewed. Its simplified version applicable 
to the 1D Lieb-Liniger Bose gas and spin-$1/2$ $XXX$ chain is also discussed.
The effects of varying the spin density on the 1D Hubbard model spectral
properties are shortly revisited. Finally, the simpler case of the 1D Bose gas dynamical correlation functions 
is used to clarify the relation between the PDT and the MQIM approaches.

%%%%%%%%%%%%%%%%%%%%%%%%%%%%%%%%%%%%%%%%%%%%%%%%%%%%%%%%%%%%%%%%%%%%%%%%%%
\section{The 1D Lieb-Liniger Bose gas}
\label{Bg}

As a first example of application of the pseudoparticle representation of the BA, we consider the 1D Bose gas with 
two-body repulsive contact interaction. It was introduced in 1963 by Lieb and Liniger \cite{Liniger-63,Lieb-63}.
The model particles satisfies Bose-Einstein statistics. It is the simplest example of BA solution. 
Its Hamiltonian describes particles interacting with each other via a two-body potential. The energy
and momentum eigenfunctions and eigenvalues can be calculated exactly by the BA. This integrable model 
helped to shape the understanding of quantum integrability \cite{Berezin-64,McGuire-64,Yang-69,Thacker-81,Korepin-93}.
It represents the non-relativistic limit of several integrable field theories \cite{Kormos-10,Pozsgay-11,Bertini-16}.
The results of Yang and Yang reported in Ref. \cite{Yang-69} were a significant step towards a deeper understanding 
of the physics of the Lieb-Liniger gas. Indeed, they presented for the first time a grand canonical description of the 
model in equilibrium.

The model seemed to be only of academic interest until with the sophisticated experimental techniques developed 
in late XX and XXI st century, it became possible to produce this kind of gas using real bosonic ultra-cold atoms as particles.
Hence in addition to being a paradigmatic example of a system of interacting bosons on the continuum, it 
became as well experimentally relevant for the physics of elongated clouds of cold atoms with contact interactions 
\cite{Amerongen-08,Fabbri-11,Fabbri-15,Meinert-15,Fang-16}.

As the temperature is lowered, a uniform gas of bosons in three dimensions will undergo a transition to a Bose-Einstein condensate
(BEC). In 1D, low-energy fluctuations prevent long-range order. For trapped gases, the situation changes.
Three regimes become possible in 1D: true condensate, quasi-condensate, and a strongly interacting regime, with BEC limited
to extremely small interaction between particles \cite{Petrov-00}. 
 
Trapped 1D gases are now accessible experimentally in all regimes \cite{Guan-13,Cazalilla-11}. The most challenging to obtain is the 
strongly interacting case. It can survive without fast decay due to a reduced three-body recombination rate,
consequence of fermionization. A natural starting point for the theoretical description of 1D atomic gases in this last regime is indeed
provided by bosons with delta-function interaction. In that case the fermionization refers to the Pauli-like zero and one allowed 
occupancies of the BA quantum numbers. As mentioned above,
the charge dynamical structure factor of that 1D Lieb-Liniger Bose gas can be probed in experimental
Bragg spectra of ultra-cold atoms on optical lattices \cite{Guan-13,Fabbri-15,Golovach-09}.

\subsection{The pseudoparticle representation of 1D Lieb-Liniger Bose gas BA solution}
\label{SoluBg}

The Hamiltonian of the 1D Bose gas with two-body repulsive interaction is in units of $\hbar =1$ 
and bare mass $m=1/2$ given by,
\begin{equation}
\hat{H} = - \sum_{j=1}^N{\partial^2\over \partial x_j^2} + 2c\sum_{j'>j}\delta (x_j - x_{j'}) - \mu\,N_b \, .
\label{HBg}
\end{equation}
Here and throughout this review, $\delta (x)$ denotes the usual Dirac delta-function distribution,
$x_j$ is the position of the $jth$ particle, $c>0$ gives the strength of the repulsive interaction,
and $\mu$ is the chemical potential. This Hamiltonian describes a set of $N_b$ particles with bosonic
statistics. All properties of the 1D Lieb-Liniger Bose gas
depend on the ratio $c/n_b$. Here $n_b$ is the particle density $n_b=N_b/L$. The limit of infinite repulsion, 
$c\rightarrow\infty$, is often called the Tonks-Girardeau limit \cite{Tonks-36,Girardeau-60}.

The {\it Bethe ansatz} method is named after the work by Hans Bethe. 
Bethe found the energy eigenfunctions and spectrum of the 1D spin-$1/2$ isotropic
Heisenberg model \cite{Bethe}. The 1D Lieb-Liniger Bose gas was the next model solved by coordinate BA 
more than 30 years later. Since it is the simplest example of BA, the BA is introduced here for that model. 
We start by considering the case of only two bosons, $N_b=2$.
The wave function is assumed to be a symmetrized product of plane waves of both bosons,
\begin{eqnarray}
\psi (x_1,x_2) & = & \psi_1 (x_1,x_2) \hspace{0.20cm}{\rm for}\hspace{0.20cm}x_1 > x_2
\nonumber \\
& = & \psi_2 (x_1,x_2) \hspace{0.20cm}{\rm for}\hspace{0.20cm}x_2 > x_1 \, .
\nonumber 
\end{eqnarray}

The functions $\psi_1 (x_1,x_2)$ and $\psi_2 (x_1,x_2)$ are expressed as what is called BA wave functions,
\begin{eqnarray}
\psi_1 (x_1,x_2) & = & A_{1\,2} (I)\,e^{i(k_1 x_1 + k_2 x_2)}  + A_{2\,1} (I)\,e^{i(k_2 x_1 + k_1 x_2)}  
\hspace{0.20cm}{\rm for}\hspace{0.20cm}x_1 > x_2
\nonumber \\
\psi_2 (x_1,x_2) & = & A_{1\,2} (II)\,e^{i(k_2 x_1 + k_1 x_2)}  + A_{2\,1} (II)\,e^{i(k_1 x_1 + k_2 x_2)}
\hspace{0.20cm}{\rm for}\hspace{0.20cm}x_2 > x_1 \, .
\nonumber 
\end{eqnarray}
By using Bose statistics, $\psi_2 (x_2,x_1)=\psi_1 (x_1,x_2)$, one finds that the amplitudes in 
these equations obey the relations,
\begin{eqnarray}
A_{1\,2} (I) = A_{1\,2} (II) \equiv A_{1\,2} 
\hspace{0.20cm}{\rm and}\hspace{0.20cm}
A_{2\,1} (I) & = & A_{2\,1} (II) \equiv A_{2\,1}  \, .
\nonumber 
\end{eqnarray}

A key feature associated with the BA wave function for this model is that it is continuous at $x_1 = x_2$ whereas
its derivative is discontinuous. The second derivative discontinuity reads,
\begin{eqnarray}
\left(- {\partial^2\over\partial x_1^2}  - {\partial^2\over\partial x_2^2}\right)\psi\vert_{x_2<x_1} -
\left(- {\partial^2\over\partial x_1^2}  - {\partial^2\over\partial x_2^2}\right)\psi\vert_{x_1<x_2} 
= 2c\,\psi\vert_{x_1 =x_2} \, .
\nonumber 
\end{eqnarray} 

It then follows from the corresponding continuity and discontinuity conditions that
the amplitudes $A_{1\,2}$ and $A_{2\,1}$ satisfy the following relation,
\begin{eqnarray}
{A_{1\,2}\over A_{2\,1}} = {k_1 - k_2 + i\,c\over k_1 - k_2 - i\,c} \equiv S_{1\,2} \, .
\nonumber 
\end{eqnarray} 
$S_{1\,2}$ is here the {\it scattering amplitude}. It corresponds to an interchange of the regions $x_1<x_2$ and $x_1>x_2$ 
that actually refers to an interchange of the particles.

The system is considered to be on a ring with length $L$. The wave function must then satisfy the boundary condition, 
\begin{eqnarray}
\psi (x_1 + L,x_2) = \psi (x_1,x_2)  \hspace{0.20cm}{\rm for}\hspace{0.20cm}x_1 < x_2  \, .
\nonumber  
\end{eqnarray} 
Its use leads to,
\begin{eqnarray}
e^{i k_j L} = \prod_{l\neq j}S_{l\,j} = \prod_{l\neq j} {k_j - k_l + i\,c\over k_j - k_l - i\,c} 
\hspace{0.20cm}{\rm for}\hspace{0.20cm}j = 1,...,N_b  \, .
\label{BAS}
\end{eqnarray} 

This is the BA equation that here was derived for $N_b=2$. However, it turns out that it
is valid for any $N_b$.  The corresponding $N_b$-boson BA wave function is then found to
be given by,
\begin{eqnarray}
\psi (x_1,...,x_{N_b}) = \sum_{\iota_j\in S_{N_b}}^{N_b!}A_{\iota}\,e^{i \sum_{j=1}^{N_b} k_{\iota_j} x_j}   \, .
\nonumber 
\end{eqnarray} 
Here $S_{N_b}$ is the permutation group of $N_b$ elements, $\iota$ is one of the elements of $S_{N_b}$,
and the amplitude reads,
\begin{eqnarray}
A_{\iota} = (-1)^{\iota} \prod_{j>l}^{N_b}\left(k_{\iota_j} - k_{\iota_l} - i\,c\,{\rm sgn}\{x_j - x_l\}\right)  \, .
\nonumber 
\end{eqnarray} 
(Further details on the $N_b$-boson BA solution can be found in Ref. \cite{Korepin-93}.)

By taking the logarithm of Eq. (\ref{BAS}), one arrives to the following form
for the BA equations of the 1D Lieb-Liniger Bose gas \cite{Liniger-63,Lieb-63,Bosegas-94},
\begin{equation}
q_j = k_j + {2\over L}\sum_{l =1}^N \arctan \left({k_j - k_l\over c}\right) 
\hspace{0.20cm}{\rm where}\hspace{0.20cm}q_j = {2\pi\over L}\,I_j 
\hspace{0.20cm}{\rm and}\hspace{0.20cm} j = 1,...,\infty \, .
\label{qjBg}
\end{equation}
The $l =1,...,N_b$ summation in this equation runs over the subset of occupied $q_{l}$ quantum numbers
out of the full $j = 1,...,\infty$ set $\{q_j\}$ and corresponding subset of rapidities $k_l$.
The different occupancy configurations of the related 
$j = 1,...,\infty$ quantum numbers $I_j$ in this equation (defined modulo $L$) generate all the model 
energy and momentum eigenstates. Such numbers are successive integers 
or half-odd integers according to the boundary conditions,
\begin{eqnarray}
I_j & = & 0,\pm 1,\pm 2,...,\pm\infty \hspace{0.20cm}{\rm for}\hspace{0.20cm}N_b\hspace{0.2cm}{\rm odd} \, ,
\nonumber \\
& = & \pm {1\over 2},\pm {3\over 2},...,\pm\infty \hspace{0.20cm}{\rm for}\hspace{0.20cm}N_b\hspace{0.2cm}{\rm even} \, .
\label{IjBg}
\end{eqnarray}

As confirmed in the following, the corresponding quantities $q_j = (2\pi/L)\,I_j$ in Eq. (\ref{qjBg}) play the role of
discrete momentum values. The quantum numbers, Eq. (\ref{IjBg}), are successive 
integers or half-odd integers. The discrete momentum values
$q_j = (2\pi/L)\,I_j$ have though the usual spacing for any energy and momentum eigenstate,
\begin{equation}
q_{j+1} - q_j =  {2\pi\over L}\hspace{0.20cm}{\rm where}\hspace{0.20cm}j = 1,...,\infty \, .
\label{qjBspacingg}
\end{equation} 

The set of $j = 1,...,\infty$ quantities $k_j$ on the right-hand side of Eq. (\ref{qjBg}) are
the BA real momentum rapidities mentioned above. The term {\it rapidity} was first used by L. Hulth\'en
in 1938 \cite{Hulthen-38} to parametrize the roots of the spin-$1/2$ isotropic Heisenberg chain.
For it and other models whose BA solution is more complex than that of the present
1D Bose gas, {\it rapidity} is actually an analogy with relativistic kinematics, in the relative motion
problem. In that case a velocity may become non-additive due to a transformation, as
the Lorentz transformation. One then introduces related suitable alternative additive parameters, called
rapidities.

The actual quantum numbers whose occupancy configurations
generate the model energy eigenstates are not the momentum
rapidities. They are rather the set of numbers $\{q_j\} = \{(2\pi/L)\,I_j\}$.
The corresponding set $j = 1,...,\infty$ of numbers $\{I_j\}$ is given in Eq. (\ref{IjBg}). 
Each energy eigenstate is defined by the subset $l = 1,...,N_b$ of numbers
$\{q_l\} = \{(2\pi/L)\,I_l\}$ that are occupied.

The BA equations, Eq. (\ref{qjBg}), define for each specific subset $l = 1,...,N_b$ of occupied 
quantum numbers $\{q_l\} = \{(2\pi/L)\,I_l\}$ the corresponding related set $j = 1,...,\infty$ of real 
momentum rapidities $k_j$ for the energy eigenstate under consideration. 
The physical meaning of these equations is thus directly related to that of
such momentum rapidities, $k_j = k (q_j)$. For instance, the energy eigenvalues depend 
on the occupancy configurations of the subset $l = 1,...,N_b$
of quantum numbers $\{q_l\}$ and thus $\{I_l\}$ through them. 

The momentum rapidities, $k_j = k (q_j)$, and thus the BA equations, Eq. (\ref{qjBg}), 
that define them, contain important physical information
beyond the mere dependence of the energy eigenvalues on the 
subset $l = 1,...,N_b$ of occupied numbers $\{q_l\}$. For instance
and as further discussed below in Section \ref{RelapsboBg}, one finds from
straightforward manipulations of these equations that the momentum rapidities $k_j = k (q_j)$
of energy eigenstates whose subset $l = 1,...,N_b$ of occupied numbers $\{q_l\} = \{(2\pi/L)\,I_l\}$ 
differs from that of a ground state in only the occupancies of $N_{\rm ex}\ll N_b$ such numbers have
a simple form. For such states, $N_{\rm ex}/N_b\rightarrow 0$ for $N_b\rightarrow\infty$.
Indeed, their momentum rapidities are within the TL expressed in terms of
the corresponding ground-state momentum rapidity $k_0 (q_j)$ as
$k (q_j) = k_0 (q_j  + 2\pi\Phi (q_j)/L)$. Here $2\pi\Phi (q_j) = \sum_{l=1}^{N{\rm ex}}\,(\alpha_l)\,2\pi\Phi (q_j,q_l)$
is a dressed phase shift. The quantity $\alpha_l$ reads $\alpha_l=-1$ and $\alpha_l=+1$ when the occupancy
of the number $q_l$ changes relative to the ground state from occupied to unoccupied and vice versa, 
respectively. Furthermore, the important two-parameter dressed phase shifts $2\pi\Phi (q_j,q_l)$
are the solution of coupled equations that are directly extracted from the BA equations, Eq. (\ref{qjBg}).

Here we use the functional representation of Ref. \cite{Bosegas-94}. Within
it, the energy and momentum eigenvalues are of the general form,
\begin{equation}
E = \sum_{l=1}^{N_b} k_l^2 - \mu\,N_b 
= \sum_{j=1}^{\infty}\,N (q_j) k^2 (q_j) - \mu\,N_b 
\hspace{0.20cm}{\rm and}\hspace{0.20cm}
P = \sum_{l=1}^{N_b} k_l = \sum_{j=1}^{\infty}\,N (q_j) k (q_j) = \sum_{j=1}^{\infty}\,N (q_j) q_j \, ,
\label{EBg}
\end{equation}
respectively. The distribution function $N (q_j)$ appearing here is such that $N (q_j)=1$ and $N (q_j)=0$ 
for occupied and unoccupied $q_j$ values, respectively. We associate one {\it pseudoparticle} with each of the 
$N_b$ occupied momentum values $q_j$ of such a distribution. That discrete momentum variable, 
Eq. (\ref{qjBg}), has the range $q_j \in [-\infty,\infty]$. 
The equality between the two last terms of the momentum expression in Eq. (\ref{EBg}) is confirmed by 
suitable manipulations of the BA equation, Eq. (\ref{qjBg}). 

Each energy eigenstate has specific values 
for the distribution function $N (q_j)$. The $j = 1,...,\infty$ rapidity momentum values $k_j$ on the
right-hand side of Eq. (\ref{qjBg}) are for each state functions of the $j = 1,...,\infty$
momentum values $q_j$, Eq. (\ref{qjBg}), $k_j = k (q_j)$. The pseudoparticles of the 1D Lieb-Liniger Bose gas
have no internal structure.  Within an alternative ``holon'' representation, the holons would be associated 
with the unoccupied momentum values $q_j$. However, for the present model there is no advantage 
in considering such a representation.

For a ground state with $N_b$ bosons, the pseudoparticle  momentum distribution function is of the form \cite{Bosegas-94},
\begin{equation}
N^0 (q_j) = \theta (q_F - \vert q_j\vert)\hspace{0.20cm}{\rm where}\hspace{0.20cm}q_F = \pi \left(n_b - {1\over L}\right) \approx \pi\,n_b 
\hspace{0.20cm}{\rm and}\hspace{0.20cm}q_{F}^{\iota} = \iota\,{2\pi\over L}\,N_{b,\iota}\hspace{0.20cm}{\rm for}\hspace{0.20cm}\iota = \pm \, .
\label{qFiota}
\end{equation}
The distribution $\theta (x)$ reads in this review $\theta (x)=1$ for $x>0$ and $\theta (x)=0$ for $x\leq 0$.
Moreover, $q_F$ is in Eq. (\ref{qFiota}) the pseudoparticle Fermi momentum. Within the TL, one can use the number $N_{b,\iota}$ of $\iota =-1$ 
left and $\iota =+1$ right pseudoparticles, with the ground-state left and right Fermi momentum value $q_{F}^{\iota}$ 
as given in Eq. (\ref{qFiota}). (In that limit, the $N_b$-odd occupancy of the $q_j=0$ momentum refers to a $1/L$ correction
to the numbers of left and right pseudoparticles that can be ignored.)

The excited states momentum distribution functions can be written as,
\begin{equation}
N (q_j) = N^0 (q_j) + \delta N (q_j) \, .
\label{NqdNqBg}
\end{equation}
Here $N^0 (q_j)$ is the ground state momentum distribution function, Eq. (\ref{qFiota}), and
$\delta N (q_j)$ is the corresponding momentum distribution function deviation.
Under transitions from the $N_b$-boson ground state to $N_b+\delta N_b$-boson excited states
for which $\delta N_b$ is an even number, the deviations, Eq. (\ref{NqdNqBg}), can have the values $0$, $+1$, and $-1$.
Under transitions to $N_b+\delta N_b$-boson excited states
for which $\delta N_b$ is an odd number, one must account for the overall $\pm\pi/L$ shifts in the set of $j=1,...,\infty$ 
discrete momentum values $q_j$, Eq.  (\ref{qjBg}). This effect is due to the corresponding
quantum numbers $I_j$ being successive half-odd integers for $N_b$ even and integers for $N_b$ odd,
as reported in Eq. (\ref{IjBg}). We denote the corresponding overall shifts by $(2\pi/L)\,\Phi^0$
where the parameter $\Phi^0$ reads,
\begin{equation}
\Phi^0 = 0\hspace{0.20cm}{\rm for}\hspace{0.20cm}\delta N_b \hspace{0.25cm} {\rm even}\hspace{0.20cm}{\rm and}\hspace{0.20cm}
\Phi^0 = \pm {1\over 2}\hspace{0.20cm}{\rm for}\hspace{0.20cm}\delta N_b \hspace{0.25cm} {\rm odd} \, .
\label{picanBg}
\end{equation}
For $\delta N_b$ odd one must then add the term $\pm 1/2$ to the above
$\delta N (q_j)$ values $0$, $+1$, or $-1$.

In spite of the bosonic nature of the present quantum problem, the BA quantum
numbers, Eq. (\ref{IjBg}), and corresponding discrete momentum values, Eq. (\ref{qjBg}), 
have Pauli-like occupancies zero and one, respectively. Due to such a fermionization,
the pseudoparticles that carry these momentum values are not simple ``dressed bosons.'' 
Indeed, addition or removal of one boson of vanishing energy to and from the ground state 
involves two pseudoparticle excitations that cannot be decomposed: (a) addition 
or removal, respectively, of one pseudoparticle of momentum $\pm q_F\approx \pm \pi n_b$
at the ground-state Fermi points
and (b) a collective excitation of {\it all} remaining pseudoparticles, each
contributing with a small fraction $\mp\pi/L$ or $\pm\pi/L$ to the momentum
of the added or removed boson, respectively. This latter excitation
results from the pseudoparticle discrete momentum values, Eq. (\ref{qjBg}), shake up.
This effect is associated with the transition between the two set of quantum number values in Eq. (\ref{IjBg}) 
upon changing $N_b$ by one. Although in the present TL
each fraction $\mp\pi/L$ or $\pm\pi/L$ is vanishing small, if we multiply by the
number of $\approx N_b$ pseudoparticles of the Fermi sea this gives 
$\mp q_F$ or $\pm q_F$, respectively. Such a collective excitation occurs
upon any transition between two arbitrary states differing in
the number of bosons by an odd number.

It is useful to classify the general deviations $\delta N (q_j)$ into deviations $\delta N^F (q_j)$ 
and $\delta N^{NF} (q_j)$ for which the quantity $q_F - \vert q_j\vert$ vanishes and remains finite
within the TL, respectively. A {\it particle subspace} (PS) is spanned by a ground state and the set
of all states generated from it by a finite number of pseudoparticle processes
such that $\sum_{j=1}^{\infty}\,\vert\delta N^{NF} (q_j)\vert/L\rightarrow 0$ as $L\rightarrow\infty$. 
The transitions from the ground state to its PS excited states include the shake-up effects. Those are associated with
the overall momentum shifts $(2\pi/L)\,\Phi^0$ of all the system pseudoparticles.

In the following we consider the pseudoparticle quantum liquid described by the present
model Hamiltonian in the PS of a ground state with arbitrary value of the boson density $n_b$. 
The excitation energy $\delta E = E_{n} - E_0$ of the corresponding PS states 
is up to ${\cal{O}} (1/L)$ given by \cite{Bosegas-94},
\begin{equation}
\delta E = \sum_{j=1}^{\infty}\varepsilon (q_j)\delta N (q_j) 
+ {1\over L}\sum_{j=1}^{\infty}\sum_{j'=1}^{\infty}
{1\over 2}\,f (q_j,q_{j'})\,\delta N (q_j)\delta N (q_{j'}) \, .
\label{DEBg}
\end{equation}
The only restriction to the applicability of the pseudoparticle energy 
functional, Eq. (\ref{DEBg}), is that associated with the PS definition, {\it i. e.}
$\lim_{L\rightarrow\infty}\sum_{j=1}^{L}\vert\delta N^{NF} (q_j)\vert/L\rightarrow 0$.

\begin{figure}
\begin{center}
\centerline{\includegraphics[width=10.50cm]{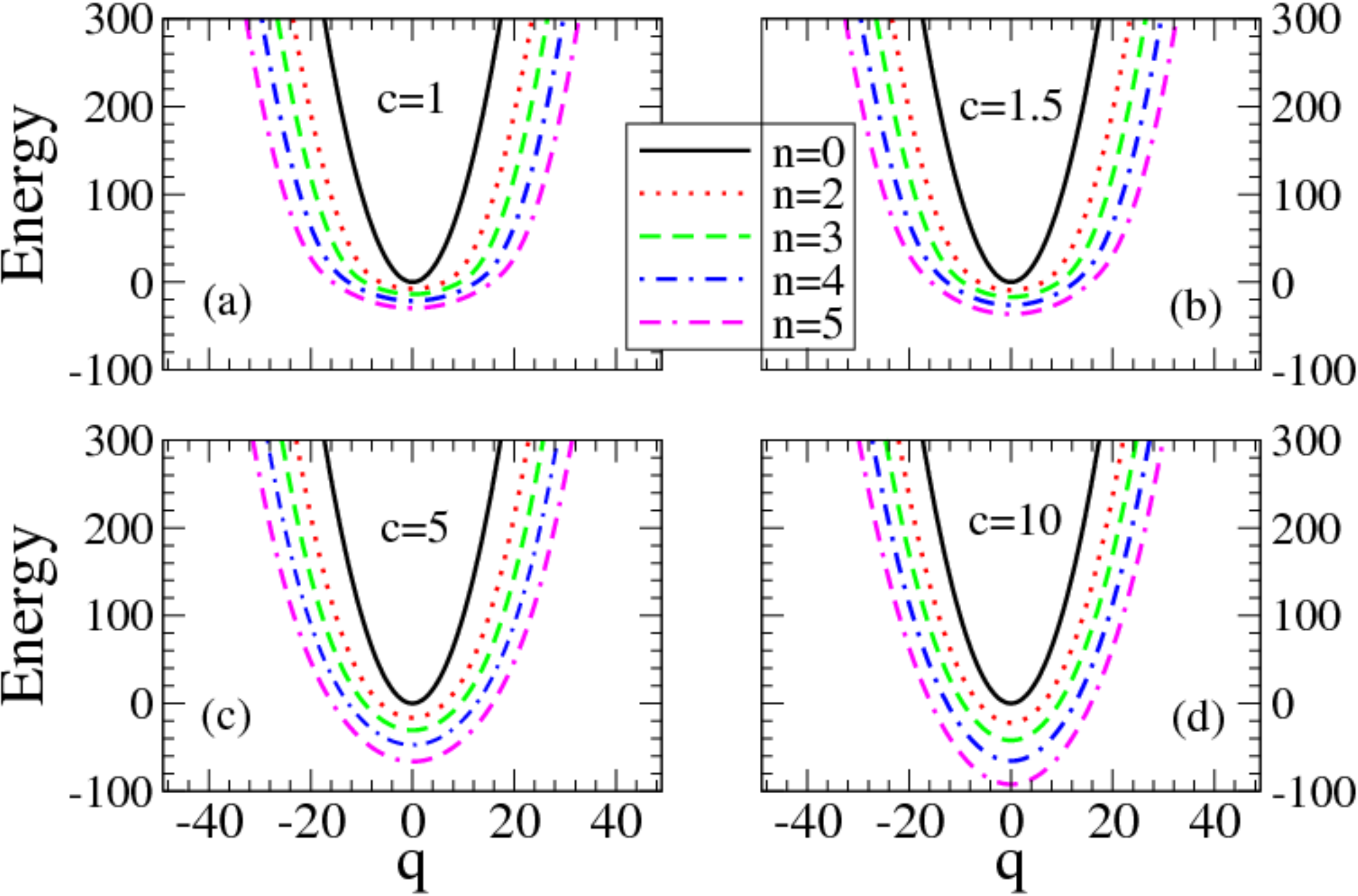}}
\caption{The pseudoparticle energy dispersion $\varepsilon (q)$, Eq. (\ref{varepsilonBg}), plotted 
as a function of the continuous momentum $q$ associated in the TL to $q_j$ such that
$q_{j+1}-q_j=2\pi/L$ for several densities $n_b$ and interaction $c$ values.
(In the figures $n_b$ is denoted by $n$.)}
\label{figure1}
\end{center}
\end{figure}

The pseudoparticle dispersion $\varepsilon (q_j)$ in the term of first order in the deviations
is plotted in Fig. \ref{figure1} as a function of the momentum $q_j$ for several densities 
$n_b$ and interaction $c$ values. It is of the form,
\begin{equation}
\varepsilon (q_j) = \varepsilon^0 (q_j) - \mu\hspace{0.20cm}{\rm where}\hspace{0.20cm}
\varepsilon^0 (q_j) = (k_0 (q_j))^2 + 2\int_{-Q}^Q dk \,k\,{\bar{\Phi}} (k,k_0 (q_j)) 
\hspace{0.20cm}{\rm and}\hspace{0.20cm}Q = \pm k_0 (\pm q_F) \, .
\label{varepsilonBg}
\end{equation}
Here $\pm Q$ are the Fermi rapidity momenta, the ground-state momentum rapidity function $k_0 (q_j)$ is 
the solution of the BA equation, Eq. (\ref{qjBg}), 
for the ground-state distribution, Eq. (\ref{qFiota}), and $\Phi (q_j,q_{j'})$ is a dressed phase shift in units of $2\pi$.
Its physical meaning is clarified below in Section \ref{RelapsboBg}. It can be written as,
\begin{equation}
\Phi (q_j,q_{j'}) = {\bar{\Phi}} (k_0 (q_j),k_0 (q_{j'})) \, .
\label{PsBg}
\end{equation}
The related momentum-rapidity phase shift ${\bar{\Phi}} (k,k')$ in units of $2\pi$ appearing here obeys the integral equation,
\begin{equation}
{\bar{\Phi}} (k,k') = - {1\over\pi}\arctan \left({k- k'\over c}\right)  
+ {1\over\pi c}\int_{-Q}^Q dk'' {{\bar{\Phi}} (k'',k')\over 1 + \left({k - k''\over c}\right)^2} \, .
\label{PsEqBg}
\end{equation}

The chemical potential $\mu$ on the right-hand side of Eq. (\ref{varepsilonBg}) and the energy bandwidth 
of the ground-state occupied pseudoparticle sea can be written as,
\begin{eqnarray}
\mu  & = & \varepsilon^0 (q_F) = Q^2 + 2\int_{-Q}^Q dk \,k\,{\bar{\Phi}} (k,Q)\hspace{0.20cm}{\rm and}
\nonumber \\
W_F & = & \varepsilon^0 (q_F) - \varepsilon^0 (0) =  Q^2 + 2\int_{-Q}^Q dk \,k\,({\bar{\Phi}} (k,Q) - {\bar{\Phi}} (k,0)) \, ,
\label{muBg}
\end{eqnarray}
respectively. Hence $\varepsilon (\pm q_F) = 0$.

The $f$ functions in the term of second order in the momentum distribution function deviations on
the right-hand side of Eq. (\ref{DEBg}) are given by,
\begin{equation}
f (q_j,q_{j'}) = v (q_{j})\,2\pi \,\Phi (q_{j},q_{j'}) + v (q_{j'})\,2\pi \,\Phi (q_{j'},q_{j}) +
{v\over 2\pi}\sum_{\iota = \pm}2\pi\Phi (\iota q_F,q_{j})\,2\pi\Phi (\iota q_F,q_{j'}) \, ,
\label{ffBg}
\end{equation}
where $f (q,q') = f (-q,-q')$. The pseudoparticle group velocities in this expression read
$v (q_j) = v (q)\vert_{q=q_j}$ where,
\begin{equation}
v (q) = {d\varepsilon (q)\over d q} 
\hspace{0.20cm}{\rm and}\hspace{0.20cm}v = v (q_F) \, ,
\label{vqBg}
\end{equation}
and $v = v (q_F)$ is the (pseudoparticle) Fermi velocity.

As the form of the expressions in Eqs. (\ref{DEBg})-(\ref{vqBg}) indicates,
for the model in a PS the present representation refers to a quantum liquid whose pseudoparticles
have residual zero-momentum forward-scattering interactions. Those are associated with the
energy terms of second order in the deviations in
the general energy spectrum, Eq. (\ref{DEBg}). The difference relative to the 
zero-momentum forward-scattering interactions of Fermi-liquid quasiparticles 
is that the latter is valid only in the limit of vanishing excitation energy. Indeed, due
to the model integrability, the pseudoparticles zero-momentum forward-scattering interactions
rather refer to all energy scales. Hence the pseudoparticle lifetime is infinite. 
Furthermore, the pseudoparticle occupancy configurations
generate all energy and momentum eigenstates from the boson vacuum. 
On the contrary, in a Fermi liquid the states generated by quasiparticle
occupancy configurations are energy eigenstates only in the limit of vanishing excitation energy.

Within two-boson excitations, the functions $f (q_j,q_{j'})$ play a role similar 
to that of the $f$ functions in Fermi-liquid theory. The following ``renormalized'' Fermi velocities 
determine the low-energy expressions of several physical quantities \cite{Bosegas-94},
\begin{equation}
v^{i} = v +  {1\over 2\pi}\sum_{\iota = \pm} (\iota)^i\,f (q_F,\iota q_F) = v\,(\xi^i)^2 \hspace{0.20cm}{\rm where}\hspace{0.20cm} i = 0,1 \, .
\label{viBg}
\end{equation}
The $f$ functions, Eq. (\ref{ffBg}), in Eq. (\ref{viBg}) involve the dressed phase shifts $2\pi\Phi (q_j,q_{j'})$, Eq. (\ref{PsBg}).
Their connection to the BA equation, Eq. (\ref{PsEqBg}), occurs through the momentum rapidity function
$k_j = k (q_j)$ of PS excited states with general distributions $N (q_j) = N^0 (q_j) + \delta N (q_j)$, Eq. (\ref{NqdNqBg}). Their
use in the BA equations leads to solutions of the form $k (q_j) = k_0 (q_j + [1/L]\sum_{j'}\delta N (q_{j'})\,2\pi\Phi (q_j,q_{j'}))$.
The quantity $\Phi (q_j,q_{j'})$ appearing here is related to the rapidity phase shift ${\bar{\Phi}} (k,k')$, Eq. (\ref{PsBg}).
The latter obeys the integral equation, Eq. (\ref{qjBg}). It emerges directly from suitable manipulations of the BA equation.

The $i=0$ and $i=1$ quantities $v^{i}/v = 1 + (1/2\pi v)\sum_{\iota = \pm} (\iota)^i\,f (q_F,\iota q_F)$ play the role of 
symmetric charge and antisymmetric current {\it Landau parameters},
respectively. The related parameters $\xi^0$ and $\xi^1$ in Eq. (\ref{viBg}) are the following simple 
symmetric and antisymmetric combinations, respectively, of dressed phase shifts in units
of $2\pi$ at the Fermi points,
\begin{equation}
\xi^i = 1 + \Phi (q_F,q_F) + (-1)^i\,\Phi (q_F,-q_F) \hspace{0.20cm}{\rm where}\hspace{0.20cm} i = 0,1 \, .
\label{xiBg}
\end{equation}
(Here in $\Phi (q_F,q_F)$ and in other phase shifts of this review whose two momenta are the same,
$(q_F,q_F)$ refers to the TL in which $(q_F,q_F\pm 2\pi/L)$ is for simplicity written as $(q_F,q_F)$.)
Manipulations of Eqs. (\ref{PsBg}), (\ref{PsEqBg}), and (\ref{xiBg})
reveal that such parameters are related as $\xi^1 = 1/\xi^0$ and that
$\xi^1= \xi^1(Q)$. Here $\xi^1(k)$ is the solution of the integral equation
$\xi^1 (k) = 1 + (c/\pi)\int_{-Q}^Q dk' \xi^1 (k')/[c^2 + (k - k')^2]$
where $\pm Q = k_0 (\pm q_F)$ and $q_F= \pi n_b$. Its solution is mathematically simplest in the
$n_b/c \ll 1$ and $n_b/c \gg 1$ limits. This leads to the following limiting expressions,
which, as all the model quantities, depend only on the ratio $n_b/c$,
\begin{equation}
\xi^1 = 1/\xi^0 \approx  1+ {2n_b\over c} \hspace{0.20cm}{\rm for}\hspace{0.20cm} n_b/c \ll 1 
\hspace{0.20cm}{\rm and}\hspace{0.20cm} 
\xi^1 = 1/\xi^0 \approx \sqrt{\pi\sqrt{n_b\over c}} = \pi^{1/2}\left({n_b\over c}\right)^{1/4} \hspace{0.20cm}{\rm for}\hspace{0.20cm} n_b/c \gg 1 \, .
\label{xi-0-1-limits}
\end{equation}

\begin{figure}
\begin{center}
\subfigure{\includegraphics[width=5.00cm]{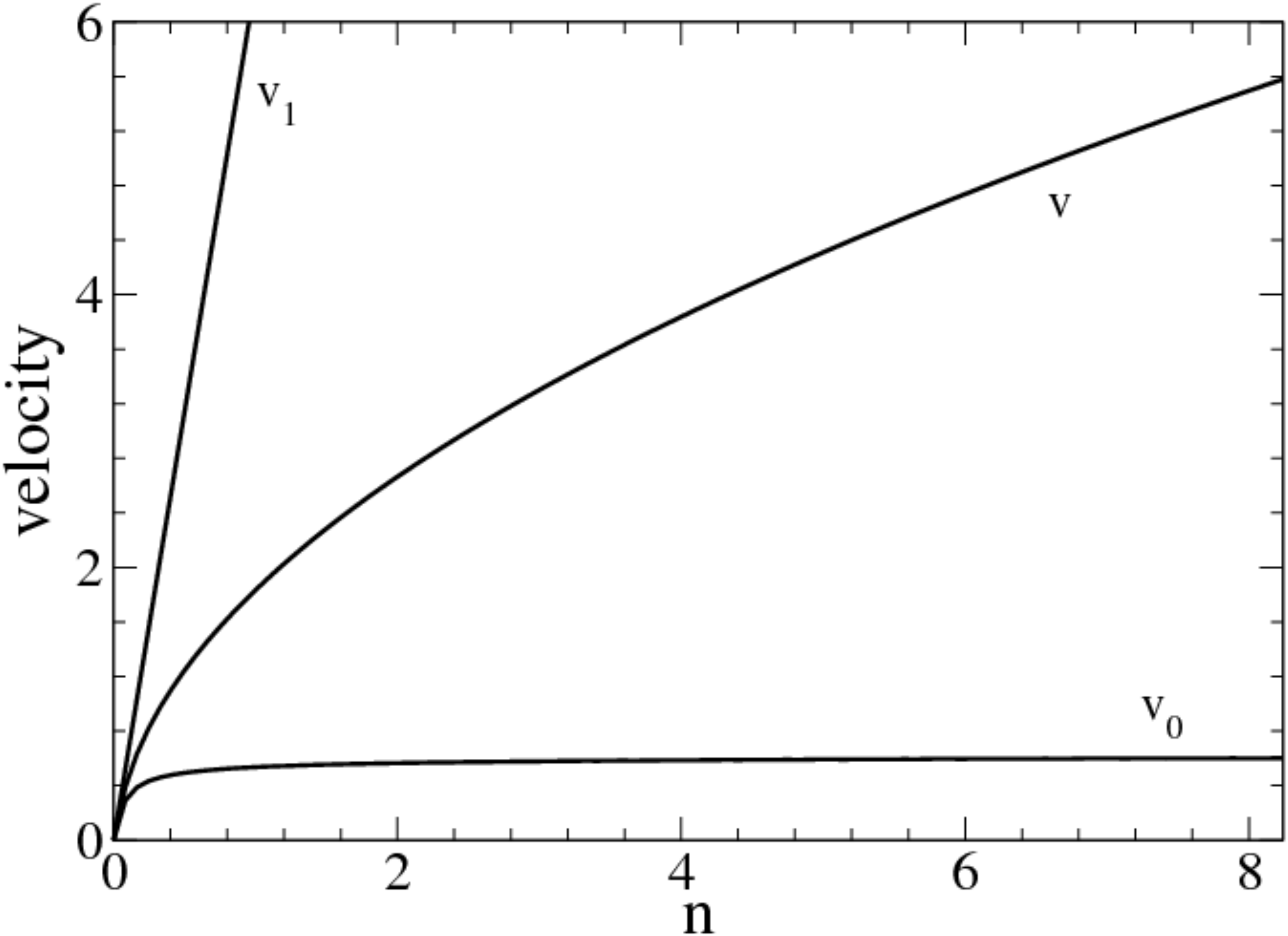}}
\hspace{0.50cm}
\subfigure{\includegraphics[width=5.00cm]{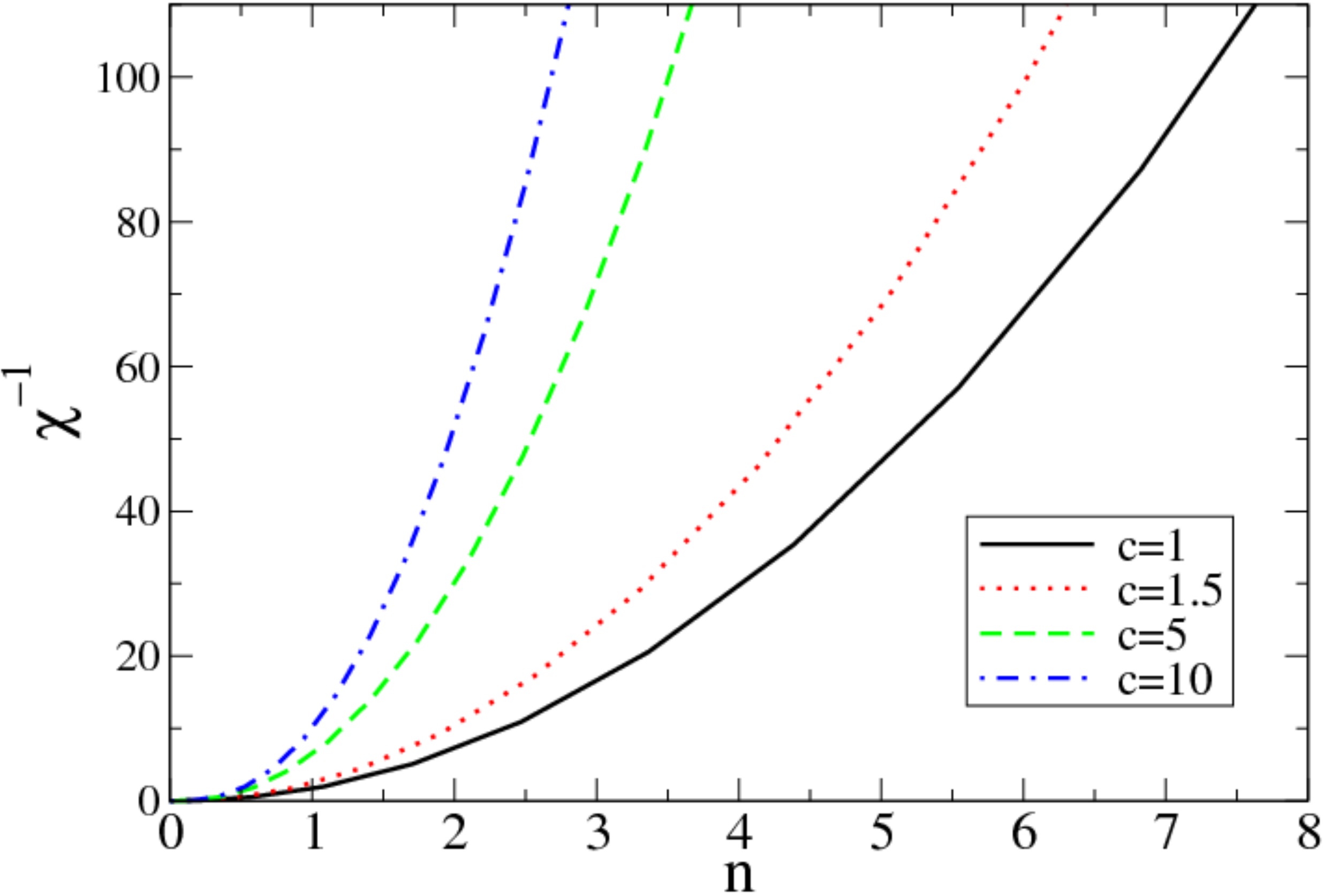}}
\caption{The group velocity and the two ``renormalized'' Fermi velocities, Eqs. (\ref{viBg}) and (\ref{vxiBg}),
(left panel) and the inverse compressibility $\chi^{-1}$, Eq. (\ref{ChiBg}), (right panel) as a function of the density $n_b$ 
for $c=1$ and several interaction $c$ values, respectively. (In the figures $n_b$ is denoted by $n$.)\\
{\it Source}: The plots of $v$, $v_0$ and $v_1$ (left) were produced using data from Fig. 3 of Ref. \cite{Bosegas-94}.}
\label{figure2}
\end{center}
\end{figure}

The related ``renormalized'' velocities $v_0$ and $v_1$, Eq. (\ref{viBg}), and the Fermi velocity $v = v (q_F)$, Eq. (\ref{vqBg}),
are given by the following expressions and obey the following relations,
\begin{equation}
v_0 = {v^2\over 2\pi n_b} \hspace{0.20cm}{\rm and}\hspace{0.20cm}v_1 = 2\pi n_b 
\hspace{0.20cm}{\rm where}\hspace{0.20cm}v = 2\pi n_b (\xi^0)^2 = {2\pi n_b\over (\xi^1)^2} \, .
\label{vxiBg}
\end{equation}
The velocities $v$, $v_0$, and $v_1$ are plotted  in Fig. \ref{figure2} as a function of the density $n_b$ for $c=1$.

The usual TLL parameter $K_0$ can be written as $K_0=(\xi^1)^2=(1/\xi^0)^2$. This is
consistent with at low energy the pseudoparticles referring to a representation of the universal TLL.
Its results can be reached by other approaches \cite{Voit,Glazman-BG-08},
as for instance conformal field theory \cite{Bogoliubov-86,Bogoliubov-87,Woy-89}.
When, as in Eq. (\ref{viBg}), the two momenta of the $f$ functions, Eq. (\ref{ffBg}), 
are at the Fermi points, they read $f (q_F,\pm q_F) = \pi [v^0 - v \pm (v^1 - v)]$.
This signals the emergence of the low-energy TLL physics. 

The system compressibility is controlled by excited states. Their momentum distribution function
deviations are given by \cite{Bosegas-94},
\begin{equation}
\delta N (q) = \delta (q_F - \vert q\vert)\,\delta q_F \, .
\label{dNdqBg}
\end{equation}
From their use in Eqs. (\ref{DEBg}) and (\ref{muBg}), one readily 
finds that $\partial\mu (n_b)/\partial n_b = - v_0\,\pi$ where $v_0$ is defined in
Eq. (\ref{viBg}). The compressibility is then easily found to read, 
\begin{equation}
\chi = - {1\over n_b^2}{1\over \partial\mu (n_b)/\partial n_b} = {1\over \pi\,n_b^2}{1\over v_0} \, .
\label{ChiBg}
\end{equation}
The inverse compressibility $\chi^{-1}$ is plotted in Fig. \ref{figure2} as a function of the density $n_b$ for
several interaction $c$ values.

An expression for the low-temperature entropy is derived by means of
simple combinatorial arguments that rely on the allowed occupancies
of the discrete momentum values $q_j$, Eq. (\ref{qjBg}), being only zero and one.
The result is,
\begin{equation}
S = -2\sum_{j=1}^{\infty}\left(N (q_j)\ln [N (q_j)] + N^h (q_j)\ln [N^h (q_j)]\right) \, .
\label{entropyBg}
\end{equation}
Here $N^h (q_j)\equiv 1 - N (q_j)$ and $N (q_j)$ is a Fermi-Dirac distribution. Its temperature-dependent
BA energy dispersion $\varepsilon (q_j)$ is obtained in a self-consistent way, as in a Fermi liquid.
In the present low-temperature limit we can use in it the $T=0$ energy dispersion, Eq.
(\ref{varepsilonBg}). Although the original particles are bosons, in connection to and due to the BA quantum
numbers $I_j$ allowed occupancies, Eq. (\ref{IjBg}), the pseudoparticles obey a Pauli-like occupancy of their momentum 
values $q_j= (2\pi/L)\,I_j$. Hence their thermal momentum distribution function deviation $\delta N (q_j)$ is given by,
\begin{equation}
\delta N (q_{j}) = {1\over 1 + e^{\varepsilon (q_j)/k_B T}} - N^0 (q_{j}) \, .
\label{dNqTBg}
\end{equation}
From the use in the energy functional, Eq. (\ref{DEBg}), of Eqs. (\ref{entropyBg}) and (\ref{dNqTBg}), provided
one accounts for that $\varepsilon (\pm q_F)=0$, it is straightforward to obtain the well known TLL low-temperature 
specific-heat leading order term \cite{Bogoliubov-86,Bogoliubov-87,Woy-89},
\begin{equation}
c_V = {L\,k_B\,\pi\over 3\,v}\,(k_B T) \, . 
\label{cVBg}
\end{equation}

The charge conductivity real part has the general form $\sigma (\omega) = 2\pi\,D\,\delta (\omega) + \sigma_{reg} (\omega)$. 
The charge stiffness or Drude weight $D$ characterizes here the response to a static field. $\sigma_{reg} (\omega)$ 
describes the absorption of light of frequency $\omega$. Both such quantities
can be expressed in terms of the charge current operator \cite{Shastry-90}. Since the system has translational invariance, for 
this model that operator commutes with the Hamiltonian. As a result, the real conductivity spectrum has no incoherent 
part. From the conductivity sum rule one then finds,
\begin{equation}
{\rm Re}\,\sigma (\omega) = 2\pi\,D\,\delta (\omega) 
\hspace{0.20cm}{\rm where}\hspace{0.20cm}2\pi\,D = v^1 = 2\pi n_b  \, . 
\label{ResigmaBg}
\end{equation}

\subsection{Pseudofermions and dynamical correlation functions}
\label{RelapsboBg}

It useful for the study of the dynamical correlation functions of the 1D Lieb-Liniger Bose gas and other integrable models 
to provide some basic information on dynamical correlation functions of general many-body quantum systems.
Time correlation functions of dynamical variables play an important role in the description of such systems. 
It is well known that transport coefficients and the cross sections for scattering of physical particles are directly 
related to time correlation functions. The simplest dynamical properties are the linear response to an external 
perturbation. Most macroscopic measurements are in the linear regime. This is because macroscopic perturbations are 
very small on the scale of microscopic forces. 

The fluctuation-dissipation theorem states that at zero and finite temperature the linear response of the system 
to a time-varying perturbation is the same as and is connected to
fluctuations, respectively, that naturally occur in statistical equilibrium \cite{Nyquist-28,Callen-51}. 
This means that the response of a system in thermodynamic equilibrium to a small applied force is connected
to its response to a spontaneous fluctuation. Therefore, the theorem connects the linear response relaxation 
of a system from a prepared non-equilibrium state to its statistical fluctuation properties in equilibrium. 
So there are two basic ways to calculate the response: either waiting for the system to fluctuate by itself or 
applying a perturbation and see what happens. The second is often referred to as the Kubo method. (In the present
case of 1D integrable models, see Ref. \cite{Mukerjee-08}.)

Within quantum field theory, the operators become functions of space and time, $\hat{O} (x,t)$. Consider two operators 
$\hat{A} (x,t)$ and $\hat{B} (x,t)$ and that the basic dynamical correlation functions,
\begin{eqnarray}
\chi (x,t;x't') = \langle\hat{A} (x,t)\,\hat{B} (x',t')\rangle \, ,
\label{AB}
\end{eqnarray}
where
\begin{eqnarray}
\hat{A} (x,t) & = & e^{i\hat{H}\,t}\hat{A} (x)e^{-i\hat{H}\,t} 
\hspace{0.20cm}{\rm and}\hspace{0.20cm}
\hat{B} (x,t) = e^{i\hat{H}\,t}\hat{B} (x)e^{-i\hat{H}\,t} \, ,
\label{AxBx}
\end{eqnarray}
are written within the Heisenberg representation. Here $\hat{H}$ is the Hamiltonian of the model
under consideration and $\langle ...\rangle$ stands for the thermal average for finite temperatures $T>0$ 
and the ground-state expectation value at $T=0$.

The dynamical correlation functions, Eq. (\ref{AB}), are the building blocks of several physically relevant 
dynamical correlation functions. This includes, for instance, the dynamical correlation functions,
\begin{eqnarray}
\chi_{\rm ret} (x,t;x't') & = & -i\,\theta (t-t')\,\langle [\hat{A} (x,t)\,\hat{B} (x',t')\pm \hat{B} (x',t')\,\hat{A} (x,t)]\rangle
\hspace{0.20cm}{\rm and}
\nonumber \\
\chi'' (x,t;x't') & = & \langle[\hat{A} (x,t),\hat{B} (x',t')]\rangle \, ,
\label{sevAB}
\end{eqnarray}
where in the retarded correlation function $\chi_{\rm ret} (x,t;x't')$ expression $+$ and $-$ refers to fermions 
and bosons, respectively. In most cases of physical interest, the correlation function $\chi'' (x,t;x't')$ plays the 
role of the quantity that describes absorption or dissipation. The related retarded correlation function $\chi_{\rm ret} (x,t;x't')$ is
characterized by the presence of the step function $\theta (t-t')$. Usually, 
when thinking about scattering amplitudes, one works with time-ordered (Feynman) correlation functions. 
Those are relevant for building perturbation theory. 

In most cases zero temperature is considered in this review, so that $\langle...\rangle$ means
ground-state expectation value $\langle GS\vert ...\vert GS\rangle$. 
The systems considered in it exhibit within the TL translational invariance in both space and time.
Hence the dynamical correlation functions depend only on the differences $x-x'$ and $t-t'$, respectively. One then
defines the dynamical correlation functions in momentum and frequency space by the Fourier transforms,
\begin{eqnarray}
\tilde{\chi} (k,\omega) = \int dx \int dt\,e^{-i(k\,x - \omega\,t)}\,\chi (x,t;0,0) \, .
\nonumber 
\end{eqnarray}

In the problems under consideration in this review one has that $\hat{A} (x,t)=\hat{O} (x,t)$ and
$\hat{B} (x,t)=\hat{O}^{\dag} (x,t)$ in Eqs. (\ref{AB}) and (\ref{sevAB}), so that for instance,
\begin{eqnarray}
\chi_{\rm ret} (x,t;x't') & = & -i\,\theta (t-t')\,\langle [\hat{O} (x,t)\,{\hat{O}}^{\dag} (x',t')\pm {\hat{O}}^{\dag} (x',t')\,\hat{O} (x,t)]\rangle
\hspace{0.20cm}{\rm and}
\nonumber \\
\tilde{\chi}_{\rm ret}  (k,\omega) & = & \int dx \int dt\,e^{-i(k\,x - \omega\,t)}\chi_{\rm ret} (x,t;0,0) \, \, .
\label{OOdag}
\end{eqnarray}
It is useful to express the dynamical correlation function $\tilde{\chi}_{\rm ret} (k,\omega)$
in this equation in a Lehmann representation. At zero temperature it reads,
\begin{eqnarray}
\tilde{\chi}_{\rm ret}  (k,\omega) & = & \sum_{\nu} {\vert\langle\nu\vert \hat{O} (k)\vert GS\rangle\vert^2
\over \omega - (E_{\nu} - E_{GS}) + i\delta}\hspace{0.20cm}{\rm so}\hspace{0.20cm}{\rm that}
\nonumber \\
- {1\over\pi}\,{\rm Im}\,\tilde{\chi}_{\rm ret}  (k,\omega) & = & \sum_{\nu} \vert\langle\nu\vert \hat{O} (k)\vert GS\rangle\vert^2
\delta (\omega - (E_{\nu} - E_{GS})) \, ,
\label{ChiT0} 
\end{eqnarray}
where $\delta$ is as usual an infinitesimal number, $\sum_{\nu}$ is a sum that runs over energy eigenstates, and,
\begin{eqnarray}
\hat{O} (k) = \int dx\,e^{-ik\,x}\,\hat{O} (x) \, .
\nonumber 
\end{eqnarray}

Correlation functions are important quantities in many branches of physics. For instance, in the framework of the physics of ultracold quantum gases, 
they provide valuable information about the quantum many-body wave function beyond the simple density profile \cite{Guan-16}.
The charge dynamical structure factor of a 1D Lieb-Liniger Bose gas is an example of a dynamical correlation function that can be probed in 
experimental Bragg spectra of ultra-cold atoms on optical lattices \cite{Fabbri-15,Golovach-09}.

There is an alternative fermionic representation for the 1D Lieb-Liniger Bose gas in terms of new
entities called pseudofermions. These emerge naturally from the pseudoparticles 
in a given PS. Such an alternative representation is particularly suitable for the study of the dynamical 
correlation functions of the integrable models reviewed in this paper. Their emergence from the pseudoparticles 
involves an important property of the momentum rapidity functions $k_j = k (q_j)$ of the excited states of any PS. 
Manipulations of the BA equations, Eq. (\ref{qjBg}), that involve the use of distributions 
of form $N (q_j) = N^0 (q_j) + \delta N (q_j)$, Eq. (\ref{NqdNqBg}), and the expansion of
these equations in such deviations, reveals that the excited-state functions $k_j = k (q_j)$
can be exactly expressed in terms of the corresponding ground-state rapidity function $k_j^0 = k_0 (q_j)$ 
as follows,
\begin{equation}
k (q_j) = k_0 (\bar{q}_j)  \hspace{0.20cm}{\rm for}\hspace{0.20cm}j=1,...,\infty \, .
\label{kjexsBg}
\end{equation}
Here,
\begin{equation}
\bar{q}_j = q_j  + {2\pi\Phi (q_j)\over L} \hspace{0.20cm}{\rm for}\hspace{0.20cm} j=1,...,\infty 
\hspace{0.20cm}{\rm where}\hspace{0.20cm}
2\pi\Phi (q_j)  = \sum_{j'=1}^{\infty}\,\delta N (q_{j'})\,2\pi\Phi (q_j,q_{j'}) \, .
\label{QPqBg}
\end{equation}

The pseudofermion representation associates each occupied discrete {\it canonical momentum value} 
${\bar{q}}_j$, Eq. (\ref{QPqBg}), with one {\it pseudofermion}. The quantity $2\pi\Phi (q_j)$ 
is below confirmed to be a scattering pseudofermion phase shift. It is such that $2\pi\Phi (q_{j+1})-2\pi\Phi (q_j)= {\cal{O}} (1/L)$.
Hence the spacing of the discrete canonical momentum values is to leading
${\cal{O}} (1/L)$ order the usual one,
\begin{equation}
{\bar{q}}_{j+1} - {\bar{q}}_j =  {2\pi\over L} + {\cal{O}} (1/L^2)\hspace{0.20cm}{\rm for}\hspace{0.20cm}j = 1,...,\infty \, .
\label{CqjBspacingg}
\end{equation} 
However, a similar equality does not hold for the discrete canonical momentum values separation for $l$ values 
such that $l/N_b$ is finite as $N_b\rightarrow\infty$, ${\bar{q}}_{j+l} - {\bar{q}}_j \neq l\,(2\pi/L)$. (It does though for $l=\infty$.)

Since $2\pi\Phi (q_{j+1})/L-2\pi\Phi (q_j)/L$ is of ${\cal{O}} (1/L^2)$ order there is no level crossing. By this it is meant that the two sets 
$\{q_j\}$ of pseudoparticle momenta and $\{{\bar{q}}_j\}$ of pseudofermion canonical momenta are similarly
ordered for $j=1,...,\infty$. For each excited state there is then a pseudoparticle-pseudofermion unitary transformation associated
with the one-to-one correspondence between $q_j$ and $\bar{q}_j = q_j  + 2\pi\Phi (q_j)/L$. The equality, 
$\bar{q}_j = q_j$, holds for the PS ground state, so that for it pseudoparticles and pseudofermions are
identical particles.

A key property of the pseudofermion representation, which renders it the most appropriate for
the study of dynamical correlation functions, is that upon expressing the energy functional, Eq. (\ref{DEBg}),
in terms of the discrete canonical momentum values ${\bar{q}}_j = {\bar{q}} (q_j)$, Eq. (\ref{QPqBg}), it simplifies
up to ${\cal{O}} (1/L)$ order to,
\begin{equation}
\delta E = \sum_{j=1}^{\infty}\varepsilon ({\bar{q}}_j )\delta {\cal{N}} ({\bar{q}}_j) \, .
\label{DEBcang}
\end{equation}
The transformation from $q_j$ to $\bar{q}_j = q_j  + 2\pi\Phi (q_j)/L$ absorbs the
energy interacting term in Eq. (\ref{DEBg}). The pseudofermion canonical momentum distribution ${\cal{N}} ({\bar{q}}_j)$ 
in Eq. (\ref{DEBcang}) is defined as ${\cal{N}} ({\bar{q}}_j)=N (q_j)$ where ${\bar{q}}_j = {\bar{q}} (q_j)$ for $j=1,...,\infty$. The pseudofermion 
energy dispersion $\varepsilon ({\bar{q}}_j)$ has exactly the same form as that given in Eq. (\ref{varepsilonBg})
with the momentum $q_j$ replaced by the corresponding canonical momentum, ${\bar{q}}_j= {\bar{q}} (q_j)$.

In contrast to the equivalent energy functional, Eq. (\ref{DEBg}), that
given in Eq. (\ref{DEBcang}) has no energy interaction terms of second-order in the deviations
$\delta {\cal{N}} ({\bar{q}}_j)$. Indeed by expanding the canonical momentum  ${\bar{q}}_j$ 
around $q_j$ in Eq. (\ref{DEBcang}) and considering all energy contributions up to ${\cal{O}} (1/L)$ order,
one recovers after some lengthy yet straightforward algebra the energy functional, Eq. (\ref{DE-fermions}), which includes terms 
of second order in the deviations $\delta N (q_j)$. Their absence from the corresponding 
energy spectrum, Eq. (\ref{DEBcang}), is a consequence of the scattering phase shift functional $2\pi\Phi (q_j)$, Eq. (\ref{QPqBg}), 
being incorporated in the pseudofermions canonical momentum, Eq. (\ref{QPqBg}).

Physically, the quantity $\pm 2\pi\Phi (q_j,q_{j'})$, Eq. (\ref{PsBg}), is the phase shift 
of a pseudofermion of ground-state momentum $q_j$. It is acquired upon it scattering off a 
pseudofermion created $(+)$ or annihilated $(-)$ at a momentum $q_{j'}$ under a transition 
from the ground state to a PS excited state. The momentum $q_{j'}$ refers to that state
canonical momentum ${\bar{q}}_{j'} = q_{j'}  + 2\pi\Phi (q_{j'})/L$. Hence 
$2\pi\Phi (q_j)  = \sum_{j'=1}^{\infty}\,\delta N (q_{j'})\,2\pi\Phi (q_j,q_{j'})$, Eq. (\ref{QPqBg}), is the overall
scattering phase shift acquired by such a pseudofermion upon scattering off all pseudofermions created or annihilated
under that transition.

The $\iota = \pm$ pseudoparticle Fermi points given in Eq. (\ref{qFiota})
for a ground state also exist for the excited states that span the corresponding PS. 
Here we denote the ground-state numbers of right $(\iota =+)$ and left
$(\iota =-)$ pseudoparticles by $N_{b,\iota}^0$. Hence the Fermi points in Eq. (\ref{qFiota})
read $q_{F}^{\iota} = \iota\,{2\pi\over L}\,N_{b,\iota}^0$. The PS excited states
pseudoparticle Fermi points are of the form $q_{F}^{\iota} = \iota\,{2\pi\over L}\,N_{b,\iota}^0+\delta q_{F}^{\iota}$ where
$\delta q_{F}^{\iota} = [2\pi/L]\,\delta N_{b,\iota}^F$ and $\delta N_{b,\iota} = \delta N_{b,\iota}^{F,0} + \Phi^0$
for $\iota = \pm$. The deviation $\delta N_{b,\iota}^{F,0}$ in this equation refers to the number of pseudoparticles (and pseudofermions) 
at the $\iota = \pm $ Fermi points. It results from the creation or annihilation pseudoparticle processes and $\Phi^0$ 
is the non-scattering phase shift, Eq. (\ref{picanBg}), in units of $2\pi$. 

Two quantities that play a key role within the pseudofermion representation of the present
model are the $\iota=\pm $ pseudofermion Fermi points deviations,
\begin{equation}
\delta {\bar{q}}_{F}^{\iota} = \delta q_{F}^{\iota} + {2\pi\Phi (q_{F}^{\iota})\over L}
 \hspace{0.20cm}{\rm for}\hspace{0.20cm} \iota = \pm \, .
\label{bardqFiota}
\end{equation}
Specifically, the square of such deviations $\delta {\bar{q}}_{F}^{\iota}$ in units of $2\pi/L$, which
is denoted by $2\Delta^{\iota}\equiv (\delta {\bar{q}}_{F}^{\iota}/[2\pi/L])^2$, corresponds to two important $\iota=\pm $ 
fluctuations functionals of the theory. Within the PDT they control the one- and two-particle matrix elements quantum overlaps.
From manipulations that rely on the form of the phase shift $2\pi\Phi (q_j)$ in the pseudoparticle 
momentum - pseudofermion canonical-momentum transformation $q_j \rightarrow q_j  + 2\pi\Phi (q_j)/L$, Eq. (\ref{QPqBg}),
one finds in the TL that,
\begin{equation}
2\Delta^{\iota} (\{q_{j'}\}) \equiv \left({\delta {\bar{q}}_{F}^{\iota}\over 2\pi/L}\right)^2 =
\left(\xi^1\,\delta J_b^F + \iota\,{\delta N_b^F\over 2\xi^1} 
+ \sum_{j'=1}^{\infty}\,\delta N^{NF} (q_{j'})\,\Phi (\iota q_F,q_{j'})\right)^2 
\hspace{0.20cm}{\rm for}\hspace{0.20cm} \iota = \pm \, .
\label{2DeltaBg}
\end{equation}
The summation $\sum_{j'=1}^{\infty}\,\delta N^{NF} (q_{j'})$ runs here over a
set of finite $j' = 1,...,N_b^{NF}$ values where $N_b^{NF} = \sum_{j'=1}^{\infty}\,\vert\delta N^{NF} (q_{j'})\vert$.
Hence $2\Delta^{\iota} (\{q_{j'}\})$ depends on a corresponding set of finite $q_{1},...,q_{N_b^{NF}}$ momentum
values.

On the one hand, the deviations $\delta N_b^F = \sum_{\iota=\pm }\delta N^F_{b,\iota}$ and 
$\delta J_b^F = {1\over 2}\sum_{\iota=\pm }(\iota)\delta N^F_{b,\iota}$ in Eq. (\ref{2DeltaBg}) refer to the low-energy part
of the excitations. On the other hand, within the TL the high-energy contributions are associated with the deviation $\delta N^{NF} (q_{j'})$
as defined above. The pseudofermion creation or annihilation at and in the vicinity of the Fermi points 
points is rather accounted for by the deviations $\delta N_b^F$ and $\delta J_b^F$.

For low-energy PS excited states for which $\delta N^{NF} (q_{j'})=0$ for all $q_{j'}$ values away from the $\iota = \pm$ 
pseudofermion Fermi points, the fluctuation functionals, Eq. (\ref{2DeltaBg}), become
the $\iota = \pm $ fields conformal dimensions of a conformal field theory,
\begin{equation}
2\Delta^{\iota}_0 \equiv \left({\delta {\bar{q}}_{F}^{\iota}\over 2\pi/L}\right)^2 =
\left(\xi^1\,\delta J^F + \iota\,{\delta N^F\over 2\xi^1}\right)^2 \hspace{0.20cm}{\rm for}\hspace{0.20cm} \iota = \pm \, .
\label{2DeltaBg0}
\end{equation}
In the low-energy limit the model can be mapped into a conformal field theory \cite{Bogoliubov-86,Bogoliubov-87,Woy-89}.
As given in Eq. (\ref{xiBg}), within the pseudofermion representation the low-energy parameter $\xi^1=1/\xi^0$
naturally emerges from the pseudofermion phase shifts at the Fermi points. $\xi^1=1/\xi^0$ is
actually the dressed charge of the conformal field theory \cite{Bogoliubov-86,Bogoliubov-87,Woy-89}.
Furthermore, since the usual low-energy TLL parameter $K_0$ \cite{Voit,Glazman-BG-08}
merely reads $K_0=(\xi^1)^2$.

As shortly reported in Section \ref{PDT}, the PDT introduced in Refs. \cite{V-1,TTF,spectral-06,VI,CarCadez-16,CarCadez-17,spectral,LE} 
for the 1D Hubbard model has been extended to simpler integrable models such as the present 1D Lieb-Liniger Bose gas \cite{DCFBo-16} 
and the spin-$1/2$ $XXX$ chain \cite{CPJD-15}. The PDT is associated with the pseudofermion representation of such models.
One of the goals of this review is to clarify the relation between the PDT and the MQIM methods \cite{Glazman-09,Glazman-12,Glazman-BG-08}.
For simplicity, below in Section \ref{RelaPDTMIM} and Appendix \ref{alphakF}
the present 1D Lieb-Liniger Bose gas is used to address that problem. 
The basic relation is qualitatively similar for the more complex models also reviewed in this paper. 
The relation of the 1D Lieb-Liniger Bose gas PDT to the MQIM of Ref. \cite{Glazman-BG-08} allows the expression of the 
general PDT $\iota = \pm$ pseudofermion Fermi points fluctuations functionals, Eq. (\ref{2DeltaBg}), in terms
of the MQIM shift function $F_B (k\vert k')$ defined in Eqs. (7) and (8) of Ref. \cite{Glazman-BG-08}.
In that reference it is called $F_B (\nu\vert\mu)$ whereas here its variables $\nu$ and 
$\mu$ are replaced by our notation for the momentum rapidities, $k$ and $k'$, respectively.
Furthermore, the corresponding limiting values $\pm q$ are replaced by our notation $\pm Q$ for them, Eq. (\ref{varepsilonBg}).
From the use of the relation $\Phi (\iota q_F,q_{j'}) = {\xi^1\over 2} - F_B (\iota k_0 (q_F),k_0 (q_{j'}))$,
Eq. (\ref{PhaseSFB}) of Appendix \ref{alphakF}, one readily finds that,
\begin{equation}
2\Delta^{\iota} (\{q_{j'}\}) = \left(\xi^1\left(\delta J_b^F + {\delta N_b^{NF}\over 2}\right) + \iota\,{\delta N_b^F\over 2\xi^1} - 
\sum_{j'=1}^{\infty}\,\delta N^{NF} (q_{j'})\,F_B (\iota Q\vert k_0 (q_{j'}))\right)^2 \hspace{0.20cm}{\rm for}\hspace{0.20cm} \iota = \pm \, ,
\label{2DeltaBgMIM}
\end{equation}
where $\delta N_b^{NF} = \sum_{j'=1}^{\infty}\,\delta N^{NF} (q_{j'})\leq N_b^{NF} = \sum_{j'=1}^{\infty}\,\vert\delta N^{NF} (q_{j'})\vert$.

By considering low-energy excited states for which $\delta N^{NF} (q_{j'})=0$ for all $q_{j'}$ values
away from the $\iota=\pm $ pseudofermion Fermi points, the functionals, Eqs. (\ref{2DeltaBg}) 
and (\ref{2DeltaBgMIM}), acquire the simplified form, Eq. (\ref{2DeltaBg0}). Hence
both the PDT and the MQIM naturally contain the present model low-energy conformal field theory. In the case of the PDT,
the link of the functionals to the conformal dimensions in the correlation functions obtained from the BA
\cite{Bogoliubov-86,Bogoliubov-87,Woy-89} can be understood as described in the following.

The property that the excitation energy spectrum, Eq. (\ref{DEBcang}), has no pseudofermion energy 
interactions simplifies the expression of the dynamical correlation functions in terms of pseudofermion spectral functions.
The lack of energy interactions achieved under the pseudoparticle momentum - pseudofermion canonical-momentum 
transformation $q_j \rightarrow q_j  + 2\pi\Phi (q_j)/L$, Eq. (\ref{QPqBg}), has though a price: The usual integer
or half-integer dynamical correlation functions exponents of non-interacting and Fermi-liquid like quantum
systems are replaced by the interaction and momentum dependent exponents whose expressions
involve the $\iota = \pm$ functional dimensions, Eqs. (\ref{2DeltaBg}) and (\ref{2DeltaBgMIM}). 

The pseudofermion representation involves a mere unitary transformation under which the integer or half-integer 
BA quantum numbers $I_j$ in $q_j = {2\pi\over L}\,I_j $, Eqs. (\ref{qjBg}) and (\ref{IjBg}), are shifted to
$I_j \rightarrow \bar{I}_j$ where $\bar{I}_j = I_j  + \Phi (q_j)$. The pseudofermion phase shift $\Phi (q_j)$ in units of $2\pi$
appearing here, Eq. (\ref{QPqBg}), is in general both interaction and momentum dependent. The dynamical correlation 
functions usual integer or half-integer dimensions are mapped under the transformation associated with such a shift, 
$I_j \rightarrow I_j  + \Phi (q_j)$, onto the exotic interaction and momentum dependent functionals, Eqs. (\ref{2DeltaBg}) and (\ref{2DeltaBgMIM}).

This is why the exponents that control the dynamical correlation 
functions line shape in the vicinity of well-defined types of $(k,\omega)$-plane singular spectral features
are interaction and momentum dependent functionals. In the low-energy limit, the general $\iota =\pm$ functional 
dimensions $2\Delta^{\iota} (\{q_{j'}\})$ in Eqs. (\ref{2DeltaBg}) and (\ref{2DeltaBgMIM}) 
where $j' = 1,...,N_b^{NF}$ lose their momentum dependence. Indeed, they refer to excited states for
which $N_b^{NF}=0$. However they remain being interaction dependent since they become the $\iota = \pm$ fields conformal 
dimensions $2\Delta^{\iota}_0$, Eq. (\ref{2DeltaBg0}). 

In the case of the more complex 1D Hubbard model, the PDT contains as well the conformal field theory 
as a limiting behavior. (For further technical information on the link of the generalized dimensions, 
Eqs. (\ref{2DeltaBg}) and (\ref{2DeltaBgMIM}), to the conformal dimensions in the correlation functions also obtained 
from the BA and how the PDT leads in the low-energy limit to exactly the same correlation functions as 
conformal-field theory, see Ref. \cite{LE}.)

The generalized functional dimensions in Eqs. (\ref{2DeltaBg}) and (\ref{2DeltaBgMIM}) correspond to an important
step beyond conformal-field theory. They apply actually both at low and high energy.
Hence one can learn from the pseudofermion representation new insights beyond the model
low-energy physics. Indeed, the additional deviations 
$\sum_{j'=1}^{\infty}\,\delta N^{NF} (q_{j'})\,\Phi (\iota q_F,q_{j'})$ in
Eq. (\ref{2DeltaBg}) actually control the high-energy regime of dynamical correlation functions.
The corresponding generalized dimensions, Eqs. (\ref{2DeltaBg}) and (\ref{2DeltaBgMIM}),
are confirmed in the following and in Sections \ref{PDT} and \ref{RelaPDTMIM} to play an important role 
in the dynamical correlation functions spectral weight distributions.
For instance, the interaction and momentum dependent exponents that control the dynamical correlation 
functions line shape in the vicinity of well-defined types of $(k,\omega)$-plane singular spectral features are
within the PDT a superposition of the such generalized dimensions. This refers to the vicinity of these 
functions $(k,\omega)$-plane lower or upper thresholds. In the case of one-particle spectral
functions, this applies as well near a particular type of singular features called within the PDT branch lines. 

Within the PDT the dynamical correlation functions are written in terms of pseudofermion 
spectral functions. Such functions spectral weights can be expressed as Slater determinants written 
in terms of anticommutators of pseudofermion operators. (For simplicity, in the case of the present model 
we do not introduce here the corresponding pseudofermion operator algebra.) The Slater determinants are 
written in terms of anticommutators of pseudofermion operators. Their expressions involve the overall phase-shift functional,
\begin{equation}
2\pi\Phi^T (q_j) = 2\pi\Phi^0 + 2\pi\Phi (q_j) = 2\pi\Phi^0 + \sum_{j'=1}^{\infty}\,\delta N (q_{j'})\,2\pi\Phi (q_j,q_{j'}) \, .
\label{QPqTBg}
\end{equation}

The corresponding dynamical correlation functions 
one- and two-boson spectral weights are written in terms of such anticommutators.
Their dependence on the phase-shift functional $2\pi\Phi^T (q_j)$ is the mechanism through
which the shake up effects occurring in the pseudofermion canonical momentum band under 
the transitions to the excited states lead to the Anderson's orthogonality catastrophes 
\cite{Anderson-67}. Such a shake up refers to the change from the ground-state momentum values
$q_j^0$ to the excited states canonical momentum values $q_j^0  + 2\pi\Phi^T (q_j)/L
= q_j  + 2\pi\Phi (q_j)/L$. (Here $q_j  = q_j^0 + 2\pi\Phi^0/L$ are the pseudoparticle
momentum values of the excited state.)

The different nature of the dynamical correlation functions of the pseudofermion quantum
liquid relative to those of a Fermi liquid originates from these Anderson's orthogonality catastrophes.
Those are associated with finite contributions to the one- and two-boson spectral weight distributions 
from a large number of low-energy and small-momentum particle-hole processes in
the pseudofermion band. The form of the functional dimensions, Eq. (\ref{2DeltaBg}) and (\ref{2DeltaBgMIM}), results
from these contributions \cite{DCFBo-16}.

In the following we provide the momentum dependent exponents obtained from the PDT. Those control the line 
shape of the one-boson addition and removal spectral functions and two-boson charge dynamical 
structure factor near their thresholds. These dynamical correlation functions for the boson problem were
studied first by the MQIM \cite{Glazman-BG-08}. Recently the use of the PDT reached exactly the same 
spectra and momentum dependent exponents for such functions \cite{DCFBo-16}.
(In Ref. \cite{Glazman-BG-08} analytical expressions for these exponents were
derived yet their momentum dependence has not been plotted.)
\begin{figure}
\begin{center}
\subfigure{\includegraphics[width=5.00cm]{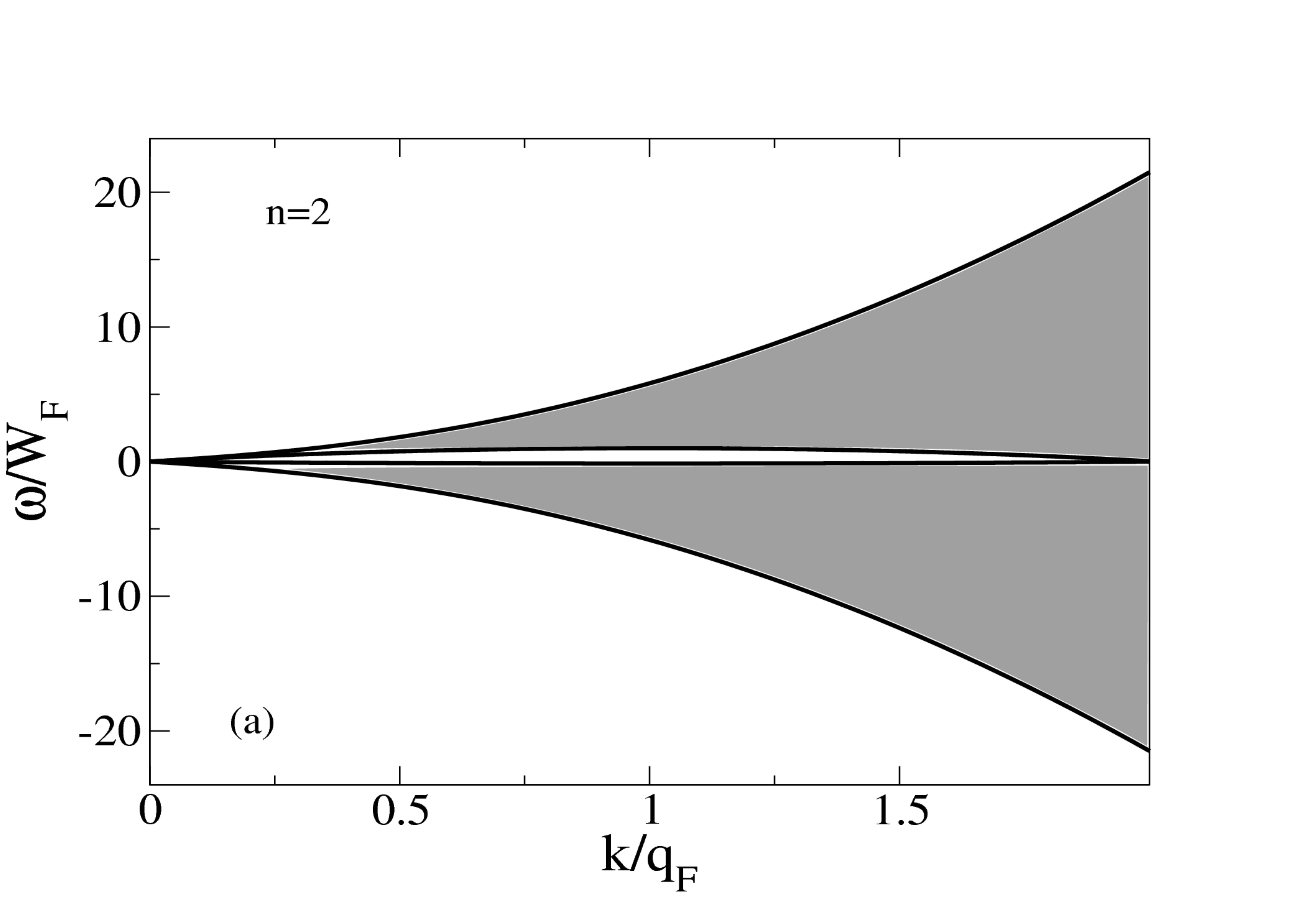}}
\hspace{0.25cm}
\subfigure{\includegraphics[width=5.00cm]{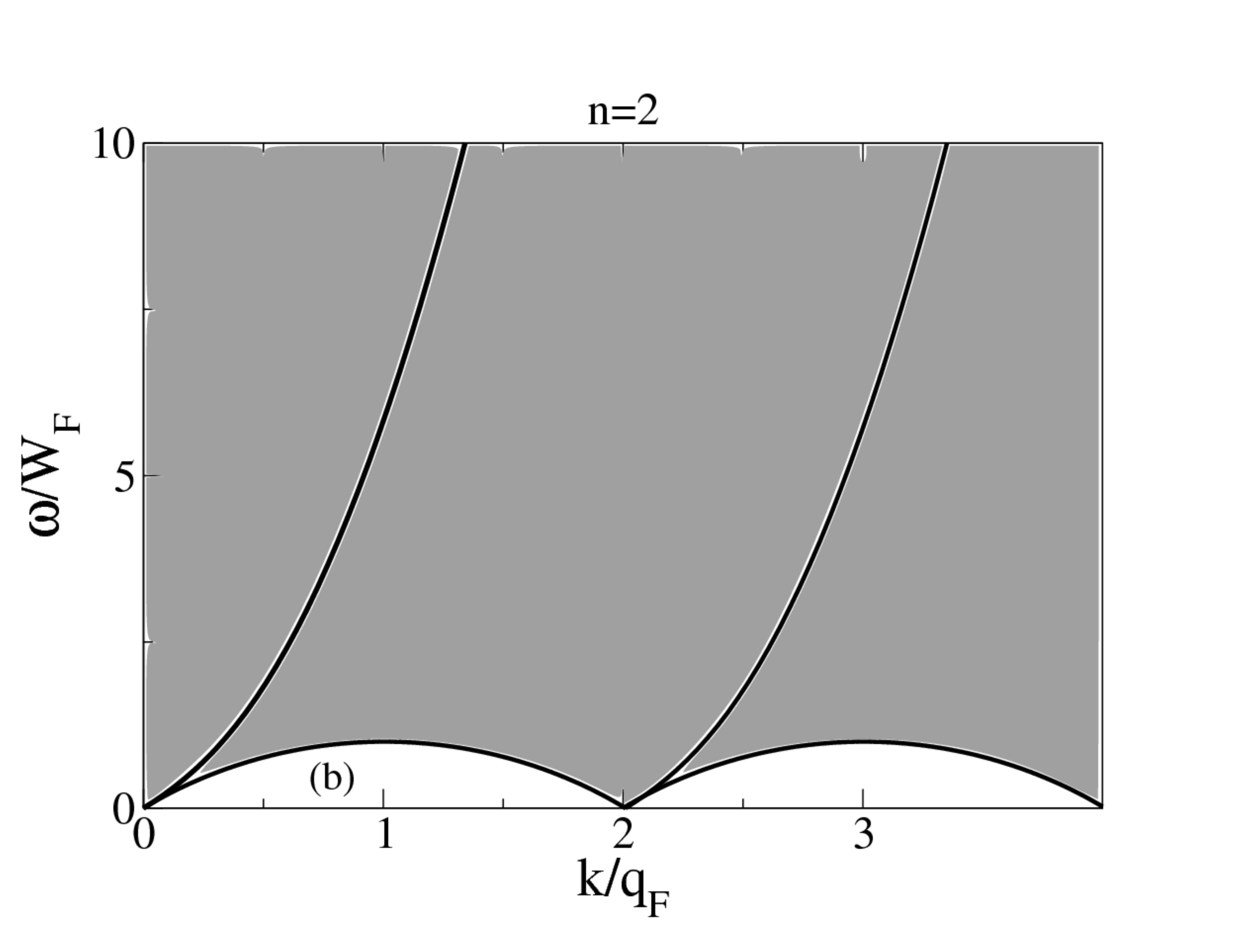}}
\caption{The spectra, Eq. (\ref{omeglBg}), of the excitations associated with the dominant contributions
to the one-boson removal $(\omega <0)$ and addition $(\omega >0)$ spectral function (left panel) and charge dynamical structure factor (right panel) 
for $c=1$ and density $n_b=2$. In the case of the latter factor, the spectrum edge for $k/q_F \in [2,4]$ and the weight
above it associated with the next-order excitations has also been included. The momentum scale $q_F$ and energy scale $W_F$ 
are the pseudoparticle Fermi momentum, Eq. (\ref{qFiota}), and energy bandwidth of the ground-state occupied 
Fermi sea, Eq. (\ref{muBg}), respectively. (As in Ref. \cite{DCFBo-16}, $n_b$ is denoted by $n$ in the figures.)\\
{\it Source}: From Ref. \cite{DCFBo-16}.}
\label{figure3}
\end{center}
\end{figure}

The one-boson removal and addition spectral functions are given by \cite{Glazman-BG-08,DCFBo-16},
\begin{equation}
S^B (k,\omega) \equiv B (k,\omega) = \sum_{f}\vert\langle f,N_b-1\vert\hat{\Psi}_k\vert GS,N_b\rangle\vert^2
\delta (\omega - \omega^B (k)) \, ,
\label{SRBg}
\end{equation}
and 
\begin{equation}
S^A (k,\omega) \equiv A (k,\omega) = \sum_{f}\vert\langle f,N_b+1\vert\hat{\Psi}_k^{\dag}\vert GS,N_b\rangle\vert^2
\delta (\omega - \omega^A (k)) \, ,
\label{SABg}
\end{equation}
respectively. Here $\omega^B (k) = E_{GS}^{N_b} - E_{f}^{N_b-1}$, $\omega^A (k) = E_{f}^{N_b+1} - E_{GS}^{N_b}$,  
$E_{n}$ is the excited states energy, $E_0$ that of the ground state,
and $\hat{\Psi}_k^{\dag}$ and $\hat{\Psi}_k$ are boson creation and annihilation operators.
Furthermore, the charge dynamical structure factor reads \cite{Glazman-BG-08,DCFBo-16,Caux-BG-06},
\begin{equation}
S^D (k,\omega) \equiv S (k,\omega) = \sum_{f}\vert\langle f\vert\hat{\rho}_k\vert GS\rangle\vert^2
\delta (\omega - \omega^D (k)) \, ,
\label{DSFBg}
\end{equation}
where $\omega^D (k) = E_{f} - E_{GS}$ and $\hat{\rho}_k$ is the Fourier transform
of the local density operator $\hat{\rho}_x$. As mentioned above, the charge dynamical structure factor,
Eq. (\ref{DSFBg}), can be probed in ultra-cold atom systems through low-momentum Bragg excitations 
\cite{Fabbri-15,Golovach-09},

The $(k,\omega)$-plane lower ($c_{\tau}=-1$) or upper ($c_{\tau}=1$) thresholds of the energy spectra 
of the above dynamical spectral functions have for the momentum range $k\in [0,2\pi n_b]$ the general form,
\begin{equation}
\omega^{\tau} = c_{\tau}\,\varepsilon (q_F-k)\hspace{0.20cm}{\rm where}\hspace{0.20cm}
k = q_F - q \in [0,2\pi n_b] \, .
\label{omeglBg}
\end{equation}
Here $\varepsilon (q)$ is the energy dispersion, Eq. (\ref{varepsilonBg}). The index $\tau$ reads
$\tau = B$ for the one-boson removal spectral function, $\tau = A$ for the one-boson addition spectral function, and
$\tau = D$ for the two-boson dynamical structure factor. The coefficient $c_{\tau}$ is given by
$c_{\tau}=1$ for $\tau = B$ and $c_{\tau}=-1$ for $\tau = A,D$. The spectra, Eq. (\ref{omeglBg}), are
shown in Fig. \ref{figure3} for $c=1$ and density $n_b=2$. 

For small energy deviations $(\omega - \omega^{\tau} (k)) >0$ in the vicinity of the $(k,\omega)$-plane lower
($c_{\tau}=-1$) or upper ($c_{\tau}=1$) thresholds, the PDT leads to the following exact line shape \cite{DCFBo-16}, 
\begin{equation}
S^{\tau} (k,\omega) = C^{\tau}\,(\omega - \omega^{\tau} (k))^{\xi_{\tau} (k)} \hspace{0.20cm}{\rm for}\hspace{0.20cm}k \in [0,2\pi n_b] 
\hspace{0.20cm}{\rm where}\hspace{0.20cm} \tau = B, A, D \, .
\label{RAD-Bg}
\end{equation}
Here $C^{\tau}$ is a coefficient whose value remains unchanged in the range of small energy deviations $(\omega - \omega^{\tau} (k)) >0$ 
for which this expression is valid. The momentum dependent exponents are given by,
\begin{equation}
\xi_{\tau} (k) = - 1 + \sum_{\iota = \pm}2\Delta^{\iota}_{\tau} =
-1 + \sum_{\iota = \pm}\left({\xi^1\over 2} +
\iota\,{b_{\tau}\over\xi^1} - \Phi (\iota q_F,q_F-k)\right)^2 \, .
\label{xilambBg}
\end{equation}
The $\iota = \pm$ functional dimensions in this expression, $2\Delta^{\iota}_{\tau}=(\delta {\bar{q}}_{F}^{\iota}/(2\pi/L))^2$,
are those given in Eq. (\ref{2DeltaBg}) for the excited states specific to each of the $ \tau = B, A, D$ dynamical
correlation functions, $b_{\tau}=0$ for $\tau = B$, $b_{\tau}=1$ for $\tau = A$, and
$b_{\tau}=1/2$ for $\tau = D$. The one-boson removal upper threshold and 
addition lower threshold are in Fig. \ref{figure3} the two boundary lines above 
and below the $\omega =0$ axis, respectively.
\begin{figure}
\begin{center}
\subfigure{\includegraphics[width=5.50cm]{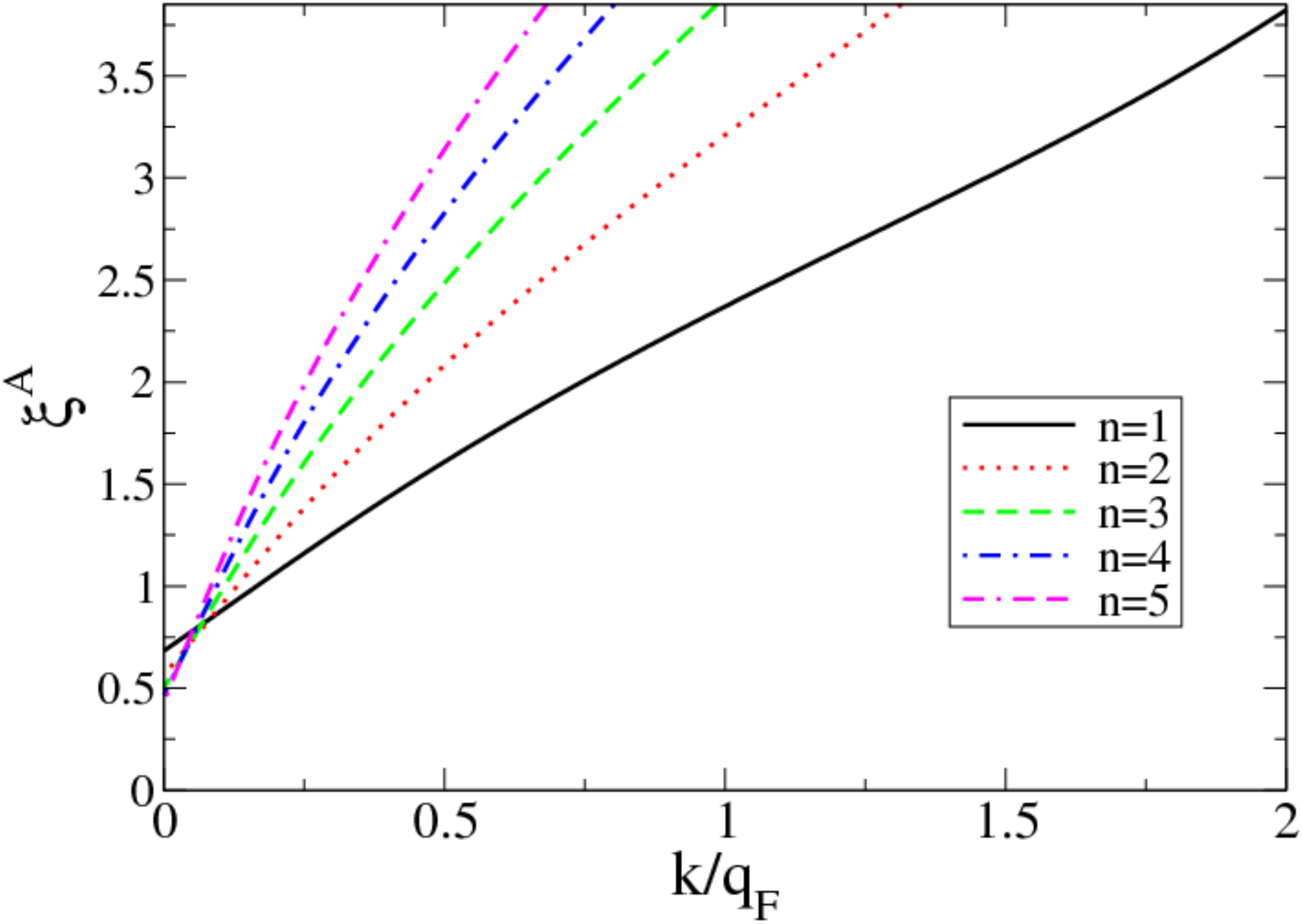}}
\hspace{0.25cm}
\subfigure{\includegraphics[width=5.50cm]{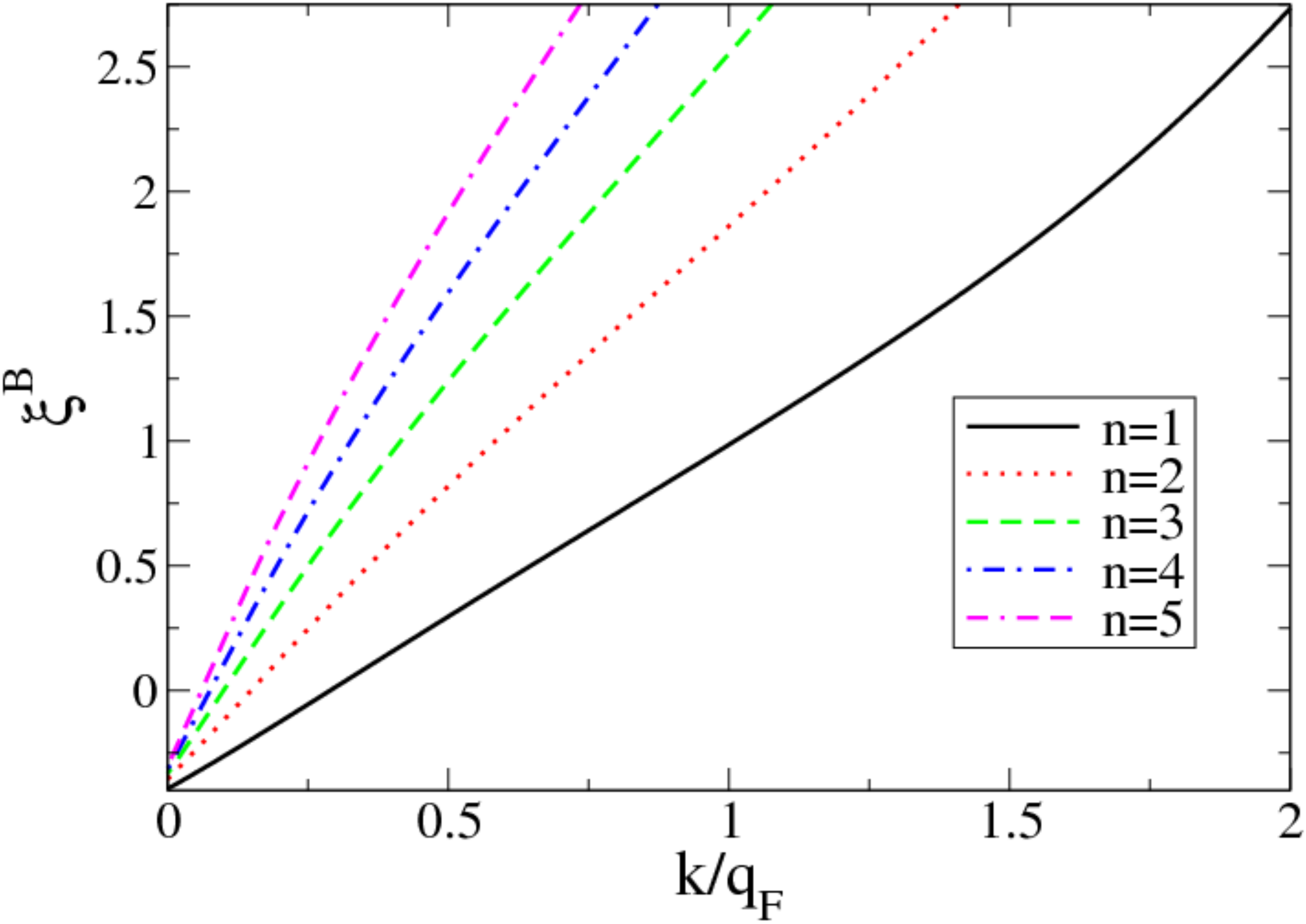}}
\hspace{0.25cm}
\subfigure{\includegraphics[width=5.50cm]{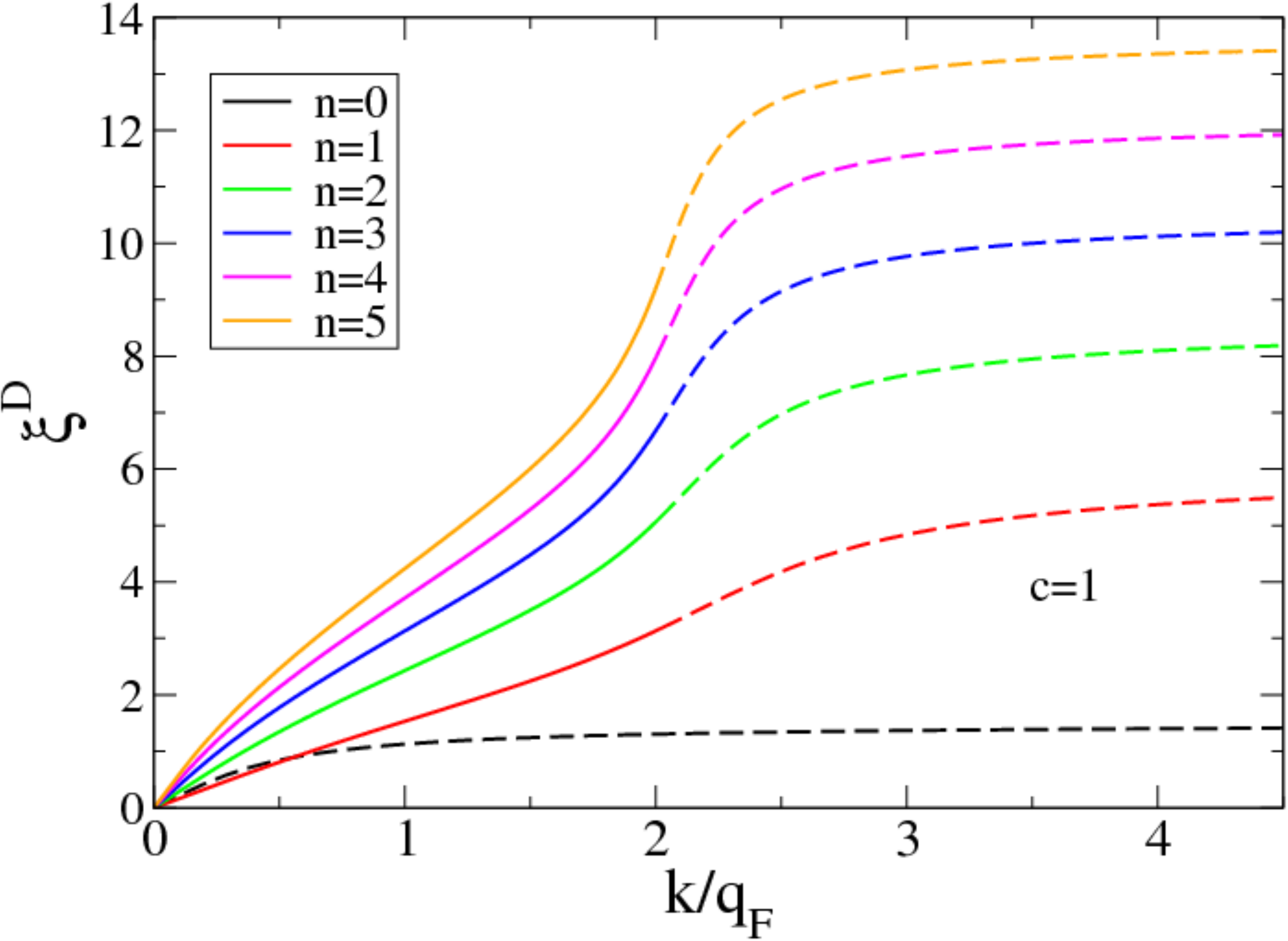}}
\caption{The exponents, Eq. (\ref{xilambBg}), $\xi_{A} (k)$ of the one-boson addition spectral function, $\xi_{B} (k)$ of 
the one-boson removal spectral function, and $\xi_{D} (k)$ of the charge dynamical structure factor
for $c=1$ plotted as a function of the momentum in units of $q_F$ for the density $n_b$ values given in the figure panels. 
(The momentum scale $q_F$ and energy scale $W_F$ are those of Fig. \ref{figure3}.) The latter exponent is exact 
for $k/q_F<2$ (solid lines) and an approximation for $k/q_F>2$ (dashed lines). For $k/q_F>2$ it is a better 
approximation for the values $c=1$ and $n_b=5$ for which $n_b/c$ is largest. Only $\xi_{B} (k)$ has negative values
for a limited momentum range associated with upper threshold singularity cusps in the one-boson removal 
spectral function. (As in Ref. \cite{DCFBo-16}, $n_b$ is denoted by $n$ in the figures.)\\
{\it Source}: From Ref. \cite{DCFBo-16}.}
\label{figure4}
\end{center}
\end{figure}

The spectral feature in Eqs. (\ref{omeglBg}) - (\ref{xilambBg}) is called a branch line \cite{DCFBo-16}.
It is generated by elementary processes where only one pseudofermion is created (and annihilated) outside 
the Fermi points. When in the $(k,\omega)$ plane
there is no spectral weight above (lower threshold) or below (upper threshold) that line, the corresponding 
dynamical correlation function analytical expression is exact. If above or below it there is a {\it very small} amount
of spectral weight, it is an approximation. 

The latter approximation is valid for one-particle spectral functions whose spectral-weight distribution
is not plateau-like. In contrast, this is the general case for two-particle dynamical correlation functions. However, in the case of the 
1D Lieb-Liniger Bose gas charge dynamical structure factor, the spectral weight between the spectrum edge line
for $k/q_F\in [2,4]$ in Fig. \ref{figure3} and the rising branch line in that figure starting at $k/q_F=2$ 
vanishes in the $n_b/c\rightarrow\infty$ limit \cite{DCFBo-16}.
Hence we consider here the charge dynamical structure factor for very large $n_b/c$ values 
in the vicinity of that branch line, which has the form,
\begin{equation}
\omega^D = \varepsilon (k-q_F) \hspace{0.20cm}{\rm where}\hspace{0.20cm}
k = q + q_F \in [2\pi n_b,\infty] \hspace{0.20cm}{\rm for}\hspace{0.20cm}q \in [\pi n_b,\infty] \, .
\label{omegDSFpBg}
\end{equation}
For small $(\omega - \omega^D (k)) >0$ values near it, an approximation for the dynamical structure factor is,
\begin{equation}
S^D (k,\omega) \approx C^{D}\,(\omega - \omega^{D} (k))^{\xi_{D} (k)}\hspace{0.20cm}{\rm for}\hspace{0.20cm}k \in [2\pi n_b,\infty] \, .
\label{Dp-Bg}
\end{equation}
The momentum dependent exponent is here of the form,
\begin{equation}
\xi_{D} (k) = -1 + \sum_{\iota = \pm}\left({\xi^1\over 2} -
{\iota\over 2\xi^1} + \Phi (\iota q_F,k-q_F)\right)^2 \, .
\label{xiDgqBg}
\end{equation}

The two $\tau = B, A$ one-boson spectral function exponents $\xi_{\tau} (k)$, Eq. (\ref{xilambBg}),
and the dynamical structure factor exponent $\xi_{D} (k)$, Eq. (\ref{xiDgqBg}), are plotted
in Fig. \ref{figure4} as a function of the momentum $k$ in units of $q_F = \pi\,n_b$ for interaction $c=1$ 
and several densities $n_b$ values. For $k/q_F>2$ the latter exponent is an approximation.
For the ranges $k/q_F \in [0,k^*/q_F]$ (where $k^*$ is such that $\xi_{B} (k^*)=0$)
in Fig. \ref{figure4} for which the exponent $\xi_{B} (k)$ is negative, there are upper threshold singularity cusps in the one-boson 
removal spectral function $S^{B} (k,\omega)$, Eq. (\ref{SRBg}). 

%%%%%%%%%%%%%%%%%%%%%%%%%%%%%%%%%%%%%%%%%%%%%%%%%%%%%%%%%%%%%%%%%%%%%%%%%%
\section{The spin-$1/2$ isotropic Heisenberg chain}
\label{Heichain}

The spin-$1/2$ $XXX$ chain is again a quantum problem of interest.
It is a paradigmatic example of an integrable strongly correlated system that
is experimentally relevant for the description of magnetic properties of spin-chain 
materials \cite{SPA-11,Ralph-11,Stone-03,Motoyama-96,Thurber-01}.
At zero magnetic field the model contains antiferromagnetic correlations 
for exchange integral $J>0$. Those have 
been observed in dynamical quantities measured in inelastic neutron scattering 
experiments on spin-chain compounds \cite{Stone-03}. It can also be prepared 
in a 1D ultra-cold atomic trap \cite{Murmann-15}.

Here we consider the spin-$1/2$ $XXX$ chain in the TL. In that limit the complex 
rapidities in the BA equations simplify in terms of the ideal strings of Ref. \cite{Takahashi-71}. 
(The deviations from such ideal strings \cite{Caux-07} do not affect in the TL the properties 
of the model revisited in this section.)

The Hamiltonian of the general spin-$1/2$ Heisenberg chain with exchange 
integral $J>0$ and anisotropy parameter $\Delta\ge 0$ in a magnetic field $H$ reads,
\begin{equation}
\hat{H}_{\Delta} = J\sum_{j=1}^{L}\left(\sum_{\tau =x,y}{\hat{S}}_j^{\tau}{\hat{S}}_{j+1}^{\tau} + \Delta\,{\hat{S}}_j^z{\hat{S}}_{j+1}^z\right) 
+ 2\mu_B\,H\sum_{j=1}^{L} \hat{S}_j^z \hspace{0.25cm}{\rm where}\hspace{0.25cm}{\hat{S}}^{\tau} = \sum_{j=1}^{L}{\hat{S}}_j^{\tau}
\hspace{0.25cm}{\rm and}\hspace{0.25cm}
{\hat{S}}^{\pm} = \sum_{j=1}^{L}{\hat{S}}_j^{\pm} \hspace{0.25cm}{\rm with}\hspace{0.25cm}\tau = x,y,z \, .
\label{HXXZ}
\end{equation}
Here $\hat{\vec{S}}_{j}$ and ${\hat{S}}^{\tau}$ are the spin-$1/2$ operators at site $j=1,...,L$ with components $\hat{S}_j^{x,y,z}$
and related total spin operators, respectively, $\hat{S}_j^{\pm} = \hat{S}_j^x \pm i \hat{S}_j^y$, and 
$\mu_B$ is the Bohr magneton.

At the isotropic point, $\Delta=1$, the model contains antiferromagnetic correlations that have been observed 
in dynamical quantities measured in experiments on spin-chain compounds \cite{SPA-11,Stone-03,Thurber-01,Motoyama-96},
\begin{equation}
\hat{H} = J\sum_{j=1}^{L} \hat{\vec{S}}_j \cdot \hat{\vec{S}}_{j+1} + 2\mu_B\,H\sum_{j=1}^{L} \hat{S}_j^z \, .
\label{Hchain}
\end{equation}
Another observable of interest for our study of this model is the $z$ component of the spin current operator,
\begin{equation}
\hat{J}^z = -i\,J\sum_{j=1}^{L}(\hat{S}_j^+\hat{S}_{j+1}^- - \hat{S}_{j+1}^+\hat{S}_j^-)  \, .
\label{c-s-currents}
\end{equation}

\subsection{The model physics: fractional excitations, spin ordering, and magnetism}
\label{HcPhysucs}

To address the phenomena brought about by quantum magnetic correlations, it is instructive to consider systems 
where the charge degrees of freedom are frozen and only spin excitations remain. Such systems are usually described
by spin-only models. They are realized, for example, in Mott insulators where magnetic interactions between the local moments of localized electrons 
are mediated by virtual exchange processes between neighboring electrons. One can describe the magnetic 
correlations through models of localized quantum spins embedded on lattices. 
At zero magnetic field local moment antiferromagnetic phases frequently occur in the arena of strongly correlated electron systems. 
The phenomenology displayed by the antiferromagnetic Hamiltonian depends sensitively on the geometry of 
the underlying lattice. 

On the one hand, on a bipartite lattice of dimension $d>1$, such as the square lattice, i.e. one in which the 
neighbors of one sub-lattice A belong to the other sub-lattice B, the zero-field ground state of the spin-$1/2$ isotropic 
Heisenberg antiferromagnet is close to a staggered spin configuration. It is known as a N\'eel state whose neighboring spins 
are antiparallel. The linear spin-wave theory for antiferromagnets \cite{Anderson-52} relies 
onto an expansion that is valid provided that ${1\over N}\sum_k\langle n_k\rangle\ll 1$. Here $n_k$ is the occupation 
number of the spin-wave state of momentum $k$. For the square lattice one finds ${1\over N}\sum_k\langle n_k\rangle \approx 0.197$, which
is indeed a rather small number. The elementary excitations above the zero-field antiferromagnetic ground
state are then spin waves of two flavors, due to the two sub-lattices. The elementary particle of a spin wave is called a magnon.
It carries a spin equal to one. 

On the other hand, for the spin chain, the lattice is trivially bipartite. However, in the 
1D case one rather finds that ${1\over N}\sum_k\langle n_k\rangle\rightarrow\infty$ at zero magnetic field \cite{Manousakis-91}. 
This is due to the long-wavelength modes. Indeed, it is well known that the zero-field ground state of the spin chain fails to 
develop long-range antiferromagnetic order. As a result, in 1D there is no zero-field linear spin-wave theory small expansion parameter. 
This is why the perturbative expansion around an ordered zero-field ground state is incorrect for the spin-$1/2$ $XXX$ chain.

The zero magnetic field spin order and associated magnetism of that spin chain is thus of a different type.
Fortunately, in spite of its non-perturbative nature and lack of zero-field ground-state long-range order, one can extract important 
information on that quantum problem physics from its exact BA solution. 
Rather than spin waves and magnons, the elementary excitations of the zero-field spin-$1/2$ $XXX$ chain, 
which correspond to well-defined BA quantum numbers distributions, are 
often associated with phenomenological nontrivially interacting spin-$1/2$ particles called spinons. Their 
energy dispersion relation is extracted from the BA \cite{Fowler-78,Faddeev-81,Anderson-87,Essler-94,Essler-94-B}.
(The phenomenological/empirical nature of the spinons stems from their precise operational relation to the model physical spins 
not being defined.)

Consider some $SU(2)$-symmetric Hamiltonian, in more than one spatial dimension, made out of spin-$1/2$ 
particles, which is in a ground state that spontaneously breaks the $SU(2)$ symmetry. If one flips a single spin, a magnon is created, 
not a spinon. Spinons that emerge in spin chains are much weirder. Since any local spin operator changes an integer amount 
of spin, one {\it cannot} create a single spinon with a local operator. Hence spinons are examples of fractionalized particles: 
They can only arise as part of a physical disturbance. 

The spin-$1/2$ $XXX$ chain whose Hamiltonian is given in Eq. (\ref{Hchain}) is a paradigmatic example of both a spin chain and 
an integrable strongly correlated quantum many-body system. 
In this paper we review the related pseudoparticle and pseudofermion representations of its exact BA solution.
Such a solution refers explicitly to the lowest-weight states (LWSs) or highest-weight states (HWSs) of the $SU(2)$ algebra for which $S=-S^z$ 
and $S=S^z$, respectively. Here (and in the following) the spin and spin projection of the spin chain, Eq. (\ref{Hchain}), 
energy eigenstates have been denoted by $S$ and $S^{z}=-(L_{\uparrow} -L_{\downarrow})/2$, respectively.
$L_{\sigma}$ such that $\sum_{\sigma=\uparrow,\downarrow}L_{\sigma}=L$ denotes the number of 
physical spins of projection $\sigma=\uparrow,\downarrow$.
(The sign choice in the expression $S^{z}=-(L_{\uparrow} -L_{\downarrow})/2$ is 
the same as in Ref. \cite{Takahashi-71}, for which $L_{\uparrow}\geq L_{\downarrow}$ for a LWS.)
The pseudoparticle and pseudofermion representations also applies to the extended Hilbert
space spanned by both the LWSs and corresponding multiplet spin $SU(2)$ towers of states.

The BA gives direct access to the energy levels of an integrable system. This allows the computation of many equilibrium 
quantities. The ground state energy of the spin-$1/2$ $XXX$ chain was
derived analytically within the TL in the early stages of its studies \cite{Hulthen-38}. However, it was not until the 1960s
and early 1970s that its excitation spectrum was computed \cite{Cloizeaux-62} and its thermodynamic properties 
derived \cite{Takahashi-71,Griffiths-64,Yang-66A,Yang-66B,Yang-66C,Gaudin-71}.
Equilibrium quantities are, nonetheless, not sufficient to completely characterize the
physics of correlated models.

The computation of dynamical quantities requires knowledge of matrix elements of spin operators between 
energy eigenstates. This goes beyond the information that can be extracted directly from the BA. At zero field 
some of such states are described by groups of real and complex BA rapidities. 
The spin dynamical structure factors are objects that motivated 
partially by experimental work have been extensively studied in the case of zero magnetic field 
\cite{Muller-81,Bougourzi-97,Bougourzi-97A,Bougourzi-96,Caux-06,Imambekov-09,Caux-11}.
In Section \ref{DSGzzxx} we thus revisit the spin dynamical structure factors of the spin-$1/2$
$XXX$ chain for the less studied case of a finite magnetic field. 

For a finite magnetic field $H>0$ the Hamiltonian term $2\mu_B\,H\sum_{j=1}^{L} \hat{S}_j^z$ in Eq. (\ref{Hchain})
does not commute with the global spin $SU(2)$ symmetry off-diagonal generators. It thus lowers
the model global symmetry to $U(1)$. However, for all spin density values $m =-2S^{z}/L\in [-1,1]$ and corresponding
fields $h \in [-H_c,H_c]$ the $S$-fixed subspaces dimensions, and thus the number of energy and momentum eigenstates
that span them, exactly equals the number of fixed-$S$ representations of the $SU(2)$ symmetry 
group. Here $\pm H_c=\pm J/\mu_B$ are the critical fields for fully polarized ferromagnetism.
Hence the $H=0$ model non-Abelian global $SU(2)$ symmetry controls and determines
the states spectrum structure for all $m$ and $H$ values.

\subsection{A functional representation of the spin-$1/2$ $XXX$ chain Bethe-ansatz solution}
\label{nbandpseudoHchain}

The model Hamiltonian, Eq. (\ref{Hchain}), is solvable by the BA. The corresponding general
BA equation is of the form \cite{Bethe,Takahashi-71},
\begin{equation}
2\arctan (\Lambda_j) = q_j + {1\over L}\sum_{\alpha\neq j}2\arctan\left({\Lambda_j-\Lambda_{\alpha}\over 2}\right) 
\hspace{0.20cm}{\rm where}\hspace{0.20cm}{\rm mod}\,2\pi \, .
\label{gen-Lambda-BA}
\end{equation}
Here the $\alpha =1,...,(L-2S)/2$ summation is over the subset of occupied $q_{\alpha}$ quantum numbers
out of the full set,
\begin{equation}
q_j = {2\pi\over L}I_j \hspace{0.20cm}{\rm for}\hspace{0.20cm}j=1,...,I_S \hspace{0.20cm}{\rm where}\hspace{0.20cm}I_S = (L+2S)/2 \, .
\label{qjR}
\end{equation}
The different occupancy configurations of the related quantum numbers $I_j$ (defined modulo $L$) 
such that $j = 1,...,I_S$ generate different energy and momentum eigenstates. The latter are successive integers 
or half-odd integers according to the boundary conditions,
\begin{eqnarray}
I_j & = & 0,\pm 1,...,\pm {I_S -1\over 2}\hspace{0.20cm}{\rm for}\hspace{0.20cm}I_S\hspace{0.2cm}{\rm odd} \, ,
\nonumber \\
& = & \pm 1/2,\pm 3/2,...,\pm {I_S -1\over 2}\hspace{0.20cm}{\rm for}\hspace{0.20cm}I_S\hspace{0.2cm}{\rm even} \, .
\label{Ij}
\end{eqnarray}

The BA equation, Eq. (\ref{gen-Lambda-BA}), explicitly refers to the LWSs. However, relying on 
the model spin $SU(2)$ symmetry one can extend its
exact solution to the non-LWSs. That global $SU(2)$ spin symmetry imposes that the energy and momentum 
eigenstates refer to state representations of the group $SU(2)$. Consistently, the LWSs and 
the non-LWSs generated from them used in our analysis are 
energy and momentum eigenstates.
They are as well eigenstates of $(\hat{\vec{S}})^2$ and $\hat{S}^z$ with eigenvalues $S(S+1)$ and $S^z$,
respectively. We thus denote all $2^{L}$ energy and momentum eigenstates by $\vert l_{\rm r},S,S^z\rangle$. Here $l_{\rm r}$ stands for all quantum 
numbers other than $S$ and $S^z$ needed to specify a
state, $\vert l_{\rm r},S,S^z\rangle$. The non-LWSs are generated from the corresponding
$n_s= S + S^z=0$ LWS $\vert l_{\rm r},S,-S\rangle$ as,
\begin{equation}
\vert l_{\rm r},S,S^z\rangle = 
\frac{1}{\sqrt{{\cal{C}}}}({\hat{S}}^{+})^{n_s}\vert l_{\rm r},S,-S\rangle\hspace{0.20cm}{\rm where}\hspace{0.20cm}
{\cal{C}} = (n_s!)\prod_{j=1}^{n_s}(\,2S+1-j\,)\hspace{0.20cm}{\rm for}\hspace{0.20cm}n_s= 1,...,2S \, .
\label{nonLWS}
\end{equation}

The BA wave functions of the LWSs $\vert l_{\rm r},S,-S\rangle$ formally vanish when two rapidities $\Lambda_j$ and $\Lambda_{j'}$ become
equal (Fermi-like statistics.) This property suggests that simply choosing $\alpha =1,...,(L-2S)/2$ distinct quantum numbers
$q_{\alpha}$ among the set of $j = 1,...,I_S$ allowed quantum numbers $q_j$, which gives a dimension
${(L+2S)/2\choose (L-2S)/2}$, would allow the reconstruction of all $2^L$ states that span the model Hilbert space. 

However such an expectation is misleading. Indeed due to the model non-Abelian global spin $SU(2)$ symmetry and 
in contrast to the simpler $U(1)$ symmetry 1D Lieb-Liniger Bose gas, only some of the solutions to the general 
BA equation, Eq. (\ref{gen-Lambda-BA}), are obtained in terms of real rapidities $\Lambda_j$. The model non-Abelian 
symmetry gives rise to new internal degrees of freedom absent from that Bose gas.
Those bring about new BA roots that involve groups of complex rapidities \cite{Bethe,Takahashi-71}.
In the context of our study, the term thermodynamic Bethe ansatz (TBA) refers to the form obtained in
Ref. \cite{Takahashi-71} for the BA equations in the TL. Within the TBA, the needed set of real and complex 
rapidities have the general form,
\begin{equation}
\Lambda_j^{n,l} = \Lambda_j^n + i (n+1-2l) \hspace{0.20cm}{\rm such}\hspace{0.20cm}{\rm that}\hspace{0.20cm} 
\Lambda_j^{n,l} = (\Lambda_j^{n,n+1-l})^*\hspace{0.20cm}{\rm where}\hspace{0.20cm} l = 1,...,n \, ,
\label{Lambda-jnl-ideal}
\end{equation}
$j = 1,...,L_n$ with $n=1,...,\infty$, and the number $L_n\geq N_n$ is defined below. 
As confirmed in the following, the extra solutions and corresponding states associated
with the quantum numbers in the complex rapidities, Eq. (\ref{Lambda-jnl-ideal}), ensure that
in each $S$-fixed subspace the number of such states equals that of state representations of
the model global spin $SU(2)$ symmetry.

For $n=1$ the rapidity, Eq. (\ref{Lambda-jnl-ideal}), is real and otherwise its imaginary part is finite. The rapidities are
roots of Eq. (\ref{gen-Lambda-BA}). In Eq. (\ref{Lambda-jnl-ideal}) they are partitioned in a configuration of strings, where a $n$-string is a group
of $n$ rapidities with the same real part $\Lambda_j^n$.
The number $n$ is often called the string length and the real part of the set 
of $n$ rapidities, $\Lambda_j^n$, is called the string center \cite{Sutherland-04}. 

After some algebra, the use of rapidities of the form given in Eq. (\ref{Lambda-jnl-ideal})
in the general BA equation, Eq. (\ref{gen-Lambda-BA}), leads to a number $n = 1,...,(L-2S)/2$ 
of TBA equations. In general we consider that $n = 1,...,\infty$ in the TL, which is
correct provided that $(1-L_{s})$ is finite. Within the momentum-distribution functional notation 
used in this review, the TBA equations read \cite{CTD-15,CT-17},
\begin{equation}
q_j = k^n_j - {1\over L}\sum_{(n',j')\neq (n,j)}N_{n'} (q_{j'})\,\Theta_{n\,n'} (\Lambda_j^n-\Lambda_{j'}^{n'}) \, .
\label{gen-Lambda}
\end{equation}
Here $\Theta_{n\,n'}(x)$ is an odd function of $x$ given in Eq. (\ref{Theta}) of Appendix \ref{TBAconfig}
where $n, n' = 1,...,\infty$. In that equation and throughout this review $\delta_{n,n'}$ is the usual Kronecker symbol.

The solutions of the TBA equations, Eq. (\ref{gen-Lambda}), define the rapidities real part, $\Lambda_j^n$. 
In these $n = 1,...,\infty$ equations,
\begin{equation}
q_j = {2\pi\over L}I_j^n \in [q_n^-,q_n^+]\hspace{0.20cm}{\rm where}\hspace{0.20cm}j=1,...,L_n 
\hspace{0.20cm}{\rm and}\hspace{0.20cm}q_n^{\pm} = \pm {\pi\over L}(L_n - 1) \, ,
\label{qj}
\end{equation}
are the momentum values of a {\it $n$-band} associated with the set of $N_n$ $n$-strings
with the same $n$ value. The quantum numbers $I_j^n$ 
are successive integers or half-odd integers according to the boundary conditions,
\begin{eqnarray}
I_j^n & = & 0,\pm 1,...,\pm {L_n -1\over 2} \hspace{0.20cm}{\rm for}\hspace{0.20cm} L_n\hspace{0.2cm}{\rm odd} \, ,
\nonumber \\
& = & \pm 1/2,\pm 3/2,...,\pm {L_n -1\over 2}\hspace{0.20cm}{\rm for}\hspace{0.20cm}L_n\hspace{0.2cm}{\rm even} \, .
\label{Ijn}
\end{eqnarray}

The distribution function $N_n (q_j)$ in Eq. (\ref{gen-Lambda}) is such that $N_n (q_j)=1$ and $N_n (q_j)=0$ 
for ``occupied'' and ``unoccupied'' $q_j$ values, respectively. Indeed, the $q_j$ values, Eq. (\ref{qj}), have 
the separation, $q_{j+1}-q_{j}=2\pi/L$, and only occupancies zero and one. For each fixed $n$ there is a number
$L_n$ of $q_j$ values given by \cite{Takahashi-71,CTD-15,CT-17},
\begin{equation}
L_n = N_n + N^h_{n} \hspace{0.20cm}{\rm where}\hspace{0.20cm}
N^h_{n} = 2S+N^{h,0}_{n} \hspace{0.20cm}{\rm and}\hspace{0.20cm}N^{h,0}_{n} = \sum_{n'=n+1}^{\infty}2(n'-n)N_{n'} \, .
\label{Mh}
\end{equation}
For consistency with the notation used for the 1D Lieb-Liniger Bose gas, here we have called $N_n$ where
$n=1,...,\infty$ the quantum numbers named $M_n$ in the TBA studies of Ref. \cite{Takahashi-71}. Otherwise we tend to use
the notations and formalism of that reference. $N_n$ and $N^h_{n}$ are in Eq. (\ref{Mh}) the numbers of 
$q_j$ values that are occupied and unoccupied, respectively, and $N^{h,0}_{n}$ is the latter number in the case of a 
$S=0$ energy and momentum eigenstate. Often an index $\alpha = 1,...,N_n$ is used to label the subset of occupied quantum numbers $I_{\alpha}^n$ 
of an energy and momentum eigenstate \cite{Takahashi-71,CTD-15,CT-17}. Moreover, in Eq. (\ref{gen-Lambda}),
\begin{equation}
k^n_j \equiv k^n (q_j) = 2\arctan \left({\Lambda_j^n\over n}\right) \, .
\label{kn-gen-Lambda}
\end{equation}
(The relation of the $n=1$ rapidity momentum 
$k^1_j  = 2\arctan (\Lambda_j^1)$, Eq. (\ref{kn-gen-Lambda}) for $n=1$, to the rapidity momentum 
$k_j$ of Ref. \cite{Takahashi-71}, such that $\Lambda_j^1 = \cot (k_j/2)$, is $k^1_j= \pi - k_j$.) 

The momentum eigenvalues $P$ and the energy eigenvalues $E$ are functionals
of the $n=1,...,\infty$ distribution functions $N_n (q_j)$ given by,
\begin{equation}
P = \pi + \sum_{n=1}^{\infty}\sum_{j=1}^{L_n}N_{n} (q_{j})\,q_j \hspace{0.20cm}{\rm and}\hspace{0.20cm}
E = - \sum_{n=1}^{\infty}\sum_{j=1}^{L_n}N_{n} (q_{j})\,{J\over n}
\left(1+ \cos k^n_j\right) - 2\mu_B\,H\,S^z \, ,
\label{PHeim}
\end{equation}
respectively.

The form of the momentum eigenvalues confirms that the quantum number variables 
$q_j$ defined in Eq. (\ref{qjR}) such that $q_{j+1}-q_{j}=2\pi/L$
play the role of $n$-band momentum values. There is one such momentum band for
each set of $N_n$ $n$-strings with the same length $n=1,...,\infty$. For  a given energy eigenstate, 
each such a $n$ band has a well-defined set of $N_n$ occupied and $N_n^h$ unoccupied momentum
values $q_j$. The full set of $L_n=N_n+N_n^h$ momentum values $q_j$ is distributed within 
an interval $q_j \in [q_n^{-},q_n^{+}]$. Its limiting values $q_n^{\pm}$ are given in Eq. (\ref{qj}) and Eq. (\ref{mmmm}) 
of Appendix \ref{TBAconfig}. 

In the ensuing section it is confirmed that the set of quantum numbers associated with the TBA equations, Eq. (\ref{gen-Lambda}), 
allows the reconstruction of the $2^L$ energy eigenstates that span the spin-$1/2$ $XXX$ chain full Hilbert space.
Out of such $\sum_{2S=0\,({\rm integers})}^{L}\,{\cal{N}}(S) = 2^{L}$ states, there is
for a given $S$ a number ${\cal{N}}(S) = (2S+1)\,{\cal{N}}_{\rm singlet} (S)$ of states. Those correspond to $(2S+1)$ 
multiplet configurations and a number ${\cal{N}}_{\rm singlet} (S)$ singlet configurations given in Eq. (\ref{NsingletS})
of Appendix \ref{HMSymmetry}.

\subsection{The $n$-pseudoparticles representation of the spin-$1/2$ $XXX$ chain Bethe-ansatz roots
and its relation to the paired physical spins $1/2$}
\label{pseudoRoots}

The pseudoparticle representation introduced in the following is in the case of the $XXX$ chain associated with 
its $L$ physical spins $1/2$ rather than with spinons. The advantage is that such a representation 
is valid for the model full Hilbert and parameter spaces. The energy eigenstates are a superposition
of lattice occupancy configurations in which the $L$ physical spins $1/2$ singly occupy
$L$ lattice sites. $L$ is even and odd when the states spin $S$ is an integer and half-odd integer number, respectively. 
For all states that span a fixed-$S$ subspace, the
corresponding lattice occupancy configurations have then a number $2S$ of sites occupied by a set of $M=2S$ physical spins $1/2$ that
participate in the multiplet configuration. The complementary set of even number $L-2S$ of sites are
singly occupied by $L-2S$ physical spins $1/2$ whose configuration forms a tensor product of singlet states. 

Since all the ${\cal N}(S)$ energy eigenstates with the same $S$ value have the same $\hat{\vec{S}}^2$ eigenvalue, 
the energy and momentum eigenstates are superpositions of such configuration terms.
Each term is characterized by a different partition of $L$ physical spins $1/2$ into $2S$ such 
physical spins that participate in a $2S+1$ multiplet, and a product of singlets involving the remaining
even number $L-2S$ of physical spins $1/2$. The latter are associated with a corresponding
number $(L-2S)/2$ of singlet pairs.

The {\it unpaired spins $1/2$} and {\it paired spins $1/2$} are the members of such two sets of $M \equiv 2S$ and 
$2\Pi\equiv L-2S$ physical spins $1/2$, respectively. For a LWS, all physical unpaired spins $1/2$
have up spin projection. The model TBA solution quantum numbers are directly related to such 
$M \equiv 2S$ up physical spins $1/2$ and different types of singlet configurations involving the 
remaining $2\Pi=L-2S$ paired physical spins $1/2$ and their $\Pi= (L-2S)/2$ singlet pairs.

A $n$-string was defined above as a group of $l = 1,...,n$ rapidities with the same real part $\Lambda_j^n$, Eq. (\ref{Lambda-jnl-ideal}).
As confirmed in the following, the set of $n$-strings of an energy eigenstate is directly related to the $\Pi\equiv (L-2S)/2$ singlet pairs
involving the $2\Pi=L-2S$ physical spins $1/2$ that participate in singlet configurations. Specifically,
each $n$-string refers to a $n$-pairs configuration within which for $n>1$ a number $n$ of singlet pairs are 
bound. Such a binding is associated with the corresponding imaginary parts, $i (n+1-2l)$, of 
the $l = 1,...,n$ rapidities $\Lambda_j^{n,l} = \Lambda_j^n + i (n+1-2l)$, Eq. (\ref{Lambda-jnl-ideal}),
with the same real part $\Lambda_j^n$. The $n>1$ singlet pairs that 
are bound within a $n$-pairs configuration associated with a string of length $n>1$
are called here {\it bound singlet pairs}. For $n=1$ the 
rapidity $\Lambda_j^{1,1}$ imaginary part vanishes because a 
$n=1$ $n$-string refers to a single singlet pair. The $N_1$ {\it unbound singlet pairs}
of an energy eigenstate are those that correspond its $N_1$ $n=1$ pair configurations.

The rapidity $\Lambda_j^{n,l}$ indexes $n$ and $l$ thus
label the $n$-pairs configuration and a specific singlet pair, respectively. Moreover,
the usual string length $n=1,...,\infty$ in Eq. (\ref{Lambda-jnl-ideal}) corresponds to 
the number of singlet pairs in each of the $N_n$ $n$-pairs configurations of a given state.
The $l = 1,...,n$ singlet pairs bound within a $n$-pairs configuration involve a number 
$2n$ of physical spins $1/2$. Those singly occupy $2n$ lattice sites.

Consistently with such a relation between the TBA $n$-strings and the $\Pi= (L-2S)/2$ singlet pairs,
the following exact TBA sum rule holds for all energy eigenstates,
\begin{equation}
\Pi = \sum_{n=1}^{\infty}n\,N_n = {1\over 2}(L-2S) \, ,
\label{Nsingletpairs}
\end{equation}
and thus $\pi = \Pi/L = \sum_{n=1}^{\infty}n\,n_n = {1\over 2}(1-m_{s})$.
Here $N_n$ is the number of $n$-pairs configurations that equals that of 
$q_j$ values that are occupied, $\pi$ is the density of singlet pairs, $m_{s} = 2S/L=M/L\geq m$
that of unpaired spins $1/2$, and $n_n = N_n/L$. The $\Pi= (L-2S)/2$ singlet pairs under
consideration involve the $2\Pi=L-2S$ physical spins $1/2$ that do not participate in multiplet configurations.

The physical spins $1/2$ configurations that generate an energy eigenstate are a superposition
of local lattice occupancy configurations. On the one hand, since each $n$-pairs configuration occupies 
a number $2n$ of lattice sites, the set of $n$-pairs configurations of an 
energy eigenstate occupy a number $2\Pi=\sum_{n=1}^{\infty}2n\,N_{n}$
of lattice sites. On the other hand, each of the $M=2S$ unpaired physical spins $1/2$ singly occupies a 
lattice site. Therefore, for an energy eigenstate of spin $S$, the following number of sites sum rule is
fulfilled,
\begin{equation}
L = M + 2\Pi = 2S + \sum_{n=1}^{\infty}2n\,N_n \, .
\label{Loccup}
\end{equation}

There is a strong requirement for each $n$-string referring to a $n$-pairs configuration that involves a number $2n$
of physical spins $1/2$ and corresponding $l = 1,...,n$ singlet pairs: That in the
dimension of any $S$-fixed subspace ${\cal{N}} (S) = (2S+1)\,{\cal{N}}_{\rm singlet} (S)$,
the number of independent singlet configurations ${\cal{N}}_{\rm singlet} (S)$ be {\it exactly} the same when 
obtained from the counting of two apparently different types of configurations. The first refers to
the counting of the $SU(2)$ group state representations associated with
the physical spins $1/2$ independent configurations with the same spin $S$, Eq. (\ref{NsingletS}) of Appendix \ref{HMSymmetry}.
The second corresponds to the counting of independent $n=1,...,\infty$ bands $\{q_j\}$ occupancy configurations
of the sets of $N_n$ $n$-strings obeying the sum rule $\sum_{n=1}^{\infty}n\,N_n = (L-2S)/2$, Eq. (\ref{Nsingletpairs}).
(The factor $(2S+1)$ in the dimension ${\cal{N}} (S) = (2S+1)\,{\cal{N}}_{\rm singlet} (S)$ refers to
the number of multiplet configurations of the $M = 2S$ unpaired physical spins $1/2$. Those are not part
of $n$-pairs configurations singlet pairs.)

As shown in Appendix A of Ref. \cite{Takahashi-71} for LWSs,
the value of the number $N_n^h = L_n - N_n$ of $n$-band holes that naturally emerges 
from the TBA, Eq. (\ref{Mh}), ensures that for each $S$-fixed subspace the singlet
dimension ${\cal{N}}_{\rm singlet} (S)$ can indeed alternatively be written as given in 
Eqs. (\ref{NsingletS}) and (\ref{Nsinglet-MM}) of Appendix \ref{HMSymmetry}, respectively. 
This also holds for the multiplet towers
of non-LWS generated from $S>0$ LWSs. Indeed, all $2S+1$ states of such a tower have
exactly the same singlet configurations as the corresponding $S>0$ LWS.

The summation $\sum_{\{N_{n}\}}$ in Eq. (\ref{Nsinglet-MM}) of Appendix \ref{HMSymmetry}
runs over all sets of $n$-strings numbers $\{N_{n}\}$ corresponding to the same 
fixed spin $S=L/2 - \sum_{n=1}^{\infty}n\,N_n$, as imposed by the exact sum rule, Eq. (\ref{Nsingletpairs}).
On the one hand, this confirms the connection between $n$-strings and the paired physical spins $1/2$.
On the other hand, it shows that the additional states described by groups of complex rapidities, absent
in the case of the 1D Lieb-Liniger Bose gas, ensure that the fixed-$S$ subspaces dimension, ${\cal{N}} (S) = (2S+1)\,{\cal{N}}_{\rm singlet} (S)$
with ${\cal{N}}_{\rm singlet} (S)$ given in Eq. (\ref{Nsinglet-MM}) of Appendix \ref{HMSymmetry}, exactly equals the corresponding
number of spin-$S$ $SU(2)$ symmetry representations, Eq. (\ref{NsingletS}) of that Appendix.

From analysis of the TBA equations given in the previous section, one finds
that there is a one-to-one correspondence between the
$N_n$ $n$-pairs configurations with the same number $n$ of singlet
pairs of an energy eigenstate and the $N_n$ occupied 
momentum values $q_j$ of the corresponding $n$-band distribution $N_n (q_j)$, respectively. 
This is consistent with the center of mass of the set of the $2n$-sites occupied by 
each $n$-pairs configuration moving with momentum $q_j$ and all its $2n$ sites singly
occupied by paired physical spins moving coherently along with it. This occurs through 
processes within which the $2n$ paired spins on such $2n$ occupied sites interchange position 
with the $M=2S$ unpaired physical spins $1/2$ that singly occupy sites.

We associate one {\it $n$-pseudoparticle} and one {\it $n$-band hole} with each of the $N_n$ occupied 
and $N_n^h$ unoccupied momentum values $q_j$, respectively, of an energy eigenstate
$n$-band. Such pseudoparticles are well defined within the TL to which the TBA applies.
Indeed, the TL ensures that the problems concerning the $n$-pseudoparticle 
internal degrees of freedom and translational degrees of freedom, respectively, separate.

On the one hand, the internal degrees of freedom of a $n$-pseudoparticle refer to
a $n$-pairs configuration. Hence there is one $n$-pseudoparticle for each $n$-pairs configuration
and corresponding BA roots. Those involve a group of $l = 1,...,n$ rapidities with the same real part, Eq. (\ref{Lambda-jnl-ideal}).
If $n>1$ the $n$-pseudoparticle has $n= 2,...,\infty$ singlet pairs bound within it.
If $n=1$, its internal degrees of freedom correspond to a single unbound singlet pair.
That a $n$-string of length $n>1$ describes the binding of $n=2,...,\infty$ singlet
pairs bound within a $n$-pseudoparticle clarifies its connection to the BA roots.

On the other hand, the momentum $q_j$, Eq. (\ref{qj}), of a $n$-pseudoparticle refers to its
translational degrees of freedom. Those are associated with its center of mass motion. The set
of $N=\sum_{n=1}^{\infty}N_n$ $n$-pseudoparticles, each carrying a momentum $q_j$, 
of a given energy eigenstate determine that state momentum eigenvalue, as given in Eq. (\ref{PHeim}). 

The magnons associated with the spin-wave representation of the spin-$1/2$ $XXX$ model
on for example a square lattice carry spin one and are associated with an antiferromagnetic long-range
order. In turn, the $l = 1,...,n$ singlet pairs bound within the $N_n$ $n$-pseudoparticles that populate 
each $n=1,...,\infty$ $n$-band of an energy eigenstate of the present spin chain have spin zero. 
The $n$-pseudoparticles are indeed spin neutral particles. The
energy eigenstates spin $S$ and spin projection $S^z$ are thus determined by the numbers
$M_{\pm 1/2}$ of unpaired spins $1/2$ with spin projection $\pm 1/2$. Specifically,
$S = (M_{+1/2}+M_{-1/2})/2$ and $S^z = - (M_{+1/2}-M_{-1/2})/2$, respectively.
The total number $L_{\pm 1/2}$ of physical spins with projection $\pm 1/2$
such that $L = L_{+1/2} + L_{-1/2}$ then reads $L_{\pm 1/2} = \Pi +  M_{\pm 1/2}$.

There is a number of pseudoparticles sum rule. It is related to that of singlet pairs, Eq. (\ref{Nsingletpairs}).
The latter sum rule implies that $N_1 = L(1-L_{s})/2 -  \sum_{n=2}^{\infty}n\,N_n$. From the use of 
this relation in the number of pseudoparticles expression, $N\equiv\sum_{n=1}^{\infty}N_n$, one confirms
that the sum rule $N = \sum_{n=1}^{\infty}N_n = {1\over 2}(L-N^h_{1})$ is obeyed. Here $N^h_{1}$ is the 
number of $n=1$ band holes, Eq. (\ref{Mh}) for $n=1$. 

The set of $\Pi=\sum_{n=1}^{\infty}n\,N_n=(L-2S)/2$ singlet pairs 
of an energy eigenstate are all bound within the set of $N=\sum_{n=1}^{\infty}N_n$ 
composite $n$-pseudoparticles that populate it. 
The question is thus which is the relation of the model physical spins $1/2$ to the
$N^h_{n} = 2S+N^{h,0}_{n}$ holes in each $n$-band for which $N_n>0$? The
$N^h_{1}$ $n=1$ band holes of the $S=0$ ground-state excited states are usually associated with
spinons \cite{Natan-94,Klauser-11}. Hence this question also refers to the relation of
spinons to the present representation in terms $n$-pseudoparticles and unpaired
spins $1/2$. This is the issue clarified in the ensuing section.

\subsection{Relation to the physical spins $1/2$ of the holes in the TBA quantum numbers distributions}
\label{relation-HS}

The spin-$1/2$ $XXX$ chain in a uniform vector potential $\Phi/L$ whose Hamiltonian is given
in Eq. (A2) of Ref. \cite{CTD-15} remains solvable by the BA. Its LWSs momentum 
eigenvalues, $P = P (\Phi/L)$, have the general form,
\begin{equation}
P (\Phi/L) = P (0) + {L-\sum_{n}2n\,N_n\over L}\Phi  = P (0) + L_{s}\,\Phi = P (0) + 2S\,{\Phi\over L} \, .
\label{Peff}
\end{equation}
The $\Phi=0$ momentum eigenvalue $P (0)$ is given in Eq. (\ref{PHeim}).

On the one hand, the current operator expectation values of the $\Phi\rightarrow 0$ LWSs can be derived from the $\Phi/L$ dependence
of the energy eigenvalues $E(\Phi/L)$ as $\langle \hat{J}^z\rangle = d E(\Phi/L)/d(\Phi/L)\vert_{\Phi=0}$.
On the other hand, $d P(\Phi/L)/d(\Phi/L)\vert_{\Phi=0}$ gives the number of spin carriers that couple to the vector potential. 
The natural candidates are the model $L$ physical spins $1/2$. However, the form of the exact momentum eigenvalues, 
Eq. (\ref{Peff}), reveals though that only the $2S$ unpaired spins $1/2$ contributing to the 
multiplet configurations couple to the vector potential $\Phi/L$. Since the $2\Pi = L-2S$ physical spins $1/2$ left over 
are those within the $\Pi = L/2 - S$ singlet pairs, this result is physically appealing.

The LWSs spin currents result from the above mentioned processes 
under which the $2n$-site configurations of the $n$-pseudoparticles
interchange position under their motion along the lattice with the set of single-site $2S$ unpaired spins. The BA 
separates each $n=1,...,\infty$ branch of $n$-pseudoparticles in a different momentum $n$ band.
The inequality $N_n^{h} \geq 2S$ follows from the BA choosing the number $N_n^{h}$ of $n$ band holes so that the fixed-$S$ subspaces dimension
obeys the sum rule, Eq. (\ref{Nsinglet-MM}) of Appendix \ref{HMSymmetry}. The additional number $\sum_{n'=n+1}^{\infty}2(n'-n)\,N_{n'}$ 
of holes in each $n$ band relative to the number $2S$ that equals that of unpaired spins $1/2$ has an interpretation in terms
of lattice sites occupancies. Indeed, locally it corresponds to $\sum_{n'=n+1}^{\infty}2(n'-n)\,N_{n'}$ sites out of the
$\sum_{n'=n+1}^{\infty}2n'\,N_{n'}$ sites occupied by bound singlet pairs within the state under consideration. 
And this refers only to $n'$-pairs configurations with $n'>n$ such pairs. Only such sites in addition to the $2S$ sites 
occupied by unpaired spins are used by the $n$ pseudoparticles as unoccupied sites. They interchange position 
under their motion along the lattice with the spins $1/2$ on such $\sum_{n'=n+1}^{\infty}2(n'-n)\,N_{n'}$ lattice sites.
They are though occupied by paired spins $1/2$ with an equal number $\sum_{n'=n+1}^{\infty}(n'-n)\,N_{n'}$
of opposite spin projections rather than by unpaired spins $1/2$. Therefore, such processes do not contribute to the spin current. 

Indeed, on average only $2S$ holes out of the $N_n^{h} = 2S + \sum_{n'=n+1}^{\infty}2(n'-n)\,N_{n'}$ holes of each $n$ band for which $N_n>0$
contribute to the spin currents. The translational degrees of freedom of the $2S$ unpaired spins $1/2$ are
described by such $n$-band holes. On average the virtual elementary currents carried by the two sets of $\sum_{n'=n+1}^{\infty}(n'-n)\,N_{n'}$
remaining holes exactly cancel each other. This is consistent with the overall spin current of $S=0$ states 
for which $N_n^{h} = N_n^{h,0} = \sum_{n'=n+1}^{\infty}2(n'-n)\,N_{n'}$ exactly vanishing.

The validity of this picture is confirmed by the form of the spin currents of the non-LWSs.
Consider a general LWS $\vert l_{\rm r},S,-S\rangle$ on the right-hand side of Eq. (\ref{nonLWS})
carrying a current $\langle  l_{\rm r},S,-S\vert\hat{J}^z\vert l_{\rm r},S,-S\rangle$. Here $\hat{J}^z$
is the $z$ component of the spin current operator, Eq. (\ref{c-s-currents}). For simplicity,
we denote that spin current by $\langle\hat{J}^z_{LWS} (l_{\rm r},S)\rangle$. 
In Appendix \ref{XXXconfig3} it is shown that its general TBA expression \cite{CTD-15,CT-17}
can be written in terms of $n$-bands holes elementary currents $j_n^h (q_j)$ as follows,
\begin{equation}
\langle\hat{J}^z_{LWS} (l_{\rm r},S)\rangle 
= \sum_{n=1}^{\infty}\sum_{j=1}^{L_n}\,N_n^h (q_j)\,\,j_n^h (q_j) \, .
\label{J-part}
\end{equation}
Here $N_n^h (q_j)=1 - N_n (q_j)$ and $l_{\rm r}$ labels the 
$\sum_{l_{\rm r}}={\cal{N}}_{\rm singlet} (S) = \sum_{\{N_n\}}\,\prod_{n =1}^{\infty} {L_n\choose N_n}$
independent singlet configurations of the $L-2S$ paired spins $1/2$. Such configurations correspond to a
well-defined set of numbers $\{N_n\}$ of $n$-pairs configurations. Those are associated with the energy and momentum 
eigenstates that span each fixed-$S$ subspace. The $n$-bands holes elementary currents $j_n^h (q_j)$
in Eq. (\ref{J-part}) are determined by the LWS rapidity functions $k^n (q_j)$ obtainable 
from solution of the TBA equations, Eq. (\ref{Theta}) of Appendix \ref{TBAconfig}. They read,
\begin{equation}
j_n^h (q_j) = {2J\sin k^n (q_j)\over 2\pi\sigma^n (k^n (q_j))} 
\hspace{0.20cm}{\rm for}\hspace{0.20cm}q_j \in [q_n^-,q_n^+] \, .
\label{jn-fn}
\end{equation}
The distribution $2\pi\sigma^n (k_j)$ appearing here is within the TL given by
$2\pi\sigma^n (k_j) \equiv 2\pi\sigma^n (k)\vert_{k=k_j}$ where $2\pi\sigma^n (k) = d q^n (k)/d k$.
$q^n (k)$ stands in that derivative for the inverse function of the $n$-band rapidity momentum function $k^n (q)$. 
\begin{figure}
\begin{center}
\centerline{\includegraphics[width=6.50cm]{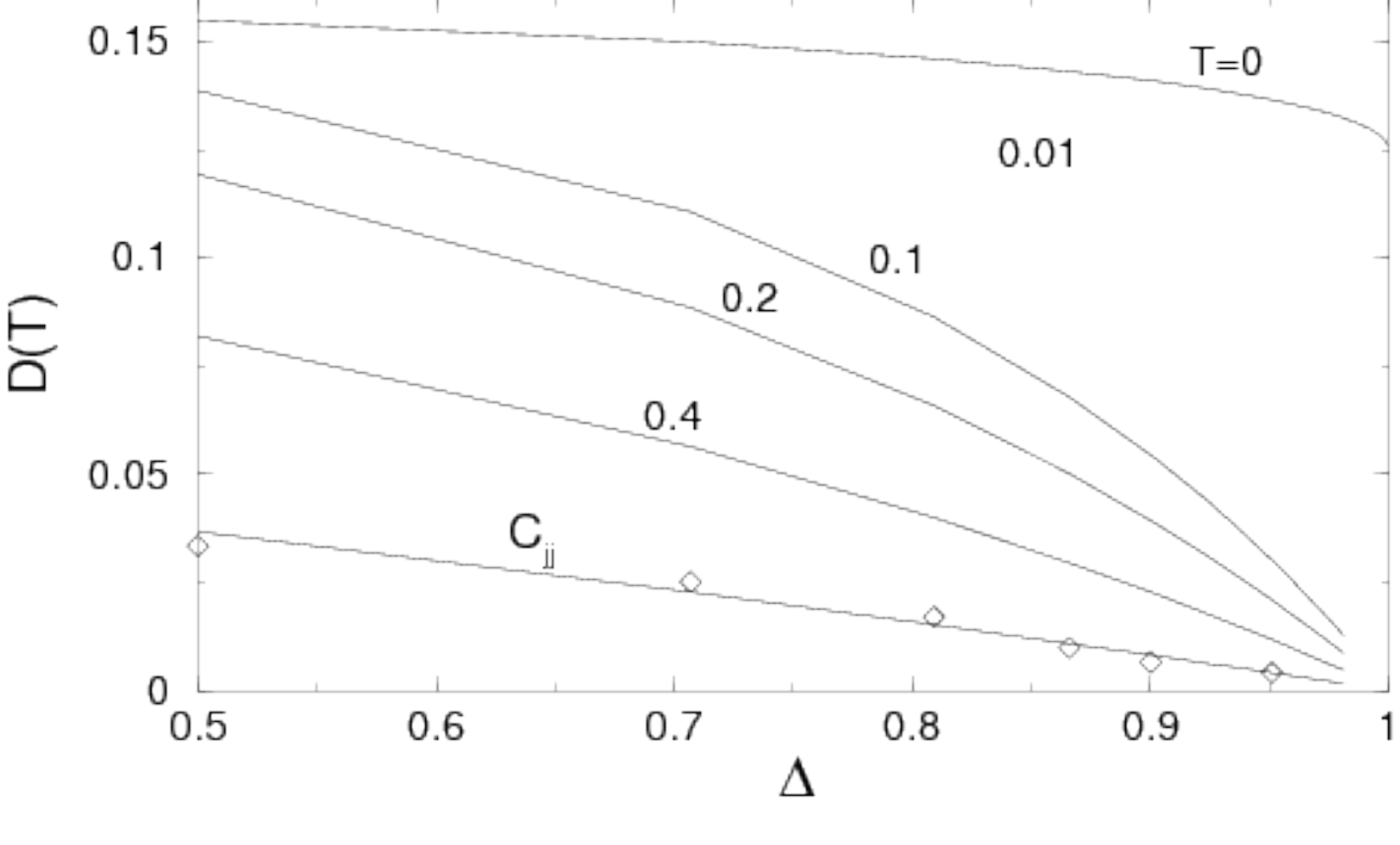}}
\caption{The spin stiffness of the spin-$1/2$ 1D Heisenberg chain, Eq. (\ref{HXXZ}), as a
function of the anisotropy parameter $\Delta $ at various temperatures.
($C_{jj}$ is a high-temperature proportionality constant \cite{Zotos-99}.)\\
{\it Source}: From Ref. \cite{Zotos-99}.}
\label{figure5}
\end{center}
\end{figure}

We consider now that $L$ is even and thus the states spin $S$ is an integer number.
However, within the TL the same results are reached for $L$ odd. 
All $2S$ unpaired spins $1/2$ of that LWS have up-spin projection. 
The following exact relation between
the spin currents of the non-LWSs belonging to the same spin $SU(2)$ tower and the spin
current of the corresponding LWS in terms of the numbers $M_{\pm 1/2}$ of
unpaired spins $1/2$ with spin projection $\pm 1/2$ holds \cite{CTD-15,CT-17}, 
\begin{equation}
\langle\hat{J}^z (l_{\rm r},M_{+1/2},M_{-1/2})\rangle = 
{(M_{+1/2} - M_{-1/2})\over 2S}\,\langle\hat{J}^z_{LWS} (l_{\rm r},S)\rangle \, .
\label{currents-genll}
\end{equation}

For each spin flip generated by application of the off-diagonal spin generator ${\hat{S}}^{+}$ in Eq. (\ref{nonLWS}) 
(and ${\hat{S}}^{-}$) onto a state with finite numbers $M_{+1/2}$ and $M_{-1/2}$, the spin current,
Eq. (\ref{currents-genll}), exactly changes by a LWS current quantum $2j_{-1/2}$ (and $2j_{+1/2}$).
The elementary currents $j_{\pm 1/2}$ in such quanta are given by,
\begin{equation}
j_{\pm 1/2} = \pm {\langle\hat{J}^z_{LWS} (l_{\rm r},S)\rangle\over 2S} = \pm {\sum_{n=1}^{\infty}\sum_{j=1}^{L_n}\,N_n^h (q_j)\,\,j_n^h (q_j)\over 2S} \, .
\label{jelementary}
\end{equation}
Hence each unpaired spin $1/2$ with spin projection $\pm 1/2$ carries such an elementary current $j_{\pm 1/2}$.
For LWSs one has that $M_{+1/2}=2S$ and $M_{-1/2}=0$ whereas $M_{+1/2} + M_{-1/2}=2S$ for their non-LWSs. 
The general expression for the number of $n$-band holes thus reads $N_n^{h} = M_{+1/2} + M_{-1/2} + \sum_{n'=n+1}^{\infty}2(n'-n)\,N_{n'}$.
For each of the $2S+1$ states in the same $SU(2)$ tower, 
an average number $M_{+1/2}$ and $M_{-1/2}$ of holes in the $n$ bands for which $N_n>0$ 
describe the translational degrees of freedom of the $M_{+1/2}$ and $M_{-1/2}$ unpaired spins $1/2$ 
of spin projection $+1/2$ and $-1/2$, respectively. 

The exact spin current expression, Eq. (\ref{currents-genll}), is proportional to $M_{+1/2} - M_{-1/2}$.
The currents of two sets of $M_{+1/2}$ and $M_{-1/2}$ 
unpaired spins of opposite spin projection then partially or totally ($M_{+1/2}=M_{-1/2}$)
cancel each other. Therefore, only an average number $\vert M_{+1/2} -  M_{-1/2}\vert$
of holes in the $n$ bands for which $N_n>0$ contribute to the spin current. The virtual 
elementary currents of a corresponding average number 
$2S_{\eta} - \vert M_{+1/2} -  M_{-1/2}\vert + \sum_{n'=n+1}^{\infty}2(n'-n)\,N_{n'}$ of holes
in these bands exactly cancel each other. For $S_{\eta}^z =0$ non-LWSs for which $M_{+1/2}=M_{-1/2}$ 
this is a total canceling. Such states have zero spin current. 

The spin stiffness $D (T)$ in the real part of the spin conductivity Drude peak,
$2\pi\,D(T)\,\delta(\omega)$, is for temperatures $T\geq 0$ an important physical quantity related to
spin ballistic transport. Indeed, a finite spin stiffness implies the occurrence of such a type of transport.
At finite temperature $T>0$ the spin stiffness can within the TL and for a fixed-$S^z$ canonical ensemble 
be expressed only in terms of the spin currents, Eq. (\ref{currents-genll}), of all 
energy and momentum with that $S^z$ value \cite{CTD-15,CT-17}. It then follows that in the TL and within 
the canonical ensemble the spin stiffness of the spin-$1/2$ $XXX$ chain vanishes
as $m^z=\vert 2S^z\vert\rightarrow 0$ for $T>0$.
At zero temperature the spin stiffness expression has additional contributions from off-diagonal matrix 
elements of the spin current operator. Some of those do not vanish in the $m^z\rightarrow 0$ limit.
The zero-temperature spin stiffness is thus finite in the TL for $m^z\rightarrow 0$, as found below in 
Section \ref{n1pseudoHchain}. (It is given in Eq. (\ref{chiD}).)

There is a direct relation between quantum spin transport and both local and quasi-local conservation laws
\cite{Prosen-11,Prosen-13,Prosen-14}. For the present model, the spin stiffness can be accessed by suitable use of the TBA \cite{Zotos-99}.
It can also be accessed employing a hydrodynamic description. Within it,
the spin stiffness is calculated from the stationary currents generated in an inhomogeneous quench from bipartitioned
initial states \cite{Ilievski-17A}. That in the $m^z\rightarrow 0$ limit the spin stiffness of the spin-$1/2$ $XXX$ chain is
within the TL finite at $T=0$ and vanishes for $T>0$ \cite{CTD-15,CT-17} is illustrated
in Fig. \ref{figure5}. The stiffness curves plotted in that figure were calculated by use of the TBA in Ref. \cite{Zotos-99} for 
the anisotropic spin-$1/2$ 1D Heisenberg chain, Eq. (\ref{HXXZ}).  (The $XXX$ chain refers to the $\Delta =1$ isotropic point in the figure.)

There is a connection between ballistic and diffusive transport in the $m^z\rightarrow 0$ limit at nonzero temperatures 
when the ballistic contribution of the spin-$1/2$ $XXX$ chain vanishes \cite{Medenjak-17}.

\subsection{The spinon representation as a limiting case of the $n$-bands hole representation}
\label{relation-spinons}

Consider LWSs for which the numbers $M=2S$ of physical unpaired spins $1/2$ and $N^h_{n} = 2S+N^{h,0}_{n}$ 
of holes in $n$ bands for which $n>1$ are within the TL finite. The densities of unpaired spins $1/2$ and $n=1$ band
holes of such states thus behave in the TL as $m_{s}=2S/L\rightarrow 0$ and $n_1^h=N_1^h/L\rightarrow 0$, respectively. 
One has then that $q_n^{\pm}\rightarrow 0$ for such $n>1$ bands whose momentum
bandwidth vanishes. For LWSs for which $m_{s}\ll 1$ and $n_1^h \ll 1$ one finds that the $n$-bands hole elementary currents
in Eq. (\ref{jn-fn}) are given by $j_n^h (q_j) = J{(n-1)\over 3n}(2\pi n_1^h)^2\sin \left({q_j\over m_n}\right)$
for $n>1$. Hence $j_n^h (q_j)\rightarrow 0$ for the above class of LWSs. 

An important quantum problem refers to the spin-$1/2$ $XXX$ chain in the reduced subspace 
spanned by the above states for which $m_s\rightarrow 0$ and $n^h_n\rightarrow 0$ for $n>1$ as
$L\rightarrow\infty$. For it the expression of the LWS spin current, Eq. (\ref{J-part}), and general current, Eq. (\ref{currents-genll}),
simplifies to, 
\begin{eqnarray}
\langle\hat{J}^z_{LWS} (l_{\rm r},S)\rangle & = & \sum_{j=1}^{L_1}\,N_1^h (q_j)\,\,j_1^h (q_j) \hspace{0.2cm} {\rm and}
\nonumber \\
\langle\hat{J}^z (l_{\rm r},M_{+1/2},M_{-1/2})\rangle & = &
{(M_{+1/2} - M_{-1/2})\over 2S}\,\sum_{j=1}^{L_1}\,N_1^h (q_j)\,\,j_1^h (q_j) 
= \sum_{\iota=\pm} M_{\iota 1/2}\times j_{\iota 1/2} \, ,
\label{jcurrentsn1}
\end{eqnarray}
respectively. Here,
\begin{equation}
j_{\pm 1/2} = \pm {\sum_{j=1}^{L_1}\,N_1^h (q_j)\,\,j_1^h (q_j)\over 2S} 
\hspace{0.20cm}{\rm and}\hspace{0.20cm}
j_1^h (q_j) = {2J\sin k^1 (q_j)\over 2\pi\sigma^1 (k^1 (q_j))} 
\hspace{0.20cm}{\rm for}\hspace{0.20cm}q_j \in [q_1^-,q_1^+] \, ,
\label{jpmn1}
\end{equation}
are elementary currents, Eq. (\ref{jelementary}), carried by a unpaired physical spin $1/2$ 
of projection $\pm 1/2$ and the $n=1$ band elementary current spectrum, Eq. (\ref{jn-fn})
for $n=1$, respectively.

The spinon representation \cite{Natan-94,Klauser-11} applies to such a quantum problem. For it the expressions in 
Eqs. (\ref{jcurrentsn1}) and (\ref{jpmn1}) are valid. Such expressions also refer to energy and momentum eigenstates 
of arbitrary spin $S$ described only by groups of real rapidities. The spinon representation also applies to them.

Within such a representation, the spinons are the $N_1^h = 2S + N^{h,0}_{1}$ holes in the $n=1$ band \cite{Natan-94,Klauser-11}. It is
assumed that each spinon carries a spin $1/2$. For LWSs described by groups of real and complex rapidities for which 
$N^{h,0}_{1} = \sum_{n=2}^{\infty}2(n-1)\,N_n$ is finite, $N^{h,0}_{1}$ gives the number of such spinons in singlet 
configurations. The elementary currents of such spinons vanish. This result is confirmed by the use of the BA solution 
for small $N^{h,0}_{1}$ values. In contrast, the currents of an average number $2S$ of remaining spinons contribute to the spin current.
The latter $2S$ spinons are intended to be a representation of the model $2S$ unpaired physical spins $1/2$.
In the case of LWSs described only by groups of real rapidities, one has that $N^{h,0}_{1}=0$ and $N_1^h=2S$.
For such states the $N_1^h=2S$ spinons are intended to describe the $2S$ unpaired physical spins $1/2$.

Each of the $N_1^h$ spinons of a LWS has been inherently constructed 
to carry an elementary current $j_1^h (q_j^h)$. This is  consistent with the LWS overall spin 
current reading $\sum_{j=1}^{L_1}\,N_1^h (q_j)\,\,j_1^h (q_j) $, Eq. (\ref{jcurrentsn1}).
The spinon representation has an empirical character. This follows for it not providing
operational relation of the spinons to the model physical spins $1/2$. A relevant question is
thus whether a spinon carrying an elementary current $j_1^h (q_j^h)$, Eq. (\ref{jpmn1}), has 
indeed internal degrees of freedom associated with a spin $1/2$ operator algebra?

This issue remains hidden in the case of LWSs. For them the spin current is the 
sum $\langle\hat{J}^z_{LWS} (l_{\rm r},S)\rangle = \sum_{j=1}^{L_1}\,N_1^h (q_j)\,\,j_1^h (q_j)$,
Eq. (\ref{jcurrentsn1}), of the elementary currents $j_1^h (q_j^h)$ associated with the $N_1^h$ spinons. It can
be clarified though if one considers the tower of non-LWSs corresponding to each LWS. It is useful
to compare the spin currents obtained for the spin LWS and HWS BA solutions. This reveals that
if a spinon with a given momentum $q_j^h$ carried one spin $1/2$, its elementary 
current would read $\pm j_1^h (q_j^h)$ in the case of spin projection $\pm 1/2$.
This would imply that one spin flip resulting from the application of the spin $SU(2)$ off-diagonal
generator ${\hat{S}}^{\pm}$ onto the non-LWS under consideration would lead to state spin current changes
$\mp 2j_1^h (q_j^h)$. 

This is though in contrast to the exact expression of the general spin current
$\langle\hat{J}^z (l_{\rm r},M_{+1/2},M_{-1/2})\rangle$ given in Eq. (\ref{jcurrentsn1}).
That expression reveals that the spin current changes under consideration rather read
$2j_{\mp 1/2} = \mp [\sum_{j=1}^{L_1}\,N_1^h (q_j)\,\,j_1^h (q_j)]/(2S)$ for
the non-LWSs belonging to the spin $SU(2)$ towers of the LWSs considered here.
This shows one cannot associate the internal degrees of freedom of one spin $1/2$ 
with a BA discrete quantum number $n=1$ band hole momentum $q_j^h$. Hence such internal degrees of freedom 
can neither be associated with the corresponding spinon.

Each of the $N_1^h$ spinons with a given momentum $q_j^h$ carrying one spin $1/2$ is appealing 
in the case of LWSs. In that case the spinon elementary currents provide a faithful representation of the state 
overall spin current. However, the $n=1$ band holes associated with the
spinons are mere neutral particles. An average number $2S$ of them merely describes the translational degrees of 
freedom of the $2S$ physical unpaired spins $1/2$. However, they 
lack their spin $1/2$ internal degrees of freedom. Such degrees of freedom are 
rather located on the $2S$ sites occupied by the physical unpaired spins $1/2$ within the local configurations whose
superposition generates a state. 

The spinon representation does not apply for general states with an arbitrary finite density $m_{s}=2S/L=M/L$ 
of unpaired spins $1/2$ described by groups of real and complex rapidities. This is
because the expressions in Eqs. (\ref{jcurrentsn1}) and (\ref{jpmn1}) do not account
for all contributions to the spin current in Eqs. (\ref{J-part})-(\ref{jelementary}) of such general states.
Indeed, such currents are not determined by the ``spinons'' occupancy 
configurations alone. For such states, a number $2S$ of holes in each $n$ band for which $N_n >0$ contributes 
to the spin current. And this applies to all $n>1$ bands with finite $n$-pseudoparticle occupancy. 
Hence the spinon representation is in this general case replaced by the extended $n$-bands hole representation considered
in Section \ref{relation-HS}. For it, the expressions in Eqs. (\ref{jcurrentsn1}) and (\ref{jpmn1}) that only involve
$n=1$ band holes are replaced by those given in Eqs. (\ref{J-part})-(\ref{jelementary}). 

An extreme example refers to the LWSs whose $\Pi=L/2-S$ singlet pairs are all bound 
within a single gigantic $n=\Pi=L/2-S$ pairs-configuration. The spin current 
$\langle\hat{J}^z_{LWS} (l_{\rm r},S)\rangle = - 2J\sin q_j$
of these LWSs stems entirely from the $2S$ holes in the $n=\Pi=L/2-S$ band.
Specifically, it results from their motion upon exchanging position with the single gigantic pseudoparticle
of momentum $q_j$. This spin current has thus no contribution whatsoever from 
the holes in the $n=1$ band. Indeed they do not exist because $N_1=0$ for such states.

\subsection{The $n=1$ band pseudoparticles quantum liquid}
\label{n1pseudoHchain}

For a LWS ground state the $n$-band limiting values $q_n^{\pm}$ in Eq. (\ref{qj}) and Eq. (\ref{mmmm}) of Appendix \ref{TBAconfig}
are given by,
\begin{equation}
q_1^{\pm} = \pm {\pi\over L}\,\left(L_1-1\right) = \pm k_{F\uparrow}\hspace{0.20cm}{\rm and}\hspace{0.20cm}
q_n^{\pm} = \pm {\pi\over L}\,\left(L_n-1\right) = \pm (k_{F\uparrow}-k_{F\downarrow}) 
\hspace{0.20cm}{\rm for}\hspace{0.20cm}n > 1 \, .
\label{mmmm0}
\end{equation}
Here,
\begin{equation}
k_{F\uparrow} = {\pi\over 2L}(L+2S-2) \approx {\pi\over 2}(1+m) \hspace{0.20cm}{\rm and}\hspace{0.20cm}
k_{F\downarrow} = {\pi\over 2L}(L-2S-2)\approx {\pi\over 2}(1-m) \, .
\label{kF}
\end{equation}
Furthermore, for such a ground state the $n$-pseudoparticle momentum distribution functions 
in Eq. (\ref{gen-Lambda}), read,
\begin{equation}
N_1^0 (q_j) = \theta (q_F - \vert q_j\vert) \hspace{0.20cm}{\rm and}\hspace{0.20cm}
N_n^0 (q_j) = 0 \hspace{0.20cm}{\rm for}\hspace{0.20cm}n > 1\, .
\label{N0qHm}
\end{equation}
The $n=1$ band Fermi momentum appearing here is given by,
\begin{equation}
q_F = k_{F\downarrow} = {\pi\over 2L}(L-2S-2)\approx {\pi\over 2}(1-m) \, .
\label{qFHm}
\end{equation}

Ground states are not populated by $n$-pseudoparticles for which $n>1$. In the case of the $n=1$ band, within the TL one
can classify the deviations $\delta N_1 (q_j)$ in Eq. (\ref{DEnHm}) as $\delta N_1^F (q_j)$ and $\delta N_1^{NF} (q_j)$,
respectively. On the one hand, for the deviations $\delta N_1^F (q_j)$ the band momentum $q_j$ is such that 
$\lim_{L\rightarrow\infty}(\vert q_j\vert - k_{F\downarrow})=0$. On the other hand, in the case of
$\delta N_1^{NF} (q_j)$ the momentum difference $\lim_{L\rightarrow\infty}(\vert q_j\vert - k_{F\downarrow})$ remains 
finite in the TL. PSs are in the present model subspaces spanned by a ground state and its excited energy eigenstates
with pseudoparticle overall deviations such that $\sum_{j=1}^{L_1}\vert\delta N_1^{NF} (q_j)\vert/L\rightarrow 0$
and $\sum_{n=2}^{\infty}\sum_{j=1}^{L_n}\vert\delta N_n (q_j)\vert/L\rightarrow 0$ as $L\rightarrow\infty$.

From the use of expansions in the deviations $\delta N_n (q_j) = N_n (q_j) - N_n^0 (q_j)$ in the TBA equations, Eq. (\ref{gen-Lambda}),
and energy eigenvalues, Eq. (\ref{PHeim}), the excitation energy $\delta E = E_{f} - E_{GS}$ 
of PS excited states is up to ${\cal{O}} (1/L)$ order found to be given by,
\begin{equation}
\delta E = \sum_{n=1}^{\infty}\sum_{j=1}^{L_n}\varepsilon_n (q_j)\delta N_n (q_j) 
+ {1\over L}\sum_{n=1}^{\infty}\sum_{n'=1}^{\infty}
\sum_{j=1}^{L_n}\sum_{j'=1}^{L_{n'}}{1\over 2}\,f_{n\,n'} (q_j,q_{j'})\,\delta N_n (q_j)\delta N_{n'} (q_{j'}) \, .
\label{DEnHm}
\end{equation}

This $n$-pseudoparticle energy functional resembles that of the low-energy Fermi liquid. The main difference is that in a Fermi liquid
the quasiparticles undergo zero-momentum forward-scattering interactions only at low energies. In contrast,
due to the present spin chain integrability, the $n$-pseudoparticle undergo 
zero-momentum forward-scattering interactions at all energy scales. This is why the energy functional,
Eq. (\ref{DEnHm}), applies at all energy scales. 

The only restriction to the applicability of the $n$-pseudoparticle energy functional, Eq. (\ref{DEnHm}), is associated
with the PS definition. It is thus such that within the TL the deviations $\delta N_1^{NF} (q_j)$ and $\delta N_n (q_j)$ 
for $n>1$ involve a finite number of $n$-pseudoparticles. This implies that,
\begin{equation}
\lim_{L\rightarrow\infty}{\left(\sum_{j=1}^{L_1}\vert\delta N_1^{NF} (q_j)\vert +
\sum_{n=2}^{\infty}\sum_{j=1}^{L_n}\vert\delta N_n (q_j)\vert\right)\over L}\rightarrow 0 \, .
\label{ConditionDEnHm}
\end{equation}

The $n$-pseudoparticles introduced in Section \ref{pseudoRoots} have internal degrees
of freedom. Those refer to a $n$-pairs configuration within which $n=2,...,\infty$ singlet pairs of $2n$ physical
spins $1/2$ are bound when $n>1$. For $n=1$ they refer to a single singlet pair.
The energy eigenstates $2\Pi=L-2S$ paired spins $1/2$ that participate in
singlet configurations are not free particles. Indeed, they interact through the Hamiltonian first-neighboring
exchange interactions, Eq. (\ref{Hchain}). All their $\Pi=(L-2S)/2$ singlet pairs 
are actually bound within $(n>1)$ or part of $(n=1)$ such states $N=\sum_{n=1}^{\infty}N_n$
composite pseudoparticles. Within the corresponding pseudoparticle representation, this refers to the $XXX$ chain
physical spins $1/2$ rather than to the usual spinons. Within it, the paired physical
spins $1/2$ exchange interactions. Those are described by the pseudoparticles zero-momentum forward-scattering 
interactions associated with the $f$ function terms in the energy functional, Eq. (\ref{DEnHm}).

The $M=2S$ physical unpaired spins $1/2$ left over are those participating in the multiplet configurations.
They are not part of composite $n$-pseudoparticles and have a free nature. 
As was discussed in Section \ref{relation-spinons}, they singly occupy
lattice sites that play the role of empty sites for the $n$-pseudoparticles. Such pseudoparticles move along
the lattice with momentum $q_j$, upon interchanging position with the unpaired spins. As reported
in that section, the spinons often used as elementary excitations of the model 
zero-field ground state, are associated with the TBA $n=1$ band holes. Within the present representation,
an average number $2S$ of such $N_1^h\geq 2S$ holes describe the translational degrees of freedom 
of the $M=2S$ physical spins $1/2$. This justifies the free fermion nature of the corresponding spinons.

The $n$-pseudoparticle dispersion $\varepsilon_n (q_j)$ in Eq. (\ref{DEnHm}) reads,
\begin{equation}
\varepsilon_n (q_j) = \varepsilon_n^0 (q_j) + 2n\mu_B\,h
\hspace{0.20cm}{\rm and}\hspace{0.20cm}
\varepsilon_n^0 (q_j) = -{J\over n}\left(1 + \cos k^n_0 (q_j) - \int_{-\pi}^{\pi} dk \,\sin k\,{\bar{\Phi}}_{1\,n} (k,k^n_0 (q_j))\right) \, .
\label{varepsilon-nHm}
\end{equation}
Here $k^n_0 (q_j)$ denotes the ground-state momentum rapidity $k^n (q_j)$.
The dressed rapidity phase shifts ${\bar{\Phi}}_{n\,n'} (k,k')$ and dressed momentum
phase shifts $\Phi_{n\,n'} (q_j,q_{j'})$ in units of $2\pi$ are defined by the following integral
equations and relation,
\begin{eqnarray}
{\bar{\Phi}}_{n\,n'} (k,k') & = & {1\over 2\pi}\Theta_{n\,n'}\left(n \tan (k/2) -n' \tan (k'/2)\right)
- {1\over 4\pi}\int_{-Q}^Q dk'' {\Theta^{[1]}_{n\,1} \left(n \tan (k/2) - \tan (k''/2)\right)\over\cos^2 (k''/2)}\,
{\bar{\Phi}}_{1\,n'} (k'',k') \, ,
\nonumber \\
\Phi_{n\,n'} (q_j,q_{j'}) & = & {\bar{\Phi}}_{n\,n'} (k^n_0 (q_j),k^{n'}_0 (q_{j'})) \, ,
\label{PsEqHm}
\end{eqnarray}
respectively. Here $Q = \pm k^1_0 (\pm q_F)$ and $\Theta^{[1]}_{n\,n'}(x)$ is the derivative of the 
function $\Theta_{n,n'}(x)$, Eq. (\ref{Theta}) of Appendix \ref{TBAconfig} given
in Eq. (\ref{The1}) of that Appendix.

The $f$ functions in Eq. (\ref{DEnHm}) read,
\begin{equation}
f_{n\,n'} (q_j,q_{j'}) = v_n (q_{j})\,2\pi \,\Phi_{n\,n'} (q_{j},q_{j'}) + v_{n'} (q_{j'})\,2\pi \,\Phi_{n'\,n} (q_{j'},q_{j}) 
+ {v\over 2\pi}\sum_{\iota = \pm}2\pi\Phi_{1\,n} (\iota q_F,q_{j})\,2\pi\Phi_{1\,n'} (\iota q_F,q_{j'}) \, ,
\label{ffnHm}
\end{equation}
where the group velocities are in the TL given by $v_n (q_{j}) = v_n (q)\vert_{q_j = q}$
with $v_n (q) = d\varepsilon_n (q)/d q$. Moreover, $v\equiv v_1 (q_F)$ is the $n=1$ pseudoparticle 
Fermi velocity.

When defined in general PSs, the spin-$1/2$ $XXX$ chain is a quantum liquid of $n=1,...,\infty$ $n$-pseudoparticle branches.
Such pseudoparticles have residual zero-momentum forward-scattering interactions associated with the term of second order
in the deviations in the energy functional, Eq. (\ref{DEnHm}). At $H=0$ the non-Abelian global spin $SU(2)$ symmetry 
renders gapless the excited energy eigenstates with $n>1$ $n$-pseudoparticle occupancy. Those are described by groups of real and 
complex rapidities. Hence the spin dynamical structure factors have contributions from transitions
from the $m=0$ ground state to these excited states. 

Turning on the magnetic field $H$, drives the system into $m\neq 0$ PSs. 
For them there emerges an energy gap $\Delta_s$ between the $m\neq 0$ ground state
and its excited states described by groups of real and complex rapidities.
Its minimum value reads $\Delta_{s}^{\rm min} = \varepsilon_{2} (0)$. 
For $H>0$ the Hamiltonian term $2\mu_B\,H\sum_{j=1}^{L} \hat{S}_j^z$ in Eq. (\ref{Hchain})
does not commute with the global spin $SU(2)$ symmetry off-diagonal generators.
Hence for excitation energy below this energy gap the physics is that of a $U(1)$ symmetry
quantum problem. Its states are described only by groups of real rapidities,
as in the case of the 1D Lieb-Liniger Bose gas. Consequently, for a finite magnetic field, $H>0$, 
the model static and low-temperature properties are determined by excitations associated with 
energy eigenstates with finite $n$-pseudoparticle occupancy $N_1 = (L-2S)/2$ only in the $n=1$ band. 
The same applies to the leading-order contributions to the longitudinal and transverse spin dynamical structure factors 
at finite energy scales below the gap $\Delta_s$.

The physical quantities considered in the following have the same values both at $H=0$
and in the $H\rightarrow 0$ limit. We thus consider finite magnetic fields in the
range $0<H< H_c$ below the critical magnetic field for fully polarized ferromagnetism $H_c=J/\mu_B$. 
Our following analysis refers to the model in $m\neq 0$ PSs. They are spanned by 
energy eigenstates with finite $n$-pseudoparticle occupancy only in the $n=1$ band.
For simplicity, often the $n=1$ pseudoparticles are called {\it pseudoparticles} and their
index $n=1$ is omitted from most quantities. For instance, the $n=1$ band is sometimes in the 
following called pseudoparticle band or simply band. Moreover, $N_1 = \Pi=(L-2S)/2$, 
$N_1^h = M=2S$, and $L_1 = (L+2S)/2$ for the PSs under consideration. 

Within the TL, the set of $j = 1,\cdots,(L+2S)/2$ momentum values $\{q_j\}$ in the pseudoparticle band
may be replaced by a continuum momentum variable, $q \in [-k_{F\uparrow},k_{F\uparrow}]$.
The set of real rapidities $\Lambda_j =\Lambda (q_j)$, Eq. (\ref{Lambda-jnl-ideal}) for $n=1$, are then
replaced by a rapidity function, $\Lambda =\Lambda (q)\in [-\infty,\infty]$,
with $\Lambda (\pm k_{F\uparrow})=\pm\infty$. The $m\in [0,1]$ ground states
pseudoparticle momentum occupancy range is $q \in [-k_{F\downarrow},k_{F\downarrow}]$.
For $m> 0$ such states are thus populated by band holes for $\vert q\vert\in [k_{F\downarrow},k_{F\uparrow}]$. 
(The BA band is full for the $m=0$ absolute ground state.)

To second order in the band momentum distribution deviations
$\delta N (q_j) = N (q_j) - N^0 (q_j)$, the energy spectrum functional 
of the excited states has for the present subspaces the general form
given in Eq. (\ref{DEBg}) with the summations $\sum_{j=1}^{\infty}$ replaced
by $\sum_{j=1}^{(L+2S)/2}$. This is as given in Eq. (\ref{DEBg}) for the 1D Lieb-Liniger 
Bose gas. Furthermore, the energy dispersion in the term of first order in the deviations
is now $\varepsilon (q_j) \equiv \varepsilon_1 (q_j)$, Eq. (\ref{varepsilon-nHm}) for $n=1$. 
The dispersion $\varepsilon^0 (q_j) \equiv \varepsilon^0_1 (q_j) $ controls the spin density curve as follows,
\begin{equation}
H (m) = - {\varepsilon^0 (k_{F\downarrow})\over 2\mu_B}\vert_{m = 1 - 2k_{F\downarrow}} \, .
\label{hm}
\end{equation}
This applies to the spin density interval $m \in ]0,1]$ and thus to the corresponding magnetic-field range $H\in ]0,H_c]$.

The energy dispersions $\varepsilon (q_j)$ and $\varepsilon^0 (q_j)$ have in the $m\rightarrow 0$ 
and $m\rightarrow 1$ limits the following analytical expressions,
\begin{eqnarray}
\varepsilon (q_j) & = & \varepsilon^0 (q_j) = -J{\pi\over 2}\cos (q_j) \hspace{0.20cm}{\rm for}\hspace{0.20cm}m\rightarrow 0 \, ,
\nonumber \\
\varepsilon (q_j) & = & -J[\cos (q_j) - 1]\hspace{0.20cm}{\rm for}\hspace{0.20cm}m\rightarrow 1 \, , 
\nonumber \\
\varepsilon^0 (q_j) & = & -J[\cos (q_j) + 1]\hspace{0.20cm}{\rm for}\hspace{0.20cm}m\rightarrow 1 \, ,
\label{varepsilon-limits}
\end{eqnarray}
respectively. Hence the corresponding group velocity $v (q) = d\varepsilon (q)/dq$ 
and Fermi velocity $v = v (q_F)$ have the limiting behaviors,
\begin{eqnarray}
v (q_j) & = & J{\pi\over 2}\sin (q_j) \hspace{0.20cm}{\rm and}\hspace{0.20cm}v = J{\pi\over 2}\hspace{0.20cm}{\rm for}\hspace{0.20cm}m\rightarrow 0 \, ,
\nonumber \\
v (q_j) & = & J\sin (q_j)\hspace{0.20cm}{\rm and}\hspace{0.20cm}v = 0\hspace{0.20cm}{\rm for}\hspace{0.20cm}m\rightarrow 1 \, .
\label{vq-limits}
\end{eqnarray}

Furthermore, in the $m\rightarrow 0$ limit, the dressed phase shift $2\pi\Phi( q_j,q_{j'}) \equiv 2\pi\Phi_{1\,1} ( q_j,q_{j'})$,
Eq. (\ref{PsEqHm}) for $n=n'=1$, has in units of $2\pi$ at $q_j =\iota\,q_F =\iota\,k_{F\downarrow}$ 
(where $\iota = \pm$) the limiting values,
\begin{equation}
\Phi (\iota\,k_{F\downarrow},q_j) =  {\iota\over 2\sqrt{2}} \hspace{0.20cm}{\rm for}\hspace{0.20cm}q_j\neq \iota\,k_{F\downarrow}
\hspace{0.20cm}{\rm and}\hspace{0.20cm}
\Phi(\iota\,k_{F\downarrow},\iota\,k_{F\downarrow}) = {\iota\over 2\sqrt{2}} (3 - 2\sqrt{2}) \, .
\label{phase-shift-kF}
\end{equation}
In the opposite $m\rightarrow 1$ limit it reads in such units,
\begin{equation}
\Phi (q_j,q_{j'}) = {1\over\pi}\arctan\left({1\over 2}\left[
\tan \left({q_j\over 2}\right)-\tan \left({q_{j'}\over 2}\right)\right]\right) \, .
\label{phi-qqprime}
\end{equation}

The $i=0,1$ ``renormalized'' Fermi velocities $v_{i}$ have the same general expression
as for the 1D Lieb-Liniger Bose gas, Eq. (\ref{viBg}). (Their specific expressions, Eq. (\ref{vxiBg}), apply
though only to that model.) The $i=0,1$ dressed phase shift parameters $\xi^i$ 
in Eq. (\ref{vxiBg}) have also the same general expression, Eq. (\ref{xiBg}).

For spin density $m>0$ the energy gap $\Delta_s$ between the ground state and the lowest-energy state with $n>1$ band
finite $n$-pseudoparticle occupancy is an increasing function of $m$. For low temperatures $T<\Delta_s/k_B$
the entropy has the form given in Eq. (\ref{entropyBg}) with the summation $\sum_{j=1}^{\infty}$ replaced by $\sum_{j=1}^{(L+2S)/2}$.
Furthermore, the thermal momentum distribution function deviation $\delta N (q_j)$ has also
the same general form as for the 1D Lieb-Liniger Bose gas, Eq. (\ref{dNqTBg}). 
The energy dispersion is however model dependent. The use of the same procedures as for
that gas then leads for spin densities not too near $m=1$ to 
the following low-temperature specific temperature leading order term,
\begin{equation}
{c_V\over L} = {k_B\,\pi\over 3\,v}\,(k_B T) \, . 
\label{cV}
\end{equation}
This is the result also reached by conformal field theory \cite{Blote-85,Affleck-85}.

The specific-heat expression, Eq. (\ref{cV}), is valid at very low temperatures $T\ll 2\mu_B(H_c-H)/k_B$
for $(H_c-H)>0$. On the one hand, at $H=0$ the low-temperature thermal excitations that contribute 
to the specific heat expression, Eq. (\ref{cV}), are singlet excited states
with $n$-strings of length $n>1$. For zero spin density they refer to gapless branches. 
On the other hand, for $H>0$ such excited states become gapped. The thermal excitations
that contribute to the low-temperature specific heat are then replaced by singlet excited states. 
They belong to a gapless branch generated by $n=1$ band low-energy and small-momentum particle-hole 
processes around that band Fermi points. Such states have no $n$-strings of length $n>1$.
For $H>0$ the specific heat, Eq. (\ref{cV}), can be expressed in terms of an effective mass $m^* = k_{F\downarrow}/v$
as $c_V/L = [2k_B\,m^*/3(1-m)]\,(k_B T)$. 

The specific-heat expression obtained for $H>0$ leads in the $H\rightarrow 0$ limit to the
correct $H=0$ expression. This is in spite of the $H>0$ and $H=0$ expressions having
contributions from the above two different types of gapless excited-state branches.
In contrast, the specific-heat expression, Eq. (\ref{cV}), is not valid in the $m\rightarrow 1$ limit.
This is because it does not describe properly the crossover to the specific heat exponential regime. 
Such a regime arises due to the gap $2\mu_B (H-H_c)$ in the excitation spectrum for $H>H_c$.
Near $H=H_c$ the minimum gap for excited energy eigenstates with $n$-strings 
of length $n>1$ reads $\Delta_{s}^{\rm min}  = 3J$. Hence for low temperatures, $T<\Delta_{s}^{\rm min}/k_B = 3J/k_B$, the 
processes contributing to the specific heat only involve $n=1$ $n$-pseudoparticles and unpaired spins $1/2$.
At low temperatures the crossover regime involves both the
above $n=1$ band gapless singlet excited states and across-gap excited 
energy eigenstates generated by elementary {\it triplet processes}.

The term triplet has been used here because the elementary processes 
under consideration lead to spin deviations $\delta S=1$ and $\delta S=-1$, 
respectively. Some authors thus associate such elementary processes with creation
and annihilation, respectively, of spin-$1$ magnons. This is mainly an
issue of wording. Nonetheless, as mentioned in Section \ref{HcPhysucs}, within the 
representation of the spin-$1/2$ $XXX$ chain 
in terms of its $L$ physical spins $1/2$ the configurations of the excited 
states under consideration lack the spin-$1$ magnons
as defined for the model on for example the square lattice.  

Specifically, within an elementary $\delta S=1$ process a transition from the 
ground state to an excited state whose minimum gap is $2\mu_B (H_c-H)$ for $H<H_c$ occurs. Within
it, one singlet pair is broken. As a consequence a deviation $\delta\Pi =-1$ occurs.
Here $\Pi$ is the number of singlet pairs, Eq. (\ref{Nsingletpairs}).
That elementary process thus corresponds to the annihilation of one $n=1$ $n$-pseudoparticle.
It is thus also associated with a deviation $\delta N_1 = -1$. The two physical spins $1/2$ that emerge 
from the broken singlet configuration join the excited state multiplet configuration. Indeed,
two initial-state paired physical spins $1/2$ become two final-state unpaired physical spins $1/2$. 
Consistently with the number of unpaired spins $1/2$ reading $M=2S$, the
emergence of the two unpaired spins $1/2$ leads to a deviation $\delta M=2$. It is behind 
the spin deviation $\delta S=1$. There is a related deviation, $\delta N_1^h=2$, in the number 
of holes in the $n=1$ band. Such two holes describe the translational degrees of freedom of the
two unpaired physical spins $1/2$ that emerge from the singlet-pair breaking.

Similarly, within an elementary $\delta S=-1$ process a transition from the ground state 
to an excited state occurs. Its minimum gap reads $2\mu_B (H-H_c)$ for $H>H_c$. Two unpaired physical spins $1/2$ 
are annihilated and one singlet pair is created under it. This gives rise to the deviations $\delta\Pi =1$, $\delta M=-2$,
and $\delta S=-1$. Within this inverse elementary process, one $n=1$ $n$-pseudoparticle is thus created, so that
$\delta N_1 = 1$. There is a related deviation $\delta N_1^h=-2$ in the number of holes in the $n=1$ band. 

Such an analysis confirms that the two opposite elementary triplet processes do not correspond to 
creation and annihilation of one spin-$1$ magnon, respectively, as defined for the spin-$1/2$ $XXX$ model 
on for example the square lattice. Rather, such processes involve the breaking and creation of one singlet pair.
Under it two paired physical spins $1/2$ are transformed into two unpaired physical spins $1/2$ and vice versa, respectively.

The effect of the temperature $T$ smooths the transition to fully polarized ferromagnetism. Therefore, the
spin density $m=1$ is only reached as $H\rightarrow\infty$ instead of for $H\rightarrow H_c=J/\mu_B$ at $T=0$. 
Nevertheless, at low temperatures the critical magnetic field $H_c$ remains a useful reference parameter.
To derive the specific heat in the close neighborhood of $H_c$, one uses a thermal momentum distribution
function deviation of the general form, Eq. (\ref{dNqTBg}). In the present case of the spin-$1/2$ $XXX$ chain, 
the ground-state distribution in it is $N^0 (q) = N_1^0 (q)$, Eq. (\ref{N0qHm}). The $n=1$ band energy dispersion 
used in such a thermal momentum distribution function deviation is that suitable for spin density $m\rightarrow 1$,
which reads $\varepsilon (q) \approx q^2/(2m^*) - 2\mu_B(H_c-H)$. The {\it effective triplet mass} $m^* = 1/J$ in
its expression refers to the $m\rightarrow 1$ limit of the above general effective mass $m^* = k_{F\downarrow}/v$. The corresponding 
thermal excited $n=1$ band momentum distribution function controls the density of singlet pairs 
in the crossover critical regime. It reads,
\begin{equation}
{\Pi\over L} = {1\over 2\pi}\int_{-\pi}^{\pi}dq\, N (q) =
{1\over 2\pi}\int_{-\pi}^{\pi}dq \,{1\over 1 + e^{\varepsilon (q)/k_B T}}
\approx {\sqrt{2m^*\,k_B T}\over\pi}\int_0^{\infty}dx \,
{1\over e^{x^2 - {2\mu_B(H_c-H)\over k_B T}} + 1} \, .
\label{PiT}
\end{equation}
Up to first order in $2\mu_B\vert H_c-H\vert/k_B T\ll 1$ and yet low temperature, one does not need to account for
the temperature dependence of the mass $m^*$ in the above energy dispersion.

One uses the thermal momentum distribution function deviation in the energy functional, Eq. (\ref{DEnHm}).
We then find that before reaching its maximum magnitude at a field slightly larger than $H_c=J/\mu_B$, 
the low-temperature specific heat behaves in a small field window $2\mu_B\vert H-H_c\vert\ll k_B T$ 
around $H_c=J/\mu_B$ as \cite{CPJD-15},
\begin{equation}
{c_V\over L}  = k_B\,c_0\,\sqrt{m^*\,k_B T\over 2}\,\left(c_1 + c_2 {2\mu_B (H_c - H)\over k_B T}\right) 
\hspace{0.20cm}{\rm for}\hspace{0.20cm}2\mu_B\vert H-H_c\vert\ll k_B T \, .
\label{cV-hc}
\end{equation}
Here,
\begin{equation}
c_0 = {(\sqrt{2}-1)\over 4\pi} \, ; \hspace{0.5cm}
c_1 = 3\sqrt{2}\,\Gamma (3/2)\zeta (3/2)  \, ; \hspace{0.5cm}
c_2 = \Gamma (1/2)\zeta (1/2) \, .
\label{c012}
\end{equation} 
$\Gamma (x)$ and $\zeta (x)$ are in this equation the usual gamma and Riemann zeta functions, respectively.
The specific heat expression, Eq. (\ref{cV-hc}), cannot be derived within conformal-field theory. 
For larger fields, $H\gg H_c + k_B T/(2\mu_B)$, the specific heat vanishes exponentially,
$c_V/L\propto e^{-2\mu_B (H - H_c)/(k_B T)}$.
\begin{figure}
\begin{center}
\centerline{\includegraphics[width=5.00cm,angle=-90]{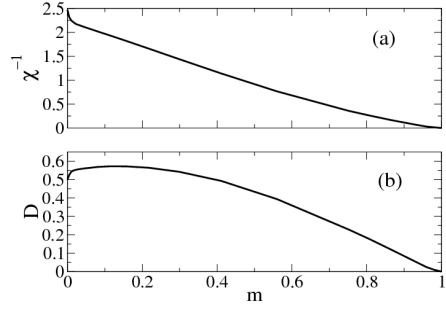}}
\caption{(a) The inverse spin susceptibility $\chi^{-1}$ and 
(b) the spin stiffness $D$ given in Eq. (\ref{chiD}) as functions of
the spin density.\\
{\it Source}: From Ref. \cite{CPJD-15}.}
\label{figure6}
\end{center}
\end{figure}

In the crossover regime defined by Eq. (\ref{cV-hc}), both the gapless and across-gap channels 
associated with singlet excitations and excitations generated by elementary spin-triplet
processes, respectively, are thermally active. That equation is only valid for a very narrow region around $H_c$.
In Appendix \ref{XXXconfig2} it is shown that in the $2\mu_B\vert H-H_c\vert\ll k_B T$ limit in which that
equation is valid it is exactly that obtained by expanding up to first order in
$2\mu_B\vert H-H_c\vert/(k_B T)$ the general scaling function of the specific heat derived in Ref. \cite{He-17}.

Procedures that resemble those of a Fermi liquid allow as well the derivation of static quantities. For instance, in 
Appendix \ref{XXXconfig3} such a type of procedures is used to show
that the spin susceptibility, $\chi = 2\mu_B/(\partial h(m)/\partial m)$, 
is fully controlled by the renormalized velocity $v_{0}$.
Similarly to Eqs. (\ref{viBg}) and (\ref{xiBg}) for the Bose gas,
$v^{i} = v +  {1\over 2\pi}\sum_{\iota = \pm} (\iota)^i\,f (k_{F\downarrow},\iota k_{F\downarrow}) = v\,(\xi^i)^2$ 
and $\xi^i = 1 + \Phi (k_{F\downarrow},k_{F\downarrow}) + (-1)^i\,\Phi (k_{F\downarrow},-k_{F\downarrow})$
where $i = 0,1$, $v \equiv v_1 (k_{F\downarrow})$, $f \equiv f_{1\,1}$, and $\Phi\equiv \Phi_{1\,1}$. 
The zero-temperature spin stiffness $D$ in the real part of the spin conductivity Drude peak,
$2\pi\,D\,\delta(\omega)$, is in that Appendix shown to be controlled by the renormalized velocity $v_{1}$. 
$D>0$ means physically the occurrence of spin ballistic transport.

The spin susceptibility and zero-temperature spin stiffness are in
Appendix \ref{XXXconfig3} found to read,
\begin{equation}
\chi = {4\mu_B^2\over\pi}{1\over v_{0}}\hspace{0.25cm}{\rm and}\hspace{0.25cm}
D = {2v_1\over\pi} \, , 
\label{chiD}
\end{equation}
respectively. 

In the $m\rightarrow 0$ limit the spin stiffness expression in Eq. (\ref{chiD}) 
recovers the stiffness found in Ref. \cite{Shastry-90}.
The inverse spin susceptibility $\chi^{-1}$ in units of $\mu_B=1$ and $J=1$
and spin stiffness $D$ in units of $J=1$ are plotted in Fig. \ref{figure6} (a) and
(b), respectively, as a function of the spin density for $m\in [0,1]$ \cite{CPJD-15}.
 
\subsection{The spin-$1/2$ $XXX$ chain longitudinal and transverse dynamical structure factors in the vicinity of their lower thresholds}
\label{DSGzzxx}

The dynamical structure factors are controlled by matrix elements of spin operators between 
energy eigenstates. At zero magnetic field this includes states described by groups of real and 
complex rapidities. Such states have more than two holes in the $n=1$ band.
Within the conventional spinon representation they are often called multi-spinon states. 
An early construction to study the zero-field dynamical structure factor is based on exact results for the 
simpler XY model, numerical computations on small chains, and known sum rules \cite{Muller-81}. It 
combines rather accurate reproduction of a number of features with its simplicity.
It leads to the exact square root singularity at the lower threshold of the spectrum. Such
a construction is commonly used in the interpretation of experimental data. However some of the expressions obtained
within it are not exact. An example is the functional form at the top of the two $n=1$ band holes 
continuum, usually called two-spinon continuum.

Mapping the infinite chain onto a relativistic quantum field theory is another successful important scheme
\cite{Luther-74,Luther-75,Kadanoff-79}. The connection of the critical exponents of the system with its behavior 
in a finite volume is achieved by finite-size scaling \cite{Blote-85,Affleck-85}. Conformal
field theory \cite{Belavin-84,Francesco-97} and bosonization allow the calculation of asymptotics of correlation
functions \cite{Affleck-89,Gogolin-98}. This includes known normalizations for the first
few leading terms in the operator expansion \cite{Affleck-98,Lukyanov-99,Lukyanov-03}.
Many studies of the dynamical structure factors in the $XXX$ chain refer to finite systems. Those rely on numerical 
diagonalizations \cite{Lefmann-96}, evaluation of matrix elements between BA states \cite{Muller-81,Karbach-00,Karbach-02}, 
and the form-factor method \cite{Biegel-02}. The latter method specifically relies on determinant representations
for matrix elements of local spin operators. They are obtained by solving the quantum inverse problem
\cite{Kitanin-99,Kitanin-00,Caux-05}. The form-factor method  applies to the spin-$1/2$ $XXZ$ chain, Eq. (\ref{HXXZ}). 
It provides the dynamical structure factor over the whole Brillouin zone \cite{Caux-06,Caux-11}. 

The isotropic $\Delta=1$ model, Eq. (\ref{Hchain}), considered here poses the most challenging technical 
problems for theory. This is because at zero magnetic field the contributions from states described by groups 
of real and complex rapidities with more than two holes in the $n=1$ band (multi-spinon states) must be 
accounted for. Indeed, at zero field the energy spectrum of such states is gapless.
As shown in Ref. \cite{Caux-11}, for anisotropy parameter $\Delta \in [0,0.8]$ in Eq. (\ref{HXXZ}) 
and zero magnetic field nearly the whole integrated spectral weight of the dynamical structure factor stems from
excited states with two holes in the $n=1$ band. Those are usually designated by two-spinon 
BA excitations. Such states are technically simpler to handle. However, as $\Delta$ increases from $0.8$ to $1.0$, 
the excited states with four holes in the $n=1$ band (often called four-spinon excitations) contribute increasingly 
as $\Delta\rightarrow 1$. This also increases the complexity of 
the quantum problem under consideration. Specifically, in that limit the contributions to it 
from the $S=1$ and $S^z=0$ excited states with two holes in the $n=1$ band
correspond to a relative integrated intensity of $\approx 0.75$ \cite{Bougourzi-97}.
If in addition one accounts for the contributions from $S=1$ and $S^z=0$ excited states with 
four holes in the $n=1$ band, the total integrated intensity increases to $\approx 0.99$. 
Nonetheless, all the singularities in the dynamical structure factor are determined by contributions 
from excited states with two holes in the $n=1$ band.

The spin dynamical structure factors have been extensively studied in the case of zero magnetic field
\cite{Muller-81,Bougourzi-97,Bougourzi-97A,Bougourzi-96,Caux-06,Imambekov-09,Caux-11}.
For instance, the square root singularity at the lower threshold exponent $-1/2$ can be shown from purely phenomenological 
considerations to be fixed at zero field by the spin $SU(2)$ symmetry invariance alone \cite{Imambekov-09}. 
In the following we revisit the longitudinal and transverse spin dynamical structure factors 
for the less studied case of finite magnetic field. The investigations of Refs. \cite{Caux-05,Rodrigo-08} on the
finite-field dynamical structure factors refer mostly 
to anisotropy parameter $\Delta < 1$ in Eq. (\ref{HXXZ}). The line shape in the vicinity of the dynamical structure 
factor thresholds was predicted in Ref. \cite{Rodrigo-08} to be controlled by momentum dependent exponents. Those
have been explicitly obtained for the spin-$1/2$ $XXX$ chain in Ref. \cite{CPJD-15}.

At finite magnetic field the longitudinal and transverse spin dynamical structure factors are different objects.
Contributions to such dynamical factors from transitions from the ground state to
excited states described by groups of real and complex rapidities are gapped for finite
magnetic field. Such states are populated by $n$-pseudoparticles for which $n>1$.
Their energy gap is for the spin density values considered in the following 
larger than the maximum lower threshold energy. Moreover, except for very small fields $H$, these excitations 
have nearly vanishing spectral weight. For instance, at $m=0.5$ their contributions correspond to a 
relative intensity of about $3\times 10^{-7}$ for anisotropy parameter $\Delta =0.3$ and 
$4\times 10^{-7}$ for $\Delta =0.7$ \cite{Caux-05}. The estimated relative intensity obtained from
the extrapolation of these results to $\Delta =1$ spin-$1/2$ $XXX$ chain is not larger than $10^{-6}$. 

For simplicity, our study focuses mainly on the spin density $m>0.15$ range. For it the contribution 
from excited states populated by $n$-pseudoparticles with $n>1$ is negligible.
Their energy gap is actually larger than the maximum lower threshold energy. Hence in the following we limit our 
analysis to finite-field subspaces spanned by energy eigenstates that are not populated by
$n$-pseudoparticles with $n>1$. The studies of Ref. \cite{Kohno-09} reveal that for
the spin density $m<0.15$ range, not considered here, decreasing the
magnetic field $H$ increases the amount of the spectral weight in the
dynamical structure factor $S^{+-} (k,\omega)$ considered below.
Such a weight stems from transitions to excited states populated by
both $n=1$ and $n=2$ $n$-pseudoparticles.
 
As in the case of the 1D Lieb-Liniger Bose gas, the pseudoparticles
can be transformed into pseudofermions. This is achieved by means of a suitable shift of their discrete momentum
values. The $n$-pseudofermions have exactly the same internal degrees of freedom
as the corresponding $n$-pseudoparticles. Indeed, they differ only in the discrete momentum
values $\bar{q}_j$ and $q_j$, respectively. Those are associated with their center of mass motion.
In the following we use the pseudofermion representation and corresponding PDT \cite{CPJD-15}
to study the line shape of the longitudinal and transverse spin dynamical structure factors 
in the vicinity of their lower thresholds. As in the case of the 1D Bose gas,
such a representation is particularly suitable to the study of high-energy dynamical correlation functions.

In the $m\neq 0$ PSs considered in Section \ref{n1pseudoHchain}, the 1D Lieb-Liniger Bose gas rapidity expression, Eq. (\ref{kjexsBg}), 
is to be replaced by $\Lambda (q_j) = \Lambda ^0 (\bar{q}_j)$. All quantities and corresponding expressions 
given in Eqs. (\ref{QPqBg})- (\ref{2DeltaBg}) for that model remain valid for the spin-$1/2$ $XXX$ chain.
This holds true though provided that in such expressions the dressed phase shift $2\pi\Phi (q_j,q_{j'})$
is now the pseudofermion phase shift $2\pi\Phi(q_j,q_{j'}) \equiv 2\pi\Phi_{1\,1} ( q_j,q_{j'})$,
Eq. (\ref{PsEqHm}) for $n=n'=1$. Moreover, the energy dispersion $\varepsilon ({\bar{q}}_j )$ in Eq. (\ref{DEBcang})
has exactly the same form as $\varepsilon (q_j) \equiv \varepsilon_1 (q_j)$, Eq. (\ref{varepsilon-nHm}) for $n=1$,
but with the momentum $q_j$ replaced by the corresponding canonical momentum, ${\bar{q}}_j= {\bar{q}} (q_j)$.

The longitudinal and transverse spin dynamical structure factors can be written as,
\begin{equation}
S^{aa} (k,\omega) = \sum_{f}\vert\langle f\vert\hat{S}^{a}_k\vert GS\rangle\vert^2
\delta (\omega - \omega^{\tau} (k)) \, .
\label{SDSF}
\end{equation}
Here $a =x,y,z$ and$\omega^{\tau} (k) = E_{f}^{\tau} - E_{GS}$ is the excitation energy. Thus $E_{f}^{\tau}$ refers to
the energies of the excited states that contribute to the longitudinal $\tau = l$ and 
transverse $\tau = t$ dynamical structure factors and $E_{GS}$ is the initial ground state energy.
Moreover, $\hat{S}^{a}_k$ are in Eq. (\ref{SDSF}) the Fourier transforms of the usual local $a =x,y,z$ 
spin operators $\hat{S}^{a}_j$, respectively. 
\begin{figure}
\begin{center}
\subfigure{\includegraphics[width=5.00cm,angle=-90]{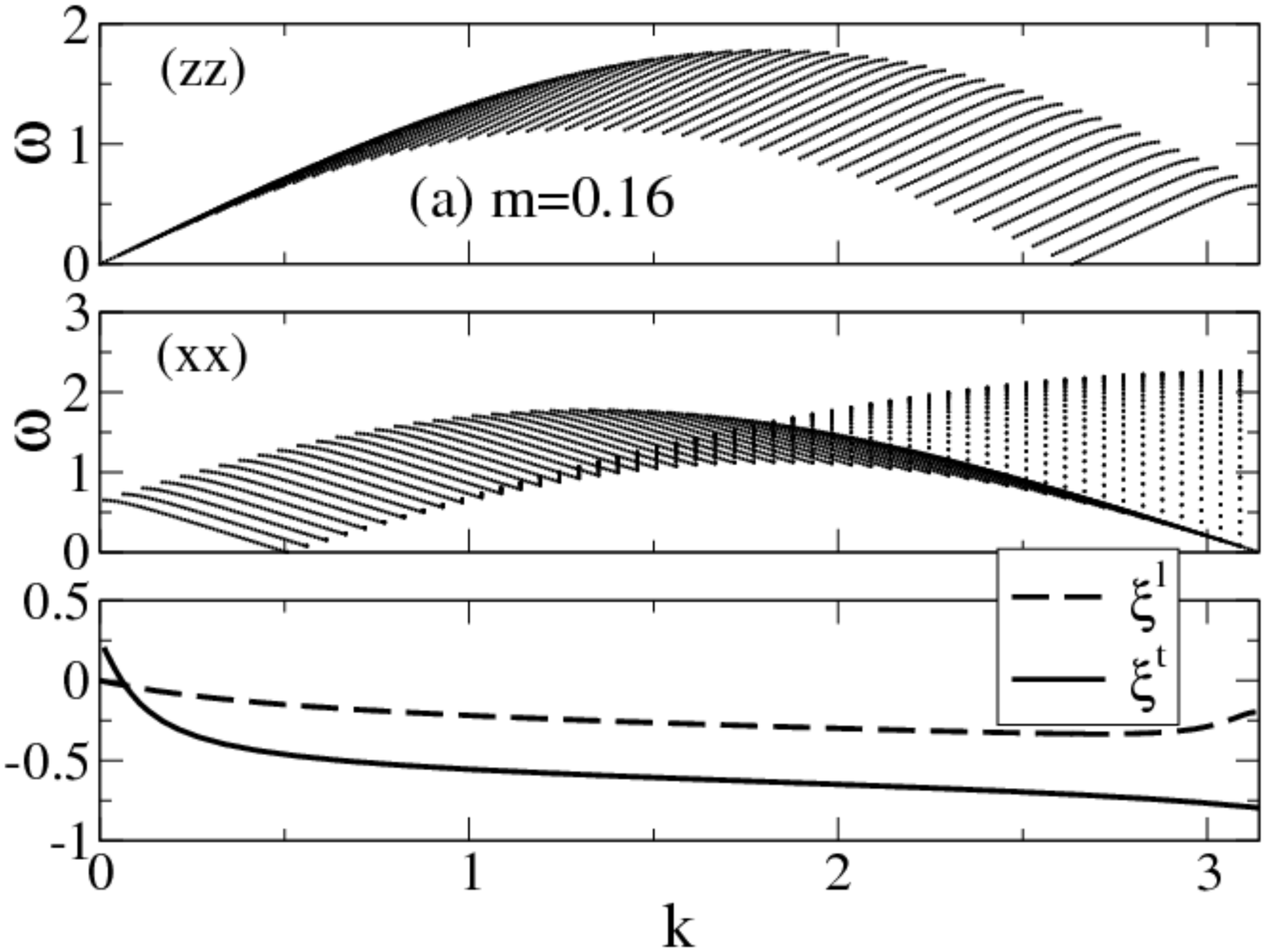}}
\hspace{0.25cm}
\subfigure{\includegraphics[width=5.00cm,angle=-90]{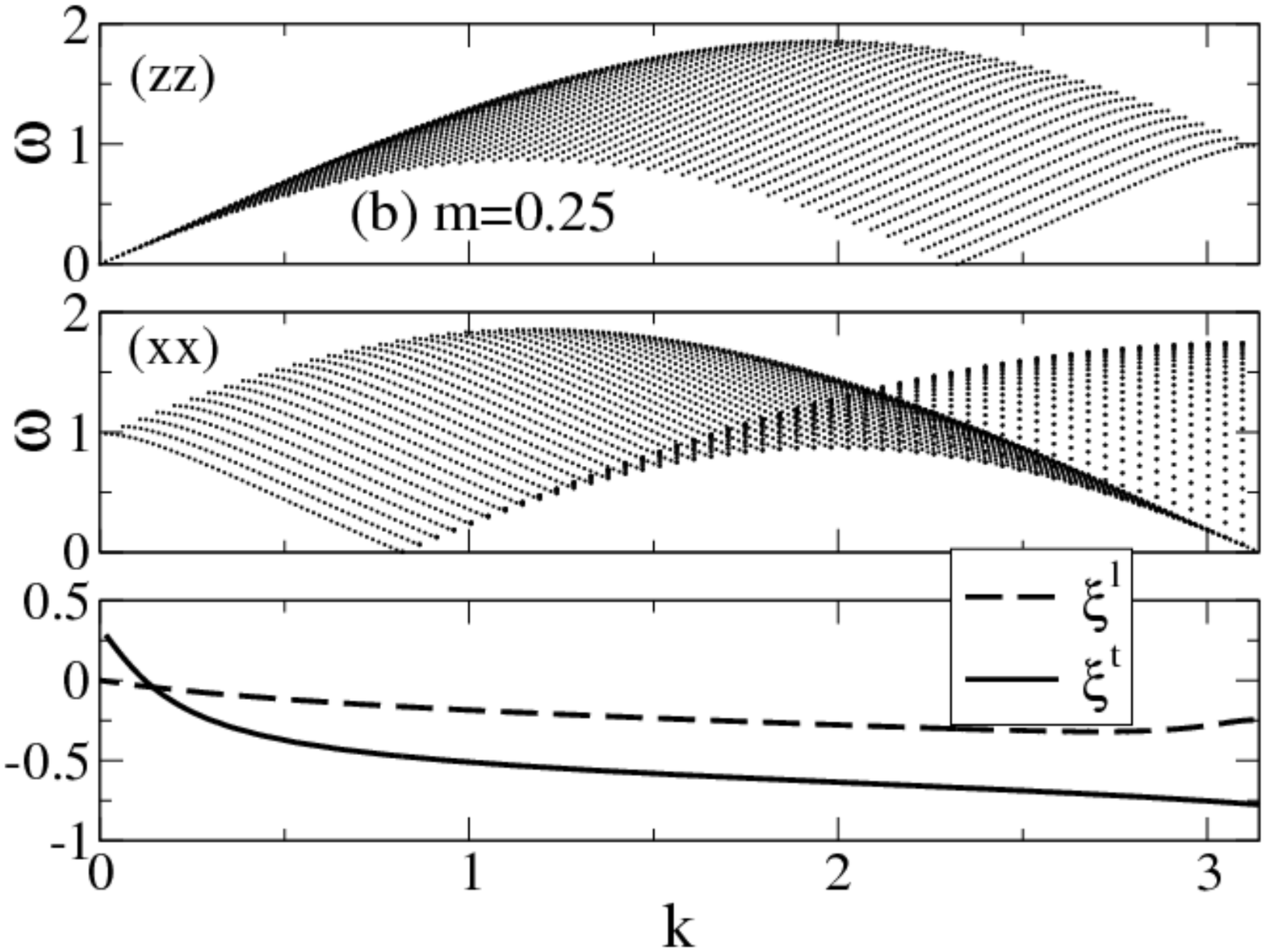}}
\caption{Two upper panels of each sub-figure (a) and (b): 
The spectra $\omega^l (k)$ and $\omega^t (k)$ for spin densities (a) $m=0.16$ and (b) $m=0.25$.
Lower panels: The exponents $\xi^{l} (k)$ and $\xi^{t} (k)$, Eq. (\ref{DSF-BL}), that control the singularities
in the vicinity of the lower thresholds of the spectra also plotted here as a function of $k\in [0,\pi]$.\\
{\it Source}: From Ref. \cite{CPJD-15}.}
\label{figure7}
\end{center}
\end{figure}

In the following we often use the classification of  Ref. \cite{Muller-81}, according to which class (i) and class (ii) 
excitations are (i) non-LWSs and non-HWSs such that $\vert S^z\vert <S$ and (ii) LWSs or/and HWSs 
such that $\vert S^z\vert =S$, respectively. 

In the case of the longitudinal dynamical structure factor $S^{zz} (k,\omega)$, the exact line shape in the vicinity
of its lower thresholds is within the PDT determined by transitions to
excited states that are generated from the $H>0$ ground state by  
high-energy one pseudofermion particle-hole elementary processes. Those
conserve the number $L_{-1/2}$ of spins $1/2$ of projection $-1/2$ \cite{CPJD-15}. 
The corresponding energy spectrum, $\omega^{l} (k) = \omega^{l} (-k)$, is for spin densities in the
interval $m\in ]0,1]$ in which the subinterval $m\in [0.15,1]$ considered here is
contained of the form,
\begin{equation}
\omega^l (k) = - \varepsilon (q_1) + \varepsilon (q_2)  \hspace{0.20cm}{\rm for}\hspace{0.20cm}
k = q_2 - q_1 \in [0,\pi] \, .
\label{dkEdP}
\end{equation}
Here $\varepsilon (q)$ is the energy dispersion, Eq. (\ref{varepsilon-nHm}) for $n=1$,
$q_1 \in [-k_{F\downarrow},k_{F\downarrow}]$, and $q_2 \in [k_{F\downarrow},k_{F\uparrow}]$.
The longitudinal dynamical structure factor line shape is determined within the PDT by a set of elementary pseudofermion particle-hole 
processes of momentum $k=\iota\,(2\pi/L)$ and energy $\omega \approx \iota\,v\,k$
Such processes occur in the vicinity of the two $\iota = \pm$ Fermi points and
dress the high-energy one-pseudofermion particle-hole processes.

For the transverse dynamical structure factor,
\begin{equation}
S^{xx} (k,\omega) = {1\over 4}\left[S^{+-} (k,\omega)+S^{-+} (k,\omega)\right] \, ,
\label{Sxx}
\end{equation}
one must consider the transitions to excited states that determine the line shape in the vicinity
of the lower thresholds of both the dynamical structure factors $S^{+-} (k,\omega)$ and $S^{-+} (k,\omega)$, respectively. Indeed, the
corresponding transverse dynamical structure factor spectrum $\omega^t (k)$,
is here expressed as the superposition of the spectra $\omega^{+-} (k)$ and $\omega^{-+} (k)$.
\begin{figure}
\begin{center}
\subfigure{\includegraphics[width=5.00cm,angle=-90]{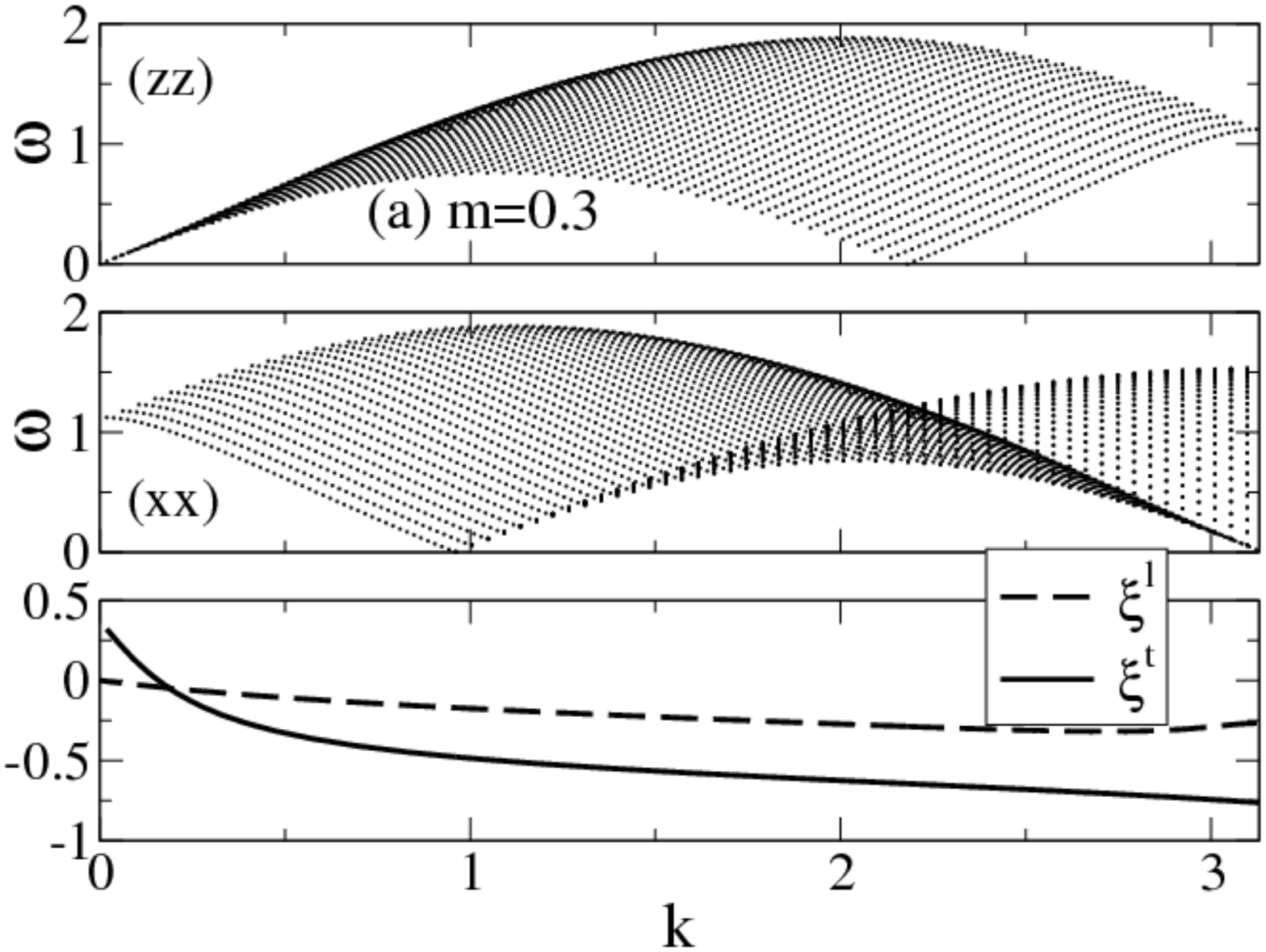}}
\hspace{0.25cm}
\subfigure{\includegraphics[width=5.00cm,angle=-90]{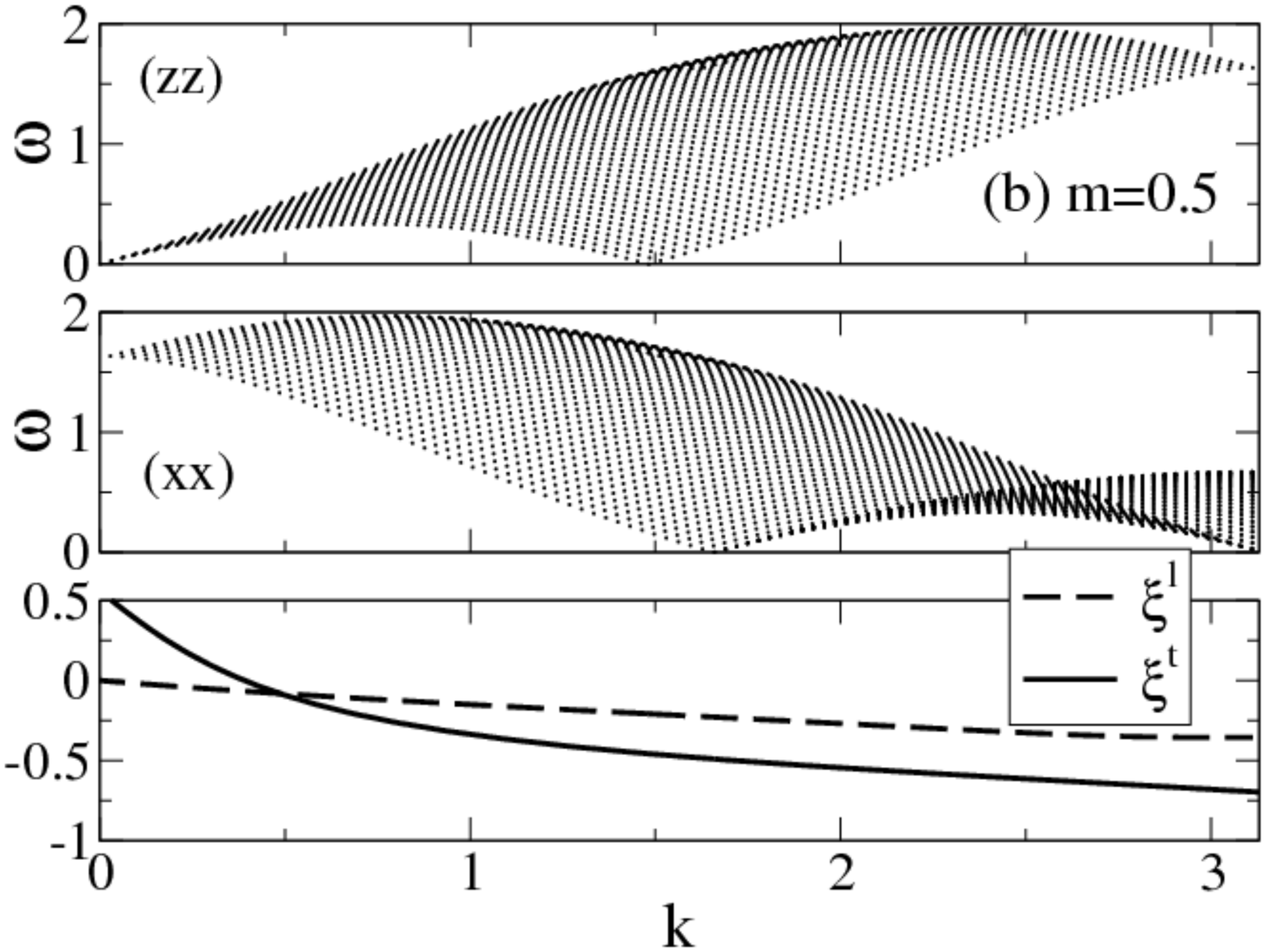}}
\caption{The same quantities as in Fig. \ref{figure7} for spin densities (a) $m=0.30$ and (b) $m=0.50$.\\
{\it Source}: From Ref. \cite{CPJD-15}.}
\label{figure8}
\end{center}
\end{figure}

The spectra $\omega^{\pm\mp} (k)$ refer to excited states that are generated 
from the $H>0$ ground state by high-energy elementary processes. Those
conserve the number $L_{-1/2}$ of spins $1/2$ of projection $-1/2$. Such states are
generated in addition by zero-energy processes that involve a $\delta L_{-1/2}=\pm 1$ deviation.
The latter processes lead as well to a related $\delta N_{\iota}^F=\pm 1$ deviation at the $\iota = \pm$ Fermi points 
and an overall band momentum shift $\delta q_j=\mp \iota\pi/L$ \cite{CPJD-15}. Here $\delta N_{\iota}^F$ is the pseudoparticle number deviation at 
the $\iota = \pm$ Fermi points. The overall zero-energy processes under consideration are a net zero-momentum process.
The high-energy elementary processes associated with the spectra $\omega^{+-} (k)$ and $\omega^{-+} (k)$
are in terms of $n=1$ band occupancies, one two-hole elementary processes and one pseudofermion 
particle-hole elementary processes, respectively.

Hence for spin densities $m\in ]0,1]$ (and thus $m\in [0.15,1]$) such spectra read,
\begin{eqnarray}
\omega^{+-} (k) & = & - \varepsilon (q_1) - \varepsilon (q_2) \hspace{0.20cm}{\rm for}\hspace{0.20cm}
k = \pi  - q_1 - q_2 \in [0,\pi] \, ;
\nonumber \\
\omega^{-+} (k) & = & \varepsilon (q_2) - \varepsilon (q_1) \hspace{0.20cm}{\rm for}\hspace{0.20cm}
k = \pi + q_2 - q_1 \in [0,\pi] \, .
\label{dkEdPxx}
\end{eqnarray}
Here $q_1 \in [-k_{F\downarrow},k_{F\downarrow}]$ for both spectra, $q_2 \in [-k_{F\uparrow},k_{F\downarrow}]$ for 
the $+-$ spectrum, and $q_2 \in [-k_{F\downarrow},-k_{F\downarrow}]$ for the $-+$ spectrum.
The line shape of the transverse dynamical structure factor is within the PDT also determined by 
a set of elementary pseudofermion particle-hole processes in the vicinity of the two $\iota = \pm$ Fermi points.
Such processes dress the above mentioned high- and zero-energy elementary processes.

A particle (and hole) branch line is a spectral feature that within the PDT is generated by high-energy elementary processes.
One pseudofermion (and pseudofermion hole) is under them created outside the Fermi points.
Such processes are dressed by pseudofermion particle-hole processes in the vicinity of such points.
If the transition to the excited states involves creation or annihilation of other pseudofermions, in the
case of a branch line it occurs at the $\iota = \pm$ Fermi points.
For both spin densities $m\rightarrow 0$ and $m>0.15$, 
the lower threshold of $\omega^l (k)$ (and $\omega^t (k)$) coincides with a hole branch line for $k \in [0,2k_{F\downarrow}]$
(and $k \in [\pi - 2k_{F\downarrow},\pi]$) and with a particle branch line for $k \in [2k_{F\downarrow},\pi]$
(and $k \in [0,\pi -2k_{F\downarrow}]$). 

The use of the PDT reveals that the lower threshold singularities of $S^{xx} (k,\omega)$
are those of $S^{-+} (k,\omega)$ near the particle branch line. Near the hole branch line
they are those of $S^{+-} (k,\omega)$. Accounting for $\varepsilon (\pm k_{F\downarrow}) =0$, the longitudinal $S^{zz} (k,\omega)$ 
and transverse $S^{xx} (k,\omega)$ hole branch lines spectra can be expressed as,
\begin{eqnarray}
\omega_h^{\tau} (k) & = & - \varepsilon (q) \hspace{0.20cm}{\rm for}\hspace{0.20cm}\tau = l,t \, ,
\nonumber \\
k & = & k_{F\downarrow} - q \in [0,2k_{F\downarrow}] \hspace{0.20cm}{\rm for}\hspace{0.20cm}\tau = l \, ,
\nonumber \\
k & = & \pi -k_{F\downarrow} - q \in [\pi - 2k_{F\downarrow},\pi]\hspace{0.20cm}{\rm for}\hspace{0.20cm}\tau = t \, .
\label{dEdP-HBL}
\end{eqnarray}
Here $q \in [-k_{F\downarrow},k_{F\downarrow}]$. The corresponding particle branch lines spectra read,
\begin{eqnarray}
\omega_p^{\tau} (k) & = & \varepsilon (q) \hspace{0.20cm}{\rm for}\hspace{0.20cm}\tau = l,t \, ,
\nonumber \\
k & = & k_{F\downarrow} + q \in [2k_{F\downarrow},\pi] \hspace{0.20cm}{\rm for}\hspace{0.20cm}\tau = l \, ,
\nonumber \\
k & = & \pi - k_{F\downarrow} + q \in [0,\pi - 2k_{F\downarrow}]\hspace{0.20cm}{\rm for}\hspace{0.20cm} \tau = t \, ,
\label{dEdP-PBL}
\end{eqnarray}
with $q \in [k_{F\downarrow},k_{F\uparrow}]$ and $q \in [-k_{F\uparrow},-k_{F\downarrow}]$ for the 
$l$ and $t$ particle branch lines, respectively.

In the present case of the longitudinal and transverse
dynamical structure factors, the use of the PDT suitable to the spin-$1/2$ $XXX$ chain 
leads to the following high-energy line shape valid for
small energy deviations $(\omega - \omega^{\tau} (k))>0$ \cite{CPJD-15},
\begin{equation}
S^{aa} (k,\omega) = C^{\tau}\,(\omega - \omega^{\tau} (k))^{\xi_{\tau} (k)}\hspace{0.20cm}{\rm for}\hspace{0.20cm}k \in [0,\pi] 
\hspace{0.20cm}{\rm where}\hspace{0.20cm}
\xi_{\tau} (k) = -1 + \sum_{\iota = \pm}2\Delta^{\iota}_{\tau} (q) \, .
\label{DSF-BL}
\end{equation}
In this equation, $a =z$ for $\tau =l$, $a =x$ for $\tau =t$, and $C^{\tau}$ is a coefficient whose value remains unchanged 
in the range of small energy deviations $(\omega - \omega^{\tau} (k))>0$ for which the present expression is valid.
The $q$ ranges are related to those of the physical momentum $k$ as
$q = k_{F\downarrow} - k$ and $q = -k_{F\downarrow} + k$ for the $S^{zz} (k,\omega)$ hole 
and particle branch lines, respectively. For the hole and particle branch line of $S^{xx} (k,\omega)$
one has that $q = \pi -k_{F\downarrow} - k$ and $q = - \pi +k_{F\downarrow} + k$, respectively.
Moreover, the functionals $2\Delta^{\iota}_{\tau} (q)$ in Eq. (\ref{DSF-BL}) are the square of the pseudofermion $\iota = \pm$ Fermi points 
deviations $(\delta {\bar{q}}_{F}^{\iota}/(2\pi/L))^2$, Eq. (\ref{2DeltaBg}), specific to the
present excitations. Those are given by \cite{CPJD-15},
\begin{equation}
2\Delta^{\iota}_{l} (q)  = \left({(\xi^1)^2 - \iota\,c\over 2\xi^1} + c\,\Phi(\iota k_{F\downarrow},q)\right)^2 
\hspace{0.20cm}{\rm and}\hspace{0.20cm}
2\Delta^{\iota}_{t} (q) = \left({\xi^1\over 2} + c\,\Phi(\iota k_{F\downarrow},q)\right)^2 \, ,
\label{deltas-lt}
\end{equation}
respectively. Here $q \in [-k_{F\downarrow},k_{F\downarrow}]$ for $c=-1$ and $\tau = l,t$, $q \in [k_{F\downarrow},k_{F\uparrow}]$ 
for $c=1$ and $\tau = l$, and $q \in [-k_{F\uparrow},-k_{F\downarrow}]$ for $c=1$ and $\tau = t$.
The band momentum $q$ is related to the physical excitation momentum $k$ as given in Eqs. (\ref{dEdP-HBL}) and (\ref{dEdP-PBL}).

\begin{figure}
\begin{center}
\subfigure{\includegraphics[width=5.00cm,angle=-90]{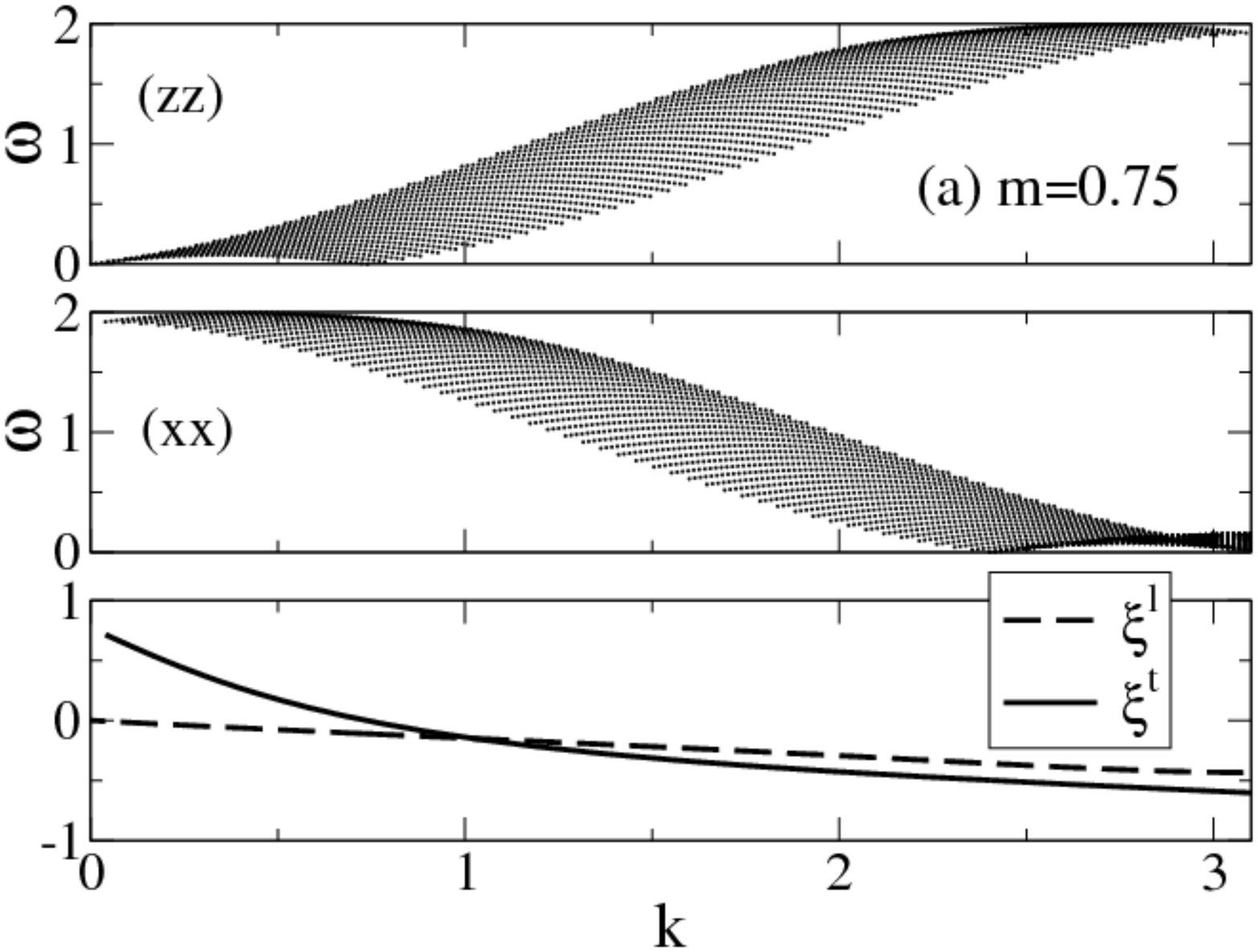}}
\hspace{0.25cm}
\subfigure{\includegraphics[width=5.00cm,angle=-90]{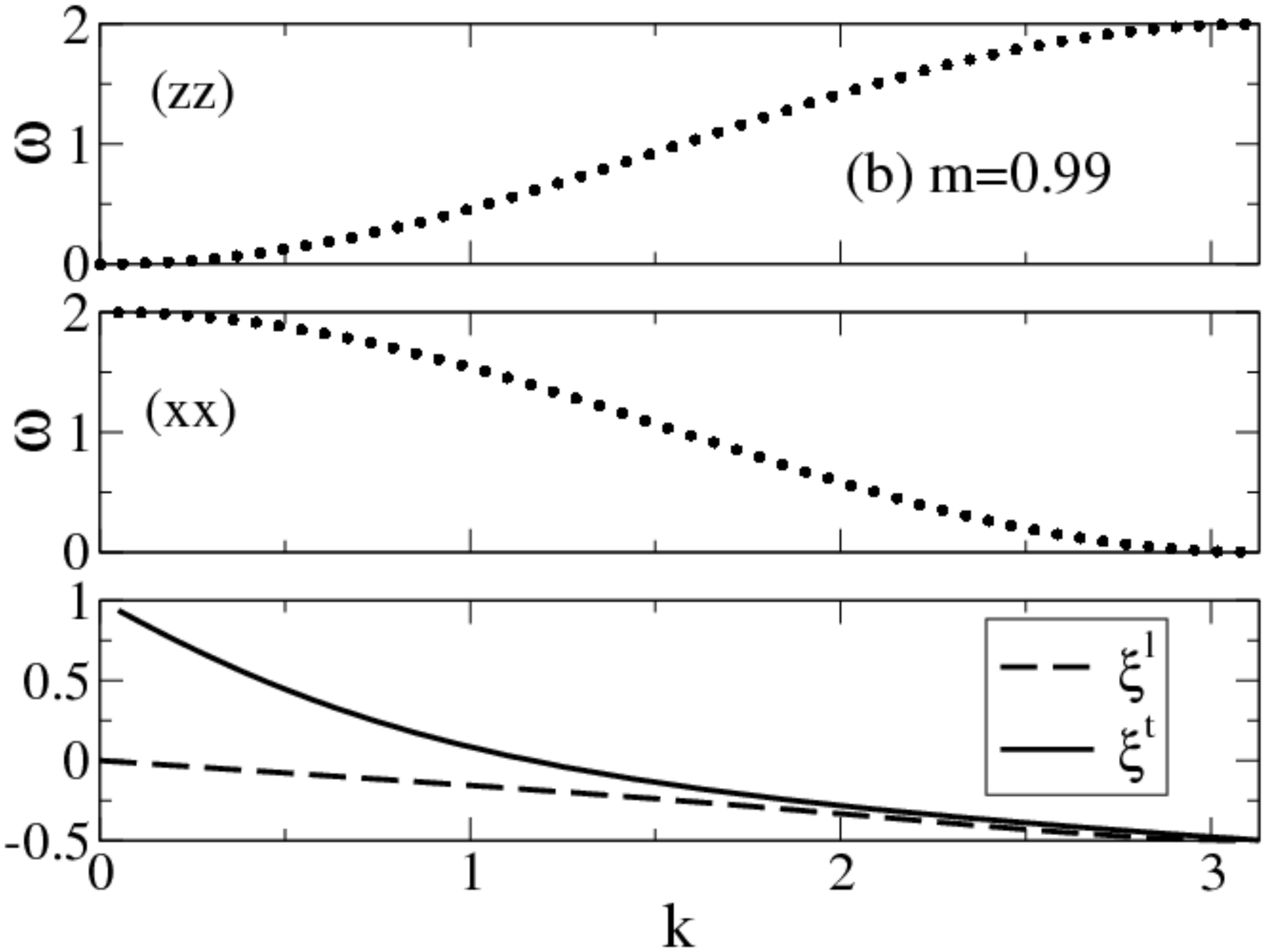}}
\caption{The same quantities as in Figs. \ref{figure7} and \ref{figure8} for spin densities (a) $m=0.75$ and (b) $m=0.99$.
Note that the spin density $m=0.99$ approaches the $m\rightarrow 1$ limit within which the spectra $\omega^l (k)$ and $\omega^t (k)$ 
become a single line.\\
{\it Source}: From Ref. \cite{CPJD-15}.}
\label{figure9}
\end{center}
\end{figure}

The longitudinal spectrum $\omega^l (k)$, Eq. (\ref{dkEdP}), and the transverse spectrum $\omega^t (k)$ that results
from combination of the spectra $\omega^{+-} (k)$ and $\omega^{-+} (k)$, Eq. (\ref{dkEdPxx}),
along with the corresponding exponents $\xi^{l} (k)$ and $\xi^{t} (k)$, respectively, given in Eq. (\ref{DSF-BL}), are
plotted in Fig. \ref{figure7} for spin densities (a) $m=0.16$ and (b) $m=0.25$. 
In Fig. \ref{figure8} they are plotted for (a) $m=0.30$ and (b) $m=0.50$ and in Fig. \ref{figure9} for (a) $m=0.75$ and (b) $m=0.99$.
On the one hand, the exponent $\xi^{l} (k)$ is negative for $k>0$ at any $m$ value. On the other hand, 
the exponent $\xi^{t} (k)$ is negative for a $m$-dependent range $k\in [k_t,\pi]$. Here $k_t$ increases from $k_t=0$ for $m\rightarrow 0$ to,
\begin{equation}
k_t = -2\arctan\left({1\over 2}\tan\left({\pi\over\sqrt{2}}\right)\right) \approx 0.37\,\pi \, ,
\label{kt}
\end{equation}
for $m\rightarrow 1$. The latter limit refers to Fig. \ref{figure9} (b).

In the $m\rightarrow 0$ limit, the spectra $\omega^l (k)$, Eq. (\ref{dkEdP}), and $\omega^{+-} (k)$, 
Eq. (\ref{dkEdPxx}), plotted in Figs. \ref{figure7} - \ref{figure9} reduce to their lower thresholds. At finite $m$, the thresholds 
of these two spectra correspond to different $(k,\omega)$-plane lines. As $m\rightarrow 0$
they become the same $(k,\omega)$-plane line. For finite $m$ values the lower threshold
of the spectrum $\omega^{+-} (k)$, Eq. (\ref{dkEdPxx}), coincides with that of $\omega^{-+} (k)$ for
$k\in [\pi-2k_{F\downarrow},\pi]$. For $k\in [0,\pi-2k_{F\downarrow}]$ it
does not exist. In the $m\rightarrow 0$ limit the lower threshold
of the spectrum $\omega^{+-} (k)$ extends to the whole $k\in [0,\pi]$ range. In that limit it coincides with those of
$\omega^l (k)$ and $\omega^{-+} (k)$. However, in contrast to the latter spectra, $\omega^{+-} (k)$
does not reduce in that limit to its lower threshold. The spectrum of the class (ii) two-hole excitations described 
by groups of real and complex rapidities is gapped for $m>0$. However, in the $m\rightarrow 0$ limit it becomes gapless and 
degenerate with that of $\omega^{-+} (k)$.

For the ranges of the momentum $k$ for which the exponent $\xi_{\tau} (k)$ is negative, there are
lower threshold singularity cusps in $S^{aa} (k,\omega)$, Eq. (\ref{DSF-BL}). Those are detectable as large 
intensity peaks in experiments. Hence analysis of Figs. \ref{figure7}-\ref{figure9} provides valuable information on
the $k$ ranges for which there are singularities in the lower thresholds of 
the dynamical structure factors $S^{zz} (k,\omega)$ and $S^{xx} (k,\omega)=S^{yy} (k,\omega)$.

In the $m\rightarrow 0$ limit, both the $\tau = l,t$ lower thresholds $\omega_{L}^{\tau} (k)$ coincide with the hole branch line
for all $k$ values. In that limit the corresponding exponents $\xi_{\tau} (k)$ are given by $\xi_{\tau} (k)=-1/2$ for 
all $k$ values, {\it i. e.},
\begin{equation}
S^{zz} (k,\omega) = S^{xx} (k,\omega) = C^0\,(\omega - \omega^0 (k))^{-1/2} \, ,
\label{DSF-BM0}
\end{equation}
for $m\rightarrow 0$ and $k \in [0,\pi]$. Here $C^0 =  C^l = C^t$ and the lower thresholds 
$\omega^{l} (k)=\omega^{t} (k)=\omega^0 (k)$ coincide with that of the $m=0$ two-hole spectrum.
Consistently, $\xi_{\tau} (k)=-1/2$ is also the value of the known
exponent that controls the line shape in the vicinity of the lower threshold of the
latter spectrum \cite{Bougourzi-97,Bougourzi-97A,Bougourzi-96,Caux-06,Imambekov-09,Caux-11}.

In the opposite limit, $m\rightarrow 1$, the lower thresholds $\omega^{\tau} (k)$ coincide with the particle branch line 
for all $k$ values. The corresponding $\tau = l,t$ exponents are given by,
\begin{equation}
\xi^l (k) = 2\Phi (0,k)[1+\Phi (0,k)] \hspace{0.20cm}{\rm and}\hspace{0.20cm}\xi^t (k) = -1/2 + 2\Phi (0,k-\pi)[1+\Phi (0,k-\pi)] \, .
\label{xilt}
\end{equation}
Here the phase shift in units of $2\pi$ is a particular case of that given in Eq. (\ref{phi-qqprime}). It reads, 
\begin{equation}
\Phi (0,q) = - {1\over\pi}\arctan\left({1\over 2}\tan \left({q\over 2}\right)\right) \, .
\label{phi-0q}
\end{equation}
In this limit, $\xi^l (k) = 0$ and $\xi^t (k) = 1$ for $k\rightarrow 0$. The values decrease to $\xi^l (k) = - 1/2$ and $\xi^t (k) = - 1/2$ 
for $k\rightarrow \pi$. The corresponding $m\rightarrow 1$ behaviors refer to a small but finite $L_{-1/2}/L$ ratio. 
Using the expression for the exponent $\xi^t (k)$ in Eq. (\ref{xilt}) leads straightforwardly to the value of the momentum $k_t$
shown in Eq. (\ref{kt}), which is reached in the $m\rightarrow 1$ limit.

Note though that $S^{zz} (k,\omega)\rightarrow 0$ as $H\rightarrow H_c$ in the TL \cite{Muller-81,CPJD-15}.
The two-component $S^{xx} (k,\omega)$ and $S^{zz} (k,\omega)$
dynamical structure factor is then dominated by $S^{xx} (k,\omega)$. At $H=H_c$, the expression
given in Eq. (\ref{DSF-BL}) is replaced by a $\delta$-function like distribution,
\begin{equation}
S^{xx} (k,\omega) = {\pi\over 2} \delta \left(\omega - J (1+\cos k)\right) 
\hspace{0.20cm}{\rm for}\hspace{0.20cm}k \in [0,\pi] \, ,
\label{Sxxm1}
\end{equation}
for $\alpha\alpha = xx$ and by $S^{zz} (k,\omega)=0$ for $\alpha\alpha = zz$.

The exponent $\xi_{\tau} (k)$ given in Eq. (\ref{DSF-BL}) does not apply near the $\omega=0$ lower threshold 
$(k,\omega)$-plane soft modes. Examples are $(k_0^{\tau},0)$ where $k_0^{l}=2k_{F\downarrow}$ and $k_0^{t}=\pi -2k_{F\downarrow}$.  
In this case the PDT reaches the same results as conformal-field theory \cite{Blote-85,Affleck-85}.
Indeed, near them the two $\iota = \pm$ functionals, Eq. (\ref{functional}), become the conformal dimensions of
the $\iota = \pm$ fields \cite{LE} given by $2\Delta^{\iota}_{l} = (\xi^1)^2$ and
$2\Delta^{\iota}_{t} = (\pm \iota/(2\xi^1) - \xi^1)^2$.
The PDT provides the corresponding low-energy $S^{aa} (k,\omega)$ behavior \cite{LE,CPJD-15}.
It is the same as that obtainable from conformal-field theory \cite{Blote-85,Affleck-85,Woy-89,Frahm-90,Frahm-91}.  

An experimental possibility is the potential observation of the theoretically predicted dynamical structure factors peaks 
in inelastic neutron scattering experiments on actual spin-chain compounds. The dynamical structure factors 
$S^{zz} (k,\omega)$ and $S^{xx} (k,\omega)$ may be investigated separately in $H>0$ experiments on spin-chain 
compounds by using a carefully oriented crystal. If the crystal is missoriented, or if a micro crystalline sample
is used, the $S^{zz} (k,\omega)$ and $S^{xx} (k,\omega)$ spectral features should appear superimposed.
Such superimposition changes the excitations lower thresholds. It leads in addition to the broadening of the singularities,
Eq. (\ref{DSF-BL}). However, this does not occur at $H=0$, since $S^{zz} (k,\omega)=S^{xx} (k,\omega)$.

These two different situations are clearly seen in the magnetic scattering intensity measured 
at zero- and finite-field inelastic neutron scattering experiments of Ref. \cite{Stone-03}, respectively,  
on Cu(C$_4$H$_4$N$_2$)(NO$_3$)$_2$. On the one hand, in Figs. 2 (a)-(c) of that reference the theoretically predicted 
sharp cusps at zero-field, Eq. (\ref{DSF-BM0}), are clearly seen
at different $k$ values. On the other hand, the $S^{zz} (k,\omega)\neq S^{xx} (k,\omega)$ spectral features appear 
superimposed in the finite-field Figs. 2 (d)-(f) of that reference. Therefore, only at $k\approx \pi$ is the theoretically predicted 
sharp cusp clearly visible. 

More demanding $H> 0$ experiments with a carefully oriented crystal to be carried out on Cu(C$_4$H$_4$N$_2$)(NO$_3$)$_2$ 
and other spin-chain compounds should yielding separately $S^{zz} (k,\omega)$ and $S^{xx} (k,\omega)$. The corresponding 
magnetic scattering intensities are expected to display the cusp singularities found theoretically reported here.

%%%%%%%%%%%%%%%%%%%%%%%%%%%%%%%%%%%%%%%%%%%%%%%%%%%%%%%%%%%%%%%%
\subsection{Outlook of the relation between the model physical spins-$1/2$ and the $n$ pseudoparticles/$n$ pseudofermions
and corresponding $n$-band holes}
\label{outlookHchain}

The spin-$1/2$ $XXX$ chain non-Abelian global spin $SU(2)$ symmetry has direct effects on its degrees of freedom. 
It gives rise to a type of solutions of the BA equation, Eq. (\ref{gen-Lambda}), in terms of groups of
both real and complex rapidities. The latter do not exist for the simpler $U(1)$ symmetry 1D Lieb-Liniger Bose gas.

For each energy and momentum eigenstate, out of the model physical $L$ spins $1/2$, there are $M = 2S$ unpaired spins
that contribute to its multiplet configuration. There are in addition $2\Pi= L-2S$ paired spins that are bound within
$\Pi\equiv (L-2S)/2$ singlet pairs. The degrees of freedom of such singlet pairs
are distributed over a set $\{N_n\}$ of $n$-pairs configurations. Here $N_n$ denotes the number of $n$-pairs configurations
within which $n=1,...,\infty$ singlet pairs are bound. The $n=1$ pair configurations contain
a single singlet pair that remains unbound.

As in the case of the 1D Lieb-Liniger Bose gas, the problem is non-perturbative.
Indeed, a spin-flip decays into a collective excitation that involves all 
pseudoparticles/pseudofermions. This follows from the boundary conditions, Eqs. (\ref{Ij}) and (\ref{Ijn}).
In contrast to the pseudoparticles of that Bose gas, for the spin-chain all $n$-pseudoparticles
have internal degrees of freedom. Those are associated with such composite pseudoparticles 
$n$-pairs configurations and the corresponding $n=1,...,\infty$ singlet pairs that for $n>1$ are
bound within them. The $n$-pseudofermions have exactly the same internal degrees of freedom
as the $n$-pseudoparticles. They differ from them in the discrete momentum
values $\bar{q}_j$ and $q_j$, respectively. Those are associated with the translational degrees
of freedom center of mass motion.

On the one hand, the $\Pi= (L-2S)/2$ singlet pairs that contain the $2\Pi= L-2S$ physical paired 
spins $1/2$ are bound within the $N = \sum_n N_n$ pseudoparticles/pseudofermions. This is consistent
with the sum rule $\Pi=\sum_n 2n\,N_n$. On the other hand, the translational degrees of freedom of the $M = 2S$ physical unpaired spins are
on average described by $2S$ holes out of the $N_n^h = 2S + N_n^{h,0}$ holes in each $n$-band for which $N_n>0$.
The additional $N_n^{h,0}$ holes occurring in states described by groups of real and complex rapidities
have a specific goal. It is to ensure that the number of TBA energy and momentum eigenstates that span each fixed-$S$ subspace, 
Eq. (\ref{Nsinglet-MM}) of Appendix \ref{HMSymmetry},
exactly equals the corresponding number of spin $SU(2)$ state representations, Eq. (\ref{NsingletS}) of that Appendix. 
The relation of the usual spinon representation for the model in well-defined subspaces to the more general $n$-bands hole representation 
for the spin-$1/2$ $XXX$ chain in its full Hilbert space has been discussed and clarified. 

Within the $n$-pseudoparticle/$n$-pseudofermion representations the spin-chain physics simplifies. 
Within them both the static and dynamical properties are controlled by pseudofermion scattering and corresponding phase shifts. 
A further simplification occurs for the model in $m\neq 0$ PSs for which there is an energy gap $\Delta_s$ between 
the $m\neq 0$ ground state and its excited states described by groups of real and complex rapidities. 
For excitation energy below this gap, the physics is that of a $U(1)$ symmetry quantum problem. Its 
states are described only by groups of real rapidities, as in the case of the 1D Lieb-Liniger Bose gas. Specifically, it is a
quantum liquid of $n=1$ pair composite pseudoparticles, which play the same role as the pseudoparticles
of that simpler gas.

%%%%%%%%%%%%%%%%%%%%%%%%%%%%%%%%%%%%%%%%%%%%%%%%%%%%%%%%%%%%%%%%%%%%%%%%%%
\section{The 1D Hubbard model: emergent fractionalized particles from rotated electrons}
\label{rot-symm-1}

The 1D Hubbard model is an integrable many-body problem that in spite of being more complex than the
1D Lieb-Liniger Bose gas and spin-$1/2$ $XXX$ chain has some common properties with both such systems.
In this section, an introductory short summary of the 1D Hubbard model development since
its coordinate BA solution by Lieb and Wu in 1968 \cite{Lieb,Lieb-03} is presented. In addition,
a uniquely defined unitary operator that transforms electrons into rotated electrons for $u= U/4t >0$ is considered.
Such rotated electrons have the same charge and spin $1/2$ as the electrons. Only their lattice occupancy configurations
and corresponding lattice degrees of freedom differ from those of the electrons. Three basic fractionalized particles 
are introduced. They naturally emerge from the rotated electrons degrees of freedom separation.
The relation of the model fractionalized particles and related exotic composite particles to the electrons is clarified.

%%%%%%%%%%%%%%%%%%%%%%%%%%%%%%%%%%%%%%%%%%%%%%%%%%%%%%%%%%%%%%%%%%%%%%%%%%
\subsection{The 1D Hubbard model: a short summary of its development}
\label{IntroHM}

Upon the 1D Hubbard model solution in 1968 by the coordinate BA \cite{Lieb,Lieb-03},
its ground state energy was derived. Moreover, the pioneering study reported in Ref. \cite{Lieb} revealed that 
the model undergoes a Mott metal-insulator transition at density $n_e =N_e/L=1$ whose corresponding critical 
onsite interaction is $U = 0$. Following that solution, the ground state properties \cite{Penc-91,Shiba-72,Takahashi-69}
and the excitation spectrum \cite{Ovchi-70,Coll-74,Woy-82,Woy,Klumper-90,Woynarovich-83,Woynarovich-83-B}
were studied by several authors. This applies as well to the preliminary version of the model pseudoparticle representation of
the BA solution reviewed here 
\cite{Carmelo-91-A,Carmelo-91,Carmelo-92,Carmelo-92-B,Carmelo-92-C,Carmelo-93,Carmelo-93-B,Carmelo-94,Carmelo-94-B,Carmelo-97,Carmelo-97-B,Carmelo-97-C,Carmelo-99,Carmelo-00}. 

In 1972 the TBA and corresponding ideal strings have been proposed in Ref. \cite{Takahashi-72} 
for the 1D Hubbard model. This has allowed the study of the model thermodynamic 
properties \cite{Kawakami-89,Usuki-90}. The energy spectra of its elementary excitations 
can be obtained from the TBA equations in the zero temperature limit \cite{Deguchi-00}. As
in the case of the spin-$1/2$ $XXX$ chain, the use
of the TBA has in the TL extended the number of pseudoparticle branches from two to infinite.
This ensures that their occupancy configurations generate all the model $4^L$ energy and momentum eigenstates
\cite{Carmelo-97-B,Carmelo-00A,Carmelo-03,Carmelo-04}.

An important property of the 1D Hubbard model is that its spectrum becomes conformally 
invariant in the low-energy limit. The corresponding finite-size corrections 
were obtained in Refs. \cite{Woynarovich-87,Woynarovich-89}. The relation between the finite-size spectrum and the 
asymptotic behavior of correlation functions was used to calculate the critical 
exponents of the model general two-point correlation functions \cite{Frahm-90,Frahm-91}.
The corresponding conformal dimensions have been expressed in terms of dressed
phase shifts associated with the pseudoparticle representation \cite{Carmelo-91,Carmelo-92}.

The conformal approach is not applicable to the zero-temperature model Mott insulating phase at half filling.
In the small-$U$ and scaling limits, dynamical correlation functions at low energies \cite{Controzzi-02,Essler-02,Essler-03,Jeckelmann-00}
can though be computed relying on methods of integrable quantum field theory \cite{Melzer-95,Woynarovich-97,Woynarovich-99}. 
Moreover, at half-filling and zero spin density the 1D Hubbard model TBA dressed phase shifts and corresponding
$S$-matrices have been associated with particles called holon, antiholon, and spinon. 
The holon and antiholon have zero spin and charge $+e$ and $-e$, respectively. The spinon has been
inherently constructed to have no charge and to have spin $1/2$ \cite{Essler-94,Essler-94-B}. The model 
$SO(4)$ symmetry group state representations were identified with occupancy configurations
of such particles.

More recently it was found in Ref. \cite{bipartite} that for $u\neq 0$ the 1D Hubbard model global symmetry is 
actually larger than $SO(4)$ and given by $[SO(4)\otimes U(1)]/Z_2$. (This applies as well to the model
on any bipartite lattice.) As further discussed below, in the case of the model full Hilbert space 
all energy eigenstates can be generated by pseudoparticle occupancy configurations. Those 
refer to the state representations of the model $[SO(4)\otimes U(1)]/Z_2$ symmetry group.

The wave functions of the energy eigenstates can be extracted from the coordinate BA solution. 
An explicit representation for the wave functions was given in Ref. \cite{Woy-82}. 
The corresponding energy and momentum eigenstates are either LWSs or HWSs
\cite{Completeness,Complete2} with respect to the model Hamiltonian $SO(4)$ symmetry 
in $[SO(4)\otimes U(1)]/Z_2$ \cite{HL,Yang,Yang-90,Lieb-89}.
The non-LWSs can be generated from the LWSs, which confirmed the
quantum problem completeness \cite{Complete3}. 

The first steps to obtain the BA solution of the 1D Hubbard model by the quantum inverse scattering method
were made in Refs.\cite{Shastry-86,Shastry-86A,CM}. The model Hamiltonian was mapped under a Jordan-Wigner 
transformation into a spin Hamiltonian. It commutes with the transfer matrix of a related covering vertex model 
\cite{Shastry-86}. The $R$-matrix of the spin model was also derived \cite{Shastry-86A,CM}. Alternative derivations 
were carried out by several authors \cite{Olmedilla-87,Olmedilla-88,Wadati-87}. 
The $R$-matrix was later shown to satisfy the Yang-Baxter equation \cite{Shiroishi-95}. 

An algebraic BA having as starting point the results of Refs.\cite{Shastry-86,Shastry-86A,CM} was afterwards constructed 
in Refs. \cite{Martins-97,Martins-98} for the 1D Hubbard model. Consistently with the model $[SO(4)\otimes U(1)]/Z_2$
symmetry, the corresponding spin and charge monodromy matrices were found to have different ABCD and ABCDF 
forms, respectively. Those are associated with the spin $SU(2)$ and charge $U(2)=SU(2)\otimes U(1)$ symmetries, respectively \cite{Martins-98}. 
The latter matrix is larger than the former and involves more fields \cite{Martins-98}. If the model global symmetry was 
only $SO(4)=[SU(2)\otimes SU(2)]/Z_2$, the charge and spin monodromy matrices would have the same traditional 
ABCD form, which is that of the spin-$1/2$ $XXX$ chain \cite{Faddeev-81}. The expressions for the eigenvalues of the transfer 
matrix of the two-dimensional statistical covering model were obtained. That problem was also addressed in Ref. \cite{Yue-97}. 

The algebraic BA introduced in Refs. \cite{Martins-97,Martins-98} allowed the quantum transfer matrix approach to 
the thermodynamics of the 1D Hubbard model \cite{Juttner-98}. Within it, the thermodynamic quantities and
correlation lengths can be calculated numerically for finite temperatures \cite{Tsunetsugu-91,Umeno-03}.
The 1D Hubbard model Hamiltonian was found in the TL to be invariant under the direct sum of two $Y(sl(2))$ 
Yangians \cite{Uglov-94}. The relation of these Yangians to the above $R$-matrix and the implications of one 
of these Yangians for the structure of the bare excitations was later clarified \cite{Mura-97,Mura-98}.

In the $u\rightarrow\infty$ limit the dynamical correlation functions can be computed at zero temperature for 
all energy scales relying on the simplified form that the BA equations acquire. This was achieved by a combination 
of analytical and numerical techniques for the whole range of electronic densities
\cite{Ogata-90,Ogata-91,Parola-90,Parola-92,Weng,Weng-94,Karlo-95,Karlo-96,Karlo-97,Gallagher-97,Gebhard-97}.
In the case of the one-electron spectral function studies of Refs. \cite{Karlo-95,Karlo-96,Karlo-97},
the method relies on the spinless-fermion phase shifts imposed by $XXX$ chain physical spins $1/2$. Such 
elementary objects naturally arise from the zero spin density and $u\rightarrow\infty$ electron wave-function factorization 
\cite{Woy,Woy-82,Ogata-90}. A related PDT \cite{V-1,TTF,spectral-06,VI,CarCadez-16,CarCadez-17} relying on a 
representation of the model BA solution in terms of
the pseudofermions generated by a unitary transformation from the corresponding pseudoparticles considered
in Ref. \cite{Carmelo-04} was introduced in Ref. \cite{V-1}. It is an extension of the $u\rightarrow\infty$ method
of Refs. \cite{Karlo-95,Karlo-96,Karlo-97} to the whole $u>0$ range of the 1D Hubbard model. 

After the PDT of the 1D Hubbard model was introduced, the MQIM methods have been developed to also
tackle the high-energy physics of both integrable and non-integrable 1D correlated quantum problems, 
beyond the low-energy TLL limit \cite{Glazman-09,Glazman-12,DSF-n1,Glazman-BG-08}.
In the case of the 1D Hubbard model, the MQIM reaches the same results as the PDT. For instance,
the momentum, electronic density, and on-site repulsion $u>0$ dependence of the exponents that control 
the line shape of the one-electron spectral function of the model at zero magnetic field
calculated in Refs. \cite{Essler-10,Seabra-14}, in the framework of the MQIM using input from the BA 
solution, is exactly the same as that obtained previously by the use of the PDT.

%%%%%%%%%%%%%%%%%%%%%%%%%%%%%%%%%%%%%%%%%%%%%%%%%%%%%%%%%%%%%%%%%%%%%%%%%%
\subsection{The Hubbard model and the infinite choices of rotated electrons}
\label{gen-rot-symm}

The Hubbard model in a chemical potential $\mu$ and magnetic field $H$
under periodic boundary conditions on a 1D lattice with an even number $L\rightarrow\infty$ 
of sites is given by,
\begin{equation}
{\hat{H}} = t\,\hat{T}+U\,\hat{V}_D + 2\mu\,{\hat{S}}_{\eta}^{z} + 2\mu_B H\,{\hat{S}}_s^{z} \, .
\label{H}
\end{equation}
Here,
\begin{equation}
\hat{T} = -\sum_{\sigma=\uparrow,\downarrow }\sum_{j=1}^{L}\left(c_{j,\sigma}^{\dag}\,
c_{j+1,\sigma} + c_{j+1,\sigma}^{\dag}\,c_{j,\sigma}\right)  
\hspace{0.20cm}{\rm and}\hspace{0.20cm}
\hat{V}_D = \sum_{j=1}^{L}\hat{\rho}_{j,\uparrow}\hat{\rho}_{j,\downarrow}
\hspace{0.20cm}{\rm with}\hspace{0.20cm}
\hat{\rho}_{j,\sigma} = c_{j,\sigma}^{\dag}\,c_{j,\sigma} -1/2 \, ,
\label{HH}
\end{equation}
are the kinetic-energy operator in units of $t$ and the electron onsite repulsion operator in units of $U$, 
respectively, and 
\begin{equation}
{\hat{S}}_{\eta}^{z}=-{1\over 2}(L-\hat{N}_e) \hspace{0.20cm}{\rm and}\hspace{0.20cm}
{\hat{S}}_s^{z}= -{1\over 2}({\hat{N}}_{e\uparrow}-{\hat{N}}_{e\downarrow}) \, ,
\label{etaSz}
\end{equation}
are the diagonal generators of the global $\eta$-spin and spin $SU(2)$ symmetry algebras, respectively.
Moreover, in Eqs. (\ref{H}) and (\ref{HH}) the operator $c_{j,\sigma}^{\dagger}$ (and $c_{j,\sigma}$)
creates (and annihilates) a spin-projection $\sigma=\uparrow,\downarrow$ electron at lattice site
$j=1,...,L$. The electron number operators read
${\hat{N}_e}=\sum_{\sigma=\uparrow ,\downarrow }\,{\hat{N}}_{e\sigma}$ and
${\hat{N}}_{e\sigma}=\sum_{j=1}^{L}\hat{N}_{e,j,\sigma}$.

The $\sigma$ electronic momentum distribution operator is given by ${\hat{N}}_{e\sigma} (k) = c_{k,\sigma }^{\dagger}\,c_{k,\sigma }$ 
where $c_{k,\sigma}^{\dagger}$ (and $c_{k,\sigma}$) creates (and annihilates) a 
$\sigma$ electron of momentum $k$. Its $z$-component $\eta$-spin/charge current operator in units of 
electronic charge $e$ ($\alpha=\eta$) and $z$-component spin current operator in units of spin $1/2$ ($\alpha =s$) read,
\begin{equation}
\hat{J}_{\alpha}^z = -i\,2t\sum_{\sigma}\sum_{j=1}^{L}(\sigma)^{\delta_{\alpha,s}}\left(c_{j,\sigma}^{\dag}\,c_{j+1,\sigma} - 
c_{j+1,\sigma}^{\dag}\,c_{j,\sigma}\right) \hspace{0.20cm}{\rm where}\hspace{0.20cm}\alpha = \eta , s \, .
\label{c-s-currents-2}
\end{equation}
It is here considered in $(\sigma)^{\delta_{\alpha,s}}$ that $\sigma=+1$ and $\sigma=-1$ for $\uparrow$ and 
$\downarrow$, respectively. These $\eta$-spin/charge and spin current operators are sometimes called 
in this paper $\alpha =\eta$ and $\alpha =s$ current operators, respectively.

The $4^L$ energy and momentum eigenstates of the 1D Hubbard model for $u>0$ can be generated by the independent
occupancy configurations of three basic fractionalized particles. They are associated with the two $SU(2)$ symmetries
and the $c$-lattice $U(1)$ symmetry, respectively, in the model global $[SU(2)\otimes SU(2)\otimes U(1)]/Z_2^2$ symmetry
mentioned in Sections \ref{elem-obj-intr} and \ref{IntroHM}.

The origin of the $u>0$ global $[SU(2)\otimes SU(2)\otimes U(1)]/Z_2^2$ symmetry is a local gauge 
$SU(2)\otimes SU(2)\otimes U(1)$ symmetry of the $U>0$ Hamiltonian $t=0$ term first identified in Ref. \cite{Stellan-91}. 
At $U=0$ and $t\neq 0$ that local gauge symmetry is unrelated to the model global $SO(4)\otimes Z_2$ symmetry. 
The $1/Z_2^2$ factor in the $u>0$ model global symmetry refers to the number $4^{L}$ of its independent representations 
being four times smaller than the dimension $4^{L+1}$ of the group $SU(2)\otimes SU(2)\otimes U(1)$. In contrast,
the factor $Z_2$ in the $U=0$ model global $SO(4)\otimes Z_2$ symmetry
corresponds to a discretely generated symmetry associated with a well-known transformation that exchanges 
spin and $\eta$-spin. It is an exact symmetry of the $U=0$ and $t\neq 0$ Hamiltonian. However, it changes the sign of 
$U$ when $U\neq 0$. 

The $c$-lattice $U(1)$ symmetry beyond $SO(4)$ found in Ref. \cite{bipartite}, which does not exist at $U=0$, 
emerges at any arbitrarily small finite-$U$ value. The related $U>0$ and $t=0$ local gauge 
$SU(2)\otimes SU(2)\otimes U(1)$ symmetry becomes for finite $U$ and $t$ a group of permissible unitary transformations. 
The corresponding local $U(1)$ canonical transformation is not the ordinary gauge $U(1)$ subgroup of electromagnetism. 
It is rather a ``nonlinear" transformation \cite{Stellan-91}.  
The $c$-lattice $U(1)$ symmetry has direct effects on the $u>0$ model BA solution 
structure. Its state representations are generated by occupancy configurations
of a specific BA quantum-number branch. 

For finite values of chemical-potential $\mu$ and magnetic field $H$ 
the corresponding operator terms in the Hamiltonian, Eq. (\ref{H}), lower the model global symmetry.
However, such terms commute with that Hamiltonian. Therefore, 
for finite $u$ and all values of the electronic density $n_e$ and spin density $m$, the quantum-numbers
occupancy configurations that generate all the model energy eigenstates from the 
electron or hole vacuum are in one-to-one correspondence to a set of independent state representations of 
the $\mu=H=0$ model Hamiltonian global symmetry algebra. Hence, for all electronic density $n_e$ and spin density 
$m$ values, that global symmetry algebra fully determines the finite-$u$ energy eigenstates spectrum structure.
Furthermore, the number of the model non-Abelian global symmetry independent state representations exactly 
equals the Hilbert-space dimension, $4^{L}$ \cite{bipartite}. 

The LWSs and HWSs of the $\eta$-spin and spin $SU(2)$ symmetry algebras have numbers 
$S_{\alpha} = - S_{\alpha}^{z}$ and $S_{\alpha} = S_{\alpha}^{z}$, respectively, where $\alpha = \eta$
and $\alpha = s$, respectively. As in the case of the spin-$1/2$ $XXX$ chain, in this review the LWS 
formulation of 1D Hubbard model BA solution is used. Here $S_{\eta}$ is the states $\eta$-spin, $S_{s}$ 
their spin, and $S_{\eta}^z = - (L-N_e)/2$ and $S_s^z = - (N_{e\uparrow}-N_{e\downarrow})/2$
are the corresponding projections, respectively. The latter are the eigenvalues of
the two $SU(2)$ algebras diagonal generators, Eq. (\ref{etaSz}). The LWSs of such
algebras considered in our studies are energy and momentum eigenstates.
Hence, as in the case of the spin-$1/2$ $XXX$ chain, they are here called LWSs.
Such LWSs have electronic densities $n_e$ and spin densities $m$ 
in the ranges $n_e \in [0,1]$ and $m \in [0,n_e]$, respectively. 

Let $\{\vert l_{\rm r},l_{\eta s},u\rangle\}$ be the complete set of $4^{L}$ energy and momentum eigenstates of the
Hamiltonian $\hat{H}$, Eq. (\ref{H}), associated with the BA solution for $u>0$. The LWSs of both $SU(2)$ symmetry algebras
are here denoted by $\vert l_{\rm r},l_{\eta s}^0,u\rangle$. The $u$-independent label $l_{\eta s}$ in them is a short notation for 
the set of quantum numbers, 
\begin{equation}
l_{\eta s} = S_{\eta},S_{s},n_{\eta},n_s \hspace{0.20cm}{\rm where}\hspace{0.20cm} n_{\alpha} = 
S_{\alpha}+S_{\alpha}^{z} = 0,1,..., 2S_{\alpha} \hspace{0.20cm}{\rm and}\hspace{0.20cm}\alpha = \eta, s \, .
\label{etas-states-ll}
\end{equation}
Furthermore, the label $l_{\rm r}$ refers to the set of all remaining $u$-independent quantum numbers needed
to uniquely specify an energy eigenstate $\vert l_{\rm r},l_{\eta s},u\rangle$.  

As for the spin-$1/2$ $XXX$ chain, the sets of $(2S_{\eta}+1)\times (2S_{s}+1)-1$ finite-$u$ energy eigenstates that
are generated from each LWS are called here non-LWSs. For the present model this applies to energy and 
momentum eigenstates that are LWSs of only one of such algebras. Indeed, such states are not LWSs as defined above.
For a LWS one then has that $n_{\eta}= n_s=0$ in Eq. (\ref{etas-states-ll}). Hence $l_{\eta s}^0$ stands for $S_{\eta},S_{s},0,0$. 
The non-LWSs $\vert l_{\rm r},l_{\eta s},u\rangle$ can be generated from the corresponding 
LWSs $\vert l_{\rm r},l_{\eta s}^0,u\rangle$ as \cite{Completeness,Complete2,Complete3},
\begin{eqnarray}
\vert l_{\rm r},l_{\eta s},u\rangle & = & \prod_{\alpha=\eta, s}\left(\frac{1}{
\sqrt{{\cal{C}}_{\alpha}}}({\hat{S}}^{+}_{\alpha})^{n_{\alpha}}\right)\vert l_{\rm r},l_{\eta s}^0,u\rangle
\hspace{0.20cm}{\rm where}\hspace{0.20cm} 
{\cal{C}}_{\alpha} = (n_{\alpha}!)\prod_{j=1}^{n_{\alpha}}(\,2S_{\alpha}+1-j\,) \, ,
\nonumber \\
n_{\alpha} & = & 1,...,2S_{\alpha} \hspace{0.20cm}{\rm where}\hspace{0.20cm}\alpha = \eta,s \, ,
\nonumber \\
{\hat{S}}^{+}_{\eta} & = & \left({\hat{S}}^{-}_{\eta}\right)^{\dag} = \sum_{j=1}^{L}(-1)^j\,c_{j,\downarrow}^{\dag}\,c_{j,\uparrow}^{\dag} 
\hspace{0.20cm}{\rm and}\hspace{0.20cm}
{\hat{S}}^{+}_{s} = \left({\hat{S}}^{-}_{s}\right)^{\dag} = \sum_{j=1}^{L}c_{j,\downarrow}^{\dag}\,c_{j,\uparrow} \, .
\label{Gstate-BAstate}
\end{eqnarray}
The model in its full Hilbert space can be described either directly within the BA solution \cite{Woy-82,Braak-99} or by application onto the 
LWSs of the $\eta$-spin and spin $SU(2)$ symmetry algebras off-diagonal generators \cite{Completeness},
as given in Eq. (\ref{Gstate-BAstate}).

There are infinite unitary transformations such that,
\begin{equation}
\hat{H} \rightarrow \hat{V}^{\dag}\hat{H}\hat{V} = e^{-\hat{S}}\hat{H}e^{\hat{S}}
\hspace{0.20cm}{\rm where}\hspace{0.20cm}
\hat{V} = e^{\hat{S}} \, .
\label{H-H}
\end{equation}
Within the physical problem studied in Refs. \cite{Stein,HO-04,Trembley-09,Mac,Mac-90,Harris},
the unitary operators $\hat{V} =e^{\hat{S}}$ transform the 
Hamiltonian $\hat{H}$, Eq. (\ref{H}), into a rotated Hamiltonian $\hat{V}^{\dag}\hat{H}\hat{V}$  
other than $\hat{H}$. For it electron double and single occupancy are good quantum numbers for finite $u$ values.
For large and intermediate $u$ values, the operator $\hat{S}$ can be expanded as 
$\hat{S} = - \sum_{i=0}^{\infty}\left({t\over U}\right)^i \hat{S}^{(i)}$. However, it is well-defined for the whole $u>0$ range. 
The infinite possible choices of operators $\hat{S}$ is an issue discussed in Ref. \cite{Stein}. (The
$\hat{S}$ expansion $\hat{S} = - \sum_{i=0}^{\infty}\left({t\over U}\right)^i \hat{S}^{(i)}$
is minus that in Eq. (57) of that reference.) 

Nonetheless, exactly the same mathematical transformation, Eq. (\ref{H-H}), can refer to the different physical
problem discussed in this review. For the latter problem, $\hat{V}^{\dag}\hat{H}\hat{V}$ is for $u>0$ the 1D Hubbard 
model written in the {\it rotated electron} representation.
The creation and annihilation operators ${\hat{V}}^{\dag}\,c_{j,\sigma}^{\dag}\,{\hat{V}}$
and ${\hat{V}}^{\dag}\,c_{j,\sigma}\,{\hat{V}}$ refer thus to rotated electrons rather than to electrons. Rotated-electron single and
double occupancies are then good quantum numbers for the finite-$u$ 1D Hubbard model. In the $u\rightarrow\infty$
limit rotated-electron single and double occupancies become electron single and double occupancies, respectively.
The global $c$-lattice $U(1)$ symmetry algebra generator beyond $SO(4)$ symmetry is the operator that
counts the number of rotated-electron singly occupied sites for $u>0$. (It can also be chosen to be
the operator that counts the number of rotated-electron unoccupied plus doubly occupied sites for $u>0$.)

The kinetic-energy operator $\hat{T}$, Eq. (\ref{HH}), can be written as $\hat{T}= \hat{T}_0 + \hat{T}_{+1} + \hat{T}_{-1}$.
The operator $\hat{T}_0$ conserves the number of rotated-electron doubly occupied sites. The
operators $\hat{T}_{+1}$ and $\hat{T}_{-1}$ enhance and lessen it by one, respectively.
The infinite electron-rotated-electron unitary transformations share an important property. It is that
although the $u$ dependent expression of the operator $\hat{S}$ in the unitary operator $\hat{V} = e^{\hat{S}}$, Eq. (\ref{H-H}),
is different for each such a transformation, for $u>0$ it always
involves {\it only} the three kinetic operators $\hat{T}_{0}$, $\hat{T}_{+1}$, and $\hat{T}_{-1}$.
Another property common to all electron-rotated-electron unitary transformations is that 
to leading order in $t/U$, the operator ${\hat{S}}$, Eq. (\ref{H-H}), has the universal form, 
${\hat{S}} = -{t\over U}\,(\hat{T}_{+1} -\hat{T}_{-1})$. From a straightforward yet cumbersome algebra, one then finds that
the momentum operator $\hat{P}$ and the six generators of the $\eta$-spin and spin $SU(2)$ symmetry
algebras commute with the three kinetic operators $\hat{T}_{0}$, $\hat{T}_{+1}$, and $\hat{T}_{-1}$.
This ensures that such operators commute with the electron-rotated-electron 
unitary operators ${\hat{V}} = e^{{\hat{S}}}$ and corresponding operators ${\hat{S}}$.

%%%%%%%%%%%%%%%%%%%%%%%%%%%%%%%%%%%%%%%%%%%%%%%%%%%%%%%%%%%%%%%%%%%%%%%%%%
\subsection{The Hubbard model BA uniquely defined rotated electrons
and corresponding $c$ pseudoparticle, rotated spin, and rotated $\eta$-spin operators}
\label{specificRE}

The pseudoparticle representation and related pseudofermion representation of the 1D Hubbard model
refer to a specific choice of the electron-rotated-electron unitary operator ${\hat{V}}$. The corresponding
rotated-electron creation and annihilation operators and the number of rotated electrons at lattice site
$j$ with spin projection $\sigma$ operator read,
\begin{equation}
{\tilde{c}}_{j,\sigma}^{\dag} =
{\hat{V}}^{\dag}\,c_{j,\sigma}^{\dag}\,{\hat{V}} \, ,
\hspace{0.20cm}
{\tilde{c}}_{j,\sigma} =
{\hat{V}}^{\dag}\,c_{j,\sigma}\,{\hat{V}} 
\hspace{0.20cm}{\rm and}\hspace{0.20cm}
{\tilde{n}}_{j,\sigma} = {\tilde{c}}_{j,\sigma}^{\dag}\,{\tilde{c}}_{j,\sigma} \, ,
\label{rotated-operators}
\end{equation} 
respectively. Such rotated electrons are generated from the electrons by a unitary transformation 
defined and performed by the BA. The corresponding electron-rotated-electron unitary operator ${\hat{V}}$ in Eq. (\ref{rotated-operators})
is uniquely defined in Ref. \cite{CarCadez-17} by its matrix elements between the model $4^L$ energy and momentum eigenstates.

In the $u\rightarrow\infty$ limit (and thus $u^{-1}\rightarrow 0$ limit)
all spin configurations and all $\eta$-spin configurations with the
same number of doubly occupied sites are degenerated. Hence there are in that limit infinite 
choices of complete sets of $4^{L}$ energy and momentum eigenstates for which electron single and
double occupancies are good quantum numbers. The unitary transformation uniquely defined in Ref. \cite{CarCadez-17}
refers to a specific set of $4^{L}$ energy and momentum eigenstates. It is that obtained from the set of $4^{L}$ finite-$u$ energy eigenstates 
$\vert l_{\rm r},l_{\eta s},u\rangle$, Eq. (\ref{Gstate-BAstate}), upon turning off adiabatically $u^{-1}$. 
The finite-$u$ energy eigenstates $\vert  l_{\rm r},l_{\eta s},u\rangle={\hat{V}}^{\dag}\vert  l_{\rm r},l_{\eta s},\infty \rangle$ 
generated from each such $u\rightarrow\infty$ energy eigenstates $\vert  l_{\rm r},l_{\eta s},\infty \rangle$
have for $u>0$ {\it exactly} the same values for all $u$-independent quantum numbers. This includes
the quantum numbers $l_{\eta s}$ given in Eq. (\ref{etas-states-ll}) and
all remaining $u$-independent quantum numbers $l_{\rm r}$ (provided below in Eq. (\ref{states-ll})) needed
to uniquely specify an energy eigenstate $\vert l_{\rm r},l_{\eta s},u\rangle$. A {\it $V(u)$-set of states}
is our designation for such continuum set of $u>0$ energy eigenstates.

Important physical information can be reached from analysis of the relation between (i) the BA quantum numbers and (ii) 
the rotated-electron occupancy configurations, respectively, that generate the finite-$u$ exact energy eigenstates 
$\vert  l_{\rm r},l_{\eta s},u\rangle={\hat{V}}^{\dag}\vert  l_{\rm r},l_{\eta s},\infty \rangle$
of any $V (u)$-set. The rotated-electron spatial occupancy configurations that generate from
the electron (and rotated-electron) vacuum the finite-$u$ energy 
eigenstates $\vert  l_{\rm r},l_{\eta s},u\rangle={\hat{V}}^{\dag}\vert  l_{\rm r},l_{\eta s},\infty \rangle$ 
of any $V (u)$-set of states are exactly the same as the electron spatial occupancy configurations that
generate from it the corresponding $u\rightarrow\infty$ energy and momentum eigenstate 
$\vert  l_{\rm r},l_{\eta s},\infty \rangle$. Hence for $u>0$ the number 
$N^R_{s,\pm 1/2}$ of spin-projection $\pm 1/2$ rotated-electron singly occupied sites, $N^R_{\eta,+1/2}$
of rotated-electron unoccupied sites, and $N^R_{\eta,-1/2}$ of rotated-electron doubly occupied sites are conserved. 

Such numbers obey the sum rules $N^R_{s,+1/2}+N^R_{\eta,-1/2}=N_{e\uparrow}$, $N^R_{s,-1/2}+N^R_{\eta,-1/2}=N_{e\downarrow}$,
$N^R_{s}+2N^R_{\eta,-1/2}=N_e$, and $N^R_{s}+N^R_{\eta}=L$. The rotated-electron numbers equal those of the electrons.
Therefore, here $N_{e\uparrow}$ and $N_{e\downarrow}$ denotes both the number of electrons and rotated electrons of
spin projection $+1/2$ and $-1/2$, respectively. However, for finite $u$ values the numbers $N^R_{s}=N^R_{s,+1/2}+N^R_{s,-1/2}$ 
of rotated-electron singly occupied sites and $N^R_{\eta}=N^R_{\eta,+1/2}+N^R_{\eta,-1/2}$ of rotated-electron doubly 
occupied plus unoccupied sites are only conserved for rotated electrons.

For any operator ${\hat{O}}$ there is a corresponding operator ${\tilde{O}}={\hat{V}}^{\dag}\,{\hat{O}}\,{\hat{V}}$ whose
expression in terms of rotated-electron creation and annihilation operators is the same as that
of ${\hat{O}}$ in terms of electron creation and annihilation operators, respectively. 
The $l = z, \pm$ local rotated spins operators ($\alpha =s$) and local rotated $\eta$-spin operators ($\alpha =\eta$), 
\begin{eqnarray}
{\tilde{S}}^{l}_{j,\alpha} & = & {\hat{V}}^{\dag}\,{\hat{S}}^{l}_{j,\alpha}\,{\hat{V}} \hspace{0.20cm}{\rm where}\hspace{0.20cm}l = z, \pm 
\hspace{0.20cm}{\rm and}\hspace{0.20cm}\alpha = \eta, s \, ,
\nonumber \\
{\tilde{S}}^{\pm}_{j,\alpha} & = & {\tilde{S}}^{x}_{j,\alpha}\pm i\,{\tilde{S}}^{y}_{j,\alpha} 
\hspace{0.20cm}{\rm where}\hspace{0.20cm}\alpha = \eta, s \, ,
\label{Slinf-exp-OS}
\end{eqnarray}
play a major role in the 1D Hubbard model pseudoparticle representation revisited below in Sections \ref{rot-symm-2} 
and \ref{exc-spectra}.

Here ${\hat{S}}^{l}_{j,s}$ and ${\hat{S}}^{l}_{j,\eta}$ are the usual unrotated local spin  
$\eta$-spin operators, respectively. The rotated local operators ${\tilde{S}}^{l}_{j,\alpha}$, Eq. (\ref{Slinf-exp-OS}),
have in terms of creation and annihilation rotated-electron operators, Eq. (\ref{rotated-operators}), 
exactly the same expressions as the corresponding unrotated local operators 
${\hat{S}}^{l}_{j,\alpha}$ in terms of creation and annihilation electron operators. 
Specifically, the spin operators ${\tilde{S}}^{l}_{j,s}$, which act onto sites singly occupied by rotated electrons, read
${\tilde{S}}^-_{j,s} = ({\tilde{S}}^+_{j,s})^{\dag} = {\tilde{c}}_{j,\uparrow}^{\dag}{\tilde{c}}_{j,\downarrow}$
and ${\tilde{S}}^z_{j,s} = ({\tilde{n}}_{j,\downarrow} - 1/2)$. Similarly, the $\eta$-spin operators 
${\tilde{S}}^{l}_{j,\eta}$, which act onto sites unoccupied by rotated electrons and sites doubly occupied by rotated electrons, are
given by ${\tilde{S}}^-_{j,\eta} = ({\tilde{S}}^+_{j,\eta})^{\dag} = (-1)^j\,{\tilde{c}}_{j,\uparrow}{\tilde{c}}_{j,\downarrow}$
and ${\tilde{S}}^z_{j,\eta} = ({\tilde{n}}_{j,\downarrow} - 1/2)$. 

For $u>0$ a non-perturbative three degrees of freedom spin - $\eta$-spin - $c$-lattice separation occurs at all energy 
scales \cite{CarCadez-17}. It naturally emerges from the independent state representations of the two $SU(2)$ symmetries 
and $c$-lattice $U(1)$ symmetry, respectively, in the model global symmetry. 
The $c$-lattice - $\eta$-spin degrees of freedom separation may be considered as a 
separation of the charge degrees of freedom. At zero temperature and energy scales lower than $2\vert\mu\vert$ 
relative to the ground state, one has that $N^R_{\eta,-1/2}=0$ (and $N^R_{\eta,+1/2}=0$) for $n_e \in [0,1[$ (and
$n_e \in ]1,2]$). Hence for such energy ranges the $\eta$-spin degrees of freedom remain hidden. The three degrees of freedom
non-perturbative $c$-lattice - $\eta$-spin - spin separation is then seen as the usual two degrees of freedom charge-spin 
separation. 

Under the three general degrees of freedom separation of the rotated-electron occupancy configurations,
their operators, Eq. (\ref{rotated-operators}), are of the form,
\begin{eqnarray}
{\tilde{c}}_{j,\uparrow}^{\dag} & = &
\left({1\over 2} - {\tilde{S}}^{z}_{j,s} - {\tilde{S}}^{z}_{j,\eta}\right)f_{j,c}^{\dag} + (-1)^j
\left({1\over 2} +{\tilde{S}}^{z}_{j,s} + {\tilde{S}}^{z}_{j,\eta}\right)f_{j,c}
\hspace{0.20cm}{\rm and}\hspace{0.20cm}{\tilde{c}}_{j,\uparrow} = ({\tilde{c}}_{j,\uparrow}^{\dag})^{\dag} \, ,
\nonumber \\
{\tilde{c}}_{j,\downarrow}^{\dag} & = &
({\tilde{S}}^{+}_{j,s} + {\tilde{S}}^{+}_{j,\eta})(f_{j,c}^{\dag} + (-1)^j\,f_{j,c}) 
\hspace{0.20cm}{\rm and}\hspace{0.20cm}
{\tilde{c}}_{j,\downarrow} = ({\tilde{c}}_{j,\downarrow}^{\dag})^{\dag} \, .
\label{c-up-c-downG}
\end{eqnarray}
Here the operators $f_{j,c}^{\dag}$ and $f_{j,c}$ defined below create and annihilate one {\it $c$ pseudoparticle}
at the $c$ effective lattice site $j=1,...,L$. That lattice is identical to the electron and rotated-electron
original lattice. Their (i) occupied and (ii) unoccupied sites are those (i) singly occupied and (ii)
unoccupied and doubly occupied by the rotated electrons, respectively. Hence the $c$ pseudoparticle local density operator
${\tilde{n}}_{j,c} \equiv f_{j,c}^{\dag}\,f_{j,c}$ and the corresponding operator $(1-{\tilde{n}}_{j,c})$ are the natural projectors 
onto the subset of $N_{R}^{s}=N_c$ original-lattice sites singly occupied by rotated electrons and
onto the subset of $N_{R}^{\eta}=N_c^h=L-N_c$ original-lattice sites unoccupied and doubly occupied by rotated electrons,
respectively. It then follows that the local operators ${\tilde{S}}^{l}_{j,\alpha}$,
Eq. (\ref{Slinf-exp-OS}), can be written as,
\begin{equation}
{\tilde{S}}^l_{j,s} = {\tilde{n}}_{j,c}\,{\tilde{q}}^l_{j} 
\hspace{0.20cm}{\rm and}\hspace{0.20cm}
{\tilde{S}}^l_{j,\eta} = (1-{\tilde{n}}_{j,c})\,{\tilde{q}}^l_{j}
\hspace{0.20cm}{\rm where}\hspace{0.20cm}l = z, \pm \, ,
\label{sir-pirG}
\end{equation}
respectively. Here the $l = z, \pm$ local {\it $\eta s$ quasi-spin} operators,
\begin{equation}
{\tilde{q}}^l_{j} = {\tilde{S}}^l_{j,s} + {\tilde{S}}^l_{j,\eta} 
\hspace{0.20cm}{\rm where}\hspace{0.20cm}l=\pm,z \, ,
\label{q-operG}
\end{equation}
such that ${\tilde{q}}^{\pm}_{j}= {\tilde{q}}^{x}_{j}\pm i\,{\tilde{q}}^{y}_{j}$, 
have the following expression in terms of rotated-electron creation and annihilation operators,
\begin{equation}
{\tilde{q}}^-_{j} = ({\tilde{q}}^+_{j})^{\dag} = 
({\tilde{c}}_{j,\uparrow}^{\dag}
+ (-1)^j\,{\tilde{c}}_{j,\uparrow})\,
{\tilde{c}}_{j,\downarrow} 
\hspace{0.20cm}{\rm and}\hspace{0.20cm}
{\tilde{q}}^{z}_{j} = ({\tilde{n}}_{j,\downarrow} - 1/2) \, .
\label{rotated-quasi-spinG}
\end{equation}

The local $c$ pseudoparticle operators $f_{j,c}^{\dag}$ and $f_{j,c}$ in Eq. (\ref{c-up-c-downG}) 
are {\it uniquely} defined for $u>0$ in terms of rotated-electron creation and 
annihilation operators, Eq. (\ref{rotated-operators}). This is achieved by combining the inversion of 
the relations, Eq. (\ref{c-up-c-downG}), with the expressions of the local operators ${\tilde{S}}^{l}_{j,s}$ and ${\tilde{S}}^{l}_{j,\eta}$ 
provided in Eqs. (\ref{sir-pirG})-(\ref{rotated-quasi-spinG}). This gives, 
\begin{equation}
f_{j,c}^{\dag} = (f_{j,c})^{\dag} = {\tilde{c}}_{j,\uparrow}^{\dag}\,
(1-{\tilde{n}}_{j,\downarrow}) + (-1)^j\,{\tilde{c}}_{j,\uparrow}\,{\tilde{n}}_{j,\downarrow} 
\hspace{0.20cm}{\rm and}\hspace{0.20cm}{\tilde{n}}_{j,c} = f_{j,c}^{\dag}\,f_{j,c} 
\hspace{0.20cm}{\rm for}\hspace{0.20cm}j = 1,...,L \, .
\label{fc+G}
\end{equation}
The operator ${\tilde{n}}_{j,\sigma}$ in this equation is the $\sigma$ rotated-electron local density operator
given in Eq. (\ref{rotated-operators}). (In the $u\rightarrow\infty$ limit, the $\eta s$ quasi-spins
associated with the operators, Eq. (\ref{rotated-quasi-spinG}), and the $c$ pseudoparticle
holes associated with operators $f_{j,c}$, Eq. (\ref{fc+G}), become the quasispins and quasicharges, respectively,
of Ref. \cite{Ostlund-06}.)

On the one hand, the {\it rotated spins $1/2$} of projection $\pm 1/2$ are the spin-$1/2$ fractionalized particles 
associated with the $l = z, \pm$ spin operators ${\tilde{S}}^l_{j,s}$ in Eq. (\ref{sir-pirG}). They refer to the 
$L_{s,\pm 1/2} = N^R_{s,\pm 1/2}$ spins $1/2$ of the rotated electrons with such a spin projection that singly occupy sites. 
On the other hand, the {\it rotated $\eta$-spins $1/2$} of projection $\pm 1/2$ are the $\eta$-spin-$1/2$ fractionalized particles 
associated with the $l = z, \pm$ $\eta$-spin operators ${\tilde{S}}^l_{j,\eta}$ in Eq. (\ref{sir-pirG}).
They refer to the $\eta$-spin degrees of freedom of the $L_{\eta,\pm 1/2} = N^R_{\eta,\pm 1/2}$ sites
unoccupied $(+1/2)$ and doubly occupied $(-1/2)$ by rotated electrons.

The charge and spin of the electrons remain invariant under the electron-rotated-electron
unitary transformation. Indeed, it only changes their spatial original lattice occupancy distributions. Therefore, the 
rotated spins $1/2$ and rotated $\eta$-spins $1/2$ are physical particles with a well-defined
relation to the rotated electrons and corresponding electrons. 
There is a rotated spin $1/2$ and rotated $\eta$-spin $1/2$ quantum problem for the 1D Hubbard model in
each fixed-$N_c$ subspace where $N_c=N^R_{s}\in [0,L]$. The reason is that only in such subspaces are the
numbers $L_{s}=N^R_{s}=N_c$ of rotated spins $1/2$ and $L_{\eta}=N^R_{\eta}=L-N_c$ of rotated $\eta$-spins $1/2$, respectively, fixed.
The rotated spin $1/2$ and rotated $\eta$-spin $1/2$ representation is well defined in such subspaces.

For simplicity, in the remaining of this review the rotated spins $1/2$ and rotated $\eta$-spins $1/2$ are called
spins $1/2$ and $\eta$-spins $1/2$, respectively. The spins $1/2$ are though only those carried by the rotated electrons
that singly occupy original lattice sites. Those within the rotated electrons doubly occupied original lattice sites
rather refer to the $\eta$-spin $SU(2)$ symmetry algebra. Indeed, such doubly occupied original lattice sites
$\eta$-spin degrees of freedom correspond to the $\eta$-spins of $\eta$-spin projection $-1/2$.
(The unoccupied original lattice sites $\eta$-spin degrees of freedom refer to the $\eta$-spins of $\eta$-spin 
projection $+1/2$.)

Within the above general separation, the (i) global $c$-lattice $U(1)$ symmetry, (ii) global $\eta$-spin $SU(2)$
symmetry, and (iii) global spin $SU(2)$ symmetry state representations are, in each subspace with a
fixed number $N^R_{s}$ of rotated-electron singly occupied sites, generated by three sets of independent 
occupancy configurations. Those involve: (i) The $N_c=N^R_{s}$ $c$ pseudoparticles without internal degrees of freedom 
and corresponding $N_c^h=N^R_{\eta}$ $c$ pseudoparticle holes;
(ii) The $L_{s,\pm 1/2} = N^R_{s,\pm 1/2}$ spins $1/2$ of projection $\pm 1/2$;
(iii) The $L_{\eta,\pm 1/2} = N^R_{\eta,+1/2}$ $\eta$-spins $1/2$ of projection $\pm 1/2$.
It then follows that their numbers are such that,
\begin{eqnarray}
&& L_{s} = L_{s,+1/2} + L_{s,-1/2} = N_c \, ,
\nonumber \\
&& L_{\eta} = L_{\eta,+1/2} + L_{\eta,-1/2} = L - N_c = N_c^h \, ,
\nonumber \\
&& L_{s,+1/2} - L_{s,-1/2} = -2S_s^z = N_{e\uparrow} - N_{e\downarrow} \, ,
\nonumber \\
&& L_{\eta,+1/2} - L_{\eta,-1/2} = -2S_{\eta}^z = L - N_e \, .
\label{severalM}
\end{eqnarray}
Here $L_{s}$ denotes the number of spins and $L_{\eta}$ that of $\eta$-spins. Those
equal the numbers $N_c$ of $c$ pseudoparticles and $N_c^h = L - N_c$ of $c$ pseudoparticle holes, respectively.

The numbers $N_c$ of $c$ pseudoparticles, $L_{\eta,\pm 1/2}$ of $\eta$-spins of projection $\pm 1/2$, 
and $L_{s,\pm 1/2}$ of spins of projection $\pm 1/2$ are fully controlled by those of rotated electrons as follows,
\begin{eqnarray}
&& N_ c = N_{R}^{s} \, , \hspace{0.20cm} N_ c^h = N_{R}^{\eta} \hspace{0.20cm}{\rm and}\hspace{0.20cm}
N_ c + N_ c^h = N_{R}^{s} + N_{R}^{\eta} = L \, ,
\nonumber \\
&& L_{\alpha,\pm 1/2} = N_{R,\pm 1/2}^{\alpha} \hspace{0.20cm}{\rm and}\hspace{0.20cm}
L_{\alpha} = L_{\alpha,+1/2} + L_{\alpha,-1/2} = N_{R}^{\alpha} \hspace{0.20cm}{\rm where}\hspace{0.20cm}\alpha=\eta, s \, .
\label{NRpm}
\end{eqnarray}
This is consistent with such fractionalized particles stemming from the rotated-electron occupancy 
configurations degrees of freedom separation.

The global three degrees of freedom rotated-electron separation leads locally in what the onsite rotated-electron 
occupancies is concerned to two degrees of freedom separation.
On the one hand, the degrees of freedom of each rotated-electron occupied site decouple into one 
spin-less $c$ pseudoparticle without internal degrees of freedom and one spin $1/2$ 
that carries its spin. On the other hand, the degrees of freedom of 
each rotated-electron unoccupied and doubly occupied site decouple into one $c$ pseudoparticle
hole and one $\eta$-spin $1/2$ of projection $+1/2$ and $-1/2$, respectively. Hence the local
two degrees of freedom separation corresponds to those of the $c$-lattice $U(1)$ symmetry and
one of the two global $SU(2)$ symmetries, respectively. 

The unitarity of the electron-rotated-electron transformation implies that the rotated-electron operators 
${\tilde{c}}_{j,\sigma}^{\dag}$ and ${\tilde{c}}_{j,\sigma}$, Eqs. (\ref{rotated-operators}) and (\ref{c-up-c-downG}),
have the same anti-commutation relations as the corresponding electron 
operators $c_{j,\sigma}^{\dag}$ and $c_{j,\sigma}$, respectively. 
Straightforward manipulations based on Eqs. (\ref{Slinf-exp-OS})-(\ref{fc+G}) then lead
to the following algebra for the local $c$ pseudoparticle creation and annihilation operators,
\begin{equation}
\{f^{\dag}_{j,c}\, ,f_{j',c}\} = \delta_{j,j'} 
\hspace{0.20cm}{\rm and}\hspace{0.20cm}
\{f_{j,c}^{\dag}\, ,f_{j',c}^{\dag}\} =
\{f_{j,c}\, ,f_{j',c}\} = 0 \, .
\label{albegra-cf}
\end{equation}

Furthermore, the local $c$ pseudoparticle operators and the local rotated quasi-spin operators ${\tilde{q}}^{l}_{j}$, 
Eq. (\ref{rotated-quasi-spinG}), commute with each other. From the use of Eqs. (\ref{Slinf-exp-OS})-(\ref{rotated-quasi-spinG})
one confirms that the $SU(2)$ algebra obeyed by the local quasi-spin operators ${\tilde{q}}^{l}_{j}$ is the usual one,
\begin{equation}
[{\tilde{q}}^{+}_{j},{\tilde{q}}^{-}_{j'}] = \delta_{j,j'}\,2\,{\tilde{q}}^{z}_{j}
\hspace{0.20cm}{\rm and}\hspace{0.20cm}
[{\tilde{q}}^{\pm}_{j},{\tilde{q}}^{z}_{j'}] = \mp \delta_{j,j'}\,{\tilde{q}}^{\pm}_{j} \, .
\label{albegra-q-com}
\end{equation}
The same applies to the $SU(2)$ algebras of the corresponding (rotated) $\eta$-spin and spin 
operators ${\tilde{s}}^{l}_{j,\eta}$ and ${\tilde{s}}^{l}_{j,s}$, respectively. Moreover, $[{\tilde{q}}^{l}_{j},{\tilde{q}}^{l}_{j'}]=0$
and $[{\tilde{s}}^{l}_{j,\alpha},{\tilde{s}}^{l}_{j',\alpha'}] = 0$. 
The $c$ pseudoparticle and $\eta s$ quasi-spin operator algebras refer to the whole Hilbert space.
In contrast, those of the (rotated) $\eta$-spin and spin operators correspond to fixed-$N_c$ subspaces. 

The degrees of freedom separation, Eq. (\ref{c-up-c-downG}), is such that the 
$c$ pseudoparticle operators, Eq. (\ref{fc+G}), (rotated) spin $1/2$ and $\eta$-spin $1/2$ operators, 
Eq. (\ref{sir-pirG}), and the related $\eta s$ quasi-spin operators, Eqs. (\ref{q-operG}) and (\ref{rotated-quasi-spinG}), 
emerge from the rotated-electron operators by an exact local transformation that {\it does not} introduce constraints. 
The expressions of the $c$ pseudoparticle, spin $1/2$, and $\eta$-spin $1/2$ operators
in terms of rotated-electron creation and annihilation operators are valid for $u>0$. The latter operators are related to the
original electron creation and annihilation operators through the transformation, Eq. (\ref{rotated-operators}). 
The unitary operator in that transformation is uniquely defined in Ref. \cite{CarCadez-17} by its $4^{L}$ matrix elements. 
Combination of all such equations thus uniquely defines for $u>0$ the $c$ pseudoparticle, spin $1/2$, and $\eta$-spin $1/2$ operators
in terms of electron creation and annihilation operators.

In the case of the spin-$1/2$ $XXX$ chain, the number $L$ of sites singly occupied by spins $1/2$ is a good quantum number. 
The emergence within the $u>0$ 1D Hubbard model rotated-electron representation of $L_s=N_{R}^{s}$ spins $1/2$ 
that singly occupy $L_s=N_c$ original-lattice sites renders the problem much similar to that of such a chain. Also
the $L_{\eta}=N_{R}^{\eta} $ $\eta$-spins $1/2$ singly occupy $L_{\eta}=N_ c^h=L-N_c$ original-lattice sites.
As justified in Appendix \ref{FEL}, the present rotated-electron representation spin-$1/2$ and $\eta$-spin-$1/2$ occupancy 
configurations that generate the two $SU(2)$ symmetries degrees of freedom of the energy and momentum eigenstates of the 1D Hubbard 
model for $u>0$ are  actually exactly {\it the same} as those that generate the energy and momentum eigenstates of a spin-$1/2$ 
and $\eta$-spin-$1/2$ $XXX$ chain with $L_s=N_c$ and $L_{\eta}=N_c^h=L-N_c$ sites, respectively.
The relation to the 1D Lieb-Liniger Bose gas is brought about by the independent occupancy 
configurations of the $c$ pseudoparticles, which have no internal structure. And as the pseudoparticles of that gas, 
they are associated with a $U(1)$ symmetry. Indeed, they generate the $c$-lattice $U(1)$ symmetry degrees of freedom of the 
energy and momentum eigenstates of the 1D Hubbard model for $u>0$.

As mentioned above, the $c$ pseudoparticles live on a $c$ effective lattice similar to the original lattice. In contrast, the (i) spins $1/2$
and (ii) $\eta$-spins $1/2$ only ``see" the sites (i) singly occupied and (ii) unoccupied and doubly occupied,
respectively, by rotated electrons. Hence for the model in fixed-$N_c$ subspaces one can define within the TL a squeezed 
spin effective lattice with $L_{s}=N^R_{s}=N_c$ sites on which the spins $1/2$ live. One can define
as well a corresponding squeezed $\eta$-spin effective lattice with $L_{\eta}=N^R_{\eta}=L-N_c$ sites for the $\eta$-spins $1/2$.
The numbers of sites of such squeezed $\eta$-spin and spin effective lattices are thus given by,
\begin{equation}
L_{\eta}= N_c^h = L - N_c \hspace{0.20cm}{\rm and}\hspace{0.20cm} L_{s} = N_c \, ,
\label{Na-eta-s}
\end{equation}
respectively. The squeezed $\eta$-spin and spin effective lattices remain the same for the $(2S_{\eta}+1)\times (2S_{s}+1)-1$ 
non-LWSs $\vert l_{\rm r},l_{\eta s},u \rangle$ generated from the LWSs, Eq. (\ref{Gstate-BAstate}). Their
configurations in such lattices are those of the non-LWSs of the corresponding $\eta$-spin-$1/2$ and spin-$1/2$ $XXX$ chains with 
$L_{\eta}$ and $L_s$ sites, respectively.

Squeezed spaces are actually well known from studies of the 1D Hubbard model in the $u\rightarrow\infty$ limit 
\cite{Woy,Ogata-90,Karlo-95,Karlo-96,Karlo-97,Zaanen-04}. Such studies have used the $u\rightarrow\infty$ energy 
and momentum eigenstates $\vert  l_{\rm r},l_{\eta s},\infty \rangle$ associated with the BA solution considered 
in this paper. In Appendix \ref{FEL} some of the $u\rightarrow\infty$ properties in terms of
electron occupancy configurations that generate the states $\vert  l_{\rm r},l_{\eta s},\infty \rangle$
are extended to the $u>0$ range in terms of rotated electrons.

The $c$ effective lattice, $\eta$-spin effective lattice, and spin effective lattice occupancy configurations are independent. The 
role of the representations of the $c$-lattice $U(1)$ symmetry generated by the
$c$ pseudoparticle occupancy configurations is indeed to store the information on the positions 
in the original lattice of the $N_{R}^{s}=N_c$ sites singly occupied by rotated electrons ($c$ pseudoparticles)
relative to the $N_{R}^{\eta}=N_c^h$ sites doubly occupied and unoccupied by rotated electrons
($c$ band holes). This ensures that the spin effective lattice occupancies of the $L_s = N_{R}^{s}=N_c$ spins $1/2$ and 
$\eta$-spin effective lattice occupancies of the $L_{\eta} = N_{R}^{\eta}=N_c^c$ $\eta$-spins $1/2$ 
associated with the spin and $\eta$-spin $SU(2)$ symmetries, respectively, are independent.

It follows from such an independence that within the TL the spin $(\alpha =s)$ and $\eta$-spin $(\alpha =\eta)$ 
effective lattice sites locations can be associated with their fixed-$N_c$ subspace average locations at
$x = a_{\alpha}\,j$ where $j=1,...,L_{\alpha}$. The spin effective lattice spacing $a_s$ and $\eta$-spin effective lattice 
spacing $a_{\eta}$ thus correspond to the average spacing between the $c$ effective lattice occupied sites and 
between such a lattice unoccupied sites, respectively, in the corresponding $N_c$-fixed subspace. This gives,
\begin{equation}
a_{\alpha} = {L\over L_{\alpha}} = {L\over L_{\alpha}}\, a
\hspace{0.20cm}{\rm where}\hspace{0.20cm}\alpha = \eta, s \, .
\label{a-alpha}
\end{equation}
This spacing ensures that the $\eta$-spin ($\alpha = \eta$) 
and spin ($\alpha = s$) effective lattices have exactly the same length as the original lattice. 
The effective lattice spacings, Eq. (\ref{a-alpha}), are in general larger than that of the original lattice. 
The exception refers to subspaces for which $n_c\rightarrow 1$ and $n_c^h\rightarrow 1$. For them
these sites numbers read $L_{s}=L\,;L_{\eta}=0$ and $L_{s}=0\,;L_{\eta}=L$, respectively. Hence in these 
two density limits the spin and $\eta$-spin effective lattice becomes the original lattice and the $\eta$-spin and spin effective 
lattice does not exist, respectively. 

\subsection{Unpaired and paired spins and $\eta$-spins }
\label{pseuOpRel2}

The energy eigenstates that span a fixed-$N_c$ subspace
are a superposition of $c$ effective lattice, spin effective lattice, and $\eta$-spin effective lattice occupancy configurations. 
As discussed in Appendix \ref{HMSymmetry}, the two degrees of freedom separation of the rotated-electron occupancies 
of each of the $L$ original lattice sites is behind the two sum rules given in Eq. (\ref{LL}) of that Appendix.

For the energy eigenstates of spin $(\alpha =s)$ or $\eta$-spin $(\alpha =\eta)$ $S_{\alpha}\leq L_{\alpha}/2$ 
in a fixed-$N_c$ subspace, the corresponding spin or $\eta$-spin effective lattice occupancy configurations
have a number $2S_{\alpha}$ of sites occupied by a set of $M_{\alpha}=2S_{\alpha}$ 
spins $1/2$ $(\alpha =s)$ or $\eta$-spins $1/2$ $(\alpha =\eta)$ that participate in the 
(spin or $\eta$-spin) multiplet configuration. They have in addition a
complementary set of even number $L_{\alpha}-2S_{\alpha}$ spin or $\eta$-spin effective lattice sites.
Those are singly occupied by $L_{\alpha}-2S_{\alpha}$ spins $1/2$ or $\eta$-spins $1/2$,
respectively, whose configuration forms a tensor product of (spin or $\eta$-spin) singlet states. 
Such results are those expected from the direct relation to the $\eta$-spin-$1/2$ and spin-$1/2$ $XXX$ chains with 
$L_{\eta}$ and $L_s$ sites, respectively.

Such an analysis applies as well in terms of the spin and $\eta$-spin
degrees of freedom of the original lattice sites rotated-electron occupancy configurations.
Indeed, the spin $1/2$ occupancy configuration order in the spin effective lattice is exactly the same as
that of the spins of the rotated electrons that singly occupy sites in the original lattice. This is independent of the positions
in it of the sites unoccupied and doubly occupied by rotated electrons. 
Similarly, the $\eta$-spin $1/2$ occupancy configuration order in the $\eta$-spin effective lattice is exactly the same as
that of the rotated-electron doubly and unoccupied sites in the original lattice. Again, this is independent of the positions
in it of the sites singly occupied by rotated electrons. 
Consistently, the spin-$1/2$ $XXX$ chain distribution of the squeezed spin wave function
$\phi^{s}_{SU(2)} (x^{s\downarrow},...)$ in Eq. (\ref{amplitude-1D}) of Appendix \ref{FEL}
does not change if the chain of rotated-electron singly occupied sites is ``diluted'' by 
rotated-electron unoccupied and doubly occupied sites. The same applies to the
$\eta$-spin-$1/2$ $XXX$ chain distribution of the squeezed $\eta$-spin wave function
$\phi^{\eta}_{SU(2)} (x^d,...)$ in that equation if the chain of rotated-electron unoccupied and doubly occupied sites is ``diluted'' by 
rotated-electron singly occupied sites. 

All the energy and momentum eigenstates with the same $S_{\alpha}$ have the same 
spin ($\alpha =s$) or $\eta$-spin ($\alpha =\eta$) $\hat{\vec{S}}_{\alpha}^2$ eigenvalue. Therefore,
the energy and momentum eigenstates are superpositions of the corresponding above two types of configuration
terms. Each term in them is characterized by a different partition of $L_{\alpha}$ spins $1/2$ ($\alpha =s$)
or $\eta$-spins $1/2$ ($\alpha =\eta$) into two types of configurations. $M_{\alpha}=2S_{\alpha}$ such spins or $\eta$-spins, respectively, 
participate in a $2S_{\alpha}+1$ (spin or $\eta$-spin) multiplet. The remaining even number $L_{\alpha}-2S_{\alpha}$ of spins 
$1/2$ or $\eta$-spins $1/2$ participate in a product of (spin or $\eta$-spin) singlets. 
The latter are associated with a corresponding number,
\begin{equation}
\Pi_{\alpha} = {1\over 2}(L_{\alpha}-2S_{\alpha})\hspace{0.20cm}{\rm where}\hspace{0.20cm}\alpha = s,\eta \, ,
\label{Pialpha}
\end{equation} 
of spin ($\alpha =s$) or $\eta$-spin ($\alpha =\eta$) singlet pairs. In the following they are
often generally called $\alpha$-singlet pairs.

The {\it unpaired spins} and {\it paired spins} ($\alpha =s$) and
{\it unpaired $\eta$-spins} and {\it paired $\eta$-spins} ($\alpha =\eta$)
are the members of such two sets of $M_{\alpha} = 2S_{\alpha}$ and 
$2\Pi_{\alpha} = L_{\alpha}-2S_{\alpha}$, respectively, spins $1/2$ and $\eta$-spins $1/2$. 
For a spin and $\eta$-spin LWS, all unpaired spins $1/2$
and unpaired $\eta$-spins $1/2$, respectively, have projection $+1/2$. 

The number of pairs $\Pi_{\alpha}$ is directly related to the spin $SU(2)$ symmetry ($\alpha =s$) and $\eta$-spin 
$SU(2)$ symmetry ($\alpha =\eta$) in the $[SO(4)\otimes U(1)]/Z_2 = [SU(2)\otimes SU(2)\otimes U(1)]/Z_2^2$
global symmetry of the $u>0$ 1D Hubbard model Hamiltonian. Indeed, the expression, Eq. (\ref{N-singlet}) of Appendix 
\ref{HMSymmetry}, of the number ${\cal{N}}_{\rm singlet} (S_{\alpha},L_{\alpha})$ of that model independent spin ($\alpha =s$) 
and $\eta$-spin ($\alpha =\eta$) $\alpha$-singlet state representations in a fixed-$N_c$ and 
fixed-$S_{\alpha}$ subspace is a function of only the number of pairs $\Pi_{\alpha}$ and of the number of 
spins $1/2$ ($\alpha =s$) and $\eta$-spins $1/2$ ($\alpha =\eta$) $L_{\alpha}$. 

For general $u>0$ LWSs and their non-LWSs one finds that the number $M_{s,\pm 1/2}$ of unpaired 
spins of projection $\pm 1/2$ and $M_{\eta,\pm 1/2}$ of unpaired $\eta$-spins of projection $\pm 1/2$ 
are good quantum numbers. They read,
\begin{equation}
M_{\alpha,\pm 1/2} = (S_{\alpha}\mp S_{\alpha}^{z}) 
\hspace{0.20cm}{\rm and}\hspace{0.20cm}M_{\alpha} = M_{\alpha,-1/2}+M_{\alpha,+1/2} = 2S_{\alpha} 
\hspace{0.20cm}{\rm where}\hspace{0.20cm}\alpha=\eta, s \, .
\label{L-L}
\end{equation}
The set of an energy and momentum eigenstate $\Pi_{\eta}$ $\eta$-spin-singlet pairs and $\Pi_{s}$ spin-singlet pairs 
contains an equal number of $\eta$-spins $1/2$ and spins $1/2$, respectively, of opposite projection. 
Hence the total number $L_{\alpha,\pm 1/2}$ of $\eta$-spins of projection $\pm 1/2$ $(\alpha =\eta)$
and spins of projection $\pm 1/2$ $(\alpha =s)$ is given by,
\begin{equation}
L_{\alpha,\pm 1/2} = \Pi_{\alpha} + M_{\alpha,\pm 1/2} 
= {1\over 2}(L_{\alpha} \mp 2S_{\alpha}^z) \hspace{0.20cm}{\rm where}\hspace{0.20cm}\alpha = \eta,s \, .
\label{Mtotal}
\end{equation}

The $\eta$-spin and spin $SU(2)$ symmetry algebras diagonal generators ${\hat{S}}_{\eta}^{z}$ and ${\hat{S}}_s^{z}$, Eq. (\ref{etaSz}), 
and off-diagonal generators ${\hat{S}}^{+}_{\eta}$, ${\hat{S}}_{\eta} = ({\hat{S}}^{+}_{\eta})^{\dag}$  and ${\hat{S}}^{+}_{s}$, 
${\hat{S}}_{s} = ({\hat{S}}^{+}_{s})^{\dag}$, Eq. (\ref{Gstate-BAstate}), commute with the electron-rotated-electron unitary operator ${\hat{V}}$, 
Eq. (\ref{rotated-operators}). Hence such operators have the same expressions
in terms of electron and rotated-electron operators. The $M_{\alpha}=2S_{\alpha}$ unpaired spins $(\alpha =s)$ and 
unpaired $\eta$-spins $(\alpha =\eta)$ multiplet configurations are as given in Eq. (\ref{Gstate-BAstate})
generated by application of the $\alpha =\eta,s$ operators ${\hat{S}}^{+}_{\alpha}$ onto the LWSs. Therefore, 
the corresponding non-LWSs original lattice spatial occupancy configurations of the unpaired spins and unpaired $\eta$-spins 
generated from the LWSs are for the whole $u>0$ range exactly the same in terms of rotated electrons and
electrons, respectively. Indeed, for $u>0$ such local configurations remain invariant under the electron-rotated-electron unitary transformation. 
Hence the unpaired spins $(\alpha =s)$ and unpaired $\eta$-spins $(\alpha =\eta)$ are for $u>0$ not rotated. They thus refer to 
electron unpaired physical spins $1/2$ and to the $\eta$-spin degrees of freedom of physical onsite spin-singlet electron pairs, respectively. 

The $\alpha$-singlet configurations of the $2\Pi_{\alpha} = L_{\alpha} - 2S_{\alpha}$ paired 
spins $1/2$ $(\alpha =s)$ and paired $\eta$-spins $1/2$ $(\alpha =\eta)$ left over
are though also physical spins $1/2$ and physical $\eta$-spins $1/2$ in what their spin and $\eta$-spin degrees of 
freedom, respectively, is concerned. Only their original lattice spatial occupancies are changed under the 
electron-rotated-electron unitary transformation.

%%%%%%%%%%%%%%%%%%%%%%%%%%%%%%%%%%%%%%%%%%%%%%%%%%%%%%%%%%%%%%%%%%%%%%%%%%
\section{The 1D Hubbard model $c$ and $\alpha n$ pseudoparticle representation}
\label{rot-symm-2}

The relation of the different types of the 1D Hubbard model pseudoparticles and band holes to its physical particles, 
the electrons, is much more involved than for the 1D Lieb-Liniger Bose gas and spin-$1/2$ $XXX$ chain. Nonetheless, the
relation of the latter model physical spins $1/2$ to its $n$-pseudoparticles and holes plays a valuable role
in the study of the corresponding more complex problem of the 1D Hubbard model.

Here the functional representation of the 1D Hubbard model TBA solution is related to
the three basic fractionalized particles that naturally arise from the rotated electrons 
degrees of freedom separation. The composite $\alpha n$ pseudoparticles emerge
from such a relation. The charge and spin current carriers and the general $c$-band
and $\alpha n$-bands hole representation is an issue also discussed in this section.

\subsection{The functional representation of the 1D Hubbard model TBA solution}
\label{repres-vacuum-GS}

Some of the 1D Hubbard model TBA solution quantities and equations introduced in Ref. \cite{Takahashi-72} 
needed for our analysis are provided here within a suitable distribution functional representation. 
The model TBA equations are within such a representation given by,
\begin{eqnarray}
q_j & = & k^c (q_j) + {2\over L}\sum_{n =1}^{\infty}
\sum_{j'=1}^{L_{s n}}\,N_{sn}(q_{j'})\arctan\left({\sin
k^c (q_j)-\Lambda^{sn}(q_{j'}) \over n u}\right)
\nonumber \\
& + & {2\over L}\sum_{n =1}^{\infty}
\sum_{j'=1}^{L_{\eta n}}\, N_{\eta n}(q_{j'}) \arctan\left({\sin
k^c (q_j)-\Lambda^{\eta n}(q_{j'}) \over n u}\right) 
\hspace{0.20cm}{\rm for}\hspace{0.20cm}j = 1,...,L \, , 
\label{Tapco1}
\end{eqnarray}
and
\begin{eqnarray}
q_j & = & \delta_{\alpha,\eta}\sum_{\iota = \pm1}\arcsin (\Lambda^{\alpha n} (q_{j}) - i\,\iota\,nu)
+ {2\,(-1)^{\delta_{\alpha,\eta}}\over L} \sum_{j'=1}^{L_c}\,
N_{c}(q_{j'})\arctan\left({\Lambda^{\alpha n}(q_j)-\sin k^c (q_{j'})\over n u}\right)
\nonumber \\
& - & {1\over L}\sum_{n' =1}^{\infty}\sum_{j'=1}^{L_{\alpha n'}}\, N_{\alpha n'}(q_{j'})\Theta_{n\,n'}
\left({\Lambda^{\alpha n}(q_j)-\Lambda^{\alpha n'}(q_{j'})\over u}\right) \hspace{0.20cm}{\rm for}\hspace{0.20cm}
j = 1,...,L_{\alpha n} \, ,
\nonumber \\
& & {\rm where}\hspace{0.20cm} \alpha = \eta, s \hspace{0.20cm}{\rm and}\hspace{0.20cm}n =1,...,\infty \, .
\label{Tapco2}
\end{eqnarray}
The BA $\beta =c,\alpha n$ branches numbers $L_{\beta}$ appearing in these equations read,
\begin{eqnarray}
L_{c} & = & N_{c} + N^h_{c} = N^R_{s} + N^R_{\eta} = L  \hspace{0.20cm}{\rm where}\hspace{0.20cm}N^h_c = L - N_c \, ,
\nonumber \\
L_{\alpha n} & = & N_{\alpha n} + N^h_{\alpha n} \hspace{0.20cm}{\rm where}\hspace{0.20cm}
N^h_{\alpha n} = 2S_{\alpha}+\sum_{n'=n+1}^{\infty}2(n'-n)N_{\alpha n'} \, , 
\nonumber \\
N_{\beta} & = & \sum_{j=1}^{L_{\beta}}\, N_{\beta}(q_{j}) 
\hspace{0.20cm}{\rm where}\hspace{0.20cm}\beta = c,\eta n, sn
\hspace{0.20cm}{\rm and}\hspace{0.20cm}n =1,...,\infty \, .
\label{N-h-an}
\end{eqnarray}
The function $\Theta_{n\,n'} (x)$ in Eqs. (\ref{Tapco1}) and (\ref{Tapco2}) is given in Eq. (\ref{Theta}) of 
Appendix \ref{TBAconfig} and the $\beta$-branch discrete quantum numbers $q_j$ read,
\begin{equation}
q_j = {2\pi\over L}\,I^{\beta}_j \hspace{0.20cm}{\rm for}\hspace{0.20cm}j=1,...,L_{\beta} 
\hspace{0.20cm}{\rm where}\hspace{0.20cm}\beta = c,\eta n,sn \hspace{0.20cm}{\rm and}\hspace{0.20cm}n =1,...,\infty \, .
\label{q-j}
\end{equation} 
Here $\{I^{\beta}_j\}$ are the $\beta$-branch $j=1,...,L_{\beta}$ quantum numbers $\{q_j\}$ in units of $2\pi/L$.
Those are either integers or half-odd integers according to the following boundary conditions \cite{Takahashi-72},
\begin{eqnarray}
I_j^{\beta} & = & 0,\pm 1,\pm 2,... \hspace{0.20cm}{\rm for}\hspace{0.15cm}I_{\beta}\hspace{0.15cm}{\rm even} \, ,
\nonumber \\
& = & \pm 1/2,\pm 3/2,\pm 5/2,... \hspace{0.20cm}{\rm for}\hspace{0.15cm}I_{\beta}\hspace{0.15cm}{\rm odd} \, ,
\label{Ic-an}
\end{eqnarray}
where,
\begin{equation}
I_c = N^{SU(2)} \equiv \sum_{\alpha =\eta,s}\sum_{n=1}^{\infty}N_{\alpha n} 
\hspace{0.20cm}{\rm and}\hspace{0.20cm}
I_{\alpha n} = L_{\alpha n} -1\hspace{0.20cm}{\rm for}\hspace{0.20cm}
\alpha = \eta, s\hspace{0.20cm}{\rm and}\hspace{0.20cm}n=1,...,\infty \, .
\label{F-beta}
\end{equation}

The $\beta = c,\alpha n$ branch successive set of discrete values $q_j$, Eq. (\ref{q-j}), 
have fixed separation, $q_{j+1}-q_{j}=2\pi/L$. In addition, they have only occupancies zero and one. 
The $\beta$-branch distribution functions $N_{\beta} (q_j)$ in Eqs. (\ref{Tapco1}) and (\ref{Tapco2}) thus read $N_{\beta} (q_j)=1$ and 
$N_{\beta} (q_j)=0$ for occupied and unoccupied such discrete values, respectively.
Each energy and momentum eigenstate is described by different occupancy configurations
of the distributions $\{N_{\beta} (q_j)\}$ corresponding to all BA $\beta =c,\eta n, sn$ branches 
where $n=1,...,\infty$. The numbers $N_{\beta}$ and $N_{\beta}^h$ defined in Eq. (\ref{N-h-an})
are thus those of occupied and unoccupied, respectively, $\beta$-branch discrete values $q_j$, Eq. (\ref{q-j}). 

Solution of the coupled TBA equations, Eqs. (\ref{Tapco1}) and (\ref{Tapco2}), provides the
real momentum rapidity function $k^c (q_j)$ and the set of $n=1,...,\infty$ real rapidity functions 
$\Lambda_{\eta n}(q_{j})$ and $\Lambda_{sn}(q_{j})$ of each energy and momentum eigenstate.
Quantities such as the energy eigenvalues given in Eqs. (\ref{E})-(\ref{2mu0}) of Appendix \ref{TBAconfig}
\cite{Lieb,Lieb-03,Ovchi-70,Takahashi-72} and the charge and spin current operators expectation values depend 
on the $\beta$-branches discrete values $q_j$ through the dependence on them of the
momentum rapidity function $k^c (q_j)$ and the $\alpha =\eta,s$ and $n=1,...,\infty$ rapidity functions $\Lambda_{\alpha n}(q_{j})$.
The latter are the real part of TBA complex rapidities of general form,
\begin{equation}
\Lambda^{\alpha n,l}(q_{j}) = \Lambda^{\alpha n} (q_{j}) + i\,(n + 1 - 2l)\,u \hspace{0.20cm}{\rm where}\hspace{0.20cm}
\alpha = \eta,s \, , \hspace{0.20cm}n = 1,...,\infty \hspace{0.20cm}{\rm and}\hspace{0.20cm}l = 1,...,n \, .
\label{complex-rap}
\end{equation}
For $n=1$ this rapidity is real and otherwise its imaginary part is finite. A TBA $\alpha n$-string is a group 
of $l = 1,...,n$ rapidities, Eq. (\ref{complex-rap}), all with the same real 
part, $\Lambda^{\alpha n} (q_{j})$. For $\alpha =s$ and $\alpha =\eta$ 
those are the spin and charge, respectively, $\alpha n$-strings \cite{Deguchi-00}.
As for the spin-$1/2$ $XXX$ chain \cite{Caux-07}, for a large finite 
system some of the 1D Hubbard model 
$\alpha n$-strings deviate from their TBA ideal form, Eq. (\ref{complex-rap}).
The effects of such string deviations \cite{Deguchi-00} are in the TL though not important
for the properties considered in this paper.

As in the case of the simpler models also reviewed in it, the discrete quantum numbers $q_j$ in Eq. (\ref{q-j}) play 
the role of $\beta = c,\alpha n$ band momentum values. Consistently, the momentum eigenvalues are additive in $q_j$
and read,
\begin{equation}
P =\sum_{j=1}^{L} q_j\, N_c (q_j)
+ \sum_{n =1}^{\infty}\sum_{j=1}^{L_{s n}}
q_{j}\, N_{sn} (q_{j}) 
+ \sum_{n =1}^{\infty}\sum_{j=1}^{L_{\eta n}}
(\pi -q_{j})\, N_{\eta n} (q_{j}) + \pi L_{\eta,-1/2} \, .
\label{P}
\end{equation}
The momentum contribution $\pi L_{\eta,-1/2}=\pi (M_{\eta} + M_{\eta,-1/2})$
involves the number $L_{\eta,-1/2}$ of $\eta$-spins of projection $-1/2$, Eq. (\ref{Mtotal}) for $\alpha =\eta$.
Such a contribution follows from the paired and unpaired spins $1/2$ and 
$\eta$-spins $1/2$ of projection $\pm 1/2$ having an {\it intrinsic momentum} given by,
\begin{equation}
q_{s,\pm 1/2} = q_{\eta,+1/2} = 0  \hspace{0.20cm}{\rm and}\hspace{0.20cm}q_{\eta,-1/2} = \pi \, .
\label{q-eta-s}
\end{equation}
The set $j=1,...,L_{\beta}$ of $\beta =c, \alpha n$ bands discrete momentum values $q_j$ belong to well-defined domains, 
$q_j\in [q_{\beta}^-,q_{\beta}^+]$. The limiting momenta $q_{\beta}^{\pm}$ appearing here are given in Eq. (\ref{qcan-range}) of Appendix
\ref{TBAconfig}.

The momentum and energy spectra, Eq. (\ref{P}) and Eq. (\ref{E}) of Appendix \ref{TBAconfig}, apply to all $4^{L}$ energy eigenstates
$\{\vert l_{\rm r},l_{\eta s},u\rangle\}$, Eq. (\ref{Gstate-BAstate}). Their label $l_{\rm r}$ 
can now be defined. It corresponds to a short notation for the following set of TBA quantum numbers, 
\begin{equation}
l_{\rm r} = \{I_j^{\beta}\}\,\,{\rm such}\,\,{\rm that}\,\,N_{\beta} (q_j) = N_{\beta} ([2\pi/L]I_j^{\beta}) =1
\hspace{0.20cm}{\rm for}\hspace{0.20cm}j = 1,...,L_{\beta} \, , \hspace{0.20cm} \beta = c, \eta n, sn
\hspace{0.20cm}{\rm and}\hspace{0.20cm}n = 1,...,\infty \, .
\label{states-ll}
\end{equation}

The TBA equations, Eqs. (\ref{Tapco1}) and (\ref{Tapco2}), refer explicitly to LWSs. However, they can be extended to non-LWSs,
Eq. (\ref{Gstate-BAstate}) \cite{Woy-82,Woy,Braak-99}.
This can be achieved by formally setting some of the rapidities $\Lambda^{\eta n}$ and $\Lambda^{sn}$ 
in such equations equal to infinity \cite{Woy-82,Woy}. For example, Eqs. (3.23b) and
(3.24b) of Ref. \cite{Woy-82} describe a $\eta$-spin non-LWS with numbers $S_{\eta} = 1$ and 
$S_{\eta}^{z} = 0$. Moreover, Eqs. (3.23a) and (3.24a) of Ref. \cite{Woy-82} describe a 
LWS with numbers $S_{\eta} = S_{\eta}^{z} = 0$, Eq. (\ref{Gstate-BAstate}). 
Alternatively, in Eq. (\ref{Gstate-BAstate}) one has combined symmetry
with the BA solution to generate the non-LWSs from the LWSs \cite{Completeness}. 

\subsection{The composite $\alpha n$ pseudoparticles associated with the paired $\eta$-spins ($\alpha =\eta$)
and paired spins ($\alpha =s$)}
\label{alphanpseudop}

It was confirmed in Refs. \cite{Completeness,Complete2,Complete3} that
the TBA quantum number configurations combined with the spin and $\eta$-spin $SU(2)$ multiplet 
configurations generate the $4^L$ energy eigenstates that span the 1D Hubbard model Hilbert
space. Beyond the analysis of Refs. \cite{Completeness,Complete2,Complete3}, the Hilbert-space dimension $4^L$ also equals 
the number of independent state representations of the 1D Hubbard model global $[SU(2)\otimes SU(2)\otimes U(1)]/Z_2^2$ 
symmetry. (In 1991 and 1992 only the $SO (4)$ symmetry in the $u>0$ model global $[SO(4)\otimes U(1)]/Z_2$ symmetry 
\cite{bipartite} was known \cite{Completeness,Complete2,Complete3}.)

The proof involves the requirement addressed in Appendix \ref{HMSymmetry} that in any 
spin ($\alpha =s$) and $\eta$-spin ($\alpha =\eta$) $S_{\alpha}$-fixed subspace the number 
of independent $\alpha$-singlet configurations ${\cal{N}}_{\rm singlet} (S_{\alpha})$ is {\it exactly} the same when 
obtained from the counting of two apparently different types of configurations. (This is similar to the
spin configurations of the $XXX$ chain.) The first type of configurations refers 
to the two $\alpha =\eta,s$ $SU(2)$ group states representations associated with
the spins $1/2$ ($\alpha =s$) and $\eta$-spins $1/2$ ($\alpha =\eta$) independent configurations with the same 
spin and $\eta$-spin, respectively, $S_{\alpha}$, Eq. (\ref{N-singlet}) of Appendix \ref{HMSymmetry}.
The second type of configurations corresponds to the independent $n=1,...,\infty$ bands $\{q_j\}$ occupancy configurations
of the sets of $N_{\alpha n}$ $\alpha n$-strings obeying the $\alpha =\eta,s$ TBA sum rules 
$\sum_{n=1}^{\infty}n\,N_{\alpha n} = (L_{\alpha}-2S_{\alpha})/2$.

It follows that the set of $\alpha n$-strings of an energy and momentum eigenstate is directly related to the set of $\Pi_{\alpha}$ spin 
($\alpha =s$) and $\eta$-spin ($\alpha =\eta$) $\alpha$-singlet pairs, Eq. (\ref{Pialpha}). 
(This is as for the spin-singlet pairs of the spin-$1/2$ $XXX$ chain.)
Such pairs involve the subset of $2\Pi_{\alpha}=L_{\alpha}-2S_{\alpha}$ 
spins $1/2$ and $\eta$-spins $1/2$, respectively, that participate in $\alpha$-singlet configurations. 
Specifically, each $\alpha n$-string refers to an $\alpha n$-pairs configuration within which a number $n>1$ of $\alpha$-singlet pairs are 
bound. For $n>1$ such a binding is associated 
with the corresponding imaginary parts, $i\,(n + 1 - 2l)\,u$, of the $l = 1,...,n$ rapidities, 
$\Lambda^{\alpha n,l}(q_{j}) = \Lambda^{\alpha n} (q_{j})+i\,(n + 1 - 2l)\,u$, Eq. (\ref{complex-rap}),
with the same real part, $\Lambda^{\alpha n} (q_{j})$. For $n=1$ an $\alpha n$-string involves a single $\alpha$-singlet pair. 

The $n>1$ $\alpha$-singlet pairs that are bound within an $\alpha n$-pairs configuration associated with a string of length $n>1$
are here called {\it bound spin-singlet pairs} ($\alpha =s$) and {\it bound $\eta$-spin-singlet pairs} ($\alpha =\eta$). 
For $n=1$ the rapidity $\Lambda^{\alpha 1,1}(q_{j})$ imaginary part vanishes because a 
$n=1$ $\alpha n$-string reduces to a single $\alpha$-singlet pair. 
The {\it unbound spin-singlet pairs} ($\alpha =s$) and {\it unbound $\eta$-spin-singlet pairs} ($\alpha =\eta$) 
of an energy eigenstate are the $N_{\alpha 1}$ $\alpha$-singlet pairs that refer to the 
$N_{\alpha 1}$ $n=1$ $\alpha n$-pairs configurations.

The numbers $\Pi_{\alpha} = (L_{\alpha} - 2S_{\alpha})/2$ of spin $(\alpha =s)$ and $\eta$-spin $(\alpha =\eta)$ $\alpha$-singlet 
pairs classify the $u>0$ energy and momentum eigenstates associated with the BA solution in two different yet 
related and complementary ways. On the one hand, since $L_{\eta} = L - N_c$ and $L_{s} = N_c$, they are
amid the quantum numbers of that solution. This follows from $N_c$, the spin $S_s$, and the $\eta$-spin $S_{\eta}$ being good 
quantum numbers that classify the corresponding $u>0$ energy and momentum eigenstates. On the other hand,
each of the numbers $l = 1,...,n$ and the number $n$ that classify a TBA $\alpha n$-string and corresponding set
of $l = 1,...,n$ rapidities, $\Lambda^{\alpha n,l}(q_{j}) = \Lambda^{\alpha n} (q_{j})+i\,(n + 1 - 2l)\,u$, 
Eq. (\ref{complex-rap}), with the same real part, $\Lambda^{\alpha n} (q_{j})$, refer to one such pairs
and to their number, respectively.

The (above mentioned) following exact TBA sum rules hold for all $u>0$ energy and momentum eigenstates,
\begin{eqnarray}
\Pi_{\alpha} & = & \sum_{n=1}^{\infty}n\,N_{\alpha n} = {1\over 2}(L_{\alpha} - 2S_{\alpha})
\hspace{0.20cm}{\rm where}\hspace{0.20cm}\alpha = s, \eta \, ,
\nonumber \\
\Pi^{SU(2)} & \equiv & \sum_{\alpha =\eta,s}\Pi_{\alpha} =
\sum_{\alpha =\eta,s}\sum_{n=1}^{\infty}n\,N_{\alpha n} = {1\over 2}(L - 2S_s - 2S_{\eta}) \, .
\label{sum-Nseta}
\end{eqnarray}
This is consistent with the relation of the set of $\sum_{n=1}^{\infty}n\,N_{\alpha n}$ TBA $\alpha n$-strings 
of all lengths $n=1,...,\infty$ of such a state to the set of $\Pi_{\alpha}$ spin ($\alpha =s$) and $\eta$-spin 
($\alpha =\eta$) $\alpha$-singlet pairs.

$\Pi^{SU(2)}$ denotes in Eq. (\ref{sum-Nseta}) the total number of both spins and $\eta$-spins singlet pairs and
$N_{\alpha n}$ is the number of $\alpha n$-pairs configurations that equals that of 
$\alpha n$-band discrete momentum values $q_j$ that are occupied. Below in Section \ref{BANTIB} it is shown that the configuration of
the two spins within one such unbound spin-singlet pair and that of the
two $\eta$-spins within one unbound $\eta$-spin-singlet pair has a binding and
anti-binding character, respectively. This applies to the internal structure of all
$\Pi_{s}$ spin-singlet pairs and all $\Pi_{\eta}$ $\eta$-spin-singlet pairs, respectively,
of a $u>0$ energy and momentum eigenstate.

There is a one-to-one correspondence between the
$N_{\alpha n}$ $\alpha n$-pairs configurations with the same number $n$ of $\alpha$-singlet
pairs of an energy eigenstate and the $N_{\alpha n}$ occupied 
momentum values $q_j$ of the corresponding $\alpha n$-band distribution $N_{\alpha n} (q_j)$, respectively. 
An $\alpha n$-pairs configuration involves a set of $2n$ paired $\eta$-spins $1/2$ ($\alpha =\eta$) or
paired spins $1/2$ ($\alpha =s$). They singly occupy a set of $2n$ original-lattice sites. The use of the TBA equations
given in the previous section reveals that their center of mass moves with momentum $q_j$. All the $2n$ 
paired spins $1/2$ ($\alpha =s$) or paired $\eta$-spins $1/2$ ($\alpha =\eta$) move coherently along with it. 
This occurs through processes within which such $2n$ paired spins $1/2$ or paired $\eta$-spins $1/2$
interchange position with the $M_{\alpha}=2S_{\alpha}$ unpaired spins $1/2$ or 
unpaired $\eta$-spins $1/2$, respectively. Between each such a elementary process,
both the latter and the $2n$ paired $\eta$-spins $1/2$ ($\alpha =\eta$) or
paired spins $1/2$ ($\alpha =s$) singly occupy original-lattice sites. 

We associate one {\it $\alpha n$ pseudoparticle} and one {\it $\alpha n$-band hole} with each of the $N_{\alpha n}$ occupied 
and $N_{\alpha n}^h$ unoccupied momentum values $q_j$, respectively, of an $u>0$ energy eigenstate
$\alpha n$-band. Such composite $\alpha n$ pseudoparticles are well defined for $u>0$ within the TL to which the TBA applies.
The TL ensures that the problems concerning the $\alpha n$ pseudoparticle 
internal degrees of freedom and translational degrees of freedom, respectively, separate.

On the one hand, the internal degrees of freedom of a composite $\alpha n$ pseudoparticle refer to
an $\alpha n$-pairs configuration. Hence there is one $\alpha n$ pseudoparticle for each $\alpha n$-pairs configuration
and corresponding BA roots that involve a group of $l = 1,...,n$ rapidities with the same real part, 
Eq. (\ref{complex-rap}). If $n>1$ the composite $\alpha n$ pseudoparticle has $n= 2,...,\infty$ $\alpha$-singlet pairs bound within it.
If $n=1$ its internal degrees of freedom correspond to a single unbound $\alpha$-singlet pair.

On the other hand, the momentum $q_j$, Eq. (\ref{q-j}), of an $\alpha n$ pseudoparticle refers to its
translational degrees of freedom. It is associated with its center of mass motion. The set
of $N_{\alpha}=\sum_{n=1}^{\infty}N_{\alpha n}$ $\alpha n$ pseudoparticles, each carrying a momentum $q_j$, 
of a given energy eigenstate determine such a state momentum eigenvalue, as given in Eq. (\ref{P}). 
That the $\eta n$ pseudoparticles contribution reads $(\pi -q_{j})$ rather than $q_j$, follows from the configuration of the two 
$\eta$-spins in each $\eta$-spin-singlet pair having an anti-binding character, as confirmed below in Section \ref{BANTIB}.

The $l = 1,...,n$ $\alpha$-singlet pairs within each of the $N_{\alpha n}$ $\alpha n$ pseudoparticles that populate 
the $n=1,...,\infty$ $\alpha n$-bands of an energy and momentum eigenstate have spin ($\alpha =s$) and 
$\eta$-spin ($\alpha =\eta$) zero. The corresponding composite $\alpha n$ pseudoparticles are thus neutral particles. 
The $c$ pseudoparticles have in turn no internal degrees of freedom. Their occupancy configurations generate the state representations
of the $c$ lattice $U(1)$ symmetry. It is independent from the model two $SU(2)$ symmetries. Hence the
energy eigenstates spin and spin projection ($\alpha =s$) and $\eta$-spin and $\eta$-spin projection ($\alpha =\eta$),
$S_{\alpha}$ and $S_{\alpha}^z$, are determined solely by their numbers of unpaired spins $1/2$ and unpaired $\eta$-spins $1/2$
of projections $\pm 1/2$. Specifically, $S_{\alpha} = (M_{\alpha,+1/2}+M_{\alpha,-1/2})/2=M_{\alpha}/2$
and $S_{\alpha}^z = - (M_{\alpha,+1/2}-M_{\alpha,-1/2})/2$, respectively.

As for the spin-$1/2$ $XXX$ chain, there is a number of $\alpha n$ pseudoparticles sum rule. It is related to 
that of $\alpha$-singlet pairs, Eq. (\ref{sum-Nseta}). The latter sum rule implies that 
$N_{\alpha 1} = L_{\alpha}/2 - S_{\alpha} - \sum_{n=2}^{\infty}n\,N_{\alpha n}$. From the use of this relation 
in the overall number of $\alpha n$ pseudoparticles expression, $N_{\alpha}=\sum_{n=1}^{\infty}N_{\alpha n}$, one confirms
that the following sum rules are obeyed,
\begin{eqnarray}
N_{s} & = & \sum_{n=1}^{\infty}N_{s n} = {1\over 2}(N_c - N_{s 1}^h) 
\hspace{0.20cm}{\rm and}\hspace{0.20cm}
N_{\eta} = \sum_{n=1}^{\infty}N_{\eta n} = {1\over 2}(N_c^h - N_{\eta 1}^h) \, ,
\nonumber \\
N^{SU(2)} & = & \sum_{\alpha =\eta,s}N_{\alpha} =
\sum_{\alpha =\eta,s}\sum_{n=1}^{\infty}N_{\alpha n} = {1\over 2}(L - N_{s 1}^h - N_{\eta 1}^h) \, . 
\label{NpsNapsSR}
\end{eqnarray}
Here $N_{\alpha 1}^h$ is the number of $\alpha 1$-band holes, 
Eq. (\ref{N-h-an}) for $\alpha = \eta,s$ and $n =1$. $N^{SU(2)}=\sum_{\alpha =\eta,s}N_{\alpha}$
is that in Eq. (\ref{F-beta}).

In contrast to the spin $1/2$-$XXX$ chain, the imaginary parts, $i\,(n + 1 - 2l)\,u$,
of each set of $l = 2,...,n$ rapidities with the same real part depend on the interaction $u=U/4t$ and thus vanish 
as $u\rightarrow 0$. Such a set of $l = 2,...,n$ rapidities describes $n>1$ $\alpha$-singlet pairs
bound within an $\alpha n$-pairs configuration. The vanishing of such rapidities imaginary parts thus gives rise to
the unbinding of all $\alpha$-singlet pairs. One finds that the two $\eta$-spins $1/2$ ($\alpha =\eta$) or
spins $1/2$ ($\alpha =s$) of each pair remain contributing to singlet configurations, yet each carries
an independent virtual elementary charge or spin current, respectively. This reveals that the corresponding composite 
$\alpha n$ pseudoparticles are only well defined for $u>0$. Such a unbinding marks for finite transfer integral 
$t$ the qualitatively different physics of the $U=0$ and $U>0$ quantum problems, 
respectively. It is associated with the rearrangement of the $\eta$-spin and spin
degrees of freedom in terms of the noninteracting electrons occupancy configurations that
generate the finite-$t$ and $U=0$ energy and momentum eigenstates.

In an extended Takahashi subspace as defined in Appendix \ref{s1pseudoOR}, the numbers
of discrete momentum values $L_{\alpha n}$ in Eq. (\ref{N-h-an}) of all $\alpha n$-bands
for which $N_{\alpha n}>0$ remain fixed. For the 1D Hubbard model in such subspaces
one associates each $j=1,...,L_{\alpha n}$ momentum values $q_j$ $\alpha n$-band, 
Eq. (\ref{q-j}), with a corresponding squeezed $\alpha n$ effective lattice with $j=1,...,L_{\alpha n}$ sites and
length $L$. Provided that the ratio $L_{\alpha n}/L$ remains finite as $L\rightarrow\infty$, in the TL such
squeezed $\alpha n$ effective lattices can be represented by 1D lattices. Their spacing
corresponds to the extended Takahashi subspace average distance of their $L_{\alpha n}$ sites,
\begin{equation}
a_{\alpha n} = {L\over L_{\alpha n}} = 
{L\over L_{\alpha n}}\, a = {L_{\alpha}\over L_{\alpha n}}\, a_{\alpha} 
\hspace{0.20cm}{\rm where}\hspace{0.20cm}\alpha =\eta ,s\hspace{0.20cm}{\rm and}\hspace{0.20cm}n = 1,...,\infty \, .
\label{a-a-nu}
\end{equation}
Therefore, the $\alpha n$ effective lattice length equals that of the original lattice.
The corresponding sites then have spatial coordinates, $a_{\alpha n}\,j$, where $j=1,...,L_{\alpha n}$.
Each composite $\alpha n$ pseudoparticle singly occupied site of the $\eta n$ (and $sn$) effective lattice describes an $\eta$-spin-singlet (or
spin-singlet) occupancy configuration. It involves a set of $2n$ paired $\eta$-spins $1/2$ (or $2n$ paired spins $1/2$)
on $2n = 2,...,\infty$ sites of the original lattice. 

For the PDT, only the $c$ and $s1$ pseudoparticle and corresponding $c$ and
$s1$ pseudofermion operator algebras are explicitly needed. Hence for simplicity
we limit our present analysis to the $\alpha n = s1$ pseudoparticle operator algebra.
The corresponding operator representation is valid for the 1D Hubbard
model in fixed-$L_{s1}$ extended Takahashi subspaces. 
In such subspaces the local $s1$ pseudoparticle operators obey a fermionic algebra, 
\begin{equation}
\{f^{\dag}_{j,s1}\, ,f_{j',s1}\} = \delta_{j,j'} 
\hspace{0.20cm}{\rm and}\hspace{0.20cm}
\{f_{j,s1}^{\dag}\, ,f_{j',s1}^{\dag}\} =
\{f_{j,s1}\, ,f_{j',s1}\} = 0 \, .
\label{ffs1}
\end{equation}
This can be confirmed in terms of their statistical interactions \cite{Haldane-91}. Such a problem is 
addressed in Appendix \ref{s1pseudoOR}. (Consistently, the TBA $\beta =c,s1$ band 
momentum value $q_j$ have only occupancies zero and one.) 
Each of the $N_{s1}$ occupied $s1$ effective lattice sites corresponds to a spin-singlet pair. It
involves two original lattice sites occupied by two paired spins $1/2$ of opposite spin projection.

The $s1$ pseudoparticle translational degrees of freedom center of mass motion are 
described by operators $f^{\dag}_{j,s1}$ (and $f_{j,s1}$). They create (and annihilate) one $s1$ pseudoparticle 
at the $s1$ effective lattice site $x_{j}=a_{s1}\,j$. Here $j = 1,...,L_{s1}$ and $L_{s1}$ is given in Eq. (\ref{N-h-an}) 
for $\alpha n=s1$. This is as for the local creation and annihilation $c$ pseudoparticle operators, 
Eq. (\ref{albegra-cf}).

The $\beta =c,s1$ pseudoparticle operators labeled by the corresponding $\beta =c,s1$ bands 
$j = 1,...,L_{\beta}$ momentum values $q_j$ defined in Eqs. (\ref{q-j}) and (\ref{Ic-an}) then read, 
\begin{equation}
f^{\dag}_{q_j,\beta} = {1\over \sqrt{L}}\sum_{j'=1}^{L_{\beta}}e^{i\,q_j\,x_{j'}}f^{\dag}_{j',\beta}
\hspace{0.20cm}{\rm and}\hspace{0.20cm} 
f_{q_j,\beta} = (f^{\dag}_{q_j,\beta})^{\dag}
\hspace{0.20cm}{\rm where}\hspace{0.20cm}j = 1,...,L_{\beta}\hspace{0.20cm}{\rm and}\hspace{0.20cm}\beta = c,s1 \, .
\label{f-f-FT}
\end{equation}
Such momentum values $q_j$ are the quantum numbers of the exact BA solution whose occupancy configurations generate 
the $u>0$ energy and momentum eigenstates,

Besides acting within fixed-$L_{s1}$ extended Takahashi subspaces, the $s1$ pseudoparticle operators
labeled by momentum $q_j$ also appear in the expressions of the shake-up effects generators. Such
generators transform extended Takahashi subspaces quantum number values into each other.

%%%%%%%%%%%%%%%%%%%%%%%%%%%%%%%%%%%%%%%%%%%%%%%%%%%%%%%%%%%%%%%%%%%%%%%%%%
\subsection{Charge (and spin) current carriers and the general $c,\eta n$ bands (and $s n$ bands) hole representation}
\label{etasnholes}

The relation of the composite $\alpha n$ pseudoparticles to the paired spins $1/2$ $(\alpha =s)$
and paired $\eta$-spins $1/2$ $(\alpha =\eta)$ was the problem revisited in Section \ref{alphanpseudop}.
A related issue whose clarification is needed for the study of the charge 
and spin currents and their carriers is addressed in this section. It refers to
the relation of the set of holes in each $\alpha n$-band populated by $\alpha n$ pseudoparticles
to the spins $1/2$ $(\alpha =s)$ and $\eta$-spins $1/2$ $(\alpha =\eta)$. There are
$N^h_{\alpha n} = 2S_{\alpha}+\sum_{n'=n+1}^{\infty}2(n'-n)N_{\alpha n'}$ such 
holes, Eq. (\ref{N-h-an}), in each $\alpha n$-band.

The 1D Hubbard model in a uniform vector potential $\Phi/L$ whose Hamiltonian is given
in Eq. (4) of Ref. \cite{GPC-07} remains solvable by the BA. 
Its coupling to the charge/$\eta$-spin and spin degrees of freedom the flux $\Phi$ reads 
$\Phi = \Phi_{\uparrow} = \Phi_{\downarrow}$ and  $\Phi = \Phi_{\uparrow} = -\Phi_{\downarrow}$,
respectively \cite{GPC-07,CNP-18}. The LWSs momentum eigenvalues, $P (\Phi_{\uparrow},\Phi_{\downarrow})$, 
have the general form \cite{CNP-18},
\begin{eqnarray}
P (\Phi/L) = P (\Phi_{\uparrow}/L,\Phi_{\downarrow}/L) & = & P (0) + (N_c-\sum_{n}2n\,N_{sn}){\Phi_{\uparrow} - \Phi_{\downarrow}\over 2L}  
- (N_c^h-\sum_{n}2n\,N_{\eta n}){\Phi_{\uparrow} + \Phi_{\downarrow}\over 2L} 
\nonumber \\
& = & P (0) + 2S_s\,{\Phi_{\uparrow} - \Phi_{\downarrow}\over 2L}  
- 2S_{\eta}\,{\Phi_{\uparrow} + \Phi_{\downarrow}\over 2L} \, .
\label{PeffU}
\end{eqnarray}
The LWSs $\Phi =0$ momentum eigenvalue $P (0)$ appearing here is that in Eq. (\ref{P}) for $L_{\eta,-1/2}=0$.

The TBA equations for the model in a uniform vector potential are given in Eq. (9) of Ref. \cite{GPC-07}.
The only difference relative to the $\Phi=0$ case, is that the $c$ band, $sn$ band, and $\eta n$ band momentum values $q_j$ are
replaced by $q_j+\Phi_{\uparrow}/L$, $q_j-n(\Phi_{\uparrow} - \Phi_{\downarrow})/L$, and
$q_j-n(\Phi_{\uparrow} + \Phi_{\downarrow})/L$, respectively. Hence concerning the coupling 
to the (i) charge/$\eta$-spin and (ii) spin degrees of freedom, this gives (i)
$q_j+\Phi/L$, $q_j$, and $q_j-2n\Phi/L$ and (ii) $q_j+\Phi/L$, $q_j-2n\Phi/L$, and $q_j$,
respectively. 

The $\alpha =\eta$ and $\alpha =s$ current operators expectation values of the $\Phi\rightarrow 0$ LWSs 
can then be derived from the $\Phi/L$ dependence of the energy eigenvalues $E  (\Phi/L)$. Specifically,
$\langle \hat{J}_{\eta}^z\rangle = d E(\Phi/L)/d(\Phi/L)\vert_{\Phi = \Phi_{\uparrow} = \Phi_{\downarrow}=0}$ 
and $\langle \hat{J}_{s}^z\rangle = d E(\Phi/L)/d(\Phi/L)\vert_{\Phi = \Phi_{\uparrow} = -\Phi_{\downarrow}=0}$,
respectively.  Moreover, $d P(\Phi/L)/d(\Phi/L)\vert_{\Phi = \Phi_{\uparrow} = \Phi_{\downarrow}=0}$ gives the
number of charge/$\eta$-spin carriers and $d P(\Phi/L)/d(\Phi/L)\vert_{\Phi = \Phi_{\uparrow} - \Phi_{\downarrow}=0}$
that of spin carriers that couple to the vector potential $\Phi/L$. 
The use of the exact momentum eigenvalues, Eq. (\ref{PeffU}), then reveals that such carriers are the $M_{\eta}=2S_{\eta}$ 
unpaired $\eta$-spins $1/2$ and $M_{s}=2S_{s}$ unpaired spins $1/2$, respectively.
(This is as for the unpaired spins of the spin-$1/2$ chain $XXX$ chain.)

It is thus useful to consider the unpaired $\eta$-spins $1/2$ $(\alpha =\eta)$ and unpaired spins $1/2$ $(\alpha =s)$ 
densities $m_{S_{\alpha}} \equiv 2S_{\alpha}/L=M_{\alpha}/L$. The energy and momentum eigenstates 
that span the subspaces with fixed values for the $\alpha = \eta,s$
numbers $L_{\alpha}$, Eq. (\ref{Na-eta-s}), have $S_{\alpha}$ values in the range $S_{\alpha} \in [0,L_{\alpha}/2]$. 
Hence, for the corresponding unpaired $\eta$-spins $1/2$ and unpaired spins densities this gives 
$m_{S_{\eta}} \in [0,n_c^h/2]$ and $m_{S_{s}} \in [0,n_c/2]$, respectively.

In the $u\rightarrow\infty$ limit, one explicitly confirms that up to first order in $\Phi/L$
the dependence of the energy eigenvalues $E(\Phi/L)$ on $\Phi/L$ can be expressed in terms of a 
dependence on $(m_{S_{\alpha}}\Phi)/L$. It refers to $(m_{S_{\eta}}\Phi)/L$
for $\Phi = \Phi_{\uparrow} = \Phi_{\downarrow}$ \cite{Carmelo-00A} and $(m_{S_{s}}\Phi)/L$
for $\Phi = \Phi_{\uparrow} = -\Phi_{\downarrow}$. This ensures that for {\it all} LWSs 
$\vert  l_{\rm r},l_{\eta s}^0,\infty \rangle$ the above $\alpha =\eta,s$ currents $\langle \hat{J}_{\alpha}^z\rangle$
have an overall factor $m_{S_{\alpha}}=2S_{\alpha}/L$.  (In Ref. \cite{Carmelo-00A} it was considered that in the $\alpha =\eta$ case
the corresponding total exact flux $2S_{\eta}\Phi$ was shared by the $N_c^h$ holes in the $c$ band whose number equals that of $\eta$-spins $1/2$,
$L_{\eta} = N_c^h$. This gives $2S_{\eta}\Phi=N_c^h\,\Phi^{eff}$ and thus $\Phi^{eff}=[2S_{\eta}/N_c^h]\,\Phi$.)

Combination of these $u\rightarrow\infty$ properties with the invariance under the electron-rotated-electron unitary transformation
of the $M_{\eta}=2S_{\eta}$ unpaired $\eta$-spins $1/2$ and $M_{s}=2S_{s}$ unpaired spins $1/2$
provides useful physical information. That invariance implies that for $u>0$ the unpaired $\eta$-spins $1/2$ and unpaired 
spins $1/2$ that populate the states $\vert  l_{\rm r},l_{\eta s}^0,u\rangle={\hat{V}}^{\dag}\vert  l_{\rm r},l_{\eta s}^0,\infty \rangle$ 
have properties similar to those that in the $u\rightarrow\infty$ limit populate the
corresponding states $\vert  l_{\rm r},l_{\eta s},\infty \rangle$ belonging the same $V (u)$-set of states. 
This reveals that in the case of coupling to charge/$\eta$-spin ($\alpha =\eta$) and spin ($\alpha =s$)
the energy eigenvalues $E(\Phi/L)$ dependence on $\Phi/L$ is up to first order in $\Phi/L$ of the
general form $C_u\,(m_{S_{\alpha}}\Phi)/L$ for $u>0$. Here $C_u$ is some $u$, $n_e$, and $m$
dependent coefficient independent of $\Phi/L$. Hence the corresponding $\alpha $
current has the general form $\langle \hat{J}_{\eta}^z\rangle = C_u\,m_{S_{\alpha}}$ for all LWSs
$\vert  l_{\rm r},l_{\eta s}^0,u\rangle={\hat{V}}^{\dag}\vert  l_{\rm r},l_{\eta s}^0,\infty \rangle$.

For simplicity, we denote the LWSs $\alpha = \eta,s$ currents by
$\langle\hat{J}^z_{\alpha, LWS} (l_{\rm r},S_{\alpha})\rangle =
\langle  l_{\rm r},S_{\alpha},-S_{\alpha}^z\vert\hat{J}_{\alpha}^z\vert  l_{\rm r},S_{\alpha},-S_{\alpha}^z\rangle$.
Here we have implicitly incorporated in $l_{\rm r}$ the spin $S_s$ when $\alpha =\eta$ and
the $\eta$-spin $S_{\eta}$ when $\alpha =s$. From the use of procedures similar to those leading to the
spin-$1/2$ $XXX$ spin currents, Eq. (\ref{currents-genll}), one finds that
the $\alpha = \eta,s$ currents carried by the non-LWSs have the following exact 
relation to that of the corresponding LWS \cite{CNP-18},
\begin{equation}
\langle\hat{J}_{\alpha}^z (l_{\rm r},M_{\alpha,+1/2},M_{\alpha,-1/2})\rangle = 
{(M_{\alpha,+1/2} - M_{\alpha,-1/2})\over 2S_{\alpha}}\,\langle\hat{J}^z_{\alpha, LWS} (l_{\rm r},S_{\alpha})\rangle 
\hspace{0.20cm}{\rm where}\hspace{0.20cm}\alpha = \eta , s \, .
\label{exp-values}
\end{equation}
The arguments already used in the case of the spin-$1/2$ $XXX$ chain, concerning the LWS currents quantum associated 
with $\alpha$ current changes generated by $\alpha$-flip processes, apply. One then finds
that the elementary current carried by a unpaired $\eta$-spin $1/2$ $(\alpha =\eta)$ of projection $\pm 1/2$ 
and unpaired spin $1/2$ of projection $\pm 1/2$ is given by,
\begin{equation}
j_{\alpha,\pm 1/2} = \pm {\langle\hat{J}^z_{\alpha, LWS} (l_{\rm r},S_{\alpha})\rangle\over 2S_{\alpha}} 
\hspace{0.20cm}{\rm where}\hspace{0.20cm}\alpha = \eta , s \, .
\label{J-eta-spin-spin}
\end{equation}
This is similar to Eq. (\ref{jelementary}) for that chain.

Here we do not address the issue of the spin currents in terms of the $s n$-bands hole representation and
corresponding spinon representation. Indeed, it is similar to that reported in Section \ref{relation-HS}
for the spin-$1/2$ $XXX$ chain. Concerning the charge/$\eta$-spin currents, the translational degrees of freedom of the
$2S_{\eta}$ unpaired $\eta$-spins that couple to the vector potential are now
described by an average number of $2S_{\eta}$ holes both in the $c$ band and 
$\eta n$ bands for which $N_{\eta n}>0$ \cite{CNP-18}. The total number of holes in such bands
can be written as $N_c^h = 2S_{\eta} + \sum_{n=1}^{\infty}2n\,N_{\eta n}$ and
$N_{\eta n}^{h} = 2S_{\eta} + \sum_{n'=n+1}^{\infty}2(n'-n)\,N_{\eta n'}$, respectively.
All processes associated with the $\eta n$ bands and $\eta$-spins $1/2$ are similar
to those described in Section \ref{relation-HS} involving spin $n$ bands and spins $1/2$. 

Some of the holon representations associate each of the $N_c^h = 2S_{\eta} + \sum_{n=1}^{\infty}2n\,N_{\eta n}$
$c$-band holes with a $\eta$-spin $1/2$ holon \cite{Essler-94,Essler-94-B}. Both the range
of validity of that representation and its relation to the extended $c$-band hole and $\eta n$-band holes
more general representation show basic similarities to the same problems for the spinon representation and extended
$n$-bands hole representation revisited in Section \ref{relation-HS} for the spin-$1/2$ $XXX$ chain.
In the case of charge currents, this applies for instance to the description of the translational degrees of freedom of 
the $M_{\eta}=2S_{\eta}$ unpaired $\eta$-spins $1/2$ of $u>0$ energy eigenstates \cite{CNP-18}. Such 
degrees of freedom are described by an average number of $2S_{\eta}$ holes out of both the
$N_c^h = 2S_{\eta} + \sum_{n=1}^{\infty}2n\,N_{\eta n}$ $c$-band holes and
$N_{\eta n}^{h} = 2S_{\eta} + \sum_{n'=n+1}^{\infty}2(n'-n)\,N_{\eta n'}$ holes of $\eta n$ bands
for which $N_{\eta n}>0$. However, their internal $\eta$-spin degrees of freedom 
cannot be associated with such $c$- and $\eta n$-bands holes. This is similar to the
spinon representation of the spin-$1/2$ $XXX$ chain. For the additional information
on the role of the $c$ band holes in charge transport see Ref. \cite{CNP-18}.

%%%%%%%%%%%%%%%%%%%%%%%%%%%%%%%%%%%%%%%%%%%%%%%%%%%%%%%%%%%%%%%%%%%%%%%%%%
\section{The 1D Hubbard model pseudoparticles quantum liquid}
\label{exc-spectra}

In this section the use of the $\beta =c,\alpha n$ pseudoparticle representation of the 1D Hubbard model 
to describe its low-energy physics is revisited. As in the case of the simpler models reviewed here, the 
$\beta =c,\alpha n$ pseudoparticle energy functional resembles that of the low-energy Fermi liquid. The 
1D Hubbard model describes interacting electrons on a lattice. It is a non-perturbative
quantum problem for which there are no quasiparticles, as defined in a Fermi liquid: The $\beta =c,\alpha n$ pseudoparticles
do not become electrons upon turning off adiabatically the interaction $U$. The one-electron physics is thus
qualitatively different from that of a Fermi liquid. However, the two-electron physics resembles that of such a liquid
\cite{Carmelo-91-A,Carmelo-91,Carmelo-92,Carmelo-92-B,Carmelo-92-C,Carmelo-03,Carmelo-04}.

Our goal here is to apply the general $\beta =c,\alpha n$ pseudoparticle energy functional introduced 
below in Section \ref{exc-spectraQL1} to the description of the model low-energy physics in
Section \ref{exc-spectraQL2}. However, that functional is valid at all energy scales. 
There is though a restriction to its applicability reported below in Section \ref{exc-spectraQL1}.
(It basically is the same as that of the spin-$1/2$ $XXX$ chain spin pseudoparticle quantum liquid.) 

There are several advantages in using the $\beta =c,\alpha n$ pseudoparticle representation in the study of the low-energy 
properties. First, since it applies to all energy scales, its use reveals that the usual low-energy spin-charge separation
results from a more general separation of the spin and charge degrees of freedom. It occurs at all energy scales
and is associated with the spin $SU(2)$ and charge $U(2)=SU(2)\otimes U(1)$ symmetries, respectively, in the 
model $[SU(2)\otimes SU(2)\otimes U(1)]/Z_2^2$ global symmetry. Second, all low-energy two-particle quantities are controlled
by simple $\beta $ pseudoparticles zero-momentum forward-scattering interactions. As in a Fermi liquid, those 
are associated with $f$ function terms in the theory energy functional. Hence all such quantities can be
computed by the familiar methods of a Fermi liquid. This simplifies their physical understanding. Third,
the form of the $\alpha 1$ pseudoparticle energy dispersions provides valuable information on the type of 
pairing of the two $\eta$-spins $1/2$ $(\alpha =\eta)$ and spins $1/2$ $(\alpha =s)$ within each
$\alpha$-singlet pair. A set of $\Pi_{\alpha} = \sum_{n=1}^{\infty}n\,N_{\alpha n} = (L_{\alpha} - 2S_{\alpha})/2$
such pairs populates each $u>0$ energy and momentum eigenstate. Furthermore, under a suitable unitary transformation that shifts 
the $\beta$ pseudoparticle momentum values $q_j$, those are mapped onto $\beta$ pseudofermions in terms 
of which the study of the model finite-energy dynamical properties simplifies.

%%%%%%%%%%%%%%%%%%%%%%%%%%%%%%%%%%%%%%%%%%%%%%%%%%%%%%%%%%%%%%%%%%%%%%%%%%
\subsection{The $c$ and $\alpha n$ pseudoparticle quantum liquid I: The general energy functional and its energy scales}
\label{exc-spectraQL1}

As in the case of the simpler models discussed in Sections \ref{Bg} and \ref{Heichain}, there is a PS for 
each ground state with electronic density $n_e$ and spin density $m$ arbitrary values. 
In this section we consider PSs whose ground states refer to electronic densities $n_e \in [0,1]$ and spin densities $m \in [0,n_e]$.
They are thus LWSs. Their $\beta$-band pseudoparticle momentum distribution functions are given by,
\begin{equation}
N_c^0 (q_j) = \theta (q_j - q_{Fc}^{-})\,\theta (q_{Fc}^{+} - q_j)  
\, , \hspace{0.20cm} 
N_{s 1}^0 (q_j) = \theta (q_j - q_{Fs1}^{-})\,\theta (q_{Fs1}^{+} - q_j)  
\hspace{0.20cm}{\rm and}\hspace{0.20cm}
N_{\alpha n}^0 (q_j) = 0 \hspace{0.20cm}{\rm for}\hspace{0.20cm}\alpha n \neq s1 \, .
\label{N0q1DHm}
\end{equation}
The $c$ and $s1$ bands Fermi momentum values $q_{F\beta}^{\pm}$ appearing here
are given in Eqs. (C.4)-(C.11) of Ref. \cite{Carmelo-04}. Ignoring ${\cal{O}} (1/L)$ corrections within the TL
simplifies the $\beta =c,s1$ distributions, Eq. (\ref{N0q1DHm}), to $N_{\beta}^0 (q_j) = \theta (q_{F\beta} - \vert q_j\vert)$. Here 
the $\beta =c,s1$ Fermi momentum $q_{F\beta}$ reads,
\begin{equation}
q_{Fc} = 2k_F = \pi\,n_e \hspace{0.20cm}{\rm and}\hspace{0.20cm}q_{Fs1} = k_{F\downarrow} = \pi\,n_{e\downarrow} \, . 
\label{q0Fcs}
\end{equation}
Hence the ground states under consideration are neither populated by composite $sn$ pseudoparticles 
with $n>1$ spin-singlet pairs nor by composite $\eta n$ pseudoparticles with any  number $n=1,...,\infty$ of $\eta$-spin-singlet pairs.
Since they are LWSs, they have no unpaired spins of projection $-1/2$ and no unpaired $\eta$-spins of projection $-1/2$. 
Their $M_{\eta} = 2S_{\eta} = -2S_{\eta}^z = L - N_e$ unpaired $\eta$-spins and 
$M_{s} = 2S_{s} = -2S_{s}^z = N_{e\uparrow}-N_{e\downarrow}$ unpaired spins 
have all projection $+1/2$ .

The PS excited states $\beta$-bands distribution functions are of the general form
$N_{\beta} (q_j) = N^{0}_{\beta} (q_j) +  \delta N_{\beta} (q_j)$.
Here $\beta = c, \alpha n$, $\alpha =\eta,s$, and $n =1,...,\infty$.
In the specific case of the $\beta = c,s1$ bands, one can classify in the TL the deviations, 
\begin{equation}
\delta N_{\beta} (q_j)  = N_{\beta} (q_j) - N^0_{\beta} (q_j) \hspace{0.20cm}{\rm for}\hspace{0.20cm}
j = 1,...,L_{\beta} \, ,
\label{DNq}
\end{equation}
as $\delta N_{\beta}^F (q_j)$ and $\delta N_{\beta}^{NF} (q_j)$,
respectively. On the one hand, for the deviations $\delta N_{\beta}^F (q_j)$ the band momentum $q_j$ is such that 
$\lim_{L\rightarrow\infty}(\vert q_j\vert - q_{F\beta})=0$. On the other hand, in the case of
$\delta N_{\beta}^{NF} (q_j)$ the momentum difference $\lim_{L\rightarrow\infty}(\vert q_j\vert - q_{F\beta})$ 
remains finite in the TL. For the excited states belonging to a PS, one as that
$\sum_{\beta =c,s1}\sum_{j=1}^{L_{\beta}}\vert\delta N_{\beta}^{NF} (q_j)\vert/L\rightarrow 0$,
$\sum_{n=1}^{\infty}\sum_{j=1}^{L_{\eta n}}\vert\delta N_{\eta n} (q_j)\vert/L\rightarrow 0$,
$\sum_{n=2}^{\infty}\sum_{j=1}^{L_{sn}}\vert\delta N_{sn} (q_j)\vert)/L\rightarrow 0$, $\delta S_{s}/L\rightarrow 0$,
and $\delta S_{\eta}/L\rightarrow 0$ as $L\rightarrow\infty$. For a PS there are though no restrictions on the value
of the excitation energy and excitation momentum. 

It is often convenient within the TL to replace the $\beta =c,\alpha n$ band discrete 
momentum values $q_j$, Eq. (\ref{q-j}), such that $q_{j+1}-q_j=2\pi/L$, by a 
corresponding continuous momentum variable, $q$. It belongs to a domain $q\in [q_{\beta}^-,q_{\beta}^+]$ whose limiting 
momentum values $q_{\beta}^{\pm }$ are given in Eq. (\ref{qcan-range}) of Appendix \ref{TBAconfig}.  
Ignoring again ${\cal{O}} (1/L)$ corrections, one finds that $q_{\beta}^{\pm }\approx \pm q_{\beta}$. For the
present PSs whose ground states are LWSs the limiting momenta $q_{\beta}$ then read,
\begin{equation}
q_{c} = \pi \, , \hspace{0.20cm} q_{s1} = k_{F\uparrow} \, , \hspace{0.20cm} q_{sn} =
(k_{F\uparrow}-k_{F\downarrow}) = \pi\,m \hspace{0.20cm}{\rm and}\hspace{0.20cm}
q_{\eta n} = (\pi -2k_F) = \pi\,(1-n_e) \, .
\label{qcanGS}
\end{equation}

Within the continuum momentum $q$ representation, the deviation values $\delta N_{\beta} (q_j)=-1$ and $\delta N_{\beta} (q_j)=+1$ in
Eq. (\ref{DNq}) become $\delta N_{\beta} (q)=-(2\pi/L)\delta (q-q_j)$ and $\delta N_{\beta} (q)=+(2\pi/L)\delta (q-q_j)$, respectively. 
According to Eqs. (\ref{q-j}) and (\ref{Ic-an}), under a
transition to an excited state, the $\beta$ band discrete momentum values $q_j = (2\pi/L)\,I_j^{\beta}$ 
may undergo a collective shift, $(2\pi/L)\,\Phi_{\beta}^0 =\pm \pi/L$. Here $\Phi_{\beta}^0$ reads,
\begin{eqnarray}
\Phi_{c}^0 & = & 0 \hspace{0.20cm}{\rm for}\hspace{0.20cm}\delta N^{SU(2)} \hspace{0.20cm} {\rm even} 
\hspace{0.20cm}{\rm and}\hspace{0.20cm}\Phi_{c}^0=\pm{1\over 2}
\hspace{0.20cm}{\rm for}\hspace{0.20cm}\delta N^{SU(2)} \hspace{0.20cm} {\rm odd} \, ; 
\nonumber \\
\Phi_{\alpha n}^0 & = & 0 \hspace{0.20cm}{\rm for}\hspace{0.20cm}\delta N_{c}+\delta N_{\alpha n}
\hspace{0.20cm}{\rm even}\hspace{0.20cm}{\rm and}\hspace{0.20cm}\Phi_{\alpha n}^0=\pm {1\over 2}
\hspace{0.20cm}{\rm for}\hspace{0.20cm} 
\delta N_{c}+\delta N_{\alpha n} \hspace{0.20cm} {\rm odd} \, ,
\label{pican}
\end{eqnarray}
where $\alpha = \eta,s$ and $n =1,...,\infty$. $\delta N^{SU(2)}$ is in this equation the deviation in the 
number $N^{SU(2)}$ in Eq. (\ref{F-beta}). For $q$ at the $\beta =c,s1$ and $\iota = \pm$
Fermi points, $\iota\,q_{F\beta}$, such a shake-up effect is captured within the continuum representation by 
additional deviations, $\pm (\pi/L)\delta (q-\iota\,q_{F\beta})$. 
For transitions to an excited state for which $\delta L_{\alpha n}\neq 0$, 
the removal or addition of BA $\alpha n$ band discrete momentum values
occurs in the vicinity of the band edges $q_{\alpha n}^-=-q_{\alpha n}^+$, Eq. (\ref{qcan-range}) of Appendix \ref{TBAconfig}. 
Those are zero-momentum and zero-energy processes. 
\begin{figure}
\begin{center}
\subfigure{\includegraphics[width=5.0cm]{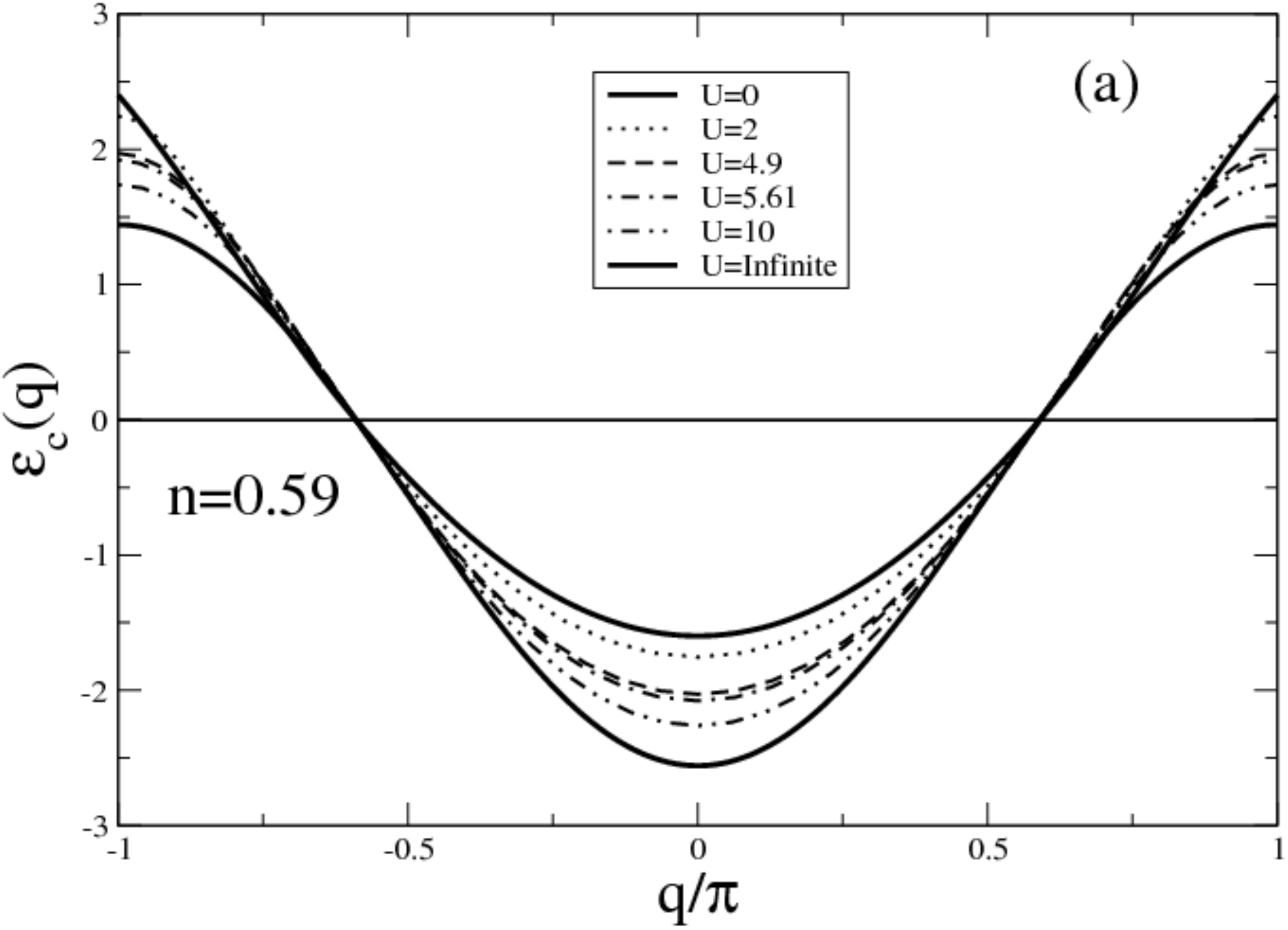}}
\hspace{0.25cm}
\subfigure{\includegraphics[width=5.0cm]{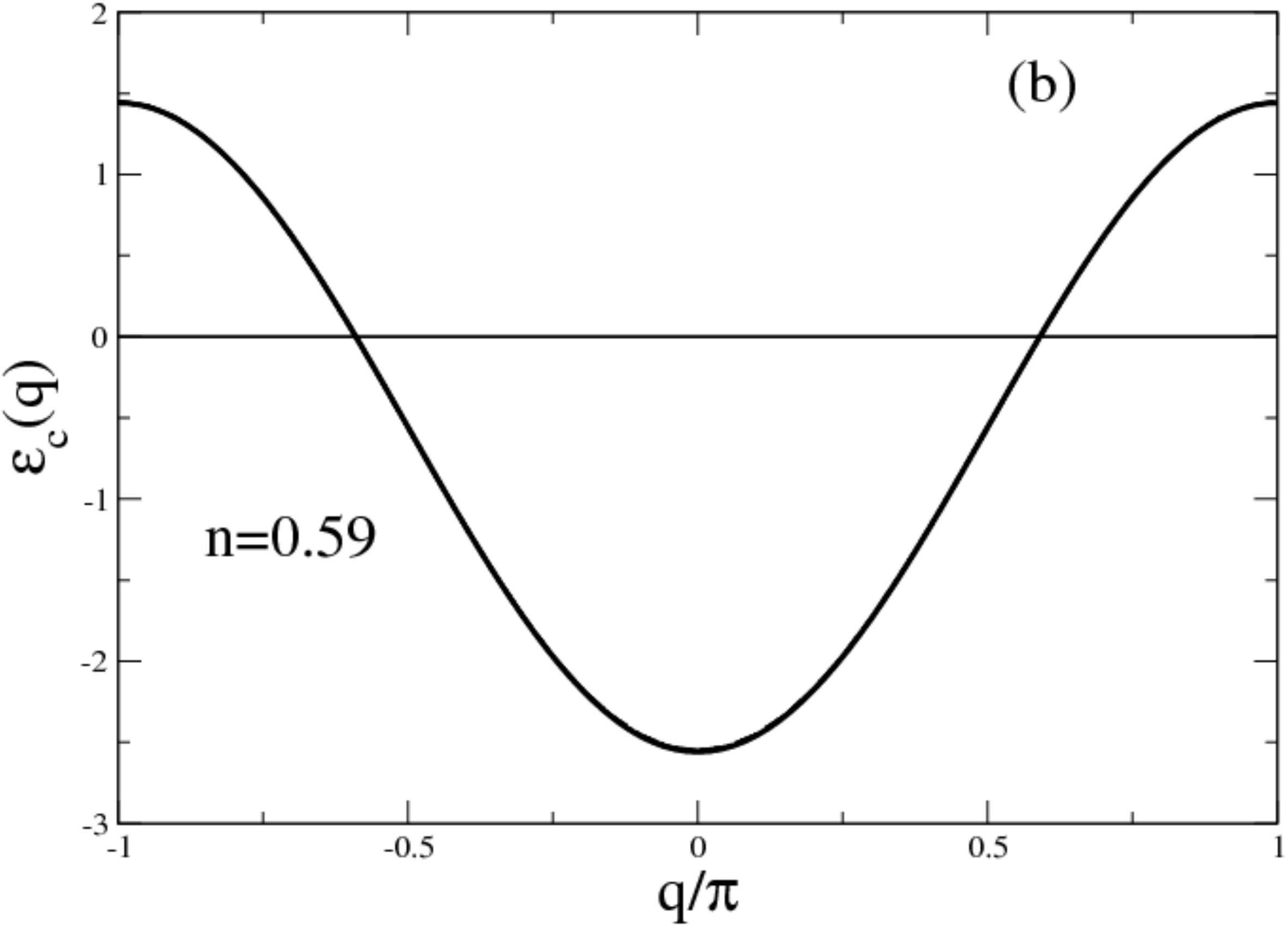}}
\caption{The $c$ band energy dispersion $\varepsilon_{c} (q)$, Eq. (\ref{epsilon-q}) for $\beta =c$, plotted as
a function of the momentum in units of $t$ for a set of $U/t$ values (in units of $t$), electronic density $n_e=0.59$, 
and spin density (a) $m=0$ and (b) $m\rightarrow n_e=0.59$.
(As in Ref. \cite{Carmelo-03}, in the figures the electronic density $n_e$ is denoted by $n$).\\
{\it Source}: The figures plots were produced using the same data as in Fig. 6 of Ref. \cite{Carmelo-03}
for other densities.}
\label{figure10}
\end{center}
\end{figure}

The PS energy functionals are derived from the use of the TBA equations,
Eqs. (\ref{Tapco1})-(\ref{Tapco2}), and general energy spectra, Eq. (\ref{E}) of Appendix \ref{TBAconfig}.
Specifically, one uses in them $\beta$-bands momentum distribution functions of form 
$N_{\beta} (q_j) = N^{0}_{\beta} (q_j) +  \delta N_{\beta} (q_j)$. Their deviations, Eq. (\ref{DNq}), play an important role.
The combined and consistent solution of those equations and spectra up to second order in such deviations
then leads to \cite{Carmelo-92-B,Carmelo-97-B},
\begin{equation}
\delta E = \sum_{\beta}\sum_{j=1}^{L_{\beta}}\varepsilon_{\beta} (q_j)\delta N_{\beta} (q_j) 
+ {1\over L}\sum_{\beta}\sum_{\beta'}\sum_{j=1}^{L_{\beta}}\sum_{j'=1}^{L_{\beta'}}
{1\over 2}\,f_{\beta\,\beta'} (q_j,q_{j'})\,\delta N_{\beta} (q_j)\delta N_{\beta'} (q_{j'})
+ \sum_{\alpha =\eta,s}\varepsilon_{\alpha,-1/2}\,M_{\alpha,-1/2} \, .
\label{DE-fermions}
\end{equation}
The $\beta =c,\alpha n$ band energy dispersions $\varepsilon_{\beta} (q_j)$ appearing here are given by,
\begin{equation}
\varepsilon_{\beta} (q_j) = E_{\beta} (q_j) + \varepsilon_{\beta}^c (q_j) 
\hspace{0.20cm}{\rm and}\hspace{0.20cm}
\varepsilon_{\beta}^c (q_j) =
{t\over \pi}\int_{-Q}^{Q}dk\,2\pi\bar{\Phi }_{c\,\beta}
\left({\sin k\over u}, {\Lambda_{0}^{\beta} (q_j)\over u}\right)\sin k  \hspace{0.20cm}{\rm for}\hspace{0.20cm}
j = 1,...,L_{\beta} \, .
\label{epsilon-q}
\end{equation} 
$E_{\beta} (q_j)$ is in this equation for $\beta = c, \eta n, s n$ the energy spectrum,
Eq. (\ref{spectra-E-an-c-0}) of Appendix \ref{TBAconfig}, with the rapidity functions 
those of the ground state,
$\Lambda_{0}^{v} (q_j) = \sin k^{c}_{0} (q_j)/u$ and $\Lambda_{0}^{\alpha n} (q_j)$.
The latter are the solutions of Eqs. (\ref{Tapco1}) and (\ref{Tapco2}) for the corresponding
distribution function distributions, Eq. (\ref{N0q1DHm}). The parameter $Q$ 
in Eq. (\ref{epsilon-q}) and related parameters $B$, $r_c^0$, and $r^s_0$ read,
\begin{equation}
Q \equiv k^{0}_c (2k_F) \, , \hspace{0.20cm} B \equiv
\Lambda_{0}^{s1}(k_{F\downarrow})  \, , \hspace{0.20cm}
r_c^0 = {\sin Q \over u}\hspace{0.20cm}{\rm and}\hspace{0.20cm}r_s^0 = {B\over u} \, .
\label{QB-r0rs}
\end{equation}

The dressed rapidity phase shifts $2\pi\bar{\Phi }_{c\,\beta} (r,r')$ in Eq. (\ref{epsilon-q}) 
are a particular case of the general dressed rapidity phase shifts $2\pi\bar{\Phi }_{\beta\,\beta'} (r,r')$.
Those are defined by the integral equations given in Appendix \ref{PSIE} \cite{Carmelo-97-B}. Such
phase shifts are associated with the following corresponding dressed phase shifts expressed in terms
of the $\beta$-band momentum $q_j$ and $\beta'$-band momentum $q_{j'}$,
\begin{equation}
2\pi\Phi_{\beta\,\beta'}(q_j,q_{j'}) = 2\pi\bar{\Phi }_{\beta\,\beta'} \left(r,r'\right) 
\hspace{0.20cm}{\rm where}\hspace{0.20cm}r = \Lambda_{0}^{\beta}(q_j)/u
\hspace{0.20cm}{\rm and}\hspace{0.20cm}r' = \Lambda_{0}^{\beta'}(q_{j'})/u \, . 
\label{Phi-barPhi}
\end{equation}

Within the continuum $q$ representation, the $\beta$ band group velocities are given by,
\begin{equation}
v_{\beta} (q_j) = {d\varepsilon_{\beta} (q)\over d q}\vert_{q=q_j} \, , \hspace{0.20cm}
v_{c} \equiv v_{c} (q_{Fc}) = v_{c} (2k_F)\hspace{0.20cm}{\rm and}\hspace{0.20cm}
v_{s1} \equiv v_{c} (q_{Fs1}) = v_{s1} (k_{F\downarrow}) \, ,
\label{vel-beta}
\end{equation}
where $\beta = c, \eta n, sn$ and $n = 1,...,\infty$. They appear in the expression of the $f$ functions in 
the second-order terms of the energy functional, Eq. (\ref{DE-fermions}), 
which reads \cite{Carmelo-92-B,Carmelo-92-C},
\begin{eqnarray}
f_{\beta\,\beta'}(q_j,q_{j'}) & = & v_{\beta}(q_{j})\,2\pi \,\Phi_{\beta\,\beta'}(q_{j},q_{j'})+
v_{\beta'}(q_{j'})\,2\pi \,\Phi_{\beta'\,\beta}(q_{j'},q_{j}) 
\nonumber \\
& + & {1\over 2\pi}\sum_{\beta''=c,s1} \sum_{\iota = \pm} v_{\beta''}\,
2\pi\Phi_{\beta''\,\beta}(\iota q_{F\beta''},q_{j})\,2\pi\Phi_{\beta''\,\beta'} (\iota q_{F\beta''},q_{j'}) \, .
\label{ff}
\end{eqnarray}
The dressed momentum phase shift $2\pi\Phi_{\beta\,\beta'}(q_j,q_{j'})$ appearing here
is given in Eq. (\ref{Phi-barPhi}).

The only restriction to the applicability of the $\beta =c,\alpha n$ pseudoparticle energy functional, Eq. (\ref{DE-fermions}),
is that associated with a PS definition. It is such that within the TL the deviations $\delta N_{c}^{NF} (q_j)$,
$\delta N_{s1}^{NF} (q_j)$, $\delta N_{sn} (q_j)$ for $n=2,...,\infty$, and $\delta N_{\eta n} (q_j)$ for $n =1,...,\infty$
involve a finite number of $\beta =c,\alpha n$ pseudoparticles. Hence such a restriction can be expressed as,
\begin{equation}
\lim_{L\rightarrow\infty}{\left(\sum_{\beta =c,s1}\sum_{j=1}^{L_{\beta}}\vert\delta N_{\beta}^{NF} (q_j)\vert +
\sum_{n=1}^{\infty}\sum_{j=1}^{L_{\eta n}}\vert\delta N_{\eta n} (q_j)\vert +
\sum_{n=2}^{\infty}\sum_{j=1}^{L_{sn}}\vert\delta N_{sn} (q_j)\vert\right)\over L}\rightarrow 0 \, .
\label{CondPDHubb}
\end{equation}
\begin{figure}
\begin{center}
\subfigure{\includegraphics[width=5.00cm]{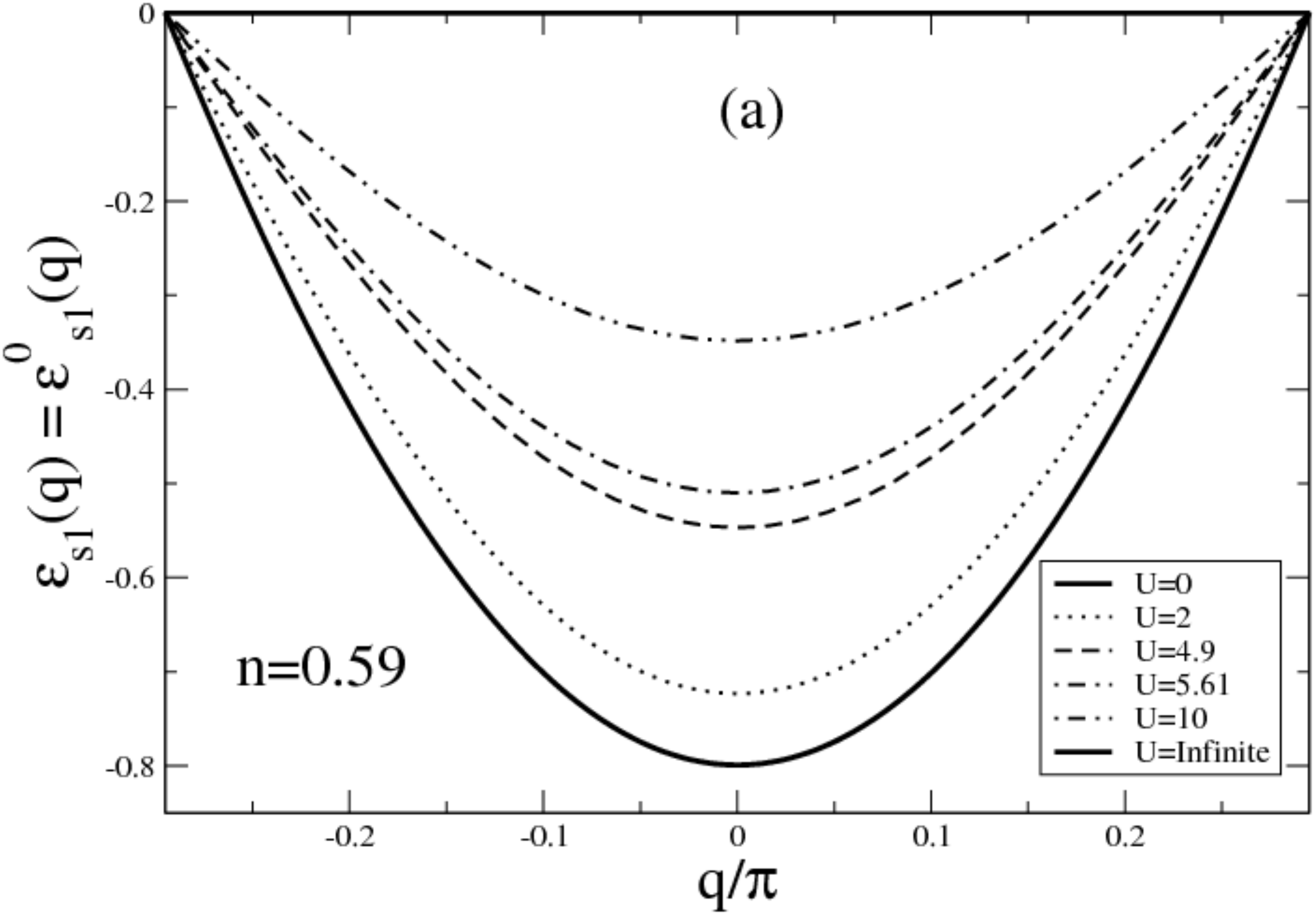}}
\subfigure{\includegraphics[width=5.00cm]{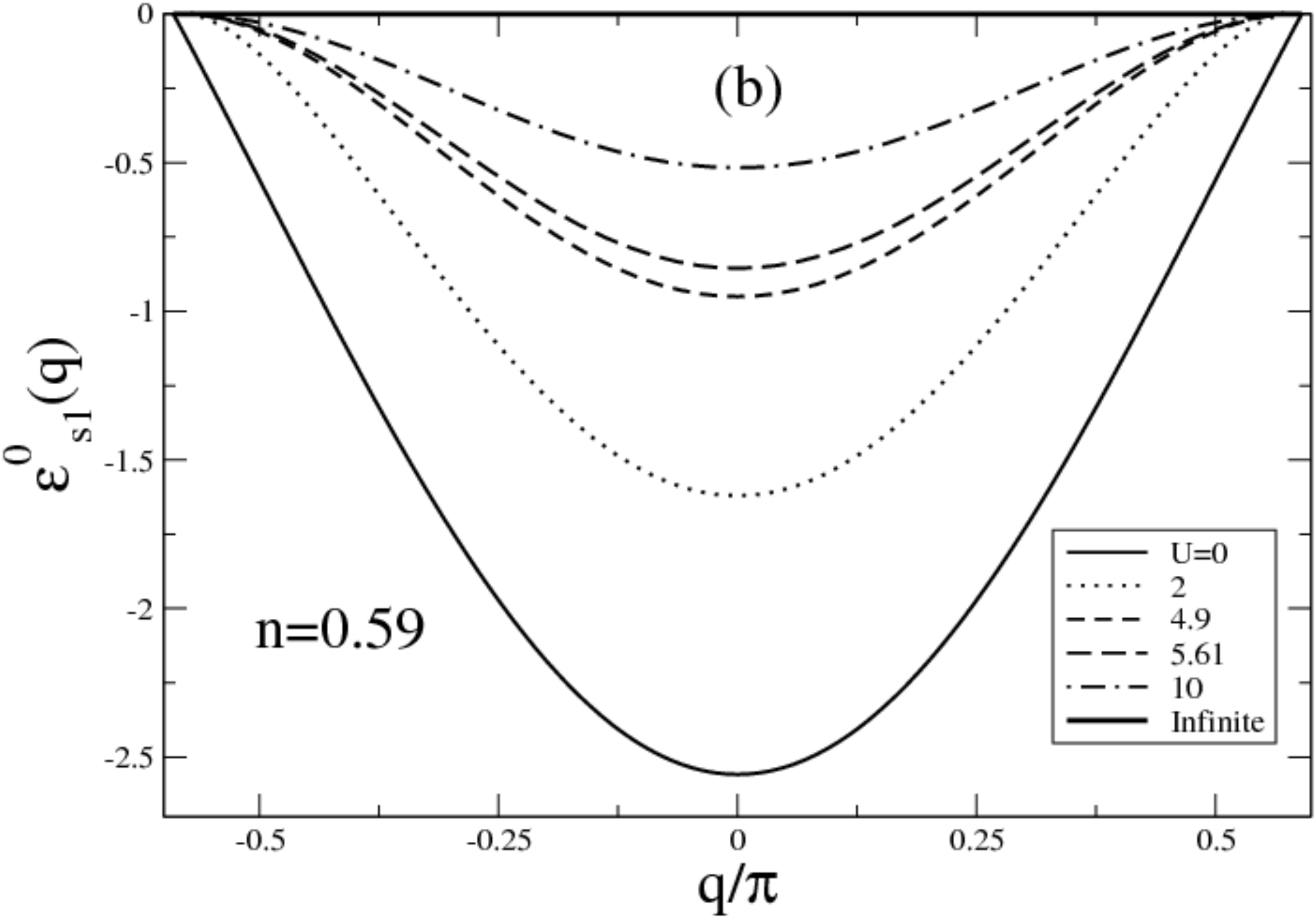}}
\caption{The $s1$ band energy dispersion $\varepsilon_{s1}^0 (q)$, Eq. (\ref{e-0-bands}) for $\alpha n =s1$,
which for $m=0$ equals the related $s1$ band energy dispersion $\varepsilon_{s1} (q)$, Eqs. (\ref{epsilon-q}) 
for $\beta =s1$, plotted as a function of the momentum in units of $t$ for the same $U/t$ and $n_e$ values 
as Fig. \ref{figure10} and (a) $m=0$ and (b) $m\rightarrow n_e=0.59$. (As in Ref. \cite{Carmelo-03}, in 
the figures the electronic density $n_e$ is denoted by $n$).\\
{\it Source}: The figures plots were produced using the same data as in Fig. 7 of Ref. \cite{Carmelo-03}
for other densities.}
\label{figure11}
\end{center}
\end{figure}

The general energy spectrum, Eq. (\ref{E}) of Appendix \ref{TBAconfig}, gives the energy eigenvalues. That in Eq. (\ref{DE-fermions})
rather provides the excited-state energy eigenvalues minus the ground state energy.
The energy dispersion term $\varepsilon_{\beta}^c (q_j)$ in Eq. (\ref{epsilon-q}) as well as the $f$-function
terms in Eq. (\ref{DE-fermions}) are absent from Eq. (\ref{E}) of Appendix \ref{TBAconfig}. Indeed, they stem from such energies differences.
This is why the expressions of the energy dispersion term $\varepsilon_{\beta}^c (q_j)$ and $f$-function 
involve dressed phase shifts. Those emerge under the transitions from the ground state to excited states. 
The one- and two-electron excited states spectra can be expressed in terms of the $\beta = c, \alpha n$ energy dispersions,
Eq. (\ref{epsilon-q}) \cite{Carmelo-91-A}. 

In the particular case of the magnetic-field energy $2\mu_B\,H = 2\mu_B\,H (m)$ and
chemical potential $\mu = \mu (n_e)$ on the right-hand side of Eq. (\ref{DE-fermions}) 
and related energy scales, we consider extended ranges of the densities 
$n_e$ and $m$. Such important energy scales appear in the following relations between energy dispersions
with different yet useful zero energy levels,
\begin{equation}
\varepsilon_{c}^0 (q_j) = \varepsilon_{c} (q_j) - \mu_{\eta} + \mu_{s}
\hspace{0.20cm}{\rm and}\hspace{0.20cm}
\varepsilon_{\alpha n}^0 (q_j) = \varepsilon_{\alpha n} (q_j) - n\,2\mu_{\alpha} \, ,
\label{e-0-bands}
\end{equation}
where $\alpha = \eta,s$ and $n = 1,..., \infty$. They are uniquely determined by the energy dispersions
$\varepsilon_{c}^0 (q_j)$ and $\varepsilon_{s1}^0 (q_j)$ at the corresponding Fermi points 
as follows \cite{Carmelo-91-A},
\begin{eqnarray}
2\mu_B\,H & = & - {\rm sgn}\{m\}\varepsilon_{s1}^0 (q_{F{s1}}) \, ,
\nonumber \\
\mu & = & - {\rm sgn}\{(1-n_e)\}\left(\varepsilon_{c}^0 (q_{Fc}) + {1\over 2}\varepsilon_{s1}^0 (q_{F{s1}})\right) 
\hspace{0.20cm}{\rm for}\hspace{0.20cm}n_e \neq 1\hspace{0.20cm}{\rm and}\hspace{0.20cm}\mu \in  [-\mu^0,\mu^0] 
\hspace{0.20cm}{\rm for}\hspace{0.20cm}n_e= 1 \, .
\label{mu-muBH}
\end{eqnarray}
Note that due to the terms in the expressions in Eq. (\ref{e-0-bands}) involving the
$\alpha = \eta,s$ energy scales $2\mu_{\alpha}$, Eq. (\ref{2mu-eta-s}) of Appendix \ref{TBAconfig}, 
the dispersions $\varepsilon_{c}^0 (q_j)$ and $\varepsilon_{\alpha n}^0 (q_j)$ are actually
independent of such energy scales. This can be confirmed by inspection of
the form of the energy dispersion term $E_{\beta} (q_j)$, Eq. (\ref{spectra-E-an-c-0}) of Appendix \ref{TBAconfig} for $\beta = c, \alpha n$,
appearing in Eq. (\ref{epsilon-q}). The zero energy level of the $\beta =c,\alpha n$ energy dispersions
$\varepsilon_{\beta} (q_j)$ is that of the ground state. In contrast, that of the $\beta =c,\alpha n$ energy dispersions
$\varepsilon_{\beta}^0 (q_j)$ refers to the BA absolute zero energy level. For the $\alpha n$ bands, the latter 
zero energy level is such that $\varepsilon_{\alpha n}^0 (q_{\alpha n}^{\pm})=0$. 

The finite-$u$ Mott-Hubbard gap $2\mu^0$, Eq. (\ref{2mu0}) of Appendix \ref{TBAconfig}, finiteness implies that the chemical potential 
curve $\mu = \mu (n_e)$ has a discontinuity at $n_e=1$. The corresponding chemical-potential dependence on the hole 
concentration, $x=(1-n_e)$, is such that $\mu (x) = -\mu (-x)$ with $\mu\in [\mu^0,\mu^1]$ for $x\geq 0$. Here
the energy scale $2\mu^1$ associated with $\mu^1 = \mp\lim_{x\rightarrow \pm 1} \mu (x)$ reads $2\mu^1 = U + 4t$. 
For $u\gg 1$ and $m=0$ the chemical-potential curve $\mu = \mu (x)$ behaves as
$2\mu (x) = {\rm sgn}\{x\}(U-4t\cos (\pi x))$ for both $x\in [-1,0]$ and $x\in [0,1]$ and
$2\mu (0) \in [-(U-4t),(U-4t)]$ at $x = 0$. Furthermore, the magnetic energy scale $2\mu_B\,H$
dependence on the spin density $m$ is such that $2\mu_B\,H (m) = -2\mu_B\,H (-m)$ with
$2\mu_B\,\vert H (m)\vert \in [0,2\mu_B\,H_c] $ for $m \in [-(1-\vert x\vert),(1-\vert x\vert)]$.
Here $2\mu_B\,H (0) = 0$ and $2\mu_B\,H_c = \pm\lim_{m\rightarrow \pm [1-\vert x\vert]} 2\mu_B\,H (m)$.
A closed-form expression for the dependence on $U$, $t$, and density $n_e$ of the energy 
scale $2\mu_B\,H_c$ where $H_c$ is the critical magnetic field for the onset of fully polarized ferromagnetism
is given below.
\begin{figure}
\begin{center}
\centerline{\includegraphics[width=5.0cm]{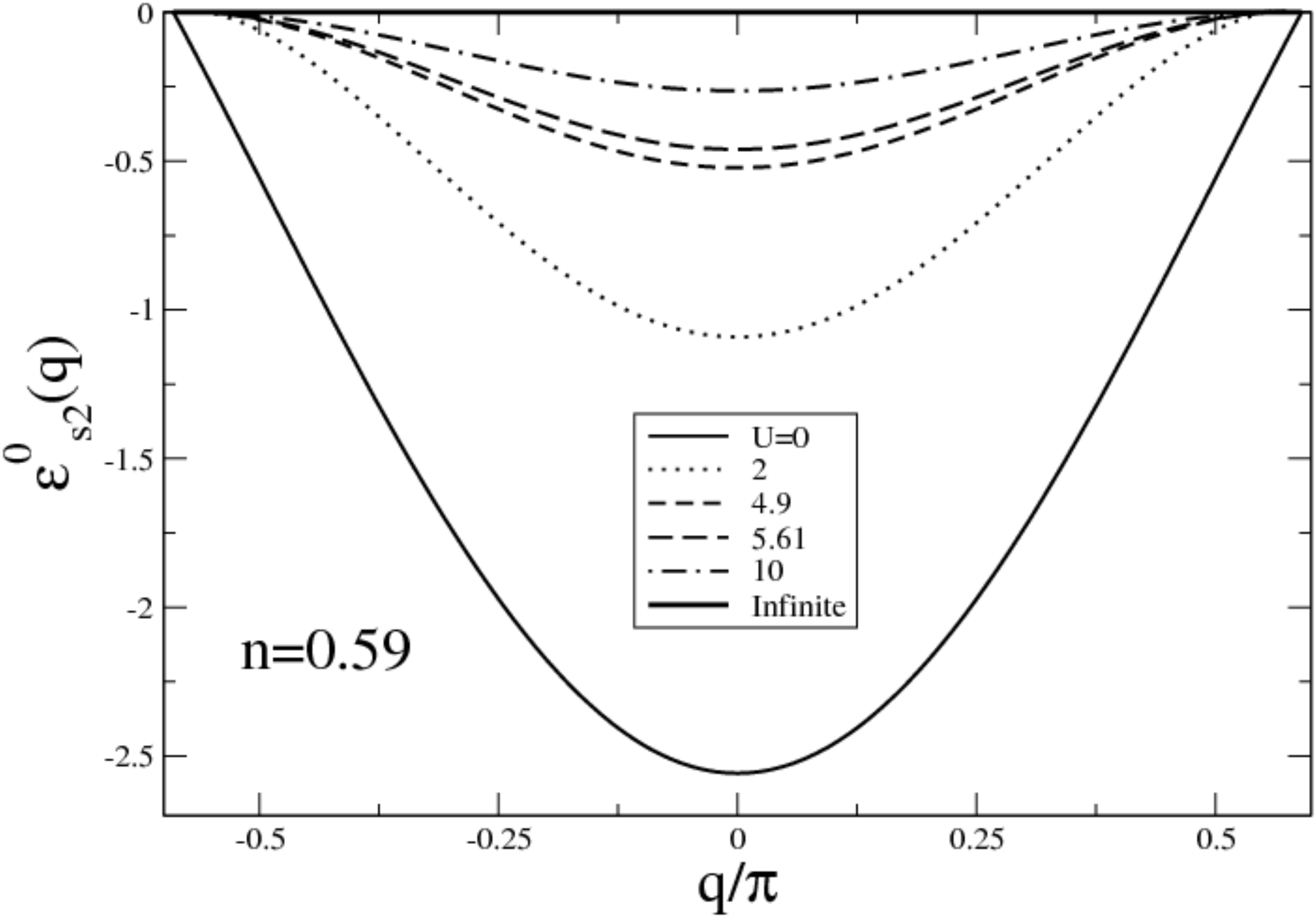}}
\caption{The $s2$ band energy dispersion $\varepsilon_{s2}^0 (q)$,
Eq. (\ref{e-0-bands}) for $\beta =s2$, plotted as
a function of the momentum in units of $t$ for the same $U/t$ and $n_e$ values 
as Fig. \ref{figure10} and $m\rightarrow n_e=0.59$. (As in Ref. \cite{Carmelo-03}, in the figure the 
electronic density $n_e$ is denoted by $n$).\\
{\it Source}: The figure plots were produced using data from Ref. \cite{Carmelo-03}.}
\label{figure12}
\end{center}
\end{figure}

The intrinsic energies $\varepsilon_{\eta,\pm1/2}$ and $\varepsilon_{s,\pm1/2}$ relative to the zero-energy ground-state
level of a unpaired $\eta$-spin $1/2$ and a unpaired spin $1/2$, respectively, of projection $\pm 1/2$ are directly
related to the energy scales $2\mu $ and $2\mu_B\,H$, respectively. Such intrinsic energies are useful 
reference scales for the analysis presented below in Section \ref{BANTIB} of the anti-binding or binding character of 
the paired spins $1/2$ and paired $\eta$-spins $1/2$ configuration within each pair. Straightforward calculations 
relying on the algebra of the $\eta$-spin and spin $SU(2)$ symmetry off-diagonal generators lead to,
\begin{eqnarray}
\varepsilon_{\eta,\pm1/2} & = & 2\vert\mu\vert \hspace{0.20cm}{\rm and}\hspace{0.20cm}\varepsilon_{\eta,\mp 1/2} = 0
\hspace{0.20cm}{\rm for}\hspace{0.20cm}{\rm sgn}\{(1-n_e)\}1 = \mp 1\hspace{0.20cm}{\rm and}\hspace{0.20cm}n_e \neq 1 \, ,
\nonumber \\
& = & (\mu^0\pm\mu)\hspace{0.20cm}{\rm for}\hspace{0.20cm}n_e=1\hspace{0.20cm}{\rm and}\hspace{0.20cm}\mu \in [-\mu^0,\mu^0] \, ,
\nonumber \\
\varepsilon_{s,\pm1/2} & = & 2\mu_B\,\vert H\vert \hspace{0.20cm}{\rm and}\hspace{0.20cm}\varepsilon_{s,\mp 1/2} = 0
\hspace{0.20cm}{\rm for}\hspace{0.20cm}{\rm sgn}\{m\}1 = \mp 1 \, .
\label{energy-eta}
\end{eqnarray}
Hence a $S_{\alpha}=1;S_{\alpha}^{z}=0$ $\alpha$-multiplet configuration of two unpaired $\eta$-spins
or two unpaired spins has an intrinsic energy given by,
\begin{equation}
\varepsilon_{\alpha+1/2} + \varepsilon_{\alpha,-1/2} = 2\mu_{\alpha} \hspace{0.20cm}{\rm where}\hspace{0.20cm}\alpha = \eta,s \, .
\label{energy-eta-s-2}
\end{equation}
The energy scales $2\mu_{\alpha}$ such that $2\mu_{\eta}=2\vert\mu\vert$ and $2\mu_{s}=2\mu_B\,\vert H\vert$
and the Mott-Hubbard gap $2\mu^0$ associated with the energy scale $\mu^0$ are
those considered above. They are given in Eqs. (\ref{2mu-eta-s}) and (\ref{2mu0}) of Appendix \ref{TBAconfig}.

\begin{figure}
\begin{center}
\subfigure{\includegraphics[width=5.0cm]{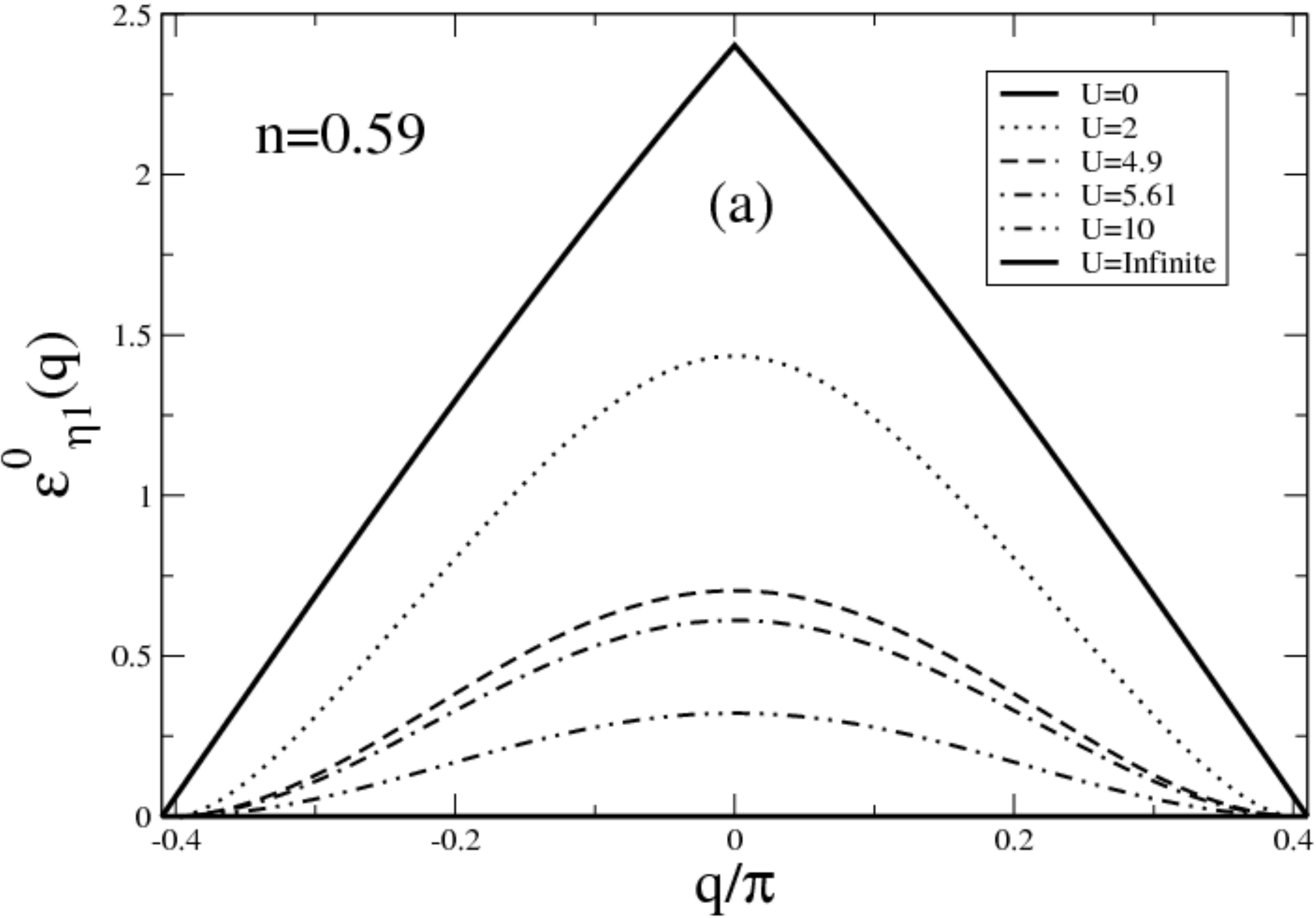}}
\subfigure{\includegraphics[width=5.0cm]{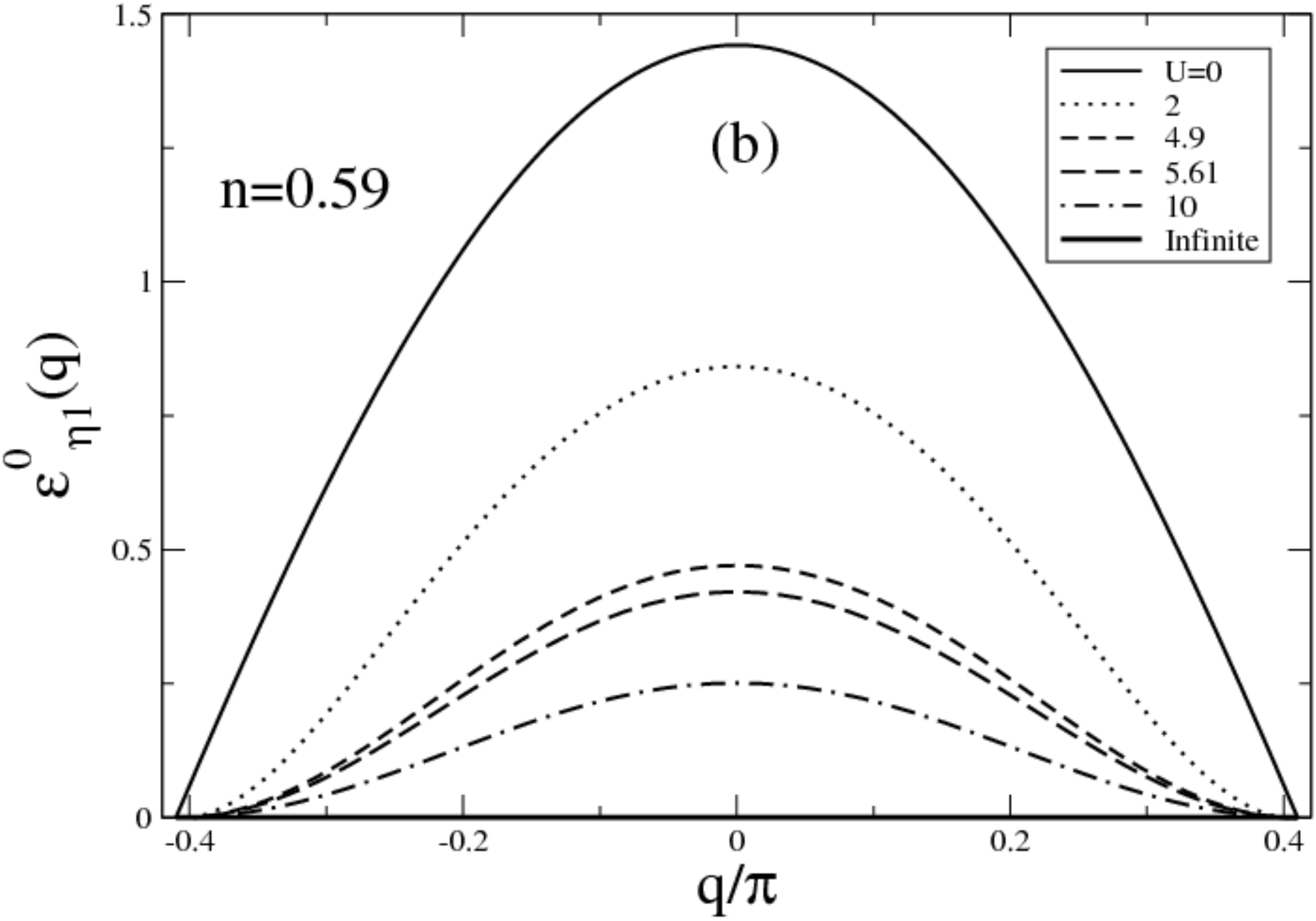}}
\caption{The $\eta 1$ band energy dispersion $\varepsilon_{\eta 1}^0 (q_j)$, Eq. (\ref{e-0-bands}) for $\beta = \eta 1$,
plotted in units of $t$ for the same $U/t$ and $n_e$ values as Fig. \ref{figure10} and (a) $m=0$ and (b) $m\rightarrow n_e=0.59$.
(As in Ref. \cite{Carmelo-03}, in the figures the electronic density $n_e$ is denoted by $n$).\\
{\it Source}: The figures plots were produced using the same data as in Figs. 8 (a) and 9 (a) of Ref. \cite{Carmelo-03}
for other densities.}
\label{figure13}
\end{center}
\end{figure}

\begin{figure}
\begin{center}
\subfigure{\includegraphics[width=5.00cm]{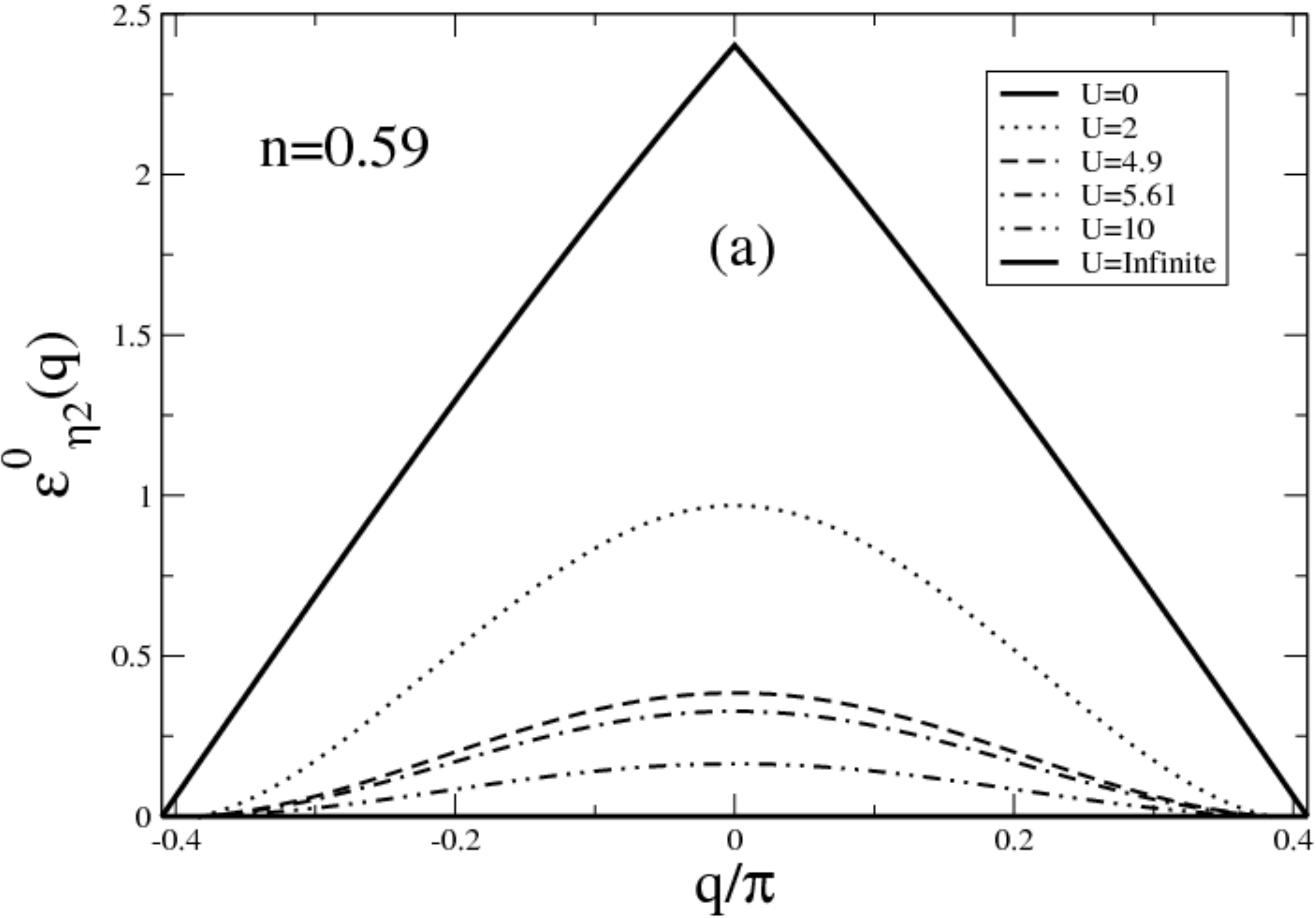}}
\subfigure{\includegraphics[width=5.00cm]{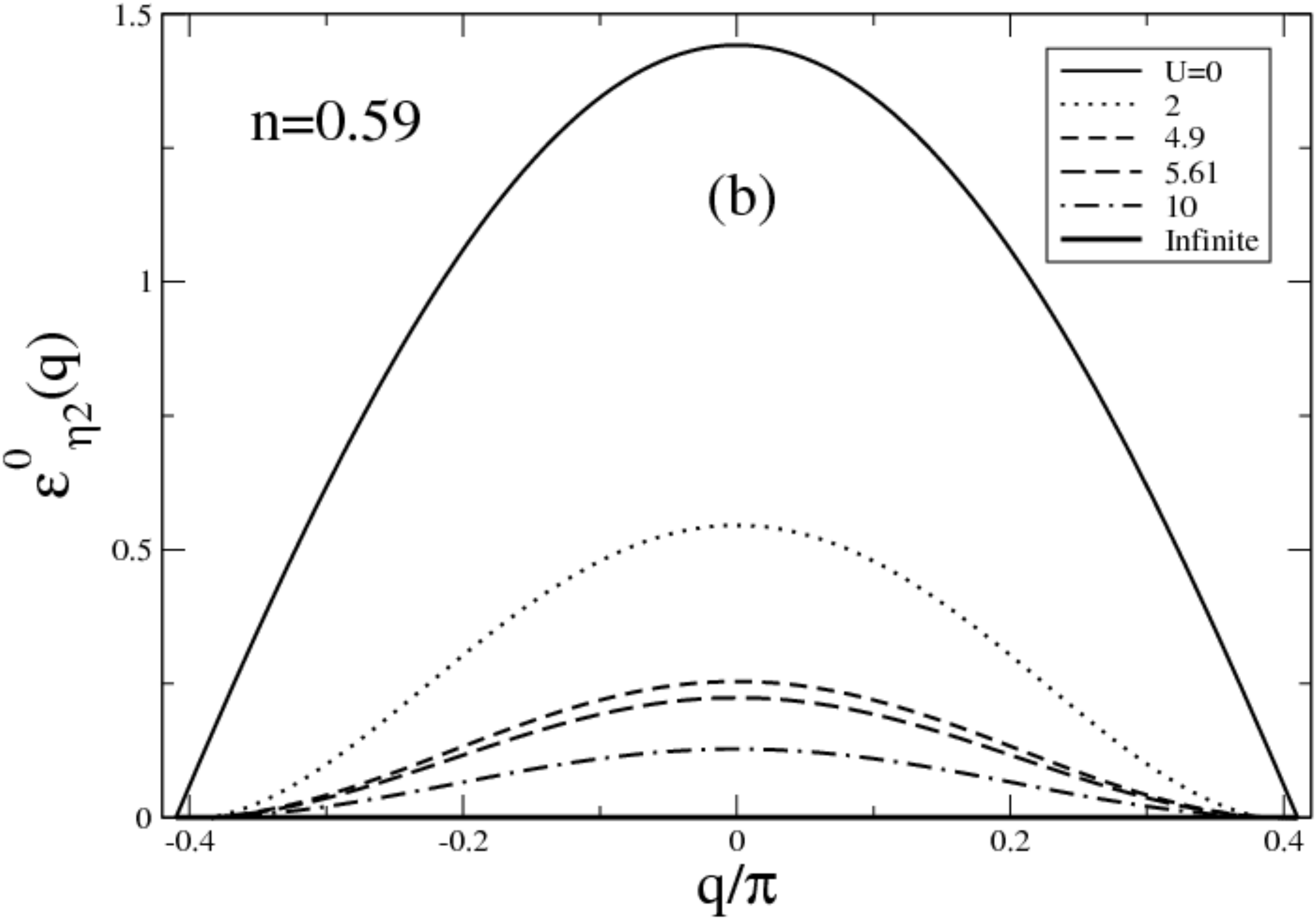}}
\caption{The $\eta 2$ band energy dispersion $\varepsilon_{\eta 2}^0 (q_j)$, Eq. (\ref{e-0-bands}) for $\beta = \eta 2$,
plotted in units of $t$ for the same $U$ and $n$ values (denoted in the figures by $n$)
as Fig. \ref{figure10} and (a) $m=0$ and (b) $m\rightarrow n_e=0.59$. 
(As in Ref. \cite{Carmelo-03}, in the figures the electronic density $n_e$ is denoted by $n$).\\
{\it Source}: The figures plots were produced using the same data as in Figs. 8 (b) and 9 (b) of Ref. \cite{Carmelo-03}
for other densities.}
\label{figure14}
\end{center}
\end{figure}

The energy dispersions $\varepsilon_{c} (q)$, $\varepsilon_{s1}^0 (q)$, $\varepsilon_{s2}^0 (q)$, $\varepsilon^0_{\eta 1} (q)$, and
$\varepsilon^0_{\eta 2} (q)$ are plotted as a function of the momentum $q$
in Figs. \ref{figure10}, \ref{figure11}, \ref{figure12}, \ref{figure13}, and \ref{figure14}, respectively, for several $U/t$ values,
electronic density $n_e=0.59$, and spin densities $m=0$ and/or $m\rightarrow n_e=0.59$. 
(The electronic density $n_e = 0.59$ is that used in Refs. \cite{TTF,spectral-06,spectral0,spectral}
for the stacks of TCNQ molecules in TTF-TCNQ.) Analysis of the figures energy-dispersions slopes reveals
that the velocity $v_{\beta} (q)$, Eq. (\ref{vel-beta}),
vanishes at $q=0$ for all $\beta =c,\alpha n$ branches. Provided that $u>0$, it also vanishes
at the limiting momentum values $q=\pm q_{\beta}$ for the $\beta\neq s1$ branches.
The energy bandwidths of the $s1$ band, $s2$ band, $\eta 1$ band, and $\eta 2$ band,
plotted in Figs. \ref{figure11}, \ref{figure12}, \ref{figure13}, and \ref{figure14}, respectively, vanish in the $u\rightarrow\infty$
limit. (In the $m\rightarrow 0$ limit, the momentum 
and energy bandwidths of the $s2$ band energy dispersion vanish for all $u$ values, so that  
it is not plotted in Fig. \ref{figure12} for $m=0$.) 
This $u\rightarrow 0$ behavior results from the degenerescence of all spin configurations and of all 
$\eta$-spin configurations with the same electron double occupancy reached in that limit.
For the $u\rightarrow 0$ and $u\gg 1$ limiting behaviors of the $\beta$ energy 
dispersions, Eqs. (\ref{epsilon-q}) and (\ref{e-0-bands}), see Ref. \cite{Carmelo-03}. 

%%%%%%%%%%%%%%%%%%%%%%%%%%%%%%%%%%%%%%%%%%%%%%%%%%%%%%%%%%%%%%%%%%%%%%%%%%
\subsection{The $c$ and $\alpha n$ pseudoparticle quantum liquid II: Applications to the low-energy physics}
\label{exc-spectraQL2}

The 1D Hubbard model in general PSs is a quantum liquid of $c$ pseudoparticles and
$n=1,...,\infty$ branches of composite $\eta n$-pseudoparticles and $sn$-pseudoparticles. 
In the following we consider again that model in its metallic-phase PSs whose ground states 
are LWSs with densities $n_e \in [0,1[$ and $m \in [0,n_e]$ for which $(\mu - \mu^0)>0$ and $H>0$.
In the case of that quantum problem, there emerge gaps $\Delta_{\eta}$ and $\Delta_s$ 
between such ground states and their PS excited energy eigenstates 
populated by $\eta n$ pseudoparticles and $n>1$ $s n$ pseudoparticles, respectively.
Such gaps minimum values are $\Delta_{\eta}^{\rm min} = \varepsilon_{\eta 1} (q_{\eta 1}^{\pm})
= 2\vert\mu\vert$ and $\Delta_{s}^{\rm min} = \varepsilon_{s 2} (0)=4\mu_B\,\vert H\vert - \vert\varepsilon_{s 2}^0 (0)\vert$. 
For excitation energy below these gaps, the physics is that of two $U(1)$ symmetry
quantum problems. The corresponding energy and momentum eigenstates are described by only groups of real rapidities.
(This is as in the case of the 1D Lieb-Liniger Bose gas.) The $\eta 1$ pseudoparticle energy spectrum is gapped, in spite of
corresponding to real rapidities. This results from creation of one $\eta 1$ pseudoparticle involving creation of one
rotated-electron doubly occupied site. (In the attractive $U<0$ 1D Hubbard model the situation
is the opposite, with the $s1$ pseudoparticle energy spectrum gaining a gap and the $\eta 1$ pseudoparticle 
spectrum being gapless.)

In the present case of chemical-potential values $(\mu - \mu^0)>0$ and magnetic fields $H>0$, 
the model static and low-temperature properties are determined by excitations associated with energy and momentum eigenstates 
with finite $c$ and $s1$ pseudoparticle occupancy $N_c=N_e$ and $N_{s1} = N_{e\downarrow}=(N_e-2S_s)/2$ only in 
the $c$ and $s1$ bands, respectively. This applies as well to the finite-energy dynamical correlation functions 
leading order contributions. Hence for such states $N_{\eta n} =0$ for all $n=1,...,\infty$ and $N_{s n} =0$ for $n>1$. 
On the one hand, due to the $n_e=1$ Mott-Hubbard gap, the physics of the model Mott-Hubbard insulator phase
is qualitatively different from that of its $n_e\neq 1$ metallic phase. On the other hand, the physical quantities have the same 
values both at $H=0$ and in the $H\rightarrow 0$ limit, respectively. 

For the quantum problem under consideration here, the following parameters involving the dressed phases 
shift $2\pi\Phi_{\beta\,\beta'}(q_j,q_{j'})$, Eq. (\ref{Phi-barPhi}), in units of $2\pi$ and with
the two momentum values $q_j$ and $q_{j'}$ at the $\beta,\beta' = c,s1$ Fermi 
points play an important role in the static and low-temperature properties
\cite{Carmelo-91-A,Carmelo-91,Carmelo-92-B,Carmelo-92-C}, 
\begin{equation}
\xi^{j}_{\beta\,\beta'} = \delta_{\beta,\beta'} 
+ \sum_{\iota = \pm} (\iota)^j\,\Phi_{\beta\,\beta'}\left(q_{F\beta},\iota q_{F\beta'}\right)
\hspace{0.20cm}{\rm where}\hspace{0.20cm}\beta, \beta' = c, s1\hspace{0.20cm}{\rm and}\hspace{0.20cm} j = 0, 1 \, .
\label{x-aa}
\end{equation}
(For $\beta =\beta'$ and $\iota=1$ in Eq. (\ref{x-aa}), the present 
TL notation assumes that the two $\beta =c,s1$ Fermi momenta in the argument of 
$\Phi_{\beta,\beta}\left(q_{F\beta},q_{F\beta}\right)$, differ by $2\pi/L$,
whereas that phase shift vanishes for identical momentum values.)

The dressed phase-shift related anti-symmetrical parameters $\xi^{1}_{\beta\,\beta'}$ and symmetrical 
parameters $\xi^{0}_{\beta\,\beta'}$ turn out to be the entries of the conformal-field theory $2\times 2$
dressed-charge matrix and of the transposition of its inverse matrix \cite{Carmelo-91,Carmelo-92-B,LE,Woy-89,Frahm-90},
\begin{equation}
Z^1 = \left[\begin{array}{cc}
\xi^{1}_{c\,c} & \xi^{1}_{c\,s1}  \\
\xi^{1}_{s1\,c}   & \xi^{1}_{s1\,s1}  
\end{array}\right]
\, ; \hspace{0.5cm}
Z^0 = \left[\begin{array}{cc}
\xi^{0}_{c\,c} & \xi^{0}_{c\,s1}  \\
\xi^{0}_{s1\,c}   & \xi^{0}_{s1\,s1} 
\end{array}\right] 
\, ; \hspace{0.5cm}
\lim_{m\rightarrow 0}\,Z^1 = \left[\begin{array}{cc}
\xi_{0} & \xi_{0}/2 \\
0 & 1/\sqrt{2} 
\end{array}\right]
\, ; \hspace{0.5cm}
\lim_{m\rightarrow 0}\,Z^0 = \left[\begin{array}{cc}
1/\xi_{0} & 0 \\
-1/\sqrt{2} & \sqrt{2} 
\end{array}\right] \, ,  
\label{ZZ-gen}
\end{equation}
respectively, where $Z^0 = ((Z^1)^{-1})^T$. (The dressed-charge matrix definition 
of Ref. \cite{Woy-89} has been used here, which is the transposition of that of Ref. \cite{Frahm-90}.)
The $m\rightarrow 0$ phase-shift parameter $\xi_0$ in Eq. (\ref{ZZ-gen})
is given by $\xi_{0} = \xi_{0} (r_c^0)$. The function $\xi_{0} (r)$ is the unique
solution of the integral equation, Eq. (74) of Ref \cite{Carmelo-92-B} for $x=r$.
That parameter has limiting values $\xi_0=\sqrt{2}$ for $u\rightarrow 0$ and $\xi_0=1$ for $u\rightarrow\infty$. 

At low energy the present quantum problem can be described by a two-component TLL
\cite{Voit,Schulz-90,Lederer-00,Tomonaga-50,Luttinger-63,Solyom-79}. The corresponding $g$ matrix \cite{Lederer-00}
can be expressed in terms of the dressed phase-shift parameters, Eq. (\ref{x-aa}).
A low-energy physical quantity that is fully controlled by the above phase-shift related parameters
in Eq. (\ref{ZZ-gen}) is the exponent in the low-energy $\omega$ power-law dependence of the electronic density 
of states suppression, $\propto\vert\omega\vert^{\alpha_0}$. In the $m\rightarrow 0$ limit its following expression
involves only the parameter $\xi_{0}$ in that equation, 
\begin{equation}
\alpha_0 = {(2-\xi_{0}^2)^2\over 8\xi_{0}^2} \in [0,1/8] \, .
\label{alphaHM}
\end{equation}
The exponent, Eq. (\ref{alphaHM}), has limiting values $\alpha_0=0$ for $u\rightarrow 0$ 
and $\alpha_0=1/8$ for $u\rightarrow\infty$, respectively. 

By use of the methods reported for the 1D Lieb-Liniger Bose gas and spin-$1/2$ $XXX$ chain, one finds that for 
(i) electronic densities $n_e$ and (ii) spin densities $m$ not too near (i) $0$ and $1$ (ii) and $n_e$, respectively, the 
low-temperature specific heat reads \cite{Carmelo-91-A},
\begin{equation}
{c_V\over L} = {k_B\,\pi\over 3}\left({1\over v_c} + {1\over v_{s1}}\right)\,(k_B T) \, ,
\label{cV-1DHm}
\end{equation}
where $v_{c}$ and $v_{s1}$ are the $\beta =c,s1$ band Fermi velocities in Eq. (\ref{vel-beta}).

On the one hand, for electronic densities $n_e \in ]0,1[$ not too close to $n_e =0$ and $n_e =1$ the 
low-temperature thermal excitations that contribute to the first term of the specific heat 
expression, Eq. (\ref{cV-1DHm}), refer to a well-defined branch of gapless excited energy and momentum eigenstates.
Their charge degrees of freedom are generated from the ground state by low-energy and small-momentum particle-hole 
processes around the $c$ band band Fermi points. 

On the other hand, as for the spin-$1/2$ $XXX$ chain, at zero magnetic field $H=0$ the spin degrees 
of freedom of such gapless branch of states that contribute to the second term of the specific heat 
expression, Eq. (\ref{cV-1DHm}), refer to $sn$-strings of lengths $n>1$. For finite magnetic field these $n>1$ 
$sn$-strings excitations become gapped. The thermal excitations that contribute to the low-temperature specific heat 
become gapless excited energy and momentum eigenstates whose charge degrees of freedom remain being
generated by low-energy and small-momentum particle-hole processes around the $c$ band band Fermi points.
Their spin degrees of freedom correspond to spin real rapidities. Those are generated from the ground state 
by low-energy and small-momentum particle-hole processes around the $s1$ band Fermi points. 

The specific-heat expression obtained for $H>0$ leads in the $H\rightarrow 0$ limit to the
correct $H=0$ expression. On the contrary, the specific heat expression, Eq. (\ref{cV-1DHm}), is not 
valid in the $m\rightarrow n_e$ limit. This is because it does not describe properly the crossover to the 
specific heat exponential regime. The latter arises due to the gap $2\mu_B (H-H_c)$ in the excitation spectrum for $H>H_c$.
More generally, the validity of that specific heat expression refers to very low temperatures $T\ll 2\mu_B(H_c-H)/k_B$
for $(H_c-H)>0$, $T\ll 2(\mu - \mu_0)/k_B$ for $(\mu - \mu_0)>0$, and $T\ll 2(\mu_1 - \mu)/k_B$ for $(\mu_1 - \mu)>0$.
In the close neighborhood of $\mu = \mu_1 = U/2 +2t$, the problem is trivial. The specific heat is then given 
by its noninteracting value. Following the technical similarities of the crossover critical regimes associated with the
$2\mu_B\vert H-H_c\vert\ll k_B T$ and $2\vert \mu - \mu_0\vert\ll k_B T$ limits \cite{Carmelo-91-A}, here 
we shortly discuss the former regime for electronic densities $n_e \in ]0,1[$ not too close to $n_e =0$ and $n_e =1$.
\begin{figure}
\begin{center}
\subfigure{\includegraphics[width=5.00cm]{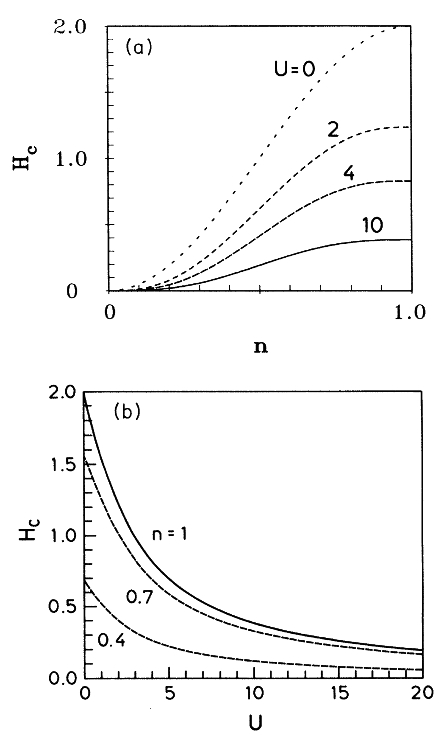}}
\hspace{0.50cm}
\subfigure{\includegraphics[width=5.00cm]{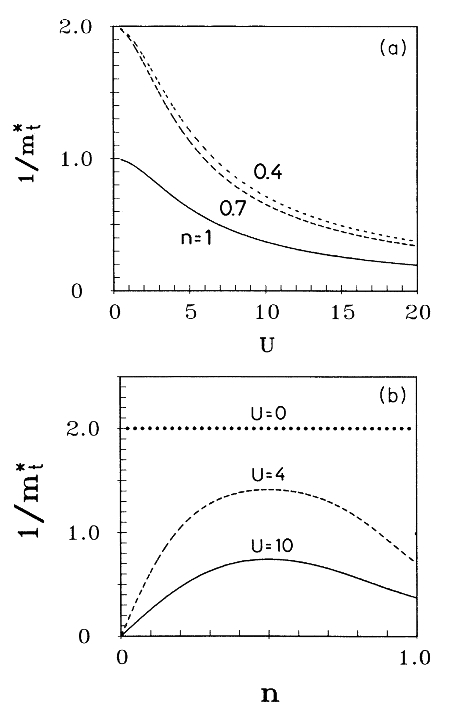}}
\caption{The critical magnetic field $H_c$ for the onset of fully polarized ferromagnetism in Eq. (\ref{muhc})
and the inverse effective spin-triplet mass $1/m_{s1}^*$, Eq. (\ref{ms1star}),
versus the electronic density $n_e$ for $U/t=0$, $U/t=2$, and $U/t=10$ and versus
$U/t$ for $n_e =0.4$, $n_e =0.7$, and $n_e =1.0$. Notice that at $n_e=1$, $1/m_{s1}^*\rightarrow t$ as $U/t\rightarrow 0$.
This singular behavior is due to the Mott-Hubbard insulator transition.
(As in Ref. \cite{Carmelo-91-A}, in the figures the electronic
density $n_e$ and effective spin-triplet mass $m_{s1}^*$ are denoted by 
$n$ and $m_{t}^*$, respectively).\\
{\it Source}: From Ref. \cite{Carmelo-91-A}.}
\label{figure15}
\end{center}
\end{figure}

Near $H=H_c$ the minimum gap for energy eigenstates with spin $s n$-strings of length $n>1$ associated with
complex rapidities is given by, 
\begin{equation}
\Delta_{s}^{\rm min} = 4\mu_B\,H_c - W_{s2} \, .
\label{Deltasmin}
\end{equation}
The dependence on $U$, $t$, and $n_e$ of the energy scale $4\mu_B\,H_c$ appearing here is given in the following.
$W_{s2}\equiv\vert\varepsilon_{s2}^0(0)\vert$ is in Eq. (\ref{Deltasmin}) the energy scale in Eq. (\ref{Ws-nu}) of Appendix \ref{TBAconfig} for $n=2$.
Its limiting behaviors for $u\rightarrow 0$ and $u \gg 1$ are,
\begin{eqnarray}
\Delta_{s}^{\rm min} & = & 2\mu_B\,H_c = 4t\sin^2\left({\pi n_e\over2}\right)\hspace{0.20cm}{\rm for}\hspace{0.20cm}u\rightarrow 0 
\nonumber \\
& = & 3\mu_B\,H_c = {12\,n_e\,t^2\over U}\left(1 - {\sin (2\pi n_e)\over 2\pi n_e}\right)
\hspace{0.20cm}{\rm for}\hspace{0.20cm}u \gg 1 \, ,
\label{Deltasminlimyts}
\end{eqnarray}
respectively.

We consider low temperatures $T< \Delta_{s}^{\rm min}/k_B$ within the critical regime
of the crossover to ferromagnetism for which $2\mu_B\vert H-H_c\vert\ll k_B T$. For such temperatures
the energy and momentum eigenstates that contribute to the specific heat have spin degrees of freedom described only 
by spin $s n$-strings of length $n=1$. They are thus only populated by $s1$ pseudoparticles and unpaired spins 
$1/2$. At low temperatures the crossover regime involves both the
above $s1$ band gapless spin-singlet excited states and across-gap excited 
energy eigenstates. The latter states spin degrees of freedom are generated by elementary spin-triplet $\delta S_s =\pm 1$ processes.
(They are similar to those considered for the spin-$1/2$ $XXX$ chain in Section \ref{n1pseudoHchain}.)

The $s1$ band energy dispersion valid for spin density $m\rightarrow n_e$ reads \cite{Carmelo-91-A},
\begin{equation}
\varepsilon_{s1} (q_j) \approx {q_j^2\over 2m_{s1}^*} - 2\mu_B(H_c-H)
\hspace{0.20cm}{\rm for}\hspace{0.20cm}n_e \in ]0,1[\hspace{0.20cm}{\rm and}\hspace{0.20cm}m\rightarrow n_e \, .
\label{vares1smallq}
\end{equation}
Here the critical magnetic energy $2\mu_B H_c$ is that associated with the zero-temperature
critical magnetic field $H_c$ for the onset of fully polarized ferromagnetism.
That energy scale and the effective spin-triplet mass $m_{s1}^*$ in Eq. (\ref{vares1smallq})
are given by \cite{Carmelo-91-A},
\begin{eqnarray}
2\mu_B H_c & = & \sqrt{(4t)^2+U^2}\,{1\over\pi}\arctan\left({\sqrt{(4t)^2+U^2}\over U}\tan (\pi n_e)\right)
- U\,n_e - 4t\cos (\pi n_e)\,{1\over\pi}\arctan\left({4t\sin (\pi n_e)\over U}\right) 
\nonumber \\
& & {\rm for}\hspace{0.20cm}n_e \in ]0,1[\hspace{0.20cm}{\rm and}\hspace{0.20cm}m\rightarrow n_e \, ,
\label{muhc}
\end{eqnarray}
and
\begin{equation}
m_{s1}^* = {U\over 4t^2}{{\sqrt{(4t)^2+U^2}\over U}{1\over\pi}\arctan\left({\sqrt{(4t)^2+U^2}\over U}\tan (\pi n_e)\right)
\over 1 - {\sqrt{(4t)^2+U^2}\over U}{1\over 1 + \left({4t\sin (\pi n_e)\over U}\right)^2}
{\sin (2\pi n_e)\over 2\arctan\left({\sqrt{(4t)^2+U^2}\over U}\tan (\pi n_e)\right)}} 
\hspace{0.20cm}{\rm for}\hspace{0.20cm}n_e \in ]0,1[\hspace{0.20cm}{\rm and}\hspace{0.20cm}m\rightarrow n_e \, ,
\label{ms1star}
\end{equation}
respectively. Their limiting behaviors for $u\rightarrow 0$ and $u \gg 1$ read,
\begin{eqnarray}
m_{s1}^* & = & {1\over 2t}\hspace{0.20cm}{\rm and}\hspace{0.20cm}
2\mu_B H_c = 4t\sin^2\left({\pi n_e\over2}\right)\hspace{0.20cm}{\rm for}\hspace{0.20cm}u\rightarrow 0 
\nonumber \\
m_{s1}^* & = & {U\over 4t^2}{n_e\over\left(1 - {\sin (2\pi n_e)\over 2\pi n_e}\right)}
\hspace{0.20cm}{\rm and}\hspace{0.20cm}
2\mu_B H_c = {8\,n_e\,t^2\over U}\left(1 - {\sin (2\pi n_e)\over 2\pi n_e}\right)
\hspace{0.20cm}{\rm for}\hspace{0.20cm}u \gg 1 \, ,
\label{ms1muhculim}
\end{eqnarray}
respectively. The critical magnetic field $H_c$ in Eq. (\ref{muhc}) and the inverse effective spin-triplet mass $1/m_{s1}^*$, 
Eq. (\ref{ms1star}), are plotted in Fig. \ref{figure15} as a function of the electronic density $n_e$
and of $U/t$ for various values of $U/t$ and $n_e$, respectively.

For the crossover critical regime under consideration, the scaling function of the specific heat is found to be given by,
\begin{eqnarray}
{c_V\over L} & = & {k_B\,\pi\over 3v_c}\,k_B T
+ \sqrt{2m_{s1}^*\,k_B T\over\pi}\left(-{3\over 8}\,f_{3/2}^s + {1\over 2}\left({2\mu_B (H_c - H)\over k_B T}\right)\,f_{1/2}^s
- {1\over 2}\left({2\mu_B (H_c - H)\over k_B T}\right)^2\,f_{-1/2}^s\right) \, ,
\nonumber \\
& & {\rm where}\hspace{0.20cm}v_c = 2t\sin (\pi n_e) \, , \hspace{0.20cm}
f^s_l = {\rm Li}_l \left(-e^{2\mu_B (H_c - H)\over k_B T}\right)
\hspace{0.20cm}{\rm and}\hspace{0.30cm}{\rm Li}_l (x) = \sum_{j=1}^{\infty} {x^j\over j^l} \, .
\label{cV-HMhcGCR}
\end{eqnarray}

Expanding this scaling function up to first order in $2\mu_B (H_c - H)/k_B T$, 
one finds that the low-temperature specific heat behaves in a small field window $2\mu_B\vert H-H_c\vert\ll k_B T$ 
around $H_c$ as \cite{Carmelo-91-A}, 
\begin{equation}
{c_V\over L} = {k_B\,\pi\over 3v_c}\,k_B T
+ k_B\,c_0\,\sqrt{m_{s1}^*\,k_B T\over 2}\,\left(c_1 + c_2 {2\mu_B (H_c - H)\over k_B T}\right)
\hspace{0.20cm}{\rm where}\hspace{0.20cm}2\mu_B\vert H-H_c\vert\ll k_B T \, .
\label{cV-HMhc}
\end{equation}
The term emerging here from the spin degrees of freedom has the same form as that
provided in Eq. (\ref{cV-hc}) and the coefficients $c_0$, $c_1$, and $c_2$ are 
thus also those given in Eq. (\ref{c012}). (Also the calculations to reach Eq. (\ref{cV-HMhc})
are similar to those presented in Appendix \ref{XXXconfig2} for the corresponding specific heat 
scaling function of the spin-$1/2$ $XXX$ chain \cite{He-17}.)

In the crossover critical regime defined by Eq. (\ref{cV-HMhc}) 
both the gapless and across-gap channels associated with spin-singlet excitations and excitations
generated by elementary spin-triplet processes, respectively, are thermally active. That equation 
is only valid for a very narrow region around $H_c$.

The charge response function or compressibility $\alpha =\eta$ and spin response function $\alpha =s$,
\begin{equation}
\chi_{\eta}\vert_{y} = - {1\over n_e^2}{1\over\partial\mu (n_e)/\partial n_e\vert_{y}} 
\hspace{0.20cm}{\rm where}\hspace{0.20cm} y = H,m \hspace{0.20cm}{\rm and}\hspace{0.20cm}
\chi_{s}\vert_{z} = - {2\mu_B\over\partial H (m)/\partial m\vert_{z}} 
\hspace{0.20cm}{\rm where}\hspace{0.20cm}z = \mu, n_e \, ,
\label{chi-eta-s}
\end{equation}
are controlled by the dressed phase-shift parameters, Eq. (\ref{x-aa}). Such functions involve partial 
derivatives of the chemical potential and magnetic field, Eq. (\ref{mu-muBH}).

By use of techniques similar to those of a Fermi liquid, which account for the $\beta=c,s1$
and $\beta'=c,s1$ pseudoparticle zero-momentum forward-scattering interactions associated
with the $f$ functions, Eq. (\ref{ff}), the studies of Ref. \cite{Carmelo-92-B} have found,
\begin{eqnarray}
\chi_{\eta}\vert_{H} & = & {1\over \pi n_e^2}\sum_{\beta =c,s1}
{(\xi^{1}_{\beta\,c})^2\over v_{\beta}} 
\hspace{0.20cm}{\rm and}\hspace{0.20cm}
\chi_{\eta}\vert_{m} = {1\over \pi n_e^2}{1\over\sum_{\beta =c,s1}
v_{\beta}(\xi^{0}_{\beta\,c}+\xi^{0}_{\beta\,s1}/2)^2} \, ,
\nonumber \\
\chi_{s}\vert_{\mu} & = & {\mu_B^2\over\pi} \sum_{\beta =c,s1}
{(\xi^{1}_{\beta\,c}-2\xi^{1}_{\beta\,s1})^2\over v_{\beta}} 
\hspace{0.20cm}{\rm and}\hspace{0.20cm}
\chi_{s}\vert_{n_e} = {\mu_B^2\over\pi} {1\over\sum_{\beta =c,s1}
v_{\beta}(\xi^{0}_{\beta\,s1}/2)^2} \, .
\label{chi-eta-s-xi}
\end{eqnarray}
\begin{figure}
\begin{center}
\subfigure{\includegraphics[width=3.50cm]{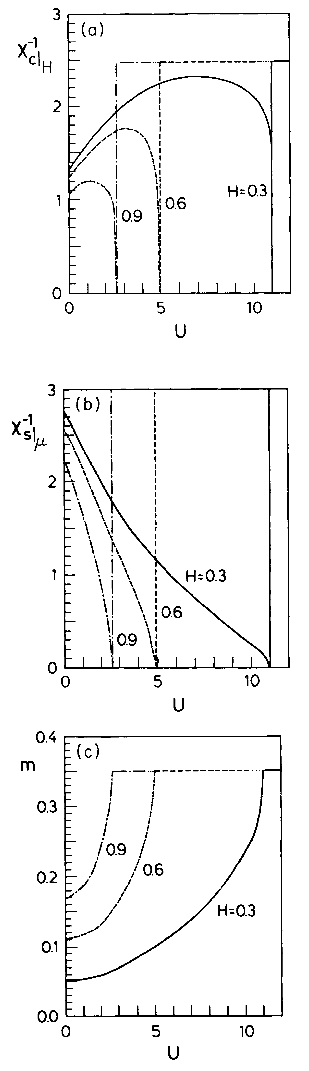}}
\hspace{0.50cm}
\subfigure{\includegraphics[width=4.00cm]{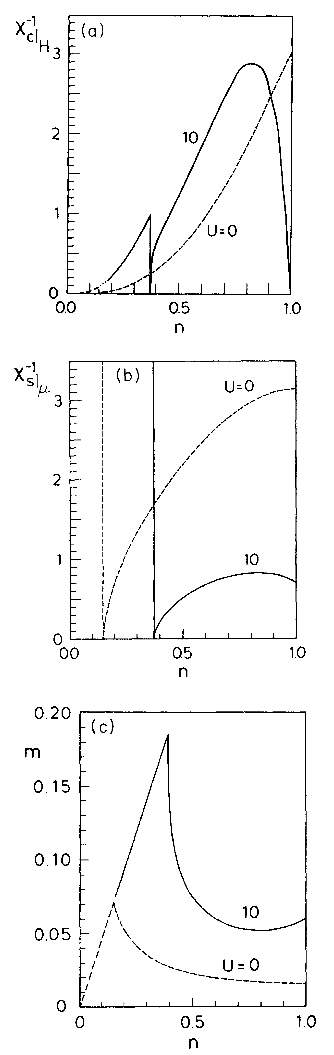}}
\caption{The inverse of the compressibility $\chi_{\eta}\vert_{H}$ and the inverse of the spin response function $\chi_{s}\vert_{\mu}$,
Eqs. (\ref{chi-eta-s}), (\ref{chi-eta-s-xi}), and (\ref{chi-eta-s-xig0}), and the spin density $m$ 
versus $U/t$ at electronic density $n_e =0.7$ and values of the magnetic field $H=0.3$
(solid line), $H=0.6$ (dashed line), and $H=0.9$ (dashed-dotted line) in suitable units
in the left panels and versus the electronic density $n_e$ for $H=0.1$ and values
of the on-site repulsion $U/t=0$ (dashed line) and $U/t=10$ (solid line) in the right panels. 
The discontinuities occur at $U/t$ or $n_e$ values at which according to Eq. (\ref{muhc})
the magnetic field value $H$ under consideration becomes $H_c$ and thus the system 
becomes ferromagnetic. Note the different behavior at $U/t=0$ and $U/t=10$ of $\chi_{\eta}\vert_{H}$ around
$n_e = 1$ due to the Mott-Hubbard insulator transition.
(The spin density $m$ of Ref. \cite{Carmelo-92-B} is is half of that considered in this
paper so that in the figures the fully polarized ferromagnetism is reached in
the limit of $m\rightarrow n_e/2$, which corresponds to $m\rightarrow n_e$ in
this review. As in that reference, in the figures the electronic density $n_e$ and compressibility $\chi_{\eta}$ 
are denoted by $n$ and $\chi_{c}$, respectively).\\
{\it Source}: From \cite{Carmelo-92-B}.}
\label{figure16}
\end{center}
\end{figure}

The derivation in Ref. \cite{Carmelo-92-B} of the charge and spin response functions, Eq. (\ref{chi-eta-s}), 
expressions provided in Eq. (\ref{chi-eta-s-xi}) uses the general procedures reported in Appendix \ref{XXXconfig3}
to obtain the corresponding expression of the spin response function of the spin-$1/2$ $XXX$ chain. 
Such two-electron response functions expressions 
can be understood as being controlled by parameters that play the same role as the Landau parameters in
a Fermi liquid. For the present $c$ and $s1$ pseudoparticle quantum liquid, the Landau parameters can
have a $i= 0$ symmetrical or $i=1$ anti-symmetrical character. They read,
\begin{eqnarray}
g_{\beta\,\beta}^i & = & 1 + {1\over 2\pi v_{\beta}}\sum_{\iota =\pm 1}(\iota)^i f_{\beta\,\beta}(q_{F\beta},\iota\,q_{F\beta}) 
\nonumber \\
& = & {1\over v_{\beta}}\left(v_c\,(\xi^i_{c\,\beta})^2 + v_{s1}\,(\xi^i_{s1\,\beta})^2\right) 
\hspace{0.20cm}{\rm for}\hspace{0.20cm} \beta = c,s1\hspace{0.20cm}{\rm and}\hspace{0.20cm}i = 0,1 \, ,
\nonumber \\
g_{\beta\,\beta'}^i & = & {1\over 2\pi v_{\beta}}\sum_{\iota =\pm 1}(\iota)^i f_{\beta\,\beta'}(q_{F\beta},\iota\,q_{F\beta'}) 
\nonumber \\
& = & {1\over v_{\beta}}\left(v_c\,\xi^i_{c\,c}\xi^i_{c\,s1} + v_{s1}\,\xi^i_{s1\,s1}\xi^i_{s1\,c}\right) 
\hspace{0.20cm}{\rm for}\hspace{0.20cm} \beta \neq \beta' = c,s1\hspace{0.20cm}{\rm and}\hspace{0.20cm}i = 0,1 \, .
\label{gbbgcs}
\end{eqnarray}
Such parameters expressions contain the $c$ and $s1$ bands velocities at a Fermi point defined in
Eq. (\ref{vel-beta}). They also contain two $f$ functions, Eq. (\ref{ff}), with the two momenta at Fermi points pointing in the same and in
opposite directions, respectively. The $i= 0$ symmetrical and $i=1$ anti-symmetrical pseudoparticle
Landau parameters involve the sum and difference of these two $f$ functions, respectively.

As given in Eq. (\ref{gbbgcs}), the pseudoparticle Landau parameters can be expressed in terms of the $\beta =c,s1$ velocities 
at a Fermi point and of the dressed phase-shift related $i=1$ anti-symmetrical and $i=0$ symmetrical parameters 
$\xi^{i}_{\beta\,\beta'}$. Those are the entries of the conformal-field theory dressed charge matrix $Z^1$ and of the matrix 
$Z^0 = ((Z^1)^{-1})^T$, respectively, Eq. (\ref{ZZ-gen}). Such entries naturally emerge within the
expressions of the $i=0,1$ and $\beta = c,s1$ renormalized velocities
$v_{\beta\,\beta}^i\equiv v_{\beta}\,g_{\beta\,\beta}^i = v_{\beta} + {1\over 2\pi}\sum_{\iota =\pm 1}(\iota)^i f_{\beta\,\beta}(q_{F\beta},\iota\,q_{F\beta})$ 
and $v_{\beta\,\beta'}^i\equiv v_{\beta}\,g_{\beta\,\beta'}^i ={1\over 2\pi}\sum_{\iota =\pm 1}(\iota)^i f_{\beta\,\beta'}(q_{F\beta},\iota\,q_{F\beta'})$
where $\beta\neq\beta'$. The pseudoparticle representation goes though beyond conformal-field theory. Indeed,
the group velocities and $f$ functions in these expressions are also well defined for arbitrary $\beta$-band
momentum values, $v_{\beta} (q_j)$ and $f_{\beta\,\beta'}(q_j,q_{j'})$ for both $\beta =\beta'$ and
$\beta \neq\beta'$. Beyond that theory, they also exist within the pseudoparticle representation for all $c$, $\eta n$, and $sn$
branches where $n=1,...,\infty$. 

The charge and spin response functions expressions, Eqs. (\ref{chi-eta-s}) and (\ref{chi-eta-s-xi}), can be expressed
in terms of the Fermi-point $c$ and $s1$ group velocities and the $i=0$ symmetrical pseudoparticle Landau parameters as follows,
\begin{eqnarray}
\chi_{\eta}\vert_{H} & = & {1\over \pi n_e^2 v_c}\,{1\over \left(g_{c\,c}^0 - {g_{c\,s}^0\,g_{s1\,c}^0\over g_{s1\,s1}^0}\right)} 
\hspace{0.20cm}{\rm and}\hspace{0.20cm} 
\chi_{\eta}\vert_{m} = {1\over \pi n_e^2 v_c}\,{1\over\left(g_{c\,c}^0 + {v_{s1}\over 4v_{c}}\,g_{s1\,s1}^0 + g_{c\,s1}^0\right)} \, ,
\nonumber \\
\chi_{s}\vert_{\mu} & = & {\mu_B^2\over\pi v_{s1}}\, 
{\left(g_{s1\,s1}^0 + {4v_c\over v_{s1}}\,g_{c\,c}^0 + 4g_{s1\,c}^0\right)\over \left({v_c\over v_{s1}}\,g_{s1\,s1}^0\,g_{c\,c}^0 - (g_{s1\,c}^0)^2\right)} 
\hspace{0.20cm}{\rm and}\hspace{0.20cm} 
\chi_{s}\vert_{n_e} = {4\mu_B^2\over\pi v_{s1}} {1\over g_{s1\,s1}^0} \, .
\label{chi-eta-s-xig0}
\end{eqnarray}

For two-electron quantities such as the charge and spin response functions,
this renormalization is qualitatively similar to that of a Fermi liquid. Indeed, that liquid 
Landau parameters, which control the effects of the electronic interactions onto the low-energy quantities, are expressed in terms 
of the $f$ functions associated with the low-energy quasiparticles residual interactions. Similarly, here the residual pseudoparticle 
interactions associated with the $f$ functions in the parameters expressions, Eq. (\ref{gbbgcs}), play exactly the same role.

In the limit of zero magnetic field and thus of zero spin density, the residual pseudoparticle interactions 
occur only through the parameter $\xi_0$. In that limit they read,
\begin{equation}
\chi_{\eta}\vert_{H} = 
\chi_{\eta}\vert_{m} = {\xi_0^2\over \pi n_e^2 v_c} \hspace{0.20cm}{\rm and}\hspace{0.20cm}  
\chi_{s}\vert_{\mu} = \chi_{s}\vert_{n_e} = {2\mu_B^2\over\pi v_{s1}} \, .
\label{chi-eta-s-xig0h0}
\end{equation}
It then follows that $\lim_{u\rightarrow 0}\chi_{\eta}\vert_{H} =2/(\pi n_e^2 v_c)$ where
$v_c = 2t\sin\left({\pi\over 2}n_e\right)$ and $\lim_{u\rightarrow\infty}\chi_{\eta}\vert_{H} =1/(\pi n_e^2 v_c)$ 
where $v_c = 2t\sin (\pi n_e)$. Moreover,
$\lim_{u\rightarrow\infty}\chi_{s}\vert_{\mu} = \lim_{u\rightarrow\infty}\chi_{s}\vert_{n_e}\rightarrow\infty$ 
because $v_{s1}\rightarrow 0$ in that limit.

In the spin density $m\rightarrow n_e$ limit of the fully polarized ferromagnetism, one finds,
\begin{eqnarray}
\chi_{\eta}\vert_{H} & = & {1\over \pi n_e^2}\,\left({1\over v_c} + {\eta_0^2\over v_{s1}}\right)\rightarrow\infty 
\hspace{0.20cm}{\rm and}\hspace{0.20cm} 
\chi_{\eta}\vert_{m} = {4\over \pi n_e^2}\,{1\over\left(v_c\,(2-\eta_0)^2 +v_{s1}\right)}\rightarrow {1\over \pi n_e^2 v_c}\,{1\over (1-\eta_0/2)^2} \, ,
\nonumber \\
\chi_{s}\vert_{\mu} & = & {\mu_B^2\over\pi v_c}\, 
\left(1 + {4v_c\over v_{s1}}(1-\eta_0/2)^2\right)\rightarrow\infty 
\hspace{0.20cm}{\rm and}\hspace{0.20cm} 
\chi_{s}\vert_{n_e} = {4\mu_B^2\over\pi} {1\over v_{s1} + v_c\,\eta_0^2}\rightarrow {4\mu_B^2\over\pi v_c} {1\over\eta_0^2} \, .
\label{chi-eta-s-xig0mne}
\end{eqnarray}
Here $\eta_ 0 = {2\over\pi}\arctan\left({\sin (\pi n_e)\over u}\right)$, $v_c = 2t\sin (\pi n_e)$, and $v_{s1}\rightarrow 0$.
The inverse of the compressibility $\chi_{\eta}\vert_{H}$ and the inverse of the spin response function $\chi_{s}\vert_{\mu}$,
Eqs. (\ref{chi-eta-s}), (\ref{chi-eta-s-xi}), and (\ref{chi-eta-s-xig0}), and the spin density $m$ are plotted
in Fig. \ref{figure16} as a function of $U/t$ at electronic density $n_e =0.7$ and various
values of the magnetic field $H$ in suitable units and as a function of $n_e$ for $H=0.1$ and
various values of $U/t$.

The 1D Hubbard model quantum phase transitions driven by a change in the chemical potential $\mu $
or the magnetic field $H$, Eq. (\ref{mu-muBH}), are marked by the leading divergences of the ground-state onsite entanglement 
entropy ${\cal{E}}$ derivatives, $\partial {\cal{E}}/\partial\mu$ and $\partial {\cal{E}}/\partial H$, respectively \cite{Johannesson-05}.
For $n_e\neq 1$, they can alternatively be signed by the related derivatives, $\partial {\cal{E}}/\partial n_e = - n_e^2\chi_{\eta}\,\partial {\cal{E}}/\partial\mu$
and $\partial {\cal{E}}/\partial m = - [\chi_s/2\mu_B]\,\partial {\cal{E}}/\partial H$ \cite{Gu-04}. ($\chi_{\eta}$ diverges at $n_e=1$.)
\begin{figure}
\begin{center}
\subfigure{\includegraphics[width=5.00cm]{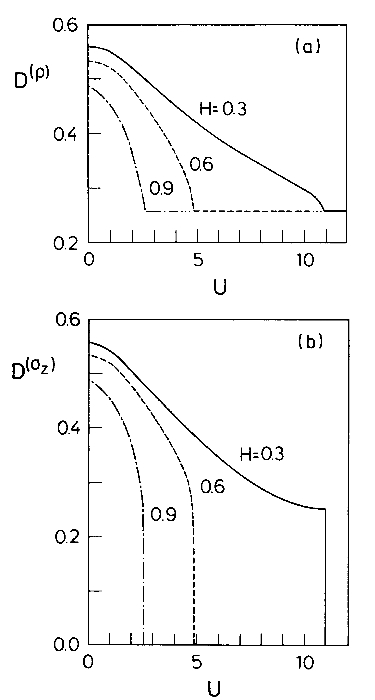}}
\hspace{0.50cm}
\subfigure{\includegraphics[width=5.00cm]{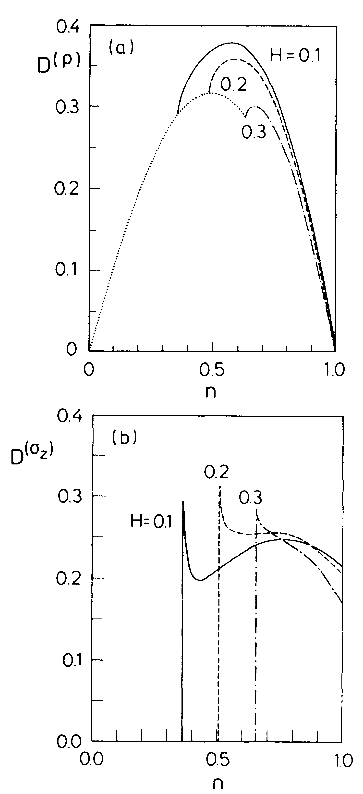}}
\hspace{0.50cm}
\subfigure{\includegraphics[width=5.00cm]{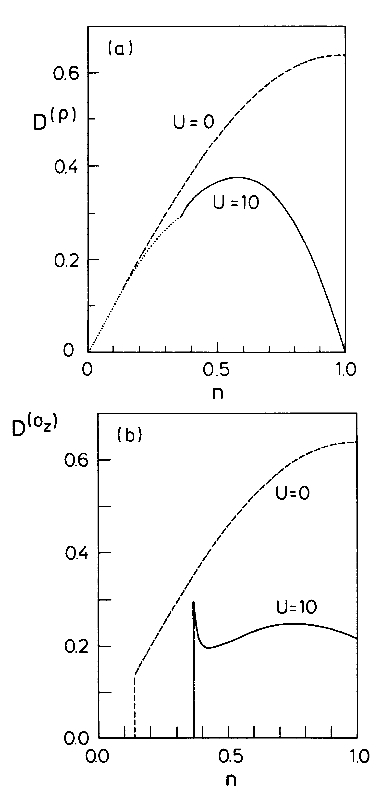}}
\caption{The zero-temperature charge stiffness $D_{\eta}$ and the spin stiffness $D_{s}$, Eq. (\ref{D-0}),
versus $U/t$ at electronic density $n_e =0.7$ and values of the magnetic field $H=0.3$
(solid line), $H=0.6$ (dashed line), and $H=0.9$ (dashed-dotted line) in suitable units
in the left panels, versus the electronic density $n_e$ for $U/t=10$ and values of the magnetic 
field $H=0.1$ (solid line), $H=0.2$ (dashed line), and $H=0.3$ (dashed-dotted line)
in the middle panels, and versus the electronic density $n_e$ for $H=0.1$ and
values of the on-site repulsion $U/t=0$ (dashed line) and $U/t=10$ (solid line) in the right panels. 
The discontinuities occur at $U/t$ or $n_e$ values at which according to Eq. (\ref{muhc})
the magnetic field value $H$ under consideration becomes $H_c$ and thus the system 
becomes ferromagnetic. Due to the Mott-Hubbard insulator transition, $D_{\eta}$ is finite at $n_e=1$ when
$U/t=0$ but vanishes when $U/t>0$. (As in Ref. \cite{Carmelo-92-C}, in the figures the
charge stiffness $D_{\eta}$, spin stiffness $D_{s}$, and electronic density $n_e$ are denoted by 
$D^{(\rho)}$, $D^{(\sigma_z)}$, and $n$, respectively).\\
{\it Source}: From Ref. \cite{Carmelo-92-C}.}
\label{figure17}
\end{center}
\end{figure}

Also charge $(\alpha =\eta)$ and spin $(\alpha =s)$ stiffnesses $D_{\alpha}$ in the corresponding
$\alpha$ conductivity real part $\sigma_{\alpha} (\omega) = 2\pi\,D_{\alpha}\,\delta (\omega) + \sigma^{reg}_{\alpha} (\omega)$
are at zero temperature fully controlled by the dressed phase-shift parameters, Eq. (\ref{x-aa}). As for the spin-$1/2$ $XXX$ chain,
those are important physical quantities. A finite charge and/or spin stiffness implies the occurrence
of charge and/or spin ballistic transport, respectively. Such ballistic transport occurs in 1D correlated 
even at finite temperatures \cite{Carmelo-00A,CNP-18}. Recently, a general formalism of hydrodynamics for the
1D Hubbard model was introduced in Ref. \cite{Ilievski-17}. By linearizing hydrodynamic equations, the
exact closed-form stiffnesses expressions valid on the hydrodynamic scale are accessed. 

For simplicity, here we focus our analysis on the zero-temperature charge and spin stiffnesses. Specifically,  
relying on charge and spin conservation laws to derive the elementary currents that contribute to 
such $\alpha =\eta,s$ stiffnesses, the studies of Ref. \cite{Carmelo-92-C} found that they read,
\begin{equation}
2\pi\,D_{\eta} = j_{c}^{\eta} \hspace{0.20cm}{\rm and}\hspace{0.20cm}  2\pi\,D_{s} = j_{c}^{s} - 2j_{s1}^{s} \, .
\label{D-0}
\end{equation}
The elementary currents in these expressions involve the $\beta =c,s1$ band Fermi velocities, Eq. (\ref{vel-beta}), 
and $\beta, \beta' =c,s1$ anti-symmetrical dressed phase-shift parameters $\xi^{1}_{\beta,\beta'}$, Eq. (\ref{x-aa}).
They are given by,
\begin{eqnarray}
j_{\beta}^{\eta} & = & v_c\,\xi^{1}_{c\,c}\,\xi^{1}_{c,\beta} + v_{s1}\,\xi^{1}_{s1\,c}\,\xi^{1}_{s1,\beta} 
\hspace{0.20cm}{\rm where}\hspace{0.20cm} \beta = c, s1 \, ,
\nonumber \\
j_{\beta}^{s} & = & v_c\,(\xi^{1}_{c\,c} - 2 \xi^{1}_{c\,s1})\,\xi^{1}_{c,\beta}
+ v_{s1}\,(\xi^{1}_{s1\,c} - 2 \xi^{1}_{s1\,s1})\,\xi^{1}_{s1,\beta} \hspace{0.20cm}{\rm where}\hspace{0.20cm} \beta = c, s1 \, .
\label{j-c-s1}
\end{eqnarray}

The $\beta =c,s1$ elementary currents $j_{\beta}^{\eta}$ and $j_{\beta}^{s}$, Eq. (\ref{j-c-s1}),
contribute both to expectation values of the charge and spin current operators, respectively, of low-energy excited energy eigenstates
and to off-diagonal matrix elements of such operators between the ground state and excited states
of vanishing energy. The derivation in Ref. \cite{Carmelo-92-C} of these elementary currents  
and charge and spin stiffnesses, Eq. (\ref{D-0}), relies on the general procedures similar to
those reported in Appendix \ref{XXXconfig3}
to obtain the corresponding expression of the zero-temperature spin stiffness of the spin-$1/2$ $XXX$ chain. 

The $\beta =c,s1$ and $\alpha =\eta,s$ elementary currents $j_{\beta}^{\alpha}$ expressions, Eq. (\ref{j-c-s1}),
can again be understood as being controlled by pseudoparticle parameters that play the same role as the Landau parameters in
a Fermi liquid. On the one hand, the two-electron static quantities are expressed in terms of the $i=0$ symmetrical pseudoparticle 
Landau parameters in Eq. (\ref{gbbgcs}). On the other hand, the elementary charge and spin currents rather involve the $i=1$ anti-symmetrical 
pseudoparticle Landau parameters also given in that equation. The residual pseudoparticle 
interactions associated with the $f$ functions in that expression control the effects of the electronic interactions 
onto such two-electron quantities associated with charge and spin ballistic transport. 

Specifically, the expression of the elementary currents, Eq. (\ref{j-c-s1}), in terms of the $\beta =c,s1$
bands group velocities at a Fermi point and $i=1$ anti-symmetrical pseudoparticle Landau parameters read,
\begin{eqnarray}
j_{c}^{\eta} & = & v_c\,g_{c\,c}^1 \hspace{0.20cm}{\rm and}\hspace{0.20cm}  j_{s1}^{\eta} = v_{s1}\,g_{s1\,c}^1 \, ,
\nonumber \\
j_{c}^{s} & = & v_c\,(g_{c\,c}^1 - 2g_{c\,s1}^1)
\hspace{0.20cm}{\rm and}\hspace{0.20cm} j_{s1}^{s} = - v_{s1}\,(2g_{s1\,s1}^1 - g_{s1\,c}^1) \, .
\label{j-c-s1LPs}
\end{eqnarray}

In the $m \rightarrow 0$ and $m \rightarrow n_e$ spin-density limits and for
electronic density in the range $n_e \in [0,1]$, these elementary currents have 
the following limiting behaviors,
\begin{eqnarray}
j_c^{\eta} & = & v_c\,\xi_0^2 
\, , \hspace{0.20cm} j_{s1}^{\eta} = v_c\,{\xi_0^2\over 2} \, ,
\hspace{0.20cm} j_c^{s} = 0\hspace{0.20cm}{\rm and}\hspace{0.20cm} 
j_{s1}^{s} = - v_{s1} 
\hspace{0.20cm}{\rm for}\hspace{0.20cm} m \rightarrow 0 \, ,
\nonumber \\
j_c^{\eta} & = & v_c \, , \hspace{0.20cm} 
j_{s1}^{\eta} = 0 \, , \hspace{0.20cm} 
j_c^{s} = v_c \hspace{0.20cm}{\rm and}\hspace{0.20cm} 
j_{s1}^{s} = 0 \hspace{0.20cm}{\rm for}\hspace{0.20cm} m \rightarrow n_e \, .
\label{j-c-s1-lim}
\end{eqnarray}
\begin{figure}
\begin{center}
\subfigure{\includegraphics[width=5.00cm]{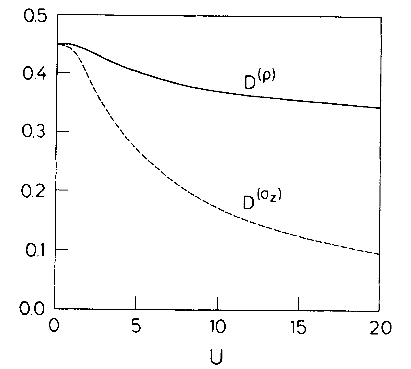}}
\hspace{0.50cm}
\subfigure{\includegraphics[width=5.70cm]{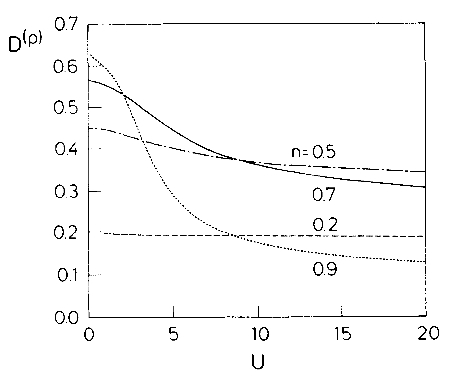}}
\caption{The zero-temperature charge stiffness $D_{\eta}$ (left and right panels) and the spin stiffness 
$D_{s}$ (left panel), Eq. (\ref{D-0}), versus $U/t$ for magnetic field $H\rightarrow 0$. The
left and right panels figures are for electronic density $n_e =0.7$ and various $n_e$ values,
respectively. Although $D_{\eta}$ has its maximum value for $n_e\rightarrow 1$, at
$n_e=1$ it vanishes when $U/t>0$, yet is finite at $U/t=0$. This strong effect of the 
Mott-Hubbard insulator transition is also present for $0<H<H_c$.
(As in Ref. \cite{Carmelo-92-C}, in the figures the charge stiffness $D_{\eta}$, spin stiffness $D_{s}$, and electronic density $n_e$ are denoted by 
$D^{(\rho)}$, $D^{(\sigma_z)}$, and $n$, respectively).\\
{\it Source}: From Ref. \cite{Carmelo-92-C}.}
\label{figure18}
\end{center}
\end{figure}

By combining the stiffnesses expressions, Eq. (\ref{D-0}), with the elementary-current limiting
behaviors, Eq. (\ref{j-c-s1-lim}), one finds that $D_{\eta} = v_c\,\xi_0^2/(2\pi)$ at $m=0$. For $n_e<1$ and $m=0$, 
the charge stiffness $D_{\eta}$ changes from $D_{\eta} = (2t/\pi)\sin (\pi n/2)$ as $u\rightarrow 0$ to
$D_{\eta} = (t/\pi)\sin (\pi n_e)$ for $u\gg 1$. For $m\rightarrow n_e$ the result is $D_{\eta} = (t/\pi)\sin (\pi n_e)$.
At $n_e=1$ and $m=0$ one finds $D_{\eta} = 2t/\pi$ at $u=0$ and $D_{\eta} = 0$ for $u>0$. 
This behavior stems from the $c$-band Fermi velocity $v_c$, Eq. (\ref{vel-beta}) for $\beta =c$, being at $n_e=1$ finite at $u=0$ and
vanishing for $u>0$. At $T=0$ the spin stiffness $D_{s}$ changes from $D_{s} = (2t/\pi)\sin (\pi n_e/2)$ for $u\rightarrow 0$ to 
$D_{s} = (t^2/U)(1-\sin (2\pi n_e)/ (2\pi n_e))$ for $u\gg 1$.
For $m\rightarrow n_e$ one finds the expected result, $D_{s} = D_{\eta} = (t/\pi)\sin (\pi n_e)$.

The charge stiffness $D_{\eta}$ and the spin stiffness $D_{s}$, Eq. (\ref{D-0}),
are plotted in Fig. \ref{figure17} as a function of $U/t$ at electronic density $n_e =0.7$
and various values of the magnetic field $H$ in suitable units and as
a function of $n_e$ for various values of the magnetic field $H$ and $U/t$.
Such quantities are also plotted in Fig. \ref{figure18} as a function of $U/t$
for magnetic field $H\rightarrow 0$ and electronic density $n_e =0.7$. In the
right panel, the charge stiffness $D_{\eta}$ curves refer to several $n_e$ values.

Below in Section \ref{PRPST} it is shown that both the static properties and the 1D Hubbard model dynamical properties are
controlled by the dressed phase shifts associated with the scattering events of the $\beta=c,s1$ pseudofermions
revisited in Section \ref{pseudofermion-R}.

\subsection{Binding and anti-binding character of the spin-singlet pairs ($\alpha =s$) and $\eta$-spin-singlet pairs ($\alpha =\eta$)}
\label{BANTIB}

From the use of the second expression in Eq. (\ref{e-0-bands}), the composite $\alpha n$ pseudoparticle energy dispersion, Eq. (\ref{epsilon-q}) 
for $\beta =\alpha n$, may be written as $\varepsilon_{\alpha n} (q_j) = n\,2\mu_{\alpha} + \varepsilon_{\alpha n}^0 (q_j)$
where $\alpha = \eta,s$ and $n = 1,..., \infty$. The term $n\,2\mu_{\alpha}$ in this energy dispersion is merely additive in the 
intrinsic energy $2\mu_{\alpha} = \varepsilon_{\alpha,-1/2} + \varepsilon_{\alpha,+1/2}$, Eq. (\ref{energy-eta-s-2}),
of two unpaired $\eta$-spins $1/2$ ($\alpha =\eta$) or two unpaired spins $1/2$ ($\alpha =s$) of opposite
$\eta$-spin and spin projection, respectively.

Our aim is clarifying the anti-binding or binding character of the $\eta$-spin ($\alpha =\eta$) and spin ($\alpha =s$)
$\alpha$-singlet configuration of two paired $\eta$-spins $1/2$ and two paired spins $1/2$, 
respectively. We first consider one single-pair $\alpha 1$ pseudoparticle
for which $\varepsilon_{\alpha n} (q_j) = 2\mu_{\alpha} + \varepsilon_{\alpha 1}^0 (q_j)$. The first energy term $2\mu_{\alpha}$ 
gives the intrinsic energy of its two $\eta$-spins $1/2$ ($\alpha =\eta$) or spins $1/2$ ($\alpha =s$) of opposite projection
if those were unpaired and in a $\alpha$-triplet $S_{\alpha}=1$ and  $S_{\alpha}^z=0$
configuration. The second energy term $\varepsilon_{\alpha 1}^0 (q_j)$ is thus
a pairing energy. It refers to a binding or anti-binding character if $\varepsilon_{\alpha 1}^0 (q_j)<0$ or $\varepsilon_{\alpha 1}^0 (q_j)>0$, respectively. 
Analysis of the form of the $\alpha 1$ pseudoparticle energy dispersion, Eq. (\ref{epsilon-q}) for $\beta =\alpha 1$, then reveals that the spin-singlet 
$s1$-pair configuration has a binding character. Indeed, it is such that $\varepsilon_{s1}^0 (q_j)<0$ for $\vert q_j\vert <q_{s 1}$, as confirmed by
inspection of Fig. \ref{figure11}. As demonstrated by Fig. \ref{figure13}, the $\eta$-spin-singlet $\eta 1$-pair 
configuration is found in turn to have an anti-binding character. For it, $\varepsilon_{\eta 1}^0 (q_j)>0$ for $\vert q_j\vert <q_{\eta 1}$.
At the $\alpha = s,\eta$ limiting momenta $q_j = \pm q_{\alpha 1}$ one has though that $\varepsilon_{\alpha 1}^0 (\pm q_{\alpha 1})=0$. 
This means that for $q_j\rightarrow q_{\alpha 1}^{\pm} =\pm q_{\alpha 1}$ the $\alpha$-singlet pair of a $\alpha 1$ pseudoparticle looses
its binding or anti-binding character as its pairing energy vanishes.

Next we consider $\alpha n$-pairs configuration with $n>1$ pairs bound within it. One finds as well that 
$\varepsilon_{s n}^0 (q_j)<0$ for $\vert q_j\vert <q_{s n}$ and $\varepsilon_{\eta n}^0 (q_j)>0$ for $\vert q_j\vert <q_{\eta n}$
and $\varepsilon_{\alpha n} (\pm q_{\alpha n})=0$ for $\alpha =\eta,s$.
(See Fig. \ref{figure12} for $\varepsilon_{s2}^0 (q_j)$ and Fig. \ref{figure14} for $\varepsilon_{\eta 2}^0 (q_j)$.)
Irrespective of its binding or anti-binding character, the energy absolute value $\vert\varepsilon_{\alpha n}^0 (q_j)\vert$
is here called {\it $\alpha n$ pseudoparticle pairing energy}. The strength of the $\alpha$-singlet pairs binding or anti-binding can be measured by the 
maximum reachable value of the $\alpha n$ pseudoparticle pairing energy upon creation of one $\alpha n$ pseudoparticle onto
the ground state. Such a maximum value is reached at $q_j = \pm q_{Fs1}$ for the $s1$ pseudoparticles
and at $q_j = 0$ for all other $\alpha n\neq s1$ pseudoparticles,
\begin{equation}
W_{s1}^{\rm pair} = \vert\varepsilon^0_{s1} (q_{Fs1})\vert = W_{s1}^h = 2\mu_B H 
\hspace{0.20cm}{\rm and}\hspace{0.20cm}
W_{\alpha n}^{\rm pair} = \vert\varepsilon^0_{\alpha n} (0)\vert = W_{\alpha n}
\hspace{0.20cm}{\rm for}\hspace{0.20cm}\alpha n \neq s1 \, .
\label{Wan-gen}
\end{equation}
Here $W_{s1}^h = 2\mu_B H$ is the ground-state energy bandwidth of the $s1$ band 
hole unoccupied part and $W_{\alpha n}$ is the energy bandwidth of $\alpha n\neq s1$ momentum bands.

The maximum pairing energy $W_{sn}^{\rm pair}$ vanishes at $H=0$. This is because then $W_{s1}^h = 2\mu_B H=0$
and the energy dispersion $\varepsilon_{sn}^0 (q_j)$ and its momentum bandwidth vanish at $H=0$ for $n>1$. 
Similarly, the maximum pairing energy $W_{\eta n}^{\rm pair}$ vanishes at electronic
density $n_e=1$. Indeed, the energy dispersion $\varepsilon_{\eta n}^0 (q_j)$ 
and its momentum bandwidth vanish at $n_e=1$. The latter maximum $\eta n$ 
pseudoparticle pairing energy is given in Eq. (\ref{Wetanu-UU}) of Appendix \ref{TBAconfig} 
in the $u\rightarrow 0$ and $u\gg 1$ limits for electronic densities $n_e\in ]0,1[$ and spin density $m=0$.
For the electronic density interval $n_e\in ]0,1[$ and spin density $m\rightarrow n_e$,
one has that $q_{Fs1}\rightarrow 0$. The maximum pairing energies 
$W_{sn}^{\rm pair} =W_{sn}= \vert\varepsilon_{sn}^0 (0)\vert$ 
and $W_{\eta n}^{\rm pair}=W_{\eta n}= \vert\varepsilon_{\eta n}^0 (0)\vert$ have 
for these densities analytical expressions that are functions of $n_e$ and $U/t$. They
are given in Eqs. (\ref{Ws-nu}) and (\ref{Weta-nu}) of Appendix \ref{TBAconfig}. In the 
$u\rightarrow 0$ and $u\gg 1$ limits these expressions simplify, as given in Eqs.
(\ref{Wsn-U0-Ul}) and (\ref{Wetanu-UUmne}) of that Appendix.
 
Actually, the suitable energy scale to measure the strength of the binding or anti-binding pairing
is the maximum value of the $\alpha n$ pseudoparticle pairing energy per
pair, $\pi_{\alpha n}^{\rm pair}\equiv W_{\alpha n}^{\rm pair}/n$. Consistently with the values and expressions
given in Eqs. (\ref{Wetanu-UU})-(\ref{Wetanu-UUmne}) of Appendix \ref{TBAconfig},
one finds that the energy per pair $\pi_{\alpha n}^{\rm pair}$ is for $n>1$ always smaller than 
$\pi_{\alpha 1}^{\rm pair}$. For densities $n_e\in ]0,1[$ and $m \in [0,n_e]$ it has the limiting behaviors,
\begin{equation}
\pi_{\alpha n}^{\rm pair} = \pi_{\alpha 1}^{\rm pair}/n 
\hspace{0.20cm}{\rm for}\hspace{0.20cm}u \rightarrow 0
\hspace{0.20cm}{\rm and}\hspace{0.20cm}
\pi_{\alpha n}^{\rm pair} = \pi_{\alpha 1}^{\rm pair}/n^2 
\hspace{0.20cm}{\rm for}\hspace{0.20cm}u \gg 1\, .
\label{ratiospis}
\end{equation}
It obeys the inequality $\pi_{\alpha 1}^{\rm pair}/n^2 \leq \pi_{\alpha n}^{\rm pair} \leq \pi_{\alpha 1}^{\rm pair}/n$
for the whole $u>0$ range. Hence the energy per pair $\pi_{\alpha n}^{\rm pair}$ decreases upon increasing $n$.
This effect is stronger upon increasing $u$. This reveals that the overall binding of the $n>1$ pairs within an
$\alpha n$-pairs configuration tends to suppress the binding $(\alpha =s)$ and anti-binding 
$(\alpha =\eta)$ energy within each such pairs. This suppression is an increasing function of both the number of pairs 
$n$ and of $u$.

In Appendix \ref{Wm*} the relation of the maximum pairing energy $W_{\alpha n}^{\rm pair}$ and 
the $\alpha n$ pseudoparticle effective mass $m_{\alpha n}^*$ in Eq. (\ref{bandssnetansq}) of that Appendix 
to $\eta$-spin ($\alpha=\eta$) and spin ($\alpha=s$) $\delta S_{\alpha} = \pm n$ multiplet excitations is discussed. 
This refers to electronic densities in the interval $n_e\in ]0,1[$, spin density $m\rightarrow n_e$, 
and the whole $u>0$ range. The quantities $W_{\alpha n}^{\rm pair}$ and $m_{\alpha n}^*$ are found to be related yet 
different quantities. 

\section{Dynamical correlation functions in the pseudofermion representation}
\label{PRPST}

Here we revisit the $\beta$ pseudofermion representation and shortly consider the corresponding PDT.
This includes its applications to the 1D Hubbard model in a magnetic field.
In addition, the relation between the PDT and the MQIM \cite{Glazman-09,Glazman-12} is
clarified. 

%%%%%%%%%%%%%%%%%%%%%%%%%%%%%%%%%%%%%%%%%%%%%%%%%%%%%%%%%%%%%%%%
\subsection{The pseudofermion representation}
\label{pseudofermion-R}

One finds from the use of the TBA  equations, Eqs. (\ref{Tapco1}) and (\ref{Tapco2}),
that for PS excited states the $\beta = c, \alpha n$ rapidity functionals $\Lambda^{\beta}(q_j)$ can be written
in terms of the corresponding ground-state rapidity functions $\Lambda^{\beta}_0(q_j)$ as follows,
\begin{equation}
\Lambda^{c}(q_j) = \Lambda_0^{c}\Bigl({\bar{q}} (q_j)\Bigr) =
\sin k_0^c\Bigl({\bar{q}} (q_j)\Bigr) \hspace{0.20cm}{\rm for}\hspace{0.20cm}  j = 1,...,L 
\hspace{0.20cm}{\rm and}\hspace{0.20cm} 
\Lambda^{\alpha n}(q_j) = \Lambda^{\alpha n}_0\Bigl({\bar{q}} (q_j)\Bigr) \hspace{0.20cm}{\rm for}\hspace{0.20cm} j = 1,...,L_{\alpha n} \, .
\label{FL}
\end{equation}
\begin{figure}
\begin{center}
\centerline{\includegraphics[width=5.00cm]{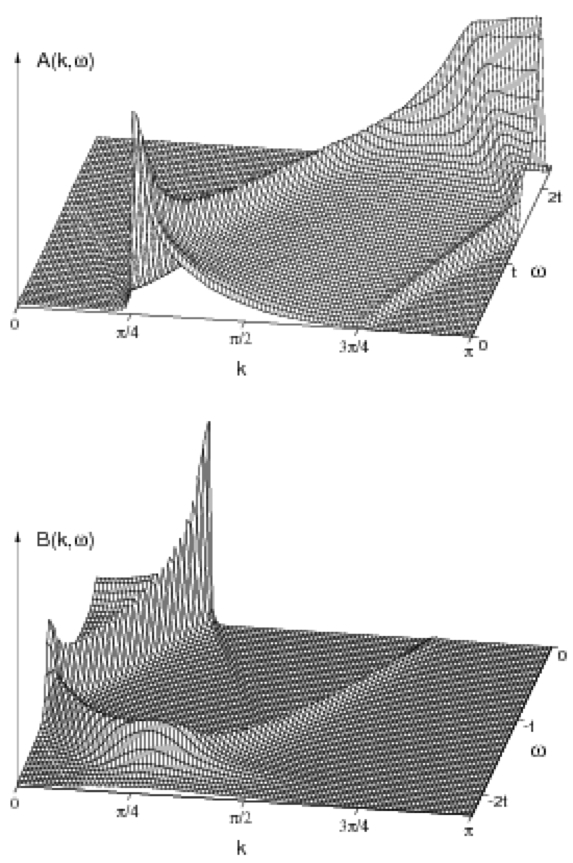}}
\caption{One-electron removal ($\omega<0$) and addition ($\omega>0$) spectral functions over the
whole $(k,\omega)$ plane for $u\gg 1$, $n_e=0.5$, and $m=0$.\\
{\it Source}: From Ref. \cite{Karlo-96}.}
\label{figure19}
\end{center}
\end{figure}
Here ${\bar{q}}_j = {\bar{q}} (q_j)$ where $j=1,...,L_{\beta}$ are the following
discrete {\it canonical momentum} values,
\begin{equation}
{\bar{q}}_j = {\bar{q}} (q_j) = q_j + {2\pi\Phi_{\beta} (q_j)\over L} = {2\pi\over
L}\,I^{\beta}_j + {2\pi\Phi_{\beta} (q_j)\over L} 
\hspace{0.20cm}{\rm for}\hspace{0.20cm} j=1,...,L_{\beta}\hspace{0.20cm}{\rm and}\hspace{0.20cm} \beta = c, \alpha n \, .
\label{barqan}
\end{equation}
(The excited-states rapidity expression, Eq. (\ref{FL}), is similar to that of the simpler models considered in Sections \ref{Bg} 
and \ref{Heichain}.) The functional $2\pi\Phi_{\beta} (q_j)$ in Eq. (\ref{barqan}) reads \cite{V-1,Carmelo-05},
\begin{equation}
2\pi\Phi_{\beta} (q_j) = \sum_{\beta'}\,
\sum_{j'=1}^{L_{\beta'}}\,2\pi\Phi_{\beta\,\beta'}(q_j,q_{j'})\, \delta N_{\beta'}(q_{j'}) \, . 
\label{qcan1j}
\end{equation}
Here the deviation $\delta N_{\beta'}(q_{j'})$ and the dressed phase shift $2\pi\Phi_{\beta\,\beta'}(q_j,q_{j'})$ are 
those in Eq. (\ref{DNq}) and Eq. (\ref{Phi-barPhi}), respectively.
The discrete canonical momentum values ${\bar{q}}_j = {\bar{q}} (q_j)$ have spacing 
${\bar{q}}_{j+1}-{\bar{q}}_{j}= 2\pi/L + {\rm h.o.}$. (h.o. stands here for terms of second order in $1/L$.) 

We associate one {\it $\beta$ pseudofermion} with each of the $N_{\beta}$ occupied $\beta$-band discrete canonical momentum values ${\bar{q}}_j$
\cite{V-1,TTF,spectral-06,VI,CarCadez-16,CarCadez-17}. We associate one {\it $\beta$ pseudofermion hole} with each of the remaining $N_{\beta}^h$ unoccupied $\beta$-band discrete canonical momentum values ${\bar{q}}_j$ of a PS excited state. There is a pseudofermion representation 
for each ground state and its PS. This holds for $u>0$ and all electronic and spin densities. The chosen initial ground state plays the
role of vacuum of the pseudofermion representation. For it one has that ${\bar{q}}_j=q_j$ in Eq. (\ref{barqan}). This also
occurs for the $\alpha n\neq s1$ bands of the PS excited states of a $S_{\alpha}=0$ ground state. For the
latter state the number of $\alpha n\neq s1$ band discrete momentum values $L_{\alpha n}$, Eq.
(\ref{N-h-an}), vanishes, $L_{\alpha n}=0$. For it the functional, Eq. (\ref{qcan1j}), also vanishes, $2\pi\Phi_{\alpha n} (q_j)=0$. 
It follows that ${\bar{q}}_j = q_j$ for the $\alpha n\neq s1$ bands of its PS excited states. 

In Sections \ref{specificRE} and \ref{alphanpseudop}, $\beta = c,s1$ pseudoparticle creation and annihilation operators
$f^{\dag}_{q_j,\beta}$ and $f_{q_j,\beta}$, respectively, have been introduced. The corresponding $\beta =c,s1$ pseudofermions
play a major role in the PDT. Their operators read,
\begin{equation}
f^{\dag}_{{\bar{q}}_j,\beta} = f^{\dag}_{q_j + 2\pi\Phi_{\beta} (q_j)/L,\beta} =
\left({\hat{S}}^{\Phi}_{\beta} \right)^{\dag}f^{\dag}_{q_j,\beta}\,{\hat{S}}^{\Phi}_{\beta} 
\hspace{0.20cm}{\rm for}\hspace{0.20cm} f_{{\bar{q}}_j,\beta} = (f^{\dag}_{{\bar{q}}_j,\beta})^{\dag} 
\hspace{0.20cm}{\rm for}\hspace{0.20cm} \beta = c,s1 \, .
\label{f-f-Q}
\end{equation}
Here ${\hat{S}}^{\Phi}_{\beta}$ denotes the $\beta$ pseudoparticle - $\beta$ pseudofermion unitary operator,
${\hat{S}}^{\Phi}_{\beta} = 
e^{\sum_{j=1}^{L_{\beta}}f^{\dag}_{q_{j} + (2\pi/L)\,\Phi_{\beta}(q_j),\beta}f_{q_{j},\beta}}$. It is such that
$\left({\hat{S}}^{\Phi}_{\beta} \right)^{\dag} = 
e^{\sum_{j=1}^{L_{\beta}}f^{\dag}_{q_{j} - (2\pi/L)\,\Phi_{\beta}(q_j),\beta}f_{q_{j},\beta}}$.

The functional $2\pi\Phi_{\beta}(q_j)$, Eq. (\ref{qcan1j}),
has an important physical meaning: It is the overall scattering phase shift acquired by a $\beta$ pseudofermion or
$\beta$ pseudofermion hole of initial-state canonical momentum $q_j$ upon scattering off all
$\beta'$ pseudofermions and/or $\beta'$ pseudofermion holes created under a transition
from the ground state to one of its PS excited states. As confirmed below in Section \ref{PDT}, 
that scattering phase shift controls the spectral weights of dynamical correlation functions.

It then follows from the form of the functional expression in Eq. (\ref{qcan1j})
that within the pseudofermion scattering theory \cite{Carmelo-05}, the function
$2\pi\Phi_{\beta\,\beta'}(q_j,q_{j'})$, Eq. (\ref{Phi-barPhi}), (and $-2\pi\Phi_{\beta\,\beta'}(q_j,q_{j'})$) 
has a well-defined physical meaning. It is the phase shift acquired by a $\beta$ pseudofermion or
$\beta$ pseudofermion hole of initial-state canonical momentum ${\bar{q}}_{j} = q_j$
upon scattering off a $\beta'$ pseudofermion (and a $\beta'$ pseudofermion hole) 
created at the canonical momentum ${\bar{q}}_{j'} = {\bar{q}} (q_{j'})$ corresponding
to the initial ground-state momentum $q_{j'}$ under a transition from that state to one of
its PS excited states. This reveals that all physical quantities whose expression 
was shown in Section \ref{exc-spectra} to depend on the phase-shift parameters, Eq. (\ref{x-aa}), are 
controlled by pseudofermion scattering events. 

Upon expressing the PS energy functional, Eq. (\ref{DE-fermions}),
in terms of the discrete canonical momentum values ${\bar{q}}_j = {\bar{q}} (q_j)$, Eq. (\ref{barqan}), it reads
$\delta E = \sum_{\beta}\sum_{j=1}^{L_{\beta}}\varepsilon_{\beta} ({\bar{q}}_j)\,\delta {\cal{N}}_{\beta}({\bar{q}}_j)
+ \sum_{\alpha =\eta,s}\varepsilon_{\alpha,-1/2}\,M_{\alpha,-1/2}$ up to ${\cal{O}}(1/L)$ order.
(This is as in the case of the 1D Lieb-Liniger Bose gas, Eq. (\ref{DEBcang}).)
The $\beta$ pseudofermion energy dispersions $\varepsilon_{\beta} ({\bar{q}}_j)$ in that functional
expression have exactly the same form as those given in Eq. (\ref{epsilon-q})
with the momentum $q_j$ replaced by the corresponding canonical momentum, ${\bar{q}}_j= {\bar{q}} (q_j)$.
The energy functional applying to the pseudofermions thus has no energy 
interaction terms of second-order in the deviations $\delta {\cal{N}}_{\beta}({\bar{q}}_j)$. 
This is in contrast to the equivalent energy functional, Eq. (\ref{DE-fermions}). Indeed,
up to ${\cal{O}}(1/L)$ order the $\beta$ pseudofermions have no such interactions. Such a property 
allows the dynamical correlation functions to be expressed as a convolution of $c$ and $s1$ 
pseudofermion spectral functions. Such functions spectral weights can be expressed as Slater determinants 
written in terms of anticommutators of pseudofermion operators.

That within the present representation the $\beta$ pseudofermion scattering phase shifts $2\pi\Phi_{\beta} (q_j)$
are incorporated in the canonical momentum, Eq. (\ref{barqan}), has though consequences on the
form of such Slater determinants. Those involve the type of 
$\beta$ pseudofermion operators anticommutators given in Eq. (\ref{pfacrGS}) of Appendix \ref{SIPDT}.
The unitarity of the $\beta$ pseudoparticle - $\beta$ pseudofermion transformation preserves the
pseudoparticle operator algebra {\it provided} that the canonical momentum values
${\bar{q}}_j$ and ${\bar{q}}_{j'}$ belong to the $\beta$ band of the same energy and momentum eigenstate. The exotic form
of the anticommutator, Eq. (\ref{pfacrGS}) of Appendix \ref{SIPDT}, follows from shake-up effects stemming 
from ${\bar{q}}_j$ and ${\bar{q}}_{j'}$ corresponding in it to the excited-state and ground-state $\beta$ band, respectively. 
Such an exotic $\beta$ pseudofermion operator algebra plays an important role in the 
one- and two-electron high-energy spectral weight distributions \cite{LE,VI} of
the PDT reviewed in the ensuing section.

\subsection{The pseudofermion dynamical theory}
\label{PDT}

The goal of this section is revisiting the PDT and thus to illustrate how the microscopic mechanisms that
control the dynamical and spectral properties are much simpler to describe in terms of pseudofermion processes than
of the underlying many-particle interactions. 

It has been difficult to apply the BA to the derivation of high-energy dynamical correlation functions.
For the 1D Hubbard model, the method employed in Refs. \cite{Karlo-95,Karlo-96,Karlo-97} 
has been the first breakthrough to address that problem in the $u\rightarrow\infty$ limit. In such references the 
one-electron spectral functions have been derived 
for the whole $(k,\omega)$ plane by accounting for the phase shifts imposed on
the spinless-fermions by the $XXX$ chain spins. (See Fig. \ref{figure19}). Such fractionalized particles naturally arise from the 
$u\rightarrow\infty$ $N_e$-electron wave-function factorization \cite{Woy,Ogata-90}. 
The related PDT relies on the corresponding factorization of the finite-$u$ $N_e$-rotated-electron wave 
function. This issue is addressed in Appendix \ref{FEL}. The PDT involves an extension of the $u\rightarrow\infty$ 
method of Refs. \cite{Karlo-95,Karlo-96,Karlo-97} to the whole finite-$u$ range. This theory has been the first 
breakthrough for the derivation of analytical expressions of the 1D Hubbard model high-energy spectral functions 
for that extended onsite repulsion range \cite{V-1,TTF,spectral-06,VI,CarCadez-16,CarCadez-17,spectral,LE}. 

In the following we consider the $\beta =c,\alpha n$ pseudofermions of the 1D Hubbard model. We indicate
as well the small differences relative to the simplified PDT suitable to the other models under review.
The aim of the theory is the evaluation of finite-$\omega$ one- and two-particle dynamical correlation functions of general form,
\begin{equation}
B (k,\omega) = \sum_{f}\, \vert\langle f\vert\,
{\hat{O}} (k) \vert GS\rangle\vert^2\,\delta\Bigl(\omega - (E_f - E_{GS})\Bigr) \hspace{0.20cm}{\rm for}\hspace{0.20cm} \omega > 0 \, . 
\label{ABON}
\end{equation}
Here ${\hat{O}} (k)$ is a one- or two-particle operator, $\vert GS\rangle$ a ground state, and $\vert f\rangle$ 
its excited states contained in ${\hat{O}} (k)\vert GS\rangle$.

The elementary processes that generate PS excited states contained in ${\hat{O}} (k)\vert GS\rangle$ from 
the ground state can be classified into the following three (A)-(B) 
classes \cite{V-1,TTF,spectral-06,VI,CarCadez-16,spectral,LE}: (A) High-energy and finite-momentum processes
that besides creation or annihilation of $c$ and $s1$ pseudofermions may involve the creation of $\alpha n\neq s1$ 
pseudofermions and/or unpaired $\eta$-spins of projection $-1/2$, (B) zero-energy 
and finite-momentum processes that conserve the
number of $\beta =c,s1$ pseudofermions yet change their number at the $\iota=+1$ right and 
$\iota=-1$ left $\beta =c,s1$ Fermi points, and (C) low-energy and small-momentum elementary $\beta =c,s1$
pseudofermion particle-hole processes in the vicinity of the
right ($\iota=+1$) and left ($\iota=+1$) $\beta =c,s1$ Fermi points onto the
momentum occupancy configurations generated by the 
elementary processes (A) and (B).

For a momentum $k$ and a given small energy range around $\omega$, the excitation ${\hat{O}} (k) \vert GS\rangle$ 
can be written as a sum of terms, $\sum_i {\hat{O}}^{\odot}_i (k) \vert GS\rangle$. Here $i=0,1,2,...$ refers 
to a suitable index. Its value for each specific operator ${\hat{O}} (k)$ can be uniquely defined in terms of the increasing number of 
pseudofermions created and annihilated by processes (A) and (B) under the transitions to the excited states. 
${\hat{O}}^{\odot}_i (k)$ is the corresponding generator onto the ground state of such processes of 
momentum $k$ and energy $\omega$. (Further information about the expansions $\sum_i {\hat{O}}^{\odot}_i (k) \vert GS\rangle$
of specific physical operators ${\hat{O}} (k)$ and the choice of the
corresponding leading-order operators ${\hat{O}}^{\odot} (k)$ can be found in Section 3.1 of Ref. \cite{CarCadez-16}
for two-particle spin operators and in Section 3.2 of Ref. \cite{CarCadez-17} for one-electron operators.)

For a well-defined small energy range around each low- or high-energy $\omega$, one approximates 
the dynamical correlation function, Eq.  (\ref{ABON}), by a corresponding leading-order term 
\cite{V-1,TTF,spectral-06,VI,CarCadez-16,CarCadez-17,spectral,LE},
\begin{eqnarray}
B (k,\omega) & \approx & B^{\odot} (k,\omega) = \sum_{f}\, \vert\langle f\vert\,
{\hat{O}}^{\odot} (k) \vert GS\rangle\vert^2\,\delta\Bigl(\omega - (E_f - E_{GS})\Bigr)
\nonumber \\
& = & \sum_{f} \Theta\Bigl(\Omega -\delta\omega_f\Bigr)\,\Theta\Bigl(\delta\omega_f\Bigr)\,
\Theta\left(\vert v_f\vert -v_{{\bar{\beta}}} \right)
{\breve{B}}^{\odot}_f (\delta\omega_f,v_f) 
\hspace{0.20cm}{\rm for}\hspace{0.20cm}  \omega > 0 \, . 
\label{ABON-odot}
\end{eqnarray}
Here
\begin{equation}
{\breve{B}}^{\odot}_{f} (\delta\omega_{f},v_{f}) = {{\rm sgn} (v_{f})\over 2\pi}\int_{0}^{\delta\omega_{f}}d\omega'\int_{-{\rm sgn}
(v_{f})\delta\omega_{f}/v_{\beta}}^{+{\rm sgn} (v_{f})\delta\omega_{f}/v_{\beta}}dk'
\,B_{Q_{{\bar{\beta}}}} (\delta\omega_{f}/v_{f} -k',\delta\omega_{f}-\omega')\,B_{Q_{\beta}} (k',\omega') \, , 
\label{B-l-i-breve}
\end{equation}
and 
\begin{eqnarray}
B_{Q_{\beta}} (k',\omega') & = & {L\over 2\pi}\sum_{m_{\beta,\,+1};m_{\beta,\,-1}}\,A^{(0,0)}_{\beta}\,a_{\beta} (m_{\beta,\,+1},\,m_{\beta,\,-1})
\nonumber \\
& \times & \delta \Bigl(\omega' -{2\pi\over L}\,v_{\beta}\sum_{\iota = \pm1} (m_{\beta,\iota}+\Delta_{\beta}^{\iota})\Bigr)\,
\delta \Bigl(k' -{2\pi\over L}\,\sum_{\iota = \pm1}\iota\,(m_{\beta,\iota}+\Delta_{\beta}^{\iota})\Bigr) 
\hspace{0.20cm}{\rm where}\hspace{0.20cm}  \beta = c,s1 \, .
\label{BQ-gen}
\end{eqnarray}
The quantities in these equations are defined below. 

In the function $B^{\odot} (k,\omega)$ initial general expression, the generator onto the particle vacuum of the ground state
$\vert GS\rangle$ is written in terms of $\beta$ pseudofermion creation operators. Their $\beta$ band  
discrete canonical momentum values, which equal the corresponding momentum 
values $q_j$, Eqs. (\ref{q-j}) and (\ref{Ic-an}), are those 
of that ground state. Both the generator onto the electron vacuum of the
PS excited states $\vert f\rangle$ and the operator
${\hat{O}}^{\odot} (k)$ are written in terms of $\beta$ pseudofermion 
operators. Their $\beta$ band discrete canonical momentum values ${\bar{q}}_j$, Eq. (\ref{barqan}),
are those of the excited states. 
\begin{figure}
\begin{center}
\centerline{\includegraphics[width=12.00cm]{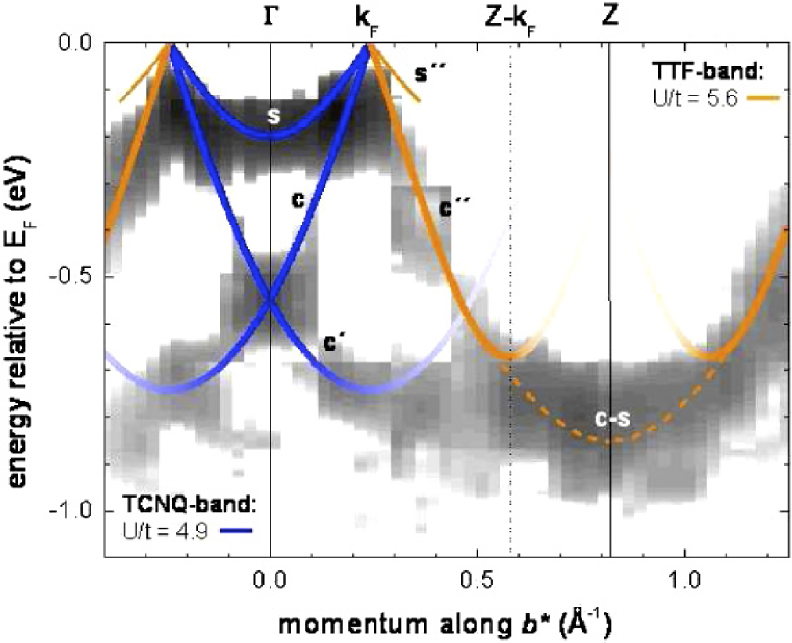}}
\caption{Experimental peak dispersions obtained by ARPES on TTF-TCNQ 
along the easy-transport axis as given in Fig. 7 of Ref. \cite{spectral0}
and matching theoretical branch and border lines, within
the 1D Hubbard model PDT. The line shape in the vicinity of the branch lines is for
that model of power-law type, with exponents that depend on
the momentum, interaction strength, and densities. (The Z-point corresponds 
in the figure to the momentum $k=\pi$.)\\
{\it Source}: From Ref. \cite{TTF}.}
\label{figure20}
\end{center}
\end{figure}

The summation $\sum_{f}$ in Eq. (\ref{ABON-odot}) runs over PS excited states generated by processes (A), (B), and (C) at fixed values 
of $k$ and $\omega$. The capital-$\Theta$ distribution $\Theta (x)$ in that equation is given here and in the
following by $\Theta (x)=1$ for $x\geq 0$ and $\Theta (x)=0$ for $x<0$.
Such states have excitation energy and momentum in the ranges $\delta E_f^{\odot}\in [\omega - \Omega,\omega]$ 
and $\delta P_f^{\odot}\in [k - \Omega/v_f,k]$, respectively, where $\Omega$ stands for the processes (C) energy range.
Moreover, in Eqs. (\ref{ABON-odot}) and (\ref{B-l-i-breve}) the fixed excitation energy $\omega$ and momentum $k$
read $\omega= \delta E_f^{\odot} + \delta\omega_f$ and $k= \delta P_f^{\odot} + \delta k_f $, respectively.
Here $\delta\omega_f = (\omega-\delta E_f^{\odot}) = (\omega - E_f^{\odot} + E_{GS})$, 
$\delta k_f = k -\delta P_f^{\odot}$, $v_f = \delta\omega_f/\delta k_f$, 
$v_{{\bar{\beta}}} = {\rm min}\{v_c,v_{s1}\}$, $v_{\beta} = {\rm max}\{v_c,v_{s1}\}$, 
and $v_c$ and $v_{s1}$ are the $\beta =c,s1$ band Fermi velocities, Eq. (\ref{vel-beta}). 
The above processes (C) energy range $\Omega$ is self-consistently determined as that for which the velocity 
$v_f$ remains nearly unchanged. 

The lack of $c$ and $s1$ pseudofermion energy interactions in their PS $u>0$ spectrum is 
behind the function ${\breve{B}}^{\odot}_{f} (\delta\omega_{f},v_{f})$ in Eq. (\ref{ABON-odot})
being expressed in Eq. (\ref{B-l-i-breve}) as a convolution of $c$ and $s1$ pseudofermion 
spectral functions \cite{V-1,LE}. Such functions expression, Eq. (\ref{BQ-gen}), involves sums 
that run over the processes (C) numbers $m_{\beta,\iota}=1,2,3,...$ \cite{TTF,LE,VI}.
In it $\Delta_{\beta}^{\iota}$ refers to the four dimensions functionals $2\Delta^{\iota}_{\beta}=(\delta {\bar{q}}_{F\beta}^{\iota}/[2\pi/L])^2$.
Those are the four $\beta=c,s1$ and $\iota = \pm$ relative weights, Eq. (\ref{a10DP-iota}) of Appendix \ref{SIPDT}.
They correspond to the smallest finite processes (C) numbers, $m_{\beta,\iota}=1$, and can be written as,
\begin{equation}
2\Delta^{\iota}_{\beta} = 
\left(\sum_{\beta'=c,s1}\left(\iota\, \xi^0_{\beta\,\beta'}\,{\delta N^F_{\beta'}\over 2} 
+ \xi^1_{\beta\,\beta'}\,\delta J^F_{\beta'}
+ \sum_{j=1}^{L_{\beta'}}\Phi_{\beta\,\beta'}(\iota q_{F\beta},q_{j})\delta N^{NF}_{\beta'} (q_{j})\right)
+ \sum_{\alpha n\neq s1}\sum_{j=1}^{L_{\alpha n}}\Phi_{\beta\,\alpha n}(\iota q_{F\beta},q_{j})\delta N_{\alpha n} (q_{j})\right)^2 \, .
\label{functional}
\end{equation}
Here $\beta =c,s1$, $\delta N^F_{\beta'}=\delta N^F_{\beta',+}+\delta N^F_{\beta',-}$, and 
$2J_{\beta'}^F=\delta N^F_{\beta',+}-\delta N^F_{\beta',-}$ for $\beta' =c,s1$.
Moreover, the $\beta =c,s1$ {\it lowest peak weight} $A^{(0,0)}_{\beta}$ in Eq. (\ref{BQ-gen}) is associated with 
transitions from the ground state to PS excited states generated by processes (A) and (B). The $\beta =c,s1$ relative weight 
$a_{\beta}=a_{\beta}(m_{\beta,\,+1},\,m_{\beta,\,-1})$ is generated by additional processes (C). 
The former weight refers to a Slater determinant that involves the
$\beta =c,s1$ pseudofermion anticommutators, Eq. (\ref{pfacrGS}) of Appendix \ref{SIPDT}. The lowest peak weight $A^{(0,0)}_{\beta}$ 
and the relative weight $a_{\beta}$ are given in Eq. (\ref{A00}) and Eqs. (\ref{aNNDP})-(\ref{a10DP-iota}) of that Appendix, respectively.

The PDT has a simplified form suitable to the 1D Lieb-Liniger Bose gas \cite{DCFBo-16}, spin-$1/2$ XXX chain  \cite{CPJD-15}, 
and spin-spin dynamical correlation functions of the 1D half-filled Hubbard model \cite{CarCadez-16}. Such functions 
involve spin excitations for which $N_c^h=0$. For that simplified dynamical theory there is no convolution, as given in Eq. (\ref{B-l-i-breve}). 
For it the function, Eq. (\ref{ABON-odot}), rather reads \cite{DCFBo-16,CPJD-15,CarCadez-16}
$B^{\odot} (k,\omega) = \sum_{f}\Theta\left(\vert v_f\vert -v\right) B_{Q} (\delta\omega_f,v_f)$ for $\delta\omega_f \in [0,\Omega]$.
Here $v = v (q_F)$ is the model-dependent ground-state single branch Fermi velocity
and $B_{Q} (k',\omega')$ is a pseudofermion spectral function. It has exactly the expression, Eq. (\ref{BQ-gen}), 
if one omits the index $\beta$. Such an omission procedure also applies to all other PDT quantities 
in the equations given in the following and in Appendix \ref{SIPDT} \cite{DCFBo-16,CPJD-15,CarCadez-16}.

The expression, Eq. (\ref{B-J-i-sum-GG}) of Appendix \ref{SIPDT}, of the $\beta =c,s1$ pseudofermion spectral function, 
Eq. (\ref{BQ-gen}), is valid in the TL. Its use in the general convolution expression, Eq. (\ref{B-l-i-breve}), of
the function ${\breve{B}}^{\odot}_{f} (\delta\omega_{f},v_{f})$ 
followed by the use of the obtained expression for such a function in the
second expression of the function $B^{\odot} (k,\omega)$ in 
Eq. (\ref{ABON-odot}), enables performing the summations 
in the latter equation for the $(k,\omega)$-plane vicinity of some singular spectral features. 
For the one-electron spectral function, these turn out to be the most important spectral features.
\begin{figure}
\begin{center}
\centerline{\includegraphics[width=6.00cm]{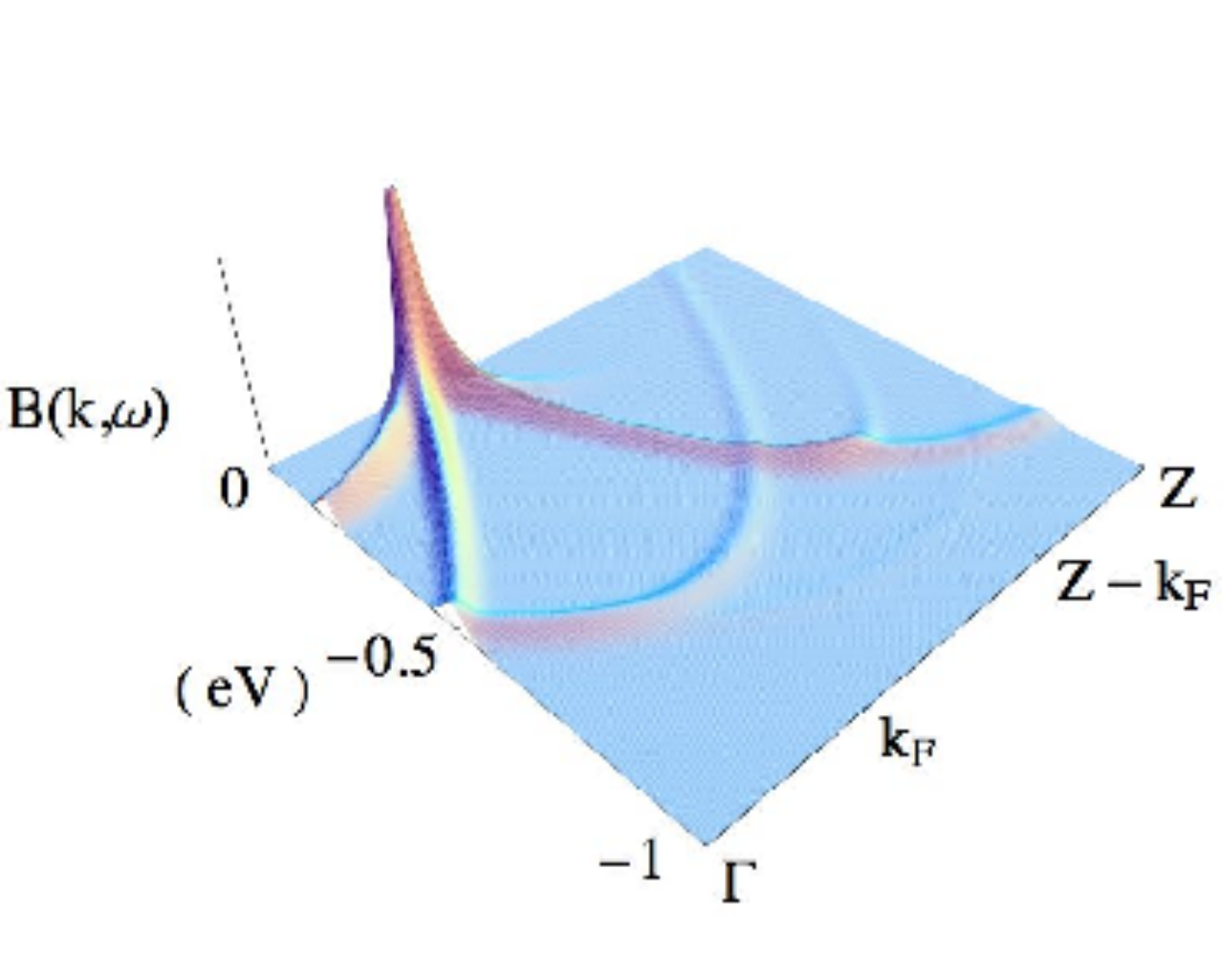}}
\caption{PDT one-electron removal spectral-weight distribution 
in the vicinity of the TTF ($n_e = 1.41$ and $U/t = 5.61$) and TCNQ ($n_e = 0.59$ and $U/t = 4.90$)
stack of molecules spectral features in Fig. \ref{figure20}.\\
{\it Source}: From Ref. \cite{TTF}.}
\label{figure21}
\end{center}
\end{figure}

The summation $\sum_{f}$ in Eq. (\ref{ABON-odot}) runs over PS excited states 
with the specific $k$ and $\omega$ values that appear in the argument of the corresponding function $B^{\odot} (k,\omega)$. 
At such fixed values, the two corresponding $\beta =c,s1$ lowest peak weights $A^{(0,0)}_{\beta}$, Eq. (\ref{A00}) of Appendix \ref{SIPDT}, 
have nearly the same magnitude for all such states. In the vicinity of the $\beta =c,s1$ branch lines whose spectrum is
defined in the following, the state summations can then be partially performed. One then finds that near them the spectral function 
behaves as \cite{V-1,TTF,spectral-06,VI,CarCadez-16,CarCadez-17,spectral,LE},
\begin{eqnarray}
B (k,\omega) & \propto & \Bigl(c_B\,\omega -\omega_{\beta} (k)\Bigr)^{\zeta_{\beta} (k)} \hspace{0.20cm}{\rm for}\hspace{0.20cm}
(c_B\,\omega -\omega_{\beta} (k)) \geq 0 \, ,
\nonumber \\
\zeta_{\beta} (k) & = & -1+\sum_{\beta' = c,s1}\sum_{\iota = \pm}2\Delta_{\beta'}^{\iota} (q_j)\vert_{q_j=c_0\,(k-k_0)} 
\hspace{0.20cm}{\rm where}\hspace{0.20cm} \beta = c,s1 \, .
\label{branch-l}
\end{eqnarray}
Here $c_B=1$ and $c_B=-1$ for $\omega\geq 0$ and $\omega\leq 0$ excitation energy, respectively. Except for the one-electron
removal spectral function, for which $c_B=-1$, the general convention is that $c_B=1$ and thus $\omega\geq 0$.
(That in Eq. (\ref{branch-l}) the $\beta =c,s1$ branch line spectrum $\omega_{\beta} (k)$ is not multiplied by $c_B$ is 
justified by it being according to Eq. (\ref{dE-dP-bl}) always such that $\omega_{\beta} (k)\geq 0$.) 

A $\beta =c,s1$ branch line in the vicinity of which the expression, Eq. (\ref{branch-l}), applies
results from transitions to excited states generated by creation ($c_0=+1$) or 
annihilation ($c_0=-1$) of one $\beta =c,s1$ pseudofermion. Its canonical momentum
${\bar{q}}_j = {\bar{q}} (q_j)$ is associated with a uniquely defined $\beta =c,s1$ band
momentum value $q_j$. A $c_0=+1$ and $c_0=-1$ branch line corresponds to the
range $\vert q_j\vert \in [q_{F\beta},q_{\beta}]$ and $q_j\in [-q_{F\beta},q_{F\beta}]$,
respectively. All remaining $\beta =c,s1$ pseudofermions are created or annihilated at the $\beta =c,s1$
Fermi points $\pm q_{F\beta}$, Eq. (\ref{q0Fcs}). The $\beta =\alpha n\neq s1$ pseudofermions (if any) are created at
the $\beta =\alpha n\neq s1$ band limiting values, $q_{\alpha n}^{\pm}=\pm q_{\alpha n}$, Eq. (\ref{qcanGS}). This gives a 
$(k,\omega)$-plane $\beta =c,s1$ branch line shape whose energy spectrum $\omega_{\beta} (k)$ 
appearing in expression, Eq. (\ref{branch-l}), reads \cite{V-1},
\begin{equation}
\omega_{\beta} (k) = \omega_0 + c_0\,\varepsilon_{\beta} (q_j)\hspace{0.20cm}{\rm and}\hspace{0.20cm}
k = k_0 + c_0\,q_j \hspace{0.20cm}{\rm where}\hspace{0.20cm} \beta = c,s1
\hspace{0.20cm}{\rm and}\hspace{0.20cm} c_0 = \pm 1 \, .
\label{dE-dP-bl}
\end{equation}
$\varepsilon_{\beta} (q_j)$ is here the $\beta =c,s1$ band energy dispersion, Eq. (\ref{epsilon-q}), and,
\begin{equation}
\omega_0 = \sum_{\alpha=\eta,s}2\mu_{\alpha}\,(L_{\alpha,-1/2} - \delta_{\alpha,s}\,N_{s1}) 
\hspace{0.20cm}{\rm and}\hspace{0.20cm} 
k_0 = \pi\,L_{\eta,-1/2} +  (\pi -2k_F) 2J_{\eta n} + \sum_{\beta=c,s1}2q_{F\beta}\,2J_{\beta}^F \, .
\label{omega0}
\end{equation}
$2J_{\eta n}=N_{\eta n,+}-N_{\eta n,-}$ where $N_{\eta n,\iota}$ is in this equation the
number of $\eta n$ pseudofermions created at the $\eta n$ band limiting momentum values 
$\iota q_{\eta n}=\iota (\pi -2k_F)$, Eq. (\ref{qcanGS}). For instance, $(N_{\eta 1,\iota}+M_{\eta,-1/2}) = 0$
and $(N_{\eta n, 1,\iota}+M_{\eta,-1/2}) = 1$ for a one-electron addition $\beta =c,s1$ branch line in the lower and upper Hubbard 
bands, respectively, defined below in Section \ref{SDIESpecPro}.

The spectral function expression, Eq. (\ref{branch-l}), is exact for $\beta =c,s1$ branch lines that
coincide with the lower thresholds ($c_B=1$) or upper thresholds ($c_B=-1$) of $(k,\omega)$-plane 
finite spectral-weight regions. For the particular case of the one-electron spectral function, Eq. (\ref{branch-l}) is a good
approximation for the $\beta =c,s1$ branch lines that have a small amount of spectral weight above ($c_B=1$) or below ($c_B=-1$) them. 
For integrable correlated problems with a single pseudofermion branch, the exponent in Eq. (\ref{branch-l})
is rather given by $-1+\sum_{\iota = \pm}2\Delta^{\iota}$. This is as in Eqs. (\ref{xilambBg}) and (\ref{DSF-BL}) for
the 1D Lieb-Liniger Bose gas and spin-$1/2$ $XXX$ chain, respectively \cite{DCFBo-16,CPJD-15,CarCadez-16}.
\begin{figure}
\begin{center}
\centerline{\includegraphics[width=5.50cm]{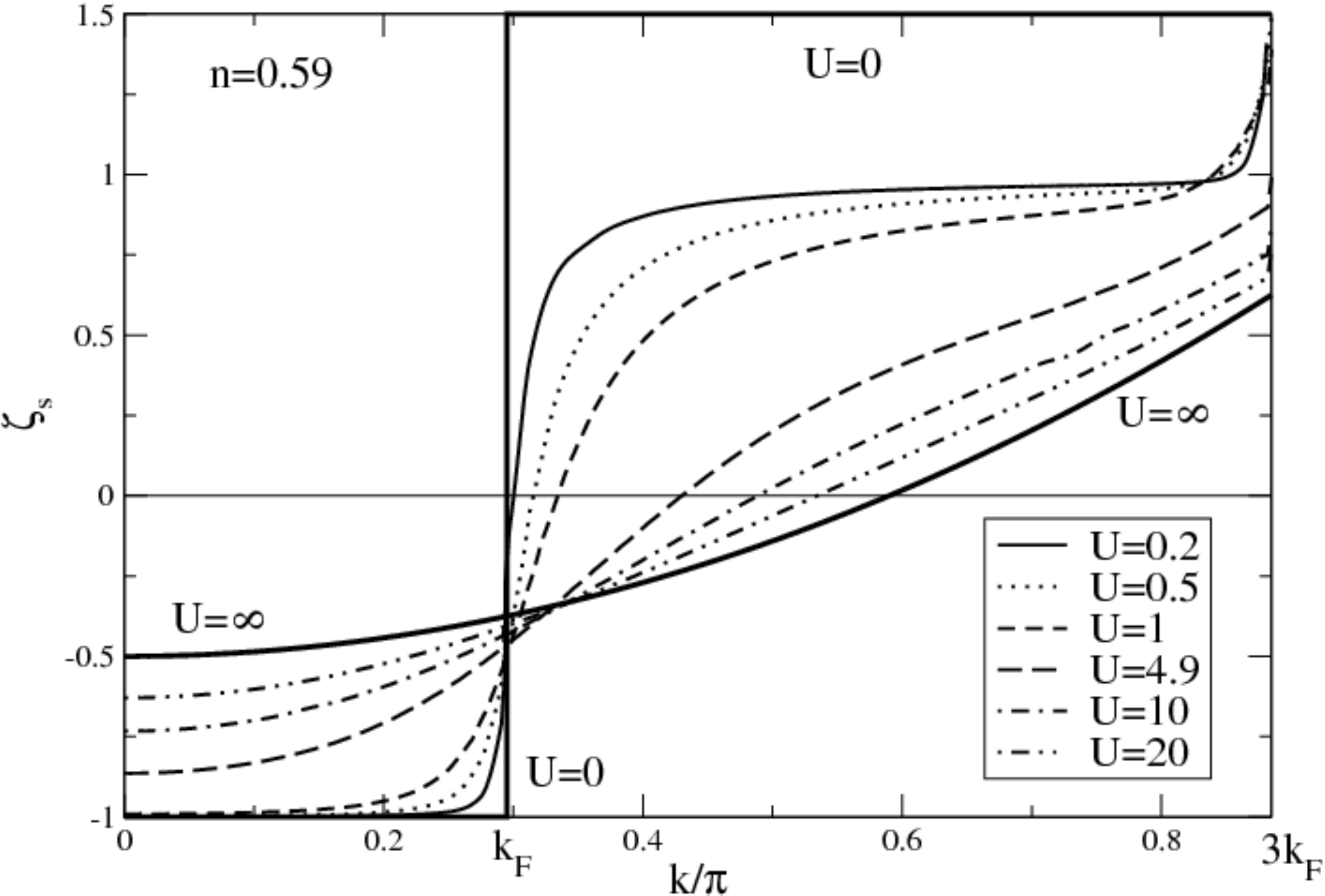}}
\caption{Momentum dependence of a one-electron spectral function $s1$ branch-line exponent, Eq. (\ref{branch-l}),
called $\zeta_{s}$ in the figure, for densities $n_e = 0.59$ and $m=0$ and several $U/t$ values. (As in
Ref. \cite{spectral}, $n_e$ is here denoted by $n$.)\\
{\it Source}: From Ref. \cite{spectral}.}
\label{figure22}
\end{center}
\end{figure}

The above dynamical correlation functions line shapes are beyond the reach
of the techniques used within the usual low-energy TLL studies. In the
limit of low-energy the PDT describes the well-known behaviors obtained by
such techniques. This refers specifically to the vicinity of $(k,\omega)$-plane points $(k_0,0)$
of which $(k_0,\omega_0)$, Eq. (\ref{omega0}), is a generalization for $\omega_0>0$.
Near them the spectral-function behavior is \cite{V-1,CarCadez-16,CarCadez-17,LE},
\begin{eqnarray}
B (k,\omega) & \propto & \Bigl(c_B\,\omega -\omega_0\Bigr)^{\zeta } \hspace{0.20cm}{\rm for}\hspace{0.20cm} 
(c_B\,\omega -\omega_0) \geq 0 \, ,
\nonumber \\
\zeta & = & -2+\sum_{\beta' = c,s1}\sum_{\iota = \pm}2\Delta_{\beta'}^{\iota} \hspace{0.20cm}{\rm for}\hspace{0.20cm} 
(c_B\,\omega - \omega_0) \neq \pm v_{\beta}\,(k-k_0) \hspace{0.20cm}{\rm where}\hspace{0.20cm} \beta =c,s1 \, ,
\nonumber \\
B (k,\omega) & \propto & \Bigl(c_B\,\omega - \omega_0 \mp v_{\beta}\,(k-k_0)\Bigr)^{\zeta^{\pm}} \hspace{0.20cm}{\rm for}\hspace{0.20cm} 
(c_B\,\omega - \omega_0 \mp v_{\beta}\,(k-k_0)) \geq 0 \, ,
\nonumber \\
\zeta^{\pm} & = & -1- 2\Delta_{\beta}^{\mp} +\sum_{\beta' = c,s1}\sum_{\iota = \pm}2\Delta_{\beta'}^{\iota}
\hspace{0.20cm}{\rm for}\hspace{0.20cm}  
(c_B\,\omega - \omega_0) \approx \pm v_{\beta}\,(k-k_0) \hspace{0.20cm}{\rm where}\hspace{0.20cm} \beta =c,s1 \, .
\label{point}
\end{eqnarray}
The expressions given here apply to the finite-weight region above ($c_B=1$) or below ($c_B=-1$) the
$(k,\omega)$-plane point. Examples of exponents $\zeta $ controlling the line shape near $(k,\omega)$-plane 
points $(k_0,0)$ are given in Ref. \cite{Carmelo-97-C}. In Ref. \cite{LE} it is confirmed that in the
limit of low excitation energy the expressions, Eq. (\ref{point}), recover those provided within the TLL
limit by conformal-field theory \cite{Woy-89,Frahm-90}. For low-energy excited eigenstates of $S_{\eta}>0$ and $S_s\geq 0$
ground states the four functionals $2\Delta^{\pm 1}_{c}$ and $2\Delta^{\pm 1}_{s1}$,
Eq. (\ref{functional}) with $\delta N^{NF}_{\beta'} (q_j)=0$ and $\delta N_{\alpha n} (q_j)=0$, are in that reference found to be
the conformal dimensions of the $c,\pm$ and $s,\pm$ primary fields, respectively. The corresponding dressed
charge matrix is in Eqs. (\ref{x-aa}) and (\ref{ZZ-gen}) expressed in terms of pseudofermion phase-shift parameters.
\begin{figure}
\includegraphics[scale=0.75]{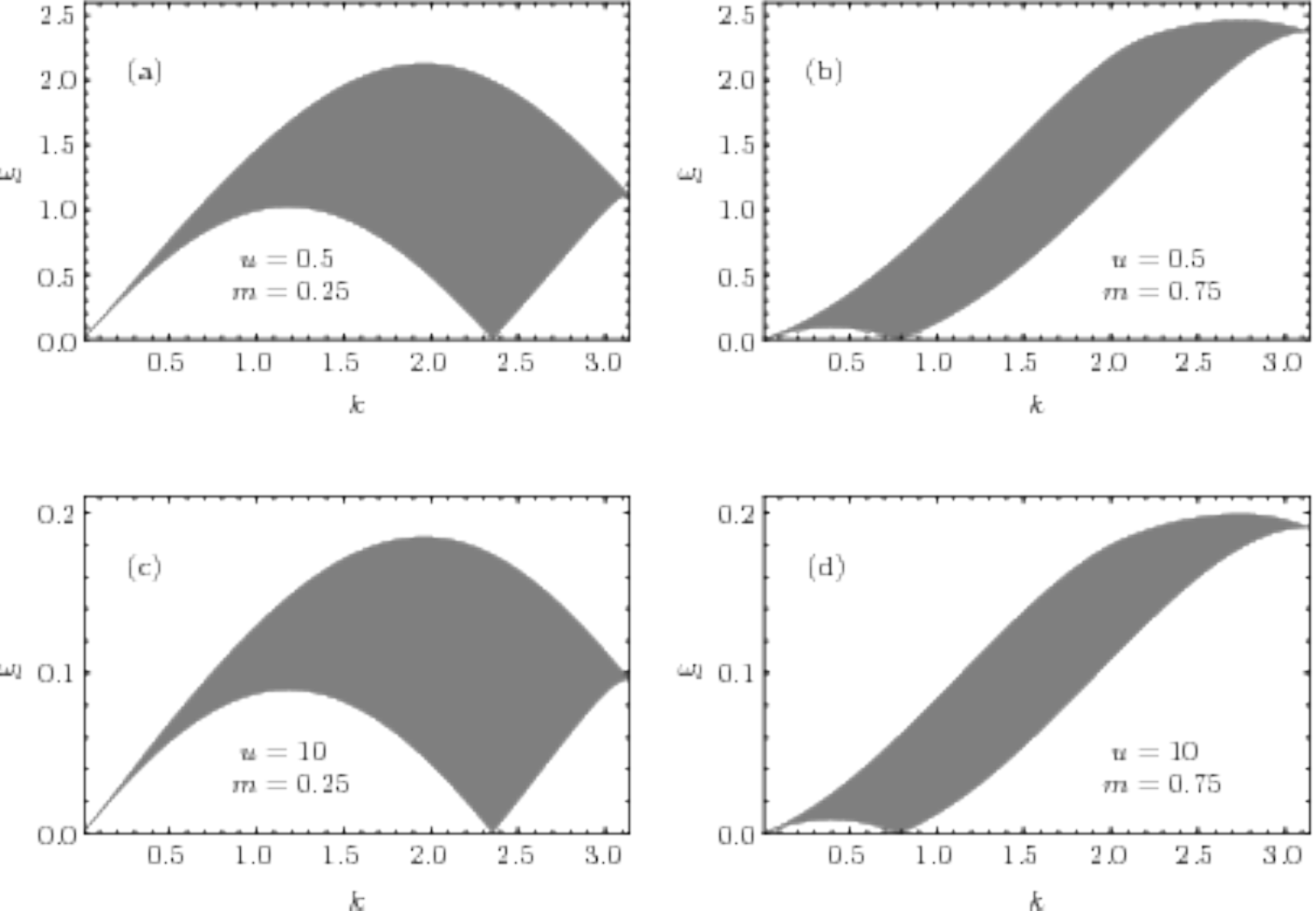}
\caption{The half-filled 1D Hubbard model longitudinal spin spectrum $\omega^l (k)$ corresponding to that
plotted in Figs. \ref{figure7} - \ref{figure9} for the spin-$1/2$ $XXX$ chain
for (a) $m=0.25$ and $u=0.5$, (b) $m=0.75$ and $u=0.5$,
(c) $m=0.25$ and $u=10.0$, and (d) $m=0.75$ and $u=10.0$. The main effect of the 
onsite repulsion is on the spectrum energy bandwidth. At fixed spin density $m$ its form remains nearly
the same for the whole $u>0$ range.\\
{\it Source}: From Ref. \cite{CarCadez-16}.}
\label{figure23} 
\end{figure}

In the particular case of the one-electron spectral function, there is a third type of high-energy 
spectral feature in the vicinity of which the PDT provides an analytical expression. 
It is generated by processes where a $c$ pseudofermion hole is annihilated (electron
addition) or created (electron removal) and a $s1$ pseudofermion hole is 
created at related momentum values $q_{j}$ and $q_{j'}$, respectively. Their relation follows from the group velocities,
Eq. (\ref{vel-beta}), obeying the equality $v_{c}(q_{j}) = v_{s1}(q_{j'})$.
The one-electron spectral feature under consideration is called a $c-s1$ border line.
Its $(k,\omega)$-plane shape is of the general form \cite{V-1,TTF,CarCadez-16,CarCadez-17},
\begin{equation}
\omega_{c-s1} (k)= \left(\omega_0 + \vert\epsilon_{c}(q_{j})\vert+\vert\epsilon_{s1}(q_{j'})\vert\right)\,\delta_{v_{c}(q_{j}) ,\,v_{s1}(q_{j'})}
\hspace{0.20cm}{\rm and}\hspace{0.20cm} 
k = k_0 + c_0\,q_{j} - q_{j'} \hspace{0.20cm}{\rm where}\hspace{0.20cm} c_0 = \pm 1 \, .
\label{dE-dP-c-s1}
\end{equation}
Near a $c-s1$ border line the spectral function has the following behavior,
\begin{equation}
B (k,\omega) \propto \Bigl(c_B\,\omega -\omega_{c-s1} (k)\Bigr)^{-1/2} \hspace{0.20cm}{\rm for}\hspace{0.20cm} 
(c_B\,\omega -\omega_{c-s1} (k)) \geq 0 \, .
\label{B-bol}
\end{equation}

Applications of the 1D Hubbard model PDT to the study of the ARPES spectral
features of actual quasi-1D materials are reported in Refs. \cite{TTF,spectral0,spectral}.
The experimental peak dispersions obtained by ARPES on 
the quasi-1D organic conductor TTF-TCNQ along the easy-transport axis \cite{spectral0} 
together with the prediction of the PDT for the 1D Hubbard model are shown in Fig. \ref{figure20}. 
The shape of the $c$, $c'$, and $c''$ spectral lines and $s$ and $s''$ spectral lines in the ARPES
spectrum plotted in such a figure is that of the $\beta =c$ and $\beta =s1$ bands energy dispersions 
$\varepsilon_{\beta} (q_j)$, Eq. (\ref{branch-l}), respectively, in the branch-line spectrum, Eq. (\ref{dE-dP-bl}). 
The indices $s$ and $s''$ read within our notation $s1$ and $s1'$, respectively.
The $c$, $c'$, and $s1$ branch lines refer to electronic densities in the range $n_e\in [0,1]$ suitable to the 
stacks of TCNQ molecules related spectral features. Their spectra expressions in terms of the energy dispersions $\varepsilon_{c} (q_j)$
and $\varepsilon_{s1} (q_j)$ and PDT momentum dependent exponents are given in 
Eqs. (\ref{3spectraOESF}) and (\ref{3expoOESF}) of Appendix \ref{SIPDT}, respectively.
Such dispersions are seen in actual experiments on quasi-1D conductors. 

The figure theoretical $c''$ and $s''$ branch lines and $c-s$ border line, Eq. (\ref{dE-dP-c-s1}), refer to the TTF 
stack of molecules spectral features derived within the 1D Hubbard model PDT for electronic density $n_e = 1.41$ 
and $U/t = 5.61$. The $c$, $s$, and $c'$ branch lines correspond to the TCNQ stack of molecules dispersions evaluated 
for electronic density $n_e = 0.59$ and $U/t = 4.90$. A corresponding approximate spectral-weight distribution in the 
vicinity of such branch lines obtained by combining the theory analytical expressions with numerical approximations is 
shown in Fig. \ref{figure21}. 
\begin{figure}
\includegraphics[scale=0.75]{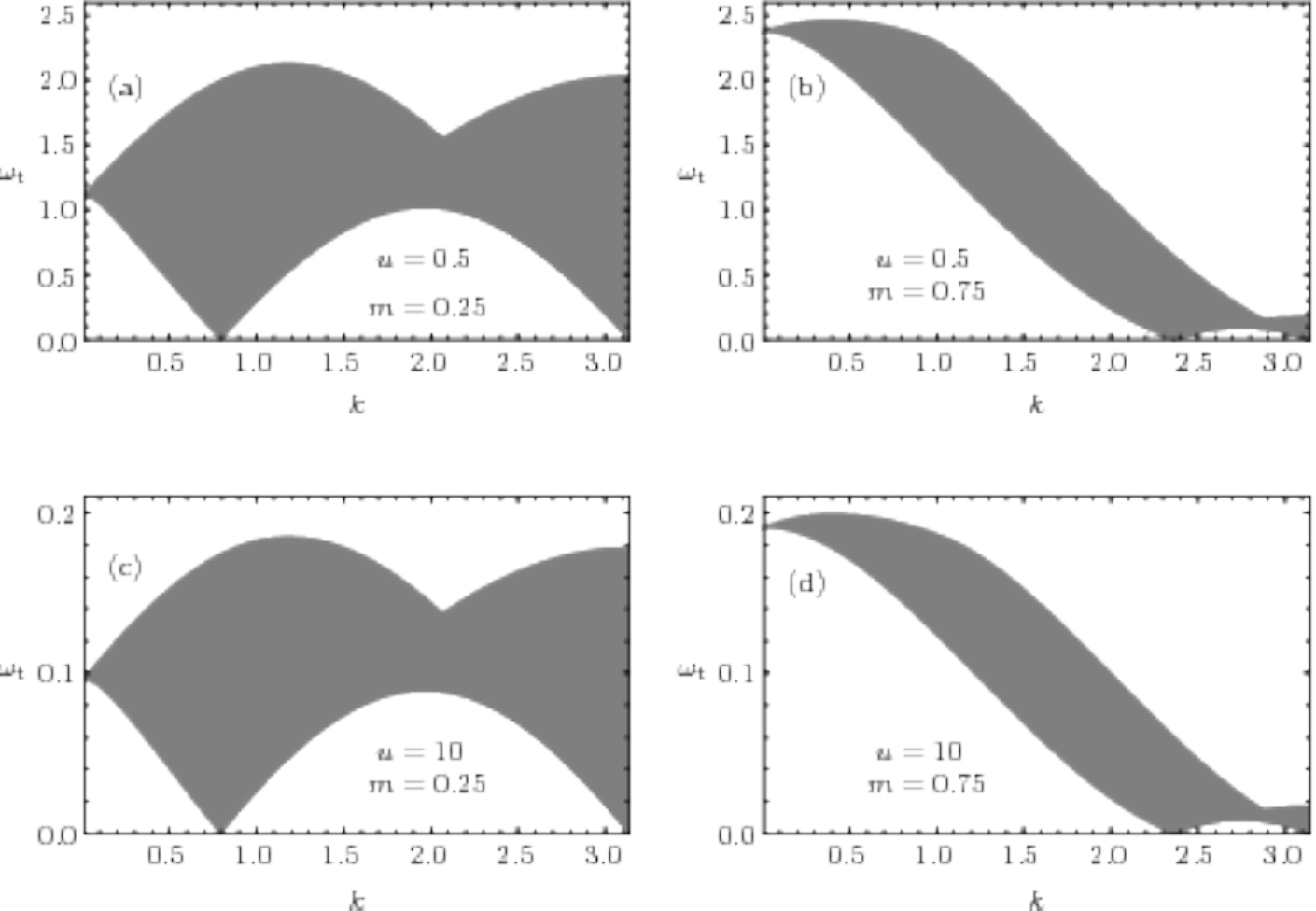}
\caption{The half-filled 1D Hubbard model transverse spin spectrum $\omega^t (k)$ 
corresponding to that plotted in Figs. \ref{figure7} - \ref{figure9} for the spin-$1/2$ $XXX$ chain
for (a) $m=0.25$ and $u=0.5$, (b) $m=0.75$ and $u=0.5$,
(c) $m=0.25$ and $u=10.0$, and (d) $m=0.75$ and $u=10.0$. As in
the case of the longitudinal spin spectrum plotted in Fig. \ref{figure23},
the main effect of the onsite repulsion is on the spectrum energy bandwidth.\\
{\it Source}: From Ref. \cite{CarCadez-16}.}
\label{figure24} 
\end{figure}

The one-electron spectral function exponent plotted in Fig. \ref{figure22} refers for $k\in [0,k_F]$ to a line shape of 
form, Eq. (\ref{branch-l}). Above it there is no one-electron removal spectral weight ($c_B=-1$). Hence it is exact for 
the present model. It corresponds to the theoretical TCNQ line called $s$ in Fig. \ref{figure20}. The one-electron 
addition exponent plotted in Fig. \ref{figure22} for $k\in [k_F,3k_F]$ and electronic density $n_e=0.59$ equals the 
corresponding one-electron removal exponent for density $n_e'=2-n_e=1.41$. The latter is that appropriate for the 
theoretical TTF $s''$ branch line plotted in Fig. \ref{figure20} for the small range of momentum values $k>k_F$ for 
which that exponent is negative. 

The experimental peak dispersions shown in Fig. \ref{figure20} show significant discrepancies from the conventional 
band-structure predictions. Figure 7 of Ref. \cite{spectral0} represents the experimental spectral features in 
that figure in comparison with the conduction band dispersions obtained by density functional theory. In contrast 
to the line shapes obtained within the 1D Hubbard model by the PDT,
those predicted by density functional theory do not agree with the experimental ARPES
features. The corresponding non-perturbative many-electron
physics justifies why standard density functional theory fails to describe
such unusual ARPES spectral-line shapes. The 
theoretical description of the microscopic mechanisms behind the spectral properties of 
1D systems and quasi-1D metals can be further improved by the use of a renormalized PDT \cite{MoSe-17}. 
It accounts for the effects of electron finite-range interactions beyond the conventional 1D Hubbard model.

The results discussed in this section and in Section \ref{exc-spectra} illustrate how the 1D Hubbard model physics 
is fully controlled by the scattering events of the pseudofermions. The model one-particle spectral
functions has also been studied by numerical methods. The authors of Ref. \cite{Jeckelmann-04} found that the 1D Hubbard model 
one-electron removal spectral function $s1$ branch line exact exponent plotted in Fig. \ref{figure22} 
for the momentum range $k\in [0,k_F]$ fully agrees with that exponent values obtained by 
the density matrix renormalization group (DMRG).
The PDT exponent for the line shape near the $c$ branch line in Fig. \ref{figure20} is not exact.
Indeed, there is some small amount of spectral weight above that line. It is though
a very good approximation. Consistently, the authors of Ref. \cite{Jeckelmann-04} have found small 
minor quantitative deviations from the DMRG values of that exponent.

The numerical results derived by the MQIM in Section VIII of Ref. \cite{Essler-10} for the momentum 
dependence of the one-electron removal spectral-function $s1$ branch line exponent of Fig. \ref{figure22} are 
in full quantitative agreement with those obtained by use of the PDT \cite{TTF,spectral,spectral-06}
for electronic density $n_e = 0.59$, interaction values $U/t = 1.00, 4.90, 10.00$. This applies to the whole range of that figure 
momentum values $k\in [0,k_F]$ associated with electron removal. Moreover, the same exponent was also calculated
in the framework of the MQIM in Ref. \cite{Seabra-14}, using input from the BA solution.
It has been plotted in that reference as a function of the momentum for densities $n_e=0.17, 0.25, $ and $U/t = 2.00, 
5.00, 10.00, 20.00$. Again, such results are in accord with those of the PDT.
\begin{figure}
\includegraphics[scale=0.85]{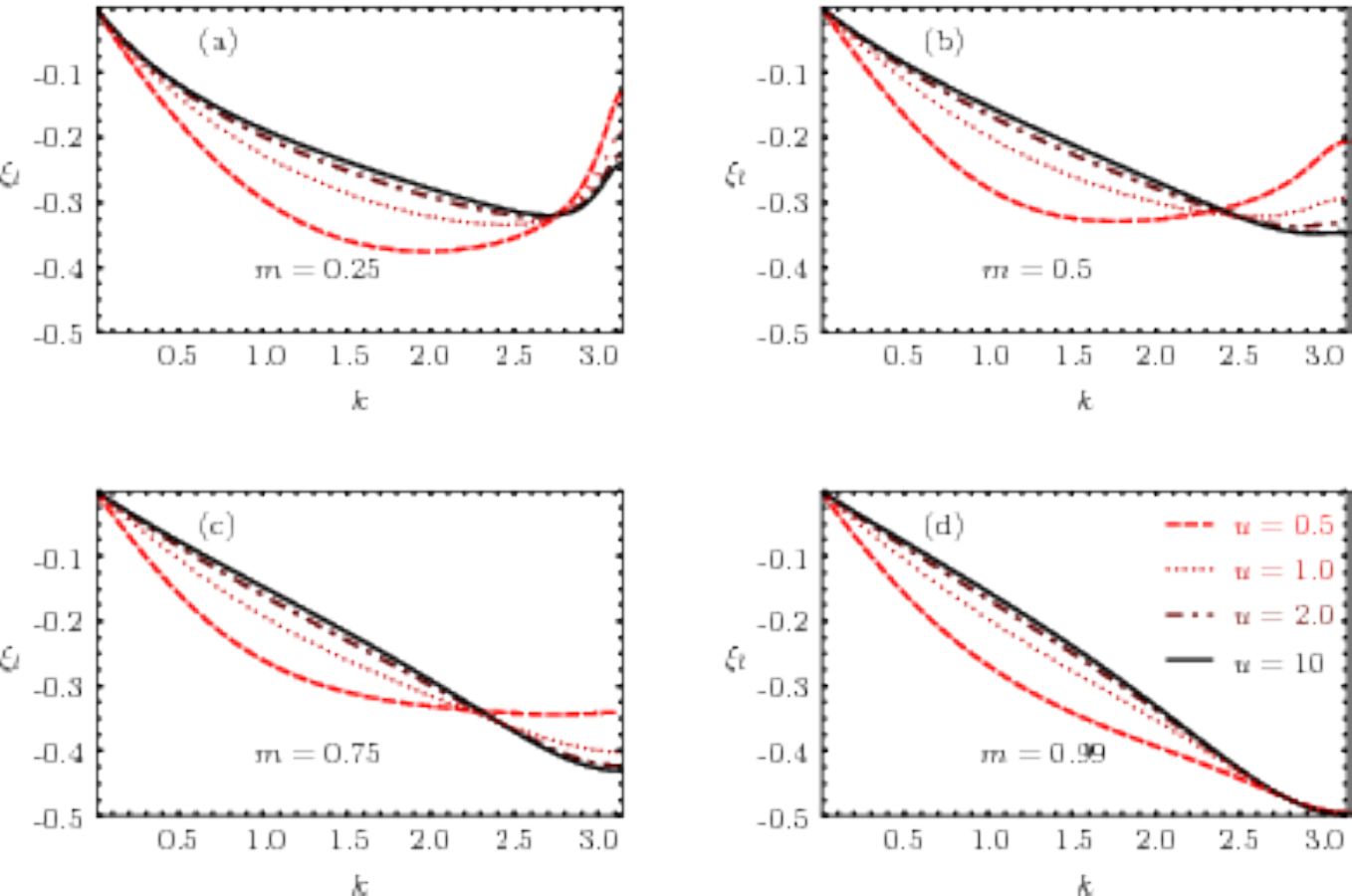}
\caption{The half-filled 1D Hubbard model exponent $\xi^{l} (k)$ corresponding to that
plotted in Figs. \ref{figure7} - \ref{figure9} for the spin-$1/2$ $XXX$ chain that controls the singularities
in the vicinity of the lower thresholds of the longitudinal spin spectrum $\omega^l (k)$ plotted 
in Fig. \ref{figure23} as a function of $k\in ]0,\pi[$ for several values of $u$ and spin densities
(a) $m=0.25$, (b) $m=0.50$, (c) $m=0.75$, and (d) $m=0.99$.\\
{\it Source}: From Ref. \cite{CarCadez-16}.}
\label{figure25} 
\end{figure}

\subsection{PDT applications to the high-energy spectral properties of the 1D Hubbard model in a magnetic field}
\label{SDIESpecPro}

The longitudinal and transverse dynamical structure factors 
studied in Section \ref{DSGzzxx} for the spin-$1/2$ $XXX$ chain 
and the one-electron spectral functions have been investigated within the framework 
of the 1D Hubbard model in a magnetic field PDT in Refs. \cite{CarCadez-16} and \cite{CarCadez-17},
respectively. The results for the $XXX$ chain correspond to those obtained for the large-$u$ half-filled 1D Hubbard model 
up to $t^2/U$ order. Previous studies of 
the factors $S^{zz} (k,\omega)$ and $S^{xx} (k,\omega)=S^{yy} (k,\omega)$ 
within the half-filled 1D Hubbard focused mainly onto magnetic fields $H=0$ for which 
$S^{zz} (k,\omega)= S^{xx} (k,\omega)=S^{yy} (k,\omega)$ \cite{Essler-99,Essler-05}.
That model lower thresholds spectra $\omega^l (k)$ and $\omega^t (k)$ of the
longitudinal and transverse, respectively, dynamical structure factors are
plotted in Figs. \ref{figure23} and \ref{figure24}, respectively. Interestingly, analysis of these
figures reveals that the main effect of $u$ on these spectra is merely on their energy bandwidth.
It increases upon decreasing $u$. Otherwise, their shape remains nearly unchanged.

\begin{figure}
\includegraphics[scale=0.85]{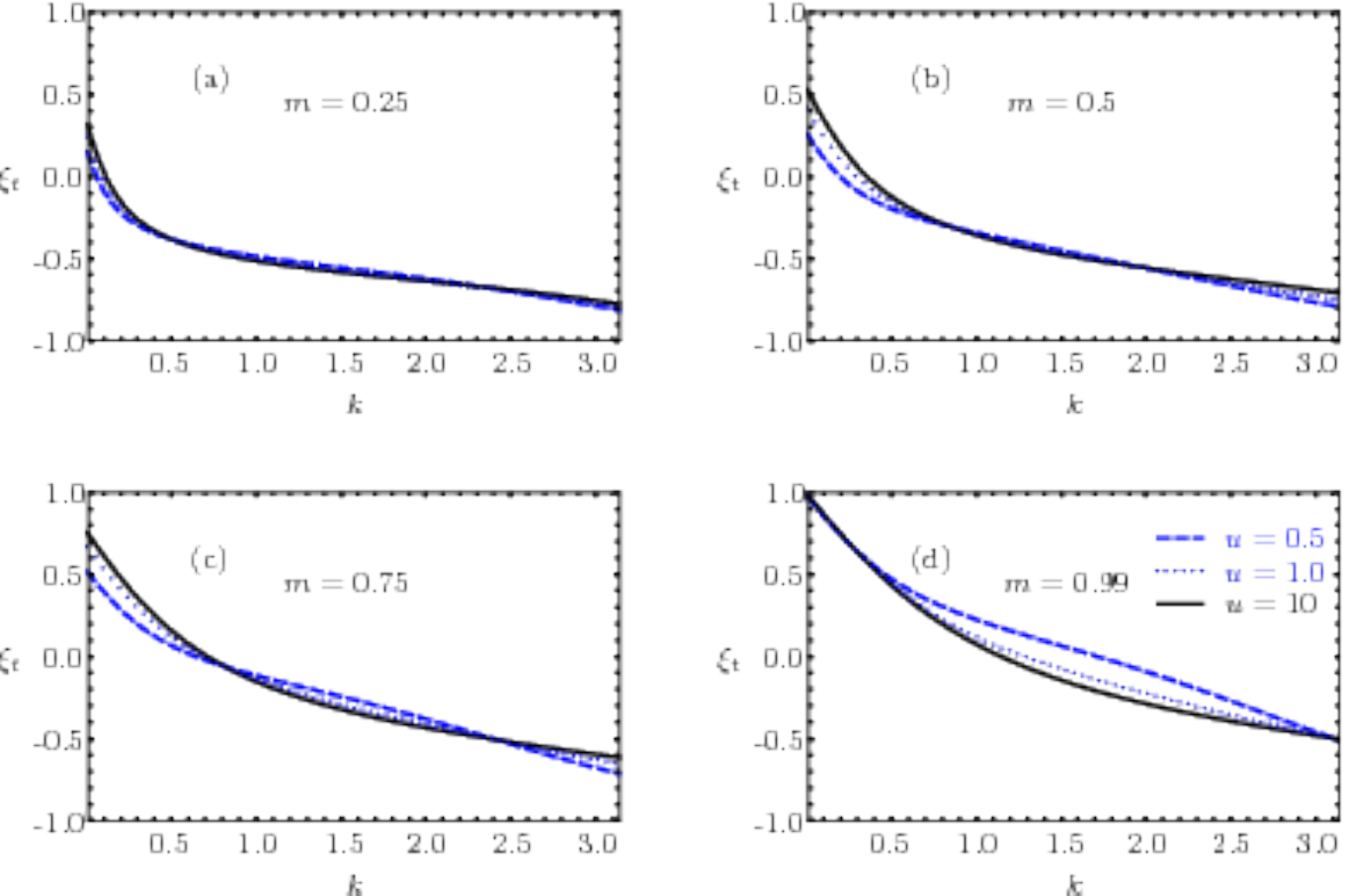}
\caption{The half-filled 1D Hubbard model  exponent $\xi^{t} (k)$  
corresponding to that plotted in Figs. \ref{figure7} - \ref{figure9} for the spin-$1/2$ $XXX$ chain
that controls the singularities in the vicinity of the lower thresholds of the transverse spin spectrum $\omega^t (k)$ plotted 
in Fig. \ref{figure24} as a function of $k\in ]0,\pi[$ for several values of $u$ and spin densities
(a) $m=0.25$, (b) $m=0.50$, (c) $m=0.75$, and (d) $m=0.99$.\\
{\it Source}: From Ref. \cite{CarCadez-16}.}
\label{figure26} 
\end{figure}

The exact behavior near the longitudinal and transverse dynamical structure factors lower thresholds refers to the
PDT general expression, Eq. (\ref{branch-l}). The corresponding singularities in the vicinity of
the thresholds of the longitudinal spin spectrum $\omega^l (k)$ in Fig. \ref{figure23}
and transverse spin spectrum $\omega^t (k)$ in Fig. \ref{figure24} are controlled by exponents $\xi^{l} (k)$ and 
$\xi^{t} (k)$, respectively. Such exponents are plotted for the half-filled 1D Hubbard model
as a function of the momentum $k\in ]0,\pi[$ for several values of $u$ and spin density $m$ in Figs. \ref{figure25} 
and \ref{figure26}, respectively.

On the one hand, the longitudinal dynamical structure factor exponent $\xi^{l} (k)$
is negative for $k>0$ at any $u$ and $m$ values. On the other hand, the
transverse dynamical structure factor exponent $\xi^{t} (k)$ 
is negative for a $u$ and $m$-dependent range $k\in [k_t,\pi]$
where the momentum $k_t$ is for $u>0$ an increasing function of $m$. Furthermore,
analysis of Fig. \ref{figure25} reveals that the negative exponent $\xi^{l} (k)$ is an
increasing and decreasing function of $u$ for the momentum ranges $k\in [0,k_l]$ and 
$k\in [k_l,\pi]$, respectively. Here $k_l$ is a spin density dependent momentum at which
the exponent $\xi^{l} (k)$ has similar value for the whole $u>0$ range.

For the intervals of the momentum $k$ for which the exponent $\xi_{\tau} (k)$ is negative, there are
lower threshold singularity cusps in the dynamical structure factors. 
Hence analysis of Figs. \ref{figure25} and \ref{figure26} provides valuable information on
the $k$ ranges for which there are singularities in the lower thresholds of 
the dynamical structure factors $S^{zz} (k,\omega)$ and $S^{xx} (k,\omega)=S^{yy} (k,\omega)$.
An interesting issue discussed in Section \ref{DSGzzxx} for the spin-$1/2$ $XXX$ chain is the potential observation of the 
theoretically predicted dynamical structure factors peaks in inelastic neutron scattering experiments. 
Analysis of Figs. \ref{figure24}, \ref{figure25}, and \ref{figure26} reveals that the effect of lessening the 
$u$ value is enhancing the $(k,\omega)$-plane energy bandwidths of the spectra edges where such 
peaks are located.

An interesting issue refers to the effects of varying the electronic density $n_e$, spin density $m$, and interaction $u$ on the momentum dependence of the exponents that control the $(k,\omega)$-plane singular features of the $\sigma =\uparrow,\downarrow$ one-electron 
spectral functions $B_{\sigma,\gamma} (k,\,\omega)$ of the 1D Hubbard model \cite{CarCadez-17}. Here
$\gamma=-1$ (and $\gamma=+1$) for one-electron removal (and addition). Such functions read,
\begin{eqnarray}
B_{\sigma,-1} (k,\,\omega) & = & \sum_{\nu^-}
\vert\langle\nu^-\vert\, c_{k,\sigma} \vert \,GS\rangle\vert^2 \,\delta \left(\omega
+ (E_{\nu^-}^{N_{\sigma}-1}-E_{GS}^{N_{\sigma}})\right) \hspace{0.20cm}{\rm for}\hspace{0.20cm}  \omega \leq 0 \, ,
\nonumber \\
B_{\sigma,+1} (k,\,\omega) & = & \sum_{\nu^+}
\vert\langle\nu^+\vert\, c^{\dagger}_{k,\sigma} \vert
\,GS\rangle\vert^2 \,\delta \left(\omega - (E_{\nu^+}^{N_{\sigma}+1}-E_{GS}^{N_{\sigma}})\right) 
\hspace{0.20cm}{\rm for}\hspace{0.20cm}  \omega \geq 0 \, .
\label{Bkomega}
\end{eqnarray}
The operators $c_{k,\sigma}$ and $c^{\dagger}_{k,\sigma}$ in this equation
annihilate and create electrons, respectively, of momentum $k$ and $\vert GS\rangle$ denotes the
initial $N_{\sigma}$-electron ground state of energy $E_{GS}^{N_{\sigma}}$. The $\nu^-$ and $\nu^+$
summations run over the $N_{\sigma}-1$ and $N_{\sigma}+1$-electron excited 
energy eigenstates, respectively, and $E_{\nu^-}^{N_{\sigma}-1}$ and 
$E_{\nu^+}^{N_{\sigma}+1}$ are the corresponding energies.

The one-electron lower Hubbard band and upper Hubbard band can be defined 
for all densities and finite repulsive onsite interaction values in terms 
of the rotated electrons associated with the model BA solution \cite{CarCadez-17}.
As discussed in Section \ref{specificRE}, the 1D Hubbard model BA quantum numbers are directly related to the
numbers of sites singly occupied, doubly occupied, and unoccupied by $\sigma$ rotated electrons. 
From the use of that relation it is found that for instance for electronic densities $n_e \in [0,1[$ and spin densities $m\in [0,n_e]$ the model ground 
states have zero rotated-electron double occupancy. The
$\sigma$ one-electron LHB addition spectral function $B^{\rm LHB}_{\sigma,+1} (k,\,\omega)$
and UHB addition spectral function $B^{\rm UHB}_{\sigma,+1} (k,\,\omega)$ are then uniquely defined
for $u>0$ as $B_{\sigma,+1} (k,\,\omega) = B^{\rm LHB}_{\sigma,+1} (k,\,\omega) + B^{\rm UHB}_{\sigma,+1} (k,\,\omega)$
where $B^{\rm LHB}_{\sigma,+1} (k,\,\omega) = \sum_{\nu^+_0}
\vert\langle\nu^+_0\vert\, c^{\dagger}_{k,\sigma} \vert
\,GS\rangle\vert^2 \,\delta (\omega - (E_{\nu^+_0}^{N_{\sigma}+1}-E_{GS}^{N_{\sigma}}))$  for $\omega \geq 0$ and
$B^{\rm UHB}_{\sigma,+1} (k,\,\omega) = \sum_{\nu^+_D}
\vert\langle\nu^+_D\vert\, c^{\dagger}_{k,\sigma} \vert
\,GS\rangle\vert^2 \,\delta (\omega - (E_{\nu^+_D}^{N_{\sigma}+1}-E_{GS}^{N_{\sigma}}))$ for $\omega \geq 0$.
Here the $\nu^+_0$ and $\nu^+_D$ summations run over the $N_{\sigma}+1$-electron excited 
energy eigenstates with zero and $D>0$, respectively, rotated-electron double occupancy 
and $E_{\nu^+_0}^{N_{\sigma}-1}$ and $E_{\nu_D^+}^{N_{\sigma}+1}$ are the corresponding energies.

\begin{figure}
\includegraphics[scale=0.85]{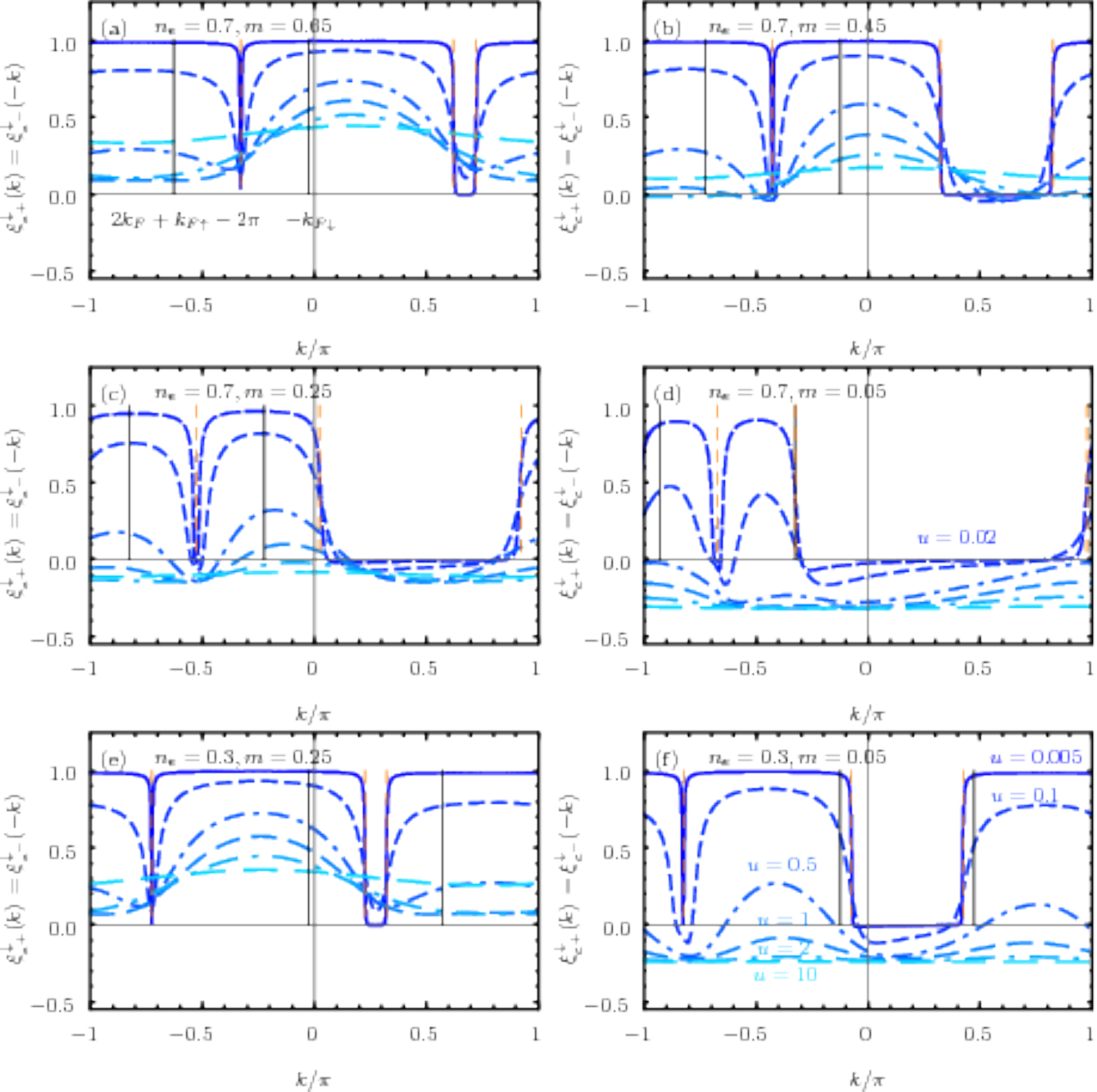}
\caption{The exponent $\xi_{c^{+}}^{\downarrow} (k)=\xi_{c^{-}}^{\downarrow} (-k)$
that controls the singularities in the vicinity of the $c^+$ branch line of the $\sigma =\downarrow$
one-electron removal and LHB addition spectral function, Eq. (\ref{Bkomega}) for $\sigma =\downarrow$ and $\gamma=-1$,
as defined in Ref. \cite{CarCadez-17}, as a function of the momentum $k/\pi\in ]-1,1[$ for several $u$ values, electronic 
density $n_e =0.7$, and spin densities (a) $m=0.65$, (b) $m=0.45$, (c) $m=0.25$, and (d)  $m=0.05$, and for electronic 
density $n_e =0.3$ and spin densities (e) $m=0.25$ and (f) $m=0.05$.\\
{\it Source}: From Ref. \cite{CarCadez-17}.}
\label{figure27} 
\end{figure}

\begin{figure}
\includegraphics[scale=0.85]{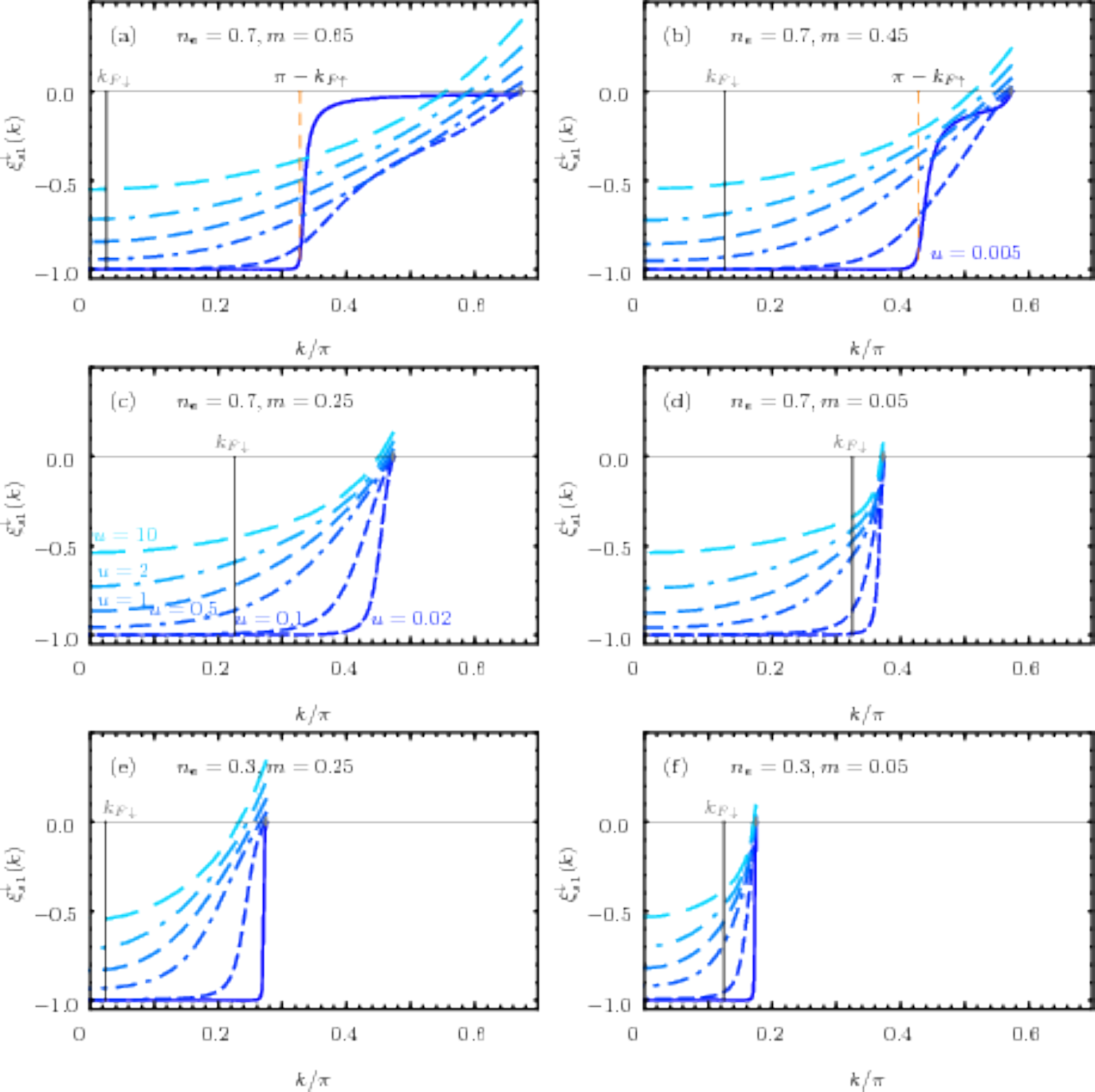}
\caption{The exponent $\xi_{s1}^{\downarrow} (k)$
that controls the singularities in the vicinity of the $s1$ branch line of the $\sigma =\downarrow$
one-electron removal and LHB addition spectral function, Eq. (\ref{Bkomega}) for $\sigma =\downarrow$ and $\gamma=-1$, 
as defined in Ref. \cite{CarCadez-17},
as a function of the momentum $k/\pi\in ]0,1[$ for the same values of $u$, electronic density $n_e$, and spin density $m$ as in 
Fig. \ref{figure27}. (For $k/\pi\in ]-1,0[$ the exponent $\xi_{s1}^{\downarrow} (k)$
is given by $\xi_{s1}^{\downarrow} (k)=\xi_{s1}^{\downarrow} (-k)$ with $-k/\pi\in ]0,1[$ as plotted here.)\\
{\it Source}: From Ref. \cite{CarCadez-17}.}
\label{figure28} 
\end{figure}

The momentum dependent exponents that control the line shape of the $\sigma =\uparrow,\downarrow$ one-electron 
spectral functions $B_{\sigma,\gamma} (k,\,\omega)$, Eq. (\ref{Bkomega}), near the main $c$ and $s1$ branch lines 
are in Ref. \cite{CarCadez-17} plotted as a function of $k$ for a large range of different $u$, $n_e$, and $m$ values. 
For simplicity, here we plot some of the momentum dependent exponents of the down-spin-one-electron removal and LHB addition spectral function,
Eq. (\ref{Bkomega}) for $\sigma =\downarrow$ and $\gamma=-1$. Such a spectral function $c^{\pm}$ branch lines
and $s1$ branch line are generated by processes that correspond to particular cases of those generated by
the leading-order operators considered in Ref. \cite{CarCadez-17}. These branch lines energy spectra are defined in that reference.
The corresponding momentum dependent exponents $\xi_{c^{+}}^{\downarrow} (k)=\xi_{c^{-}}^{\downarrow} (-k)$
and $\xi_{s1}^{\downarrow} (k)$ that control the spectral-function singularities in the vicinity of the $c^+$ and $s1$ branch lines
are plotted in Figs. \ref{figure27} and \ref{figure28}, respectively. The effects of varying $u$ on these one-electron exponents
are stronger than in the case of the exponents of the spin dynamical structure factors plotted in Figs. \ref{figure25} and \ref{figure26}.

\subsection{Relation between the pseudofermion dynamical theory and the mobile quantum impurity model}
\label{RelaPDTMIM}

For simplicity, here we use the PDT expressions for the dynamical correlation functions 
of the 1D Lieb-Liniger Bose gas discussed in Section \ref{RelapsboBg} to clarify the relation between the PDT and the MQIM 
\cite{Glazman-09,Glazman-12}. The basic relation found in the following is qualitatively similar to that of the more complex models 
also reviewed in this paper.

Within the MQIM, the pseudoparticles in the vicinity of the $\iota = \pm$ Fermi
points are called {\it particles}. The MQIM relies on an effective Hamiltonian 
of general form \cite{Glazman-09,Glazman-12},
\begin{eqnarray}
\hat{H} & = & \hat{H}_0 + \hat{H_d} + \hat{H}_{\rm int} \hspace{0.20cm}{\rm where}\hspace{0.20cm} 
\hat{H_0} = {v\over 2\pi}\int dx\left(K_0(\nabla\theta (x))^2 + {1\over K_0}(\nabla\phi (x))^2\right) \, ,
\nonumber \\
\hat{H_d} & = & \int dx\,d^{\dag} (x)\left(\varepsilon_1 (k) - i {\partial\varepsilon_1 (k)\over \partial k}{\partial\over \partial x}\right)d (x) \, ,
\nonumber \\
\hat{H}_{\rm int} & = & \int dx\left(V_R\nabla{\theta (x) - \phi (x)\over 2\pi} - V_L\nabla{\theta (x) + \phi (x)\over 2\pi}\right)d^{\dag} (x)d (x) \, .
\label{HMIM}
\end{eqnarray}
Here $\hat{H}_0$ in whose expression $v$ is the particles Fermi velocity describes their kinetic energy near the BA Fermi points.
The {\it mobile impurity} motion is described by $\hat{H}_d$. Furthermore, $\hat{H}_{\rm int}$ contains the density-density
interactions between the impurity and the particles in the vicinity of the Fermi points.

The operator $d^{\dag}$ creates the mobile impurity with momentum near $k$ and
energy near $\varepsilon_1 (k)$. This is the excitation energy $\omega^{\tau} = c_{\tau}\,\varepsilon (q_F-k)$, 
Eq. (\ref{omeglBg}). Its velocity is ${\partial\varepsilon_1 (k)\over \partial k}{\partial\over \partial x}$.
The quantities $\theta (x)$ and $\phi (x)$ in Eq. (\ref{HMIM}) are conventional bosonic fields. They
satisfy the canonical commutation relation,
\begin{equation}
[\phi (x),\nabla\theta (x')] = i\pi\delta (x-x') \, .
\end{equation}
The particles annihilation operator reads $\Psi_B (x,t) \approx e^{i\theta (x,t)}$.
Moreover, $\nabla{\theta (x) - \phi (x)\over 2\pi}$ and $-\nabla{\theta (x) + \phi (x)\over 2\pi}$ are in the
effective Hamiltonian $\hat{H}_{\rm int}$ expression the densities of right and left movers, respectively.
The interactions described by that effective Hamiltonian lead to the formation
of low-energy particle-hole pairs. Those are crucial for the line shape of the dynamical
correlation functions.

A key step of the MQIM method is that the interaction Hamiltonian term $\hat{H}_{\rm int}$
can be removed by a unitary transformation, 
$\hat{U}^{\dag}(\hat{H}_0 + \hat{H}_d + \hat{H}_{\rm int})\hat{U}$, where \cite{Glazman-09,Glazman-12},
\begin{equation}
\hat{U}^{\dag} = e^{i\int dx\left\{{\delta_+\over 2\pi}[\tilde{\theta} (x) - \tilde{\phi} (x)]
- {\delta_-\over 2\pi}[\tilde{\theta} (x) + \tilde{\phi} (x)]\right\}d^{\dag} (x)\,d (x)}
\hspace{0.20cm}{\rm where}\hspace{0.20cm}\delta_{\pm} = 2\pi F_B (\pm Q\vert k_0 (q_{F}-k)) \, .
\label{UdMIM}
\end{equation}
Here $\delta_{\pm}$ is a function defined in Eq. (15) of Ref. \cite{Glazman-BG-08}. It is a particular case of the
MQIM shift function $F_B (k\vert k')$ defined in Eqs. (7) and (8) of that reference.
(As in Section \ref{RelapsboBg}, the MQIM shift-function $F_B (\nu\vert\mu)$ variables $\nu$ and 
$\mu$ of Ref. \cite{Glazman-BG-08} are here denoted by $k$ and $k'$, respectively, and the corresponding
limiting values $\pm q$ by $\pm Q$, Eq. (\ref{varepsilonBg}).)

We start by discussing the technical equivalence of the line shapes near the branch lines, as defined within the PDT 
and the MQIM. The relation between PDT and MQIM is addressed afterwards, from the point of view
of the physical processes under consideration. A first issue to be clarified is the relation of the PDT $\tau = B, A, D$ 
exponents, Eq. (\ref{xilambBg}), to those derived by the MQIM in Ref. \cite{Glazman-BG-08}. In Appendix \ref{alphakF} 
it is rigorously shown that the following equalities {\it exactly} hold for the momentum range $k\in [0,2\pi n_b]$ considered
in that reference,
\begin{equation}
\xi_{B} (k) = - {\bar{\mu}}_{-} (k) \, ,
\hspace{0.20cm}
\xi_{A} (k) = - {\underline{\mu}}_{+} (k) 
\hspace{0.20cm}{\rm and}\hspace{0.20cm} 
\xi_{D} (k) = - \mu_2 (k) \, .
\label{3exp}
\end{equation}
Such exact relations hold in spite of the apparent different expressions given in Eqs. (16)-(18) of Ref. \cite{Glazman-BG-08} for 
the edge exponents on the right-hand side of the three equations in Eq. (\ref{3exp}). The PDT exponent expressions,
\begin{equation}
\xi_{\tau} (0) = {(2b_{\tau}-1)\over\xi^1} \left(1 + {(2b_{\tau}-1)\over 2\xi^1}\right) 
\hspace{0.20cm}{\rm and}\hspace{0.20cm} 
\xi_{\tau} (2\pi n) = -2\xi^1 + 2(\xi^1)^2 +
{(2b_{\tau}-1)\over\xi^1} \left(1 + {(2b_{\tau}-1)\over 2\xi^1}\right) \, ,
\label{xilamb02qFBg}
\end{equation}
refer to the limiting momenta $k\rightarrow 0$ and $k\rightarrow 2\pi n$, 
respectively. It follows from the equalities in Eq. (\ref{3exp}) that they
are fully equivalent to those provided for ${\bar{\mu}}_{-} (k) = -\xi_{B} (k)$, ${\underline{\mu}}_{+} (k) = -\xi_{A} (k)$, and
$\mu_2 (k) = -\xi_{D} (k)$ in Eqs. (19), (22), and (23) of Ref. \cite{Glazman-BG-08} for
$k\rightarrow 0$ and Eqs. (20), (24), and (25) of such a reference for $k\rightarrow 2\pi n$.

On the one hand, the dynamical correlation functions exponents are within the PDT controlled by the pseudofermion
phase shifts $2\pi\Phi (q_j,q_{j'}) = 2\pi{\bar{\Phi}} (k_0 (\bar{q}_j),k_0 (\bar{q}_{j'}))$, Eq. (\ref{PsBg}).
They are thus also controlled by the related momentum rapidity phase shifts $2\pi{\bar{\Phi}} (k,k')$, Eq. (\ref{PsEqBg}).
Specifically, the edge singularities exponents are fully controlled by the pseudofermion phase shifts $2\pi\Phi (\pm q_F,q') = 
2\pi{\bar{\Phi}} (\pm Q,k_0 (q'))$. 

On the other hand, within the MQIM used in Ref. \cite{Glazman-BG-08} to derive
the same exponents, those are determined by the MQIM shift function $F_B (k\vert k')$. The two MQIM 
$\iota =\pm$ shift functions $F_B (\pm Q\vert k')$
play the major role, since they fully control the edge singularities exponents.

That the exact equalities in Eq. (\ref{3exp}) hold for $k\in [0,2\pi n_b]$ is an issue addressed in Appendix \ref{alphakF}.
It is shown to result from the MQIM shift function $F_B (k\vert k')$ of Ref. \cite{Glazman-BG-08} being
exactly related to the momentum rapidity phase shift ${\bar{\Phi}} (k,k')$ in units of $2\pi$, Eq. (\ref{PsEqBg}), as 
given in Eq. (\ref{PhiFB}) of that Appendix. Hence the MQIM $\iota = \pm$ shift functions $F_B (\pm Q\vert k')= \delta_{\pm}/2\pi$ that in 
Ref. \cite{Glazman-BG-08} control the edge singularities exponents are simply related to the $\iota = \pm$ 
pseudofermion phase shifts $\Phi (\iota q_F,q') = {\bar{\Phi}} (\iota Q,k_0 (q'))$ in units of $2\pi$ as follows,
\begin{equation}
F_B (\iota Q\vert k') = {\delta_{\iota}\over 2\pi} =  {\xi^1\over 2} - {\bar{\Phi}} (\iota Q,k') 
= {\sqrt{K_0}\over 2} - {\bar{\Phi}} (\iota Q,k') \hspace{0.20cm}{\rm for}\hspace{0.20cm} \iota = \pm \, .
\label{PhiFBQ}
\end{equation}

Within the PDT, the $\iota = \pm$ pseudofermion phase shifts $2\pi\Phi (\iota q_F,q') = 2\pi{\bar{\Phi}} (\iota Q,k_0 (q'))$ 
control the important exponent functional, Eq. (\ref{2DeltaBg}). In addition, they determine the parameter 
$\xi^1 = 1 + \Phi (q_F,q_F) - \Phi (q_F,-q_F)$, Eq. (\ref{xiBg}), that also appears 
in that exponent functional expression. This thus also applies the TLL parameter, 
$K_0=(\xi^1)^2=(1 + \Phi (q_F,q_F) - \Phi (q_F,-q_F))^2$. The exact relation between the MIM
shift functions $F_B (\iota Q\vert k')$ and the pseudofermion phase shifts $2\pi\Phi (\iota q_F,q') = 2\pi{\bar{\Phi}} (\iota Q,k_0 (q'))$ 
given in Eq. (\ref{PhiFBQ}) clarifies the basic relation between the PDT and the MIM.
(It plays the key role in the rigorous proof presented in Appendix \ref{alphakF}.)

The physical processes described by the MQIM are fully equivalent to those of the PDT
that control the line shape of the dynamical correlation functions near the branch lines, 
Eqs. (\ref{dE-dP-bl})-(\ref{branch-l}). Specifically:
\vspace{0.25cm}

(I) The unitary transformation, Eq. (\ref{UdMIM}), that removes the interaction Hamiltonian term $\hat{H}_{\rm int}$
is equivalent to the pseudoparticle-pseudofermion unitary transformation, $q_j \rightarrow q_j + 2\pi\Phi (q_j)/L$.
Such a transformation removes the pseudoparticle energy spectrum interactions and introduces
shake-up effects. Those result from the discrete canonical momentum value shifts $2\pi\Phi (q_j)/L$, 
Eq. (\ref{QPqBg}), under the transitions to the excited states. Such shake-up effects are behind 
a large number of small-momentum and low-energy pseudofermion-pseudofermion-hole processes (C). 
They occur in the linear part of the pseudofermion energy dispersions and lead to finite spectral-weight 
contributions. Similarly, in the case of the MQIM the interactions described by the effective Hamiltonian 
$\hat{H}_{\rm int}$ lead to the formation of low-energy particle-hole pairs that are crucial for the line 
shape of the dynamical correlation functions.
\vspace{0.25cm}

(II) The mobile quantum impurity described by the effective Hamiltonian $\hat{H}_d$ in Eq. (\ref{HMIM})
corresponds to the pseudofermion or pseudofermion hole with canonical momentum away from the 
Fermi points. It is created within the PDT by processes (A). They occur under the transitions to excited 
states associated with the corresponding pseudofermion branch line.
\vspace{0.25cm}

(III) The effective Hamiltonian $\hat{H}_0$ in Eq. (\ref{HMIM}) describes the same processes
near the Fermi points as the PDT processes of classes (B) and (C).  
\vspace{0.25cm}

Hence although relying on apparently different physical starting points, the MQIM describes
exactly the same processes as the PDT. This applies in the particular case of line shapes 
near the branch lines, as defined within the latter dynamical theory.

%%%%%%%%%%%%%%%%%%%%%%%%%%%%%%%%%%%%%%%%%%%%%%%%%%%%%%%%%%%%%%%%
\section{General Outlook and future developments}
\label{outlook}

The static properties of the 1D Lieb-Liniger Bose gas, spin-$1/2$ isotropic Heisenberg chain, and 1D Hubbard model
have been revisited in this review in terms of quantum liquids of pseudoparticles. The static quantities of these
integrable systems are controlled by Landau parameters associated with the pseudoparticles residual 
interactions $f$ functions. The similarities to and differences from the usual Fermi liquid have been 
discussed.

The high-energy dynamical correlation functions of these integrable systems have been studied
in the suitable and related pseudofermion representation of the BA solutions. The line
shape in the vicinity of the high-energy one-particle spectral functions of the
1D Lieb-Liniger Bose gas and 1D Hubbard model is controlled by momentum dependent
exponents. Such exponents have simple expressions in terms of the pseudofermions scattering phase shifts.
For all models under review, the same applies to the line shape near the spectra edges of
the two-particle dynamical correlation functions.

One of the goals of this review is to contribute to the further understanding of
the fractionalized particles microscopic mechanisms that control the low-
and high-energy properties of 1D correlated systems.
In the case of the spin-$1/2$ $XXX$ chain, an exact expression of the spin currents of
non-LWSs, Eq. (\ref{jcurrentsn1}), was used to study the elementary currents $j_1^h (q_j)$
in Eq. (\ref{jpmn1}) that are conventionally associated with spinons. It has been found that the latter elementary currents 
describe the translational degrees of freedom of the model unpaired physical spins $1/2$ whose occupancies
generate the energy eigenstates multiplet configurations. Our study includes an analysis of the exact elementary spin currents 
$j_{\pm 1/2}$ in Eq. (\ref{jpmn1}) carried by such physical spins $1/2$. It is found from it that 
one cannot associate the internal spin degrees of freedom of physical spins $1/2$ with the BA quantum numbers $q_j$ in the argument
of the elementary current spectrum $j_1^h (q_j)$. Hence such internal spin degrees of freedom
can neither be associated with the corresponding conventional spinons.

This reveals that spinons, as defined within integrable models, are not spin-$1/2$ particles.  
They do not contain the internal spin degrees of the physical spins whose translational degrees 
of freedom they describe. This clarification is important, because spinons are conventional fractionalized particles
widely used in the description of the spin degrees of freedom of integrable systems. 
It applies to all integrable models with $SU(2)$ symmetries associated for instance with spin or $\eta$-spin 
degrees of freedom. However, conventional spinons describe correctly the translational
degrees of freedom of the multiplets physical spins $1/2$. They thus can be successfully used
in the description of some of the properties of 1D spin chains and electronic correlated models.
  
Moreover, the results under review confirm that the non-perturbative relation of the pseudoparticles 
to the models physical particles (bosons, spins $1/2$, and electrons) is more involved for
models with more types of degrees of freedom and thus of increasing complexity. 
Specifically, it becomes more involved as one goes from the Abelian global $U(1)$ symmetry 1D Lieb-Liniger 
Bose gas, to the non-Abelian global $SU(2)$ symmetry spin-$1/2$ $XXX$ chain, and further to the much more involved
non-Abelian global $[SU(2)\otimes SU(2)\otimes U(1)]/Z_2^2$ symmetry 1D Hubbard model. The
discussion of the increase in complexity and of the role played in it by the interplay of the BA pseudoparticles
with the global symmetries representations, has unified the results being reviewed. 
It thus contributed to the further understanding of their physical meaning.

Concerning future studies in this field, an interesting future development would 
be to extend the computations of the PDT high-energy dynamical correlation functions
beyond the vicinity of the one- and two-particle spectra edges and one-particle 
singular branch and border lines at finite $u$. The derivation of 
accurate finite-$u$ line shapes over the whole $(k,\omega)$ plane, as 
those obtained in Refs. \cite{Karlo-95,Karlo-96,Karlo-97} 
for the $u\rightarrow\infty$ limit, would require more demanding numerical computations 
than those employed in these references. Future developments 
and improvements in numerical techniques may allow such computations 
of the finite-$u$ one- and two-electron spectral-weight distributions.

A development that also deserves future studies, refers to the need of accounting for
electron finite-range and long-range interactions beyond the 1D Hubbard model in the description
of the microscopic mechanisms behind the ARPES in 1D and quasi-1D metallic states of actual physical
systems. Indeed, within that integrable model the parameter $\xi_0$ and related charge TLL parameter  
$K_{\rho}^0\equiv \xi_{0}^2/2$ \cite{Schulz-90,Lederer-00,Blumenstein-11} can vary in the intervals
$\xi_{0} \in [1,\sqrt{2}]$ and $K_{\rho}^0 \in [1/2,1]$, respectively.
(The $m\rightarrow 0$ parameter $\xi_{0}$ is related to phase shifts, as given Eq. (\ref{ZZPS}) of Appendix \ref{PSIE}.)
Through Eq. (\ref{alphaHM}), such intervals correspond to a suppression of density of states exponent 
range $\alpha_0 \in [0,1/8]$.

In contrast, the density of states suppression exponent $\alpha$ experimentally measured for instance 
in 1D metallic states of Bi/InSb(001) \cite{Ohtsubo-15}, 1D line defects in transition dichalcogenides such 
as MoSe$_2$ \cite{MoSe-17}, and quasi-1D conductors \cite{Blumenstein-11}
belongs typically to the interval $\alpha\in [0.50,0.80]$. This implies that for such systems
the charge TLL parameter $K_{\rho}$ and the related parameter $\xi_0 = \sqrt{2K_{\rho}}$ have
values $K_{\rho}<1/2$ and ${\tilde{\xi}}_0 <1$, respectively. As discussed in Ref. \cite{Schulz-90},
such values result from electron interactions beyond onsite, {\it i.e.}
finite-range interactions (of at least one lattice spacing) or long-range interactions that must 
be accounted for. (Here $K_{\rho}$ and ${\tilde{\xi}}_{0}$ is our notation for the parameters corresponding to the 1D Hubbard model 
parameters $K_{\rho}^0 \in [1/2,1]$ and $\xi_{0} \in [1,\sqrt{2}]$ 
in the general case of models with electron finite-range or long-range interactions for which
they have values within the extended intervals $K_{\rho} \in[1/8,1]$
and ${\tilde{\xi}}_{0} \in [1/2,\sqrt{2}]$, respectively.)

The electron finite-range renormalized theory introduced in Ref. \cite{MoSe-17} involves
the transformation of the 1D Hubbard model into non-integrable models.
The $c$ pseudofermions are not well defined in such models except at and
near the $c$ band Fermi points. The transformation involves gently turning on the
finite-range part of electronic potentials $V_{\rm el} (r)$. This leads to a renormalized effective potential between the
$c$ pseudofermions at and near the $c$ band Fermi points and the $c$-band hole created 
in that band under one-electron removal excitations. Within the MQIM, that $c$-band hole
refers to the mobile quantum impurity. The corresponding interaction
between such $c$ pseudofermions and the emerging $c$ band hole associated with
their renormalized potential has an effective range equal or larger than one lattice spacing. 
The effective range plays an important role for instance in scattering of atoms \cite{Flambaum-99}. 

The renormalized theory of Ref. \cite{MoSe-17} is applied in that reference to the 
metallic 1D line defects in MoSe$_2$ for which the effects of the $c$ pseudofermions 
and $c$ band-hole effective range are small. Specifically, the effects of such an
effective range give rise to small changes in the spectral-function momentum dependent exponents that
control the spectral peaks distribution that lay within the ARPES experimental uncertainty.
Therefore, they have been neglected in the studies of Ref. \cite{MoSe-17}. An interesting development 
is to account for such interaction effective-range effects within the $c$ pseudofermions phase-shift
renormalization. Indeed, this is needed in the case of the description of larger
finite-range or long-range microscopic mechanisms behind the ARPES of other 1D metallic states as those in 1D Bi/InSb(001) \cite{Ohtsubo-15}
and quasi-1D conductors \cite{TTF,Ralph-02,spectral0,spectral,Blumenstein-11}.

Another interesting future development would involve further advances in the 
use of ultra-cold atoms in optical lattices to simulate the 1D correlated models
under review here and related models \cite{Guan-13,Cazalilla-11,Fabbri-15,Golovach-09,Zoller-05,Massel-05,Huo-12,Cheneau-12}.
This would provide complementary information on both the spectral-functions line 
shapes over the whole $(k,\omega)$ plane and the exotic fractionalized particles
and related composite particles reviewed in this paper. 

Furthermore, spectral signatures of fractionalized particles have been clearly seen in quantum wires 
\cite{Jompol-09}. This makes them potential candidates for technological applications in quantum computers.
As mentioned in Section \ref{elem-obj-intr}, the further understanding the properties of the fractionalized particles 
may be important for such quantum technologies. It may as well as helping to develop more complete theories 
of superconductivity and conduction in low-dimensional condensed-matter systems.

Whether fractionalized particles and their composite particles also emerge
in two-dimensional correlated problems is an open problem of high scientific interest.
The global $[SU(2)\otimes SU(2)\otimes U(1)]/Z_2^2$ symmetry found in Ref. 
\cite{bipartite} applies to the Hubbard model on any bipartite lattice. Hence 
rotated-electron representations associated with the model energy 
eigenstates, as those reviewed in Section \ref{specificRE} for the 1D 
lattice, apply as well to the model on the square and other bipartite 
lattices. Whether the rotated-electron degrees of freedom separation 
reported in this paper partially survives for the 
Hubbard model on the square lattice or other low-dimensional bipartite 
lattices is a problem of physical interest that deserves further 
investigations.

A possible scenario is that for the Hubbard model on the square 
lattice fractionalized particles and/or their composite particles emerge from a rotated-electron degrees 
of freedom separation. However, their interactions would not be of the simple zero-momentum forward-scattering 
type. This is a property specific to the pseudofermions of integrable 1D models. It results from the occurrence 
of an infinite number of conservation laws \cite{Mura-97,Mura-98,Prosen}, which is 
associated with their integrability. The Hubbard model on the square lattice is not integrable.
Nonetheless, a scenario within which there are energy and momentum
exchanges among the charge-like and spin-like fractionalized particles and/or their composite particles is a possible interesting future development. 

The interactions of such elementary fractionalized particles and their possible 
composite particles could be a simpler problem to handle than that of the underlying many-electron 
interactions. This research direction could be of interest for developing a further understanding of the 
over 30 years old unsolved problem of the microscopic mechanisms behind the cuprates superconductivity 
\cite{ARPES-review,Kresin-06,2D-MIT,two-gaps,Scalapino-12} 
and its relation to the properties of the undoped Mott-Hubbard insulators parent compounds
\cite{Phillips-10,LCO-12}.

%%%%%%%%%%%%%%%%%%%%%%%%%%%%%%%%%%%%%%%%%%%%%%%%%%%%%%%%%%%%%%%%%%%%%%%%%%
\begin{acknowledgments}
We thank M. A. N. Ara\'ujo, D. Baeriswyl, P.-A. Bares, D. Bozi, D. K. Campbell,  A. H. Castro Neto, T. \v{C}ade\v{z}, R. G. Dias, 
J. M. E. Guerra, F. Guinea, P. Horsch, H. Q. Lin, A. Luther, L. M. Martelo, A. Moreno, S. \"Ostlund, 
K. Penc, R. G. Pereira, N. M. R. Peres, T. Prosen, J. M. Rom\'an, M. J. Sampaio, and J. M. P. L. Santos
for illuminating discussions and their contributions to common collaborations
that led to some of the results on the issues being reviewed. We also thank N. Andrei, E. Castro, and H. Johannesson
for illuminating discussions, and M. Belsley for the critical reading of a preliminary version of the review 
manuscript and useful discussions. We acknowledge our former collaborator, 
the late S.-J. Gu, for his important contributions to the success of our common research related to 
the topics reviewed here. Over the long course of his study of this problem, J. M. P. C. has benefited 
from discussions with P. W. Anderson, M. C. Asensio, M. Batzill, L. Carlos,
Y.-H. Chen, R. Claessen, F. Essler, J. Ferrer, X.-W. Guan, E. Jeckelmann, S.-i. Kimura, V. E. Korepin, 
P. A. Lee, R. Micnas, S. Nemati, Y. Ohtsubo, T. Ribeiro, A. W. Sandvik, M. Sing, A. L. L. Videira, J. Voit, X.-G. Wen, S. R. White, 
and X. Zotos. He especially wishes to acknowledge his former collaborators, the late K. Maki, A. Muramatsu, and A. A. Ovchinnikov,
for illuminating discussions on 1D correlated systems and their contributions to his understanding of the 
Hubbard model. He also acknowledges the late A. Imambekov for discussions that were helpful in 
writing this review. P. D. S. thanks K. -J. -B. Lee, J. W. Rasul, and P. Schlottmann for discussions on integrable systems. 
We thank the FEDER through the COMPETE Program and the Portuguese FCT in the framework of the 
Strategic Projects UID/FIS/04650/2013 and UID/CTM/04540/2013 and the support of the Beijing Computational Science Research Center 
where part of this review was written. 
\end{acknowledgments}

%%%%%%%%%%%%%%%%%%%%%%%%%%%%%%%%%%%%%%%%%%%%%%%%%%%%%%%%%%%%%%%%%%%%%%%%%%
\appendix

%%%%%%%%%%%%%%%%%%%%%%%%%%%%%%%%%%%%%%%%%%%%%%%%%%%%%%%%%%%%%%%%
\section{Equality of the Bose gas edge singularities exponents as derived by the PDT and MIM}
\label{alphakF}

The goals of this Appendix are to obtain the 1D Bose gas pseudofermion phase shift relation 
to the MQIM shift function $\Phi (\iota q_F,q_{j'}) = {\xi^1\over 2} - F_B (\iota k_0 (q_F),k_0 (q_{j'}))$ used to derive
expression, Eq. (\ref{2DeltaBgMIM}), and to provide a rigorous proof of the equalities given in Eq. (\ref{3exp}).
As a first step to reach such goals, the relation provided in Eq. (\ref{PhiFBQ}) is derived.
As in Sections \ref{RelapsboBg} and \ref{RelaPDTMIM}, here we denote the MQIM shift function $F_B (\nu\vert\mu)$ variables $\nu$ and 
$\mu$ of Ref. \cite{Glazman-BG-08} by $k$ and $k'$, respectively. The corresponding limiting values 
$\pm q$ are denoted by $\pm Q$, Eq. (\ref{varepsilonBg}). Moreover, as elsewhere in this paper the Tomonaga-Luttinger 
liquid parameter is denoted by $K_0$. (In Ref. \cite{Glazman-BG-08} it is denoted by $K$.) 

The MQIM shift function $F_B (k\vert k')$ has been defined in Ref. \cite{Glazman-BG-08} as the solution of the integral equation,
\begin{equation}
F_B (k\vert k') = {1\over 2} - {1\over\pi}\arctan \left({k- k'\over c}\right)  
+ {1\over\pi c}\int_{-Q}^Q dk'' {F_B (k\vert k')\over 1 + \left({k - k''\over c}\right)^2} \, .
\label{FBintEq}
\end{equation}

The momentum rapidity phase shift ${\bar{\Phi}} (k,k')$ in units of $2\pi$ and the MQIM shift function $F_B (k\vert k')$ obey integral equations, Eqs. (\ref{PsEqBg})
and (\ref{FBintEq}), respectively, with the same kernel. Hence their sum ${\bar{\Phi}} (k,k')+F_B (k\vert k')$ 
obeys an integral equation with such a kernel and whose free term is merely the sum of those of
Eqs. (\ref{PsEqBg}) and (\ref{FBintEq}). This gives,
\begin{equation}
{\bar{\Phi}} (k,k')+F_B (k\vert k') = {1\over 2}  
+ {1\over\pi c}\int_{-Q}^Q dk'' {{\bar{\Phi}} (k,k')+F_B (k\vert k') \over 1 + \left({k - k''\over c}\right)^2} \, .
\label{PhiFBintEq}
\end{equation}

Next we show that the function ${\bar{\Phi}} (k,k')+F_B (k\vert k')$ does not depend on $k'$.
Indeed, one finds from the use of Eq. (\ref{PhiFBintEq}) that the derivative $\partial ({\bar{\Phi}} (k,k')+F_B (k\vert k'))/\partial k'$
obeys the equation,
\begin{equation}
{\partial ({\bar{\Phi}} (k,k')+F_B (k\vert k'))\over\partial k'} = 
{1\over\pi c}\int_{-Q}^Q dk'' {{\partial ({\bar{\Phi}} (k,k')+F_B (k\vert k'))\over\partial k'}\over 1 + \left({k - k''\over c}\right)^2} \, .
\label{DerivPhiFBintEq}
\end{equation}
Since the free term of this equation vanishes, it follows from known properties of this type of integral equations that 
its unique solution is $\partial ({\bar{\Phi}} (k,k')+F_B (k\vert k'))/\partial k'=0$.
This confirms that $\alpha (k)\equiv {\bar{\Phi}} (k,k')+F_B (k\vert k')$ is an even function, $\alpha (k)=\alpha (-k)$,
of only the variable $k$. As given in Eq. (\ref{PhiFBintEq}), it obeys the integral equation, 
\begin{equation}
\alpha (k) = {1\over 2}
+ {1\over\pi c}\int_{-Q}^Q dk' {\alpha (k')\over 1 + \left({k - k'\over c}\right)^2} \, .
\label{alphak}
\end{equation}

One finds from simple manipulations of the integral equation, Eq. (\ref{PsEqBg}),
obeyed by the momentum rapidity phase shift ${\bar{\Phi}} (k,k')$ in units of $2\pi$ that the function,
\begin{equation}
\xi^1 (k) = 1 + {\bar{\Phi}} (k,Q) - {\bar{\Phi}} (k,-Q) \, ,
\label{xi1k}
\end{equation}
is the unique solution of the integral equation,
\begin{equation}
\xi^1 (k) = 1
+ {1\over\pi c}\int_{-Q}^Q dk' {\xi^1 (k')\over 1 + \left({k - k'\over c}\right)^2} \, .
\label{x1kEq}
\end{equation}
Since it is an even function, $\xi^1 (k)=\xi^1 (-k)$, one finds that $\xi^1 = \sqrt{K_0} = \xi^1 (\iota Q)$
where $\xi^1 = 1 + \Phi (q_F,q_F) - \Phi (q_F,-q_F)$ is the phase-shift parameter, Eq. (\ref{xiBg}). It is
related to the TLL parameter as $K_0=(\xi^1)^2=(1 + \Phi (q_F,q_F) - \Phi (q_F,-q_F))^2$.

The functions $\alpha (k)$ and $\xi^1 (k)$ obey again integral equations, Eqs. (\ref{alphak})
and (\ref{x1kEq}), respectively, with the same kernel. Hence the difference function
$\xi^1 (k)-\alpha (k)$ obeys an integral equation with that kernel and whose free term is
the difference of those of Eqs. (\ref{x1kEq}) and (\ref{alphak}). The latter free term
reads $1/2$. It follows that the difference function $\xi^1 (k)-\alpha (k)$ obeys the same integral equation, Eq. (\ref{alphak}),
as the function $\alpha (k)$. Since the solution of that integral equation is unique, one
arrives to the exact relations,
\begin{equation}
\alpha (k) = {\xi^1 (k)\over 2}  \hspace{0.20cm}{\rm and}\hspace{0.20cm} \alpha (\iota Q) = {\xi^1\over 2} \, .
\label{ALxi1kx1}
\end{equation}

By combining these relations with the expressions $\alpha (k)\equiv {\bar{\Phi}} (k,k')+F_B (k\vert k')$ and
$\alpha (\iota Q)\equiv {\bar{\Phi}} (\iota Q,k')+F_B (\iota Q\vert k')$, one readily finds that,
\begin{equation}
F_B (k\vert k') = {1\over 2}[1 + {\bar{\Phi}} (k,Q) - {\bar{\Phi}} (k,-Q)] - {\bar{\Phi}} (k,k') \, ,
\label{PhiFB}
\end{equation}
and thus $F_B (\iota Q\vert k') = \xi^1/2 - {\bar{\Phi}} (\iota Q,k')$. By accounting for the $\xi^1 (k)$ expression, 
Eq. (\ref{xi1k}), the latter is indeed the expression, Eq. (\ref{PhiFBQ}).

The important PDT $\iota = \pm$ $c$ pseudofermion Fermi points fluctuations functionals, Eq. (\ref{2DeltaBg}),
can be expressed in terms of the MQIM shift function $F_B (k\vert k')$ as given in Eq. (\ref{2DeltaBgMIM}).
To reach that expression, one uses the relation $F_B (\iota Q\vert k') = \xi^1/2 - {\bar{\Phi}} (\iota Q,k')$
that can be written as ${\bar{\Phi}} (\iota Q,k') = \xi^1/2 - F_B (\iota Q\vert k')$ so that,
\begin{equation}
\Phi (\iota q_F,q_{j'}) = {\xi^1\over 2} - F_B (\iota k_0 (q_F)\vert k_0 (q_{j'})) \, .
\label{PhaseSFB}
\end{equation}
Here $k_0 (q_{j'})$ is the ground-state momentum rapidity function. The use of this relation in
Eq. (\ref{2DeltaBg}) readily leads to the functional dimensions expression, Eq. (\ref{2DeltaBgMIM}).

It follows from Eq. (\ref{PhiFBQ}) that the $\tau = B,A,D$ exponents in Eq. (\ref{xilambBg}) can
be expressed in terms of the $\iota = \pm$ shift functions $F_B (\iota Q\vert k')$ as follows,
\begin{equation}
\xi_{\tau} (k) = -1 + \sum_{\iota = \pm}\left({\xi^1\over 2} +
\iota\,{b_{\tau}\over\xi^1} - {\bar{\Phi}} (\iota Q,k_0 (q_{F}-k))\right)^2 
= -1 + \sum_{\iota = \pm}\left(F_B (\iota Q\vert k_0 (q_{F}-k)) + \iota\,{b_{\tau}\over\xi^1}\right)^2 \, .
\label{exPhiFBBg}
\end{equation}
The function $k_0 (q_{F}-k)$ in this expression is the ground-state momentum rapidity function $k_0 (q)$ at
$q=q_{F}-k$. (Such a function is the solution of the ground-state BA equation.)

It is useful to express the second expression in Eq. (\ref{exPhiFBBg}) in terms of the functions $\delta_{\pm}$ in Eq. (\ref{UdMIM}),
defined in Eq. (15) of Ref. \cite{Glazman-BG-08}. This gives,
\begin{eqnarray}
\xi_{\tau} (k) & = & -1 + \left({\delta_+\over 2\pi} + {b_{\tau}\over\xi^1}\right)^2 + \left({\delta_-\over 2\pi} - {b_{\tau}\over\xi^1}\right)^2 
= -1 + {\delta_+^2 + \delta_-^2 \over (2\pi)^2} + {b_{\tau}(\delta_+ -\delta_-)\over 2\pi \xi^1} + {2b_{\tau}^2\over (\xi^1)^2}
\nonumber \\
& = & -1 + {\delta_+^2 + \delta_-^2 \over (2\pi)^2} + {b_{\tau}(\delta_+ -\delta_-)\over 2\pi\sqrt{K_0}} + {2b_{\tau}^2\over K_0} \, .
\label{exPhideltaBg}
\end{eqnarray}

Or specifically for each of the three dynamical correlation functions under consideration,
\begin{eqnarray}
\xi_{B} (k) & = & -1 + {\delta_+^2 + \delta_-^2 \over (2\pi)^2} 
\hspace{0.20cm}{\rm and}\hspace{0.20cm} 
\xi_{A} (k) = -1 + {\delta_+^2 + \delta_-^2 \over (2\pi)^2} + {\delta_+ -\delta_-\over 2\pi\sqrt{K_0}} + {2\over K_0} \, ,
\nonumber \\
\xi_{D} (k) & = & -1 + {\delta_+^2 + \delta_-^2 \over (2\pi)^2} + {\delta_+ -\delta_-\over 4\pi\sqrt{K_0}} + {1\over 2K_0} \, ,
\label{exBADBg}
\end{eqnarray}
where we used that $b_{B}=0$, $b_{A}=1$, and $b_{D}=1/2$. 

The exponents $-{\bar{\mu}}_{-}$, $-{\underline{\mu}}_{+}$, and $-\mu_2$ in Eq. (\ref{3exp})
were found in Ref. \cite{Glazman-BG-08} to read,
\begin{eqnarray}
- {\bar{\mu}}_{-} & = & -1 + {1\over 2}\left({\delta_+ - \delta_-\over 2\pi}\right)^2 + {1\over 2}\left({\delta_+ + \delta_-\over 2\pi}\right)^2 \, ,
\nonumber \\
- {\underline{\mu}}_{+} & = &  -1 + {1\over 2}\left({2\over \sqrt{K_0}} + {\delta_+ - \delta_-\over 2\pi}\right)^2 
+ {1\over 2}\left({\delta_+ + \delta_-\over 2\pi}\right)^2 \, ,
\nonumber \\
-\mu_2 & = & -1 + {1\over 2}\left({1\over \sqrt{K_0}} + {\delta_+ - \delta_-\over 2\pi}\right)^2 
+ {1\over 2}\left({\delta_+ + \delta_-\over 2\pi}\right)^2 \, ,
\label{exGlazmanBg}
\end{eqnarray}
as given in Eqs. (16)-(18) of that reference. It is a simple exercise to show that these expressions can be rewritten as,
\begin{eqnarray}
- {\bar{\mu}}_{-} & = & -1 + {\delta_+^2 + \delta_-^2 \over (2\pi)^2} \, ; \hspace{0.75cm}
- {\underline{\mu}}_{+} = -1 + {\delta_+^2 + \delta_-^2 \over (2\pi)^2} + {\delta_+ -\delta_-\over 2\pi\sqrt{K_0}} + {2\over K_0} \, ,
\nonumber \\
-\mu_2 & = & -1 + {\delta_+^2 + \delta_-^2 \over (2\pi)^2} + {\delta_+ -\delta_-\over 4\pi\sqrt{K_0}} + {1\over 2K_0} \, ,
\label{exGlazman0Bg}
\end{eqnarray}
respectively.

Finally, comparison of the expressions provided in Eqs. (\ref{exBADBg}) and (\ref{exGlazman0Bg}) confirms the validity
of the equalities given in Eq. (\ref{3exp}).

\section{Some additional results on the spin-$1/2$ $XXX$ chain and 1D Hubbard model TBA solutions}
\label{TBAconfig}

The function $\Theta_{n\,n'}(x)$ appearing in Eq. (\ref{gen-Lambda}) for the spin-$1/2$ $XXX$ chain
and in Eq. (\ref{Tapco2}) for the 1D Hubbard model is given by,
\begin{eqnarray}
\Theta_{n\,n'}(x) & = & \delta_{n,n'}\Bigl\{2\arctan\Bigl({x\over 2n}\Bigl) 
+ \sum_{l=1}^{n -1}4\arctan\Bigl({x\over 2l}\Bigl)\Bigr\} 
\nonumber \\
& + & (1-\delta_{n,n'})\Bigl\{ 2\arctan\Bigl({x\over \vert\,n-n'\vert}\Bigl)
+ 2\arctan\Bigl({x\over n+n'}\Bigl) 
+ \sum_{l=1}^{{n+n'-\vert\,n-n'\vert\over 2} -1}4\arctan\Bigl({x\over \vert\, n-n'\vert +2l}\Bigl)\Bigr\} \, ,
\label{Theta}
\end{eqnarray}
where $n, n' = 1,...,\infty$ and $\delta_{n,n'}$ is the usual Kronecker symbol. Its derivative reads,
\begin{eqnarray}
\Theta^{[1]}_{n\,n'}(x) & = & {d\Theta_{n,n'}(x)\over dx} 
= \delta_{n,n'}\Bigl\{{1\over n\,(1+({x\over 2n})^2)}
+ \sum_{l=1}^{n -1}{2\over l(1+({x\over 2l})^2)}\Bigr\} + (1-\delta_{n,n'})\Bigl\{{2\over |n-n'|(1+({x\over
|n-n'|})^2)} 
\nonumber \\
& + & \sum_{l=1}^{{n+n'-|n-n'|-2\over 2}}{4\over
(|n-n'|+2l)(1+({x\over |n-n'|+2l})^2)} + {2\over (n+n')(1+({x\over n+n'})^2)}\Bigr\} \, .
\label{The1}
\end{eqnarray}

The BA momentum bands have often exotic limiting values, in some cases dependent
on the densities. In the case of the spin-$1/2$ $XXX$ chain, each $n$-band, such that $q_{j+1}-q_j=2\pi/L$, 
has a momentum range $q_j \in [q_n^{-},q_n^{+}]$ whose limiting momentum values $q_n^{\pm}$ are
given in Eq. (\ref{qj}). Within the TL they read,
\begin{equation}
q_n^{\pm} = \pm {\pi\over L}\,\left(L_n-1\right) \approx \pm \pi\,m_n  \hspace{0.20cm}{\rm where}\hspace{0.20cm} 
m_n = n_n + n_n^h = L_n/L  \hspace{0.20cm}{\rm and}\hspace{0.20cm} n_n^h = N_n^h/L \, .
\label{mmmm}
\end{equation}

The 1D Hubbard model set $j=1,...,L_{\beta}$ of $\beta =c, \alpha n$ bands discrete momentum values $q_j$ belong to well-defined domains, 
$q_j\in [q_{\beta}^-,q_{\beta}^+]$, where 
\begin{eqnarray}
q_{c}^{\pm} & =  & \pm {\pi\over L}(L-1) \approx \pm\pi \hspace{0.20cm}{\rm for}\hspace{0.20cm}N^{SU(2)}\hspace{0.20cm}{\rm odd} \, ;
\hspace{0.20cm}
q_{c}^{\pm} = \pm {\pi\over L}(L-1\pm 1) \approx \pm\pi \hspace{0.20cm}{\rm for}\hspace{0.20cm}N^{SU(2)}\hspace{0.20cm}{\rm even} \, ,
\nonumber \\
q_{\alpha n}^{\pm} & = & \pm {\pi\over L}(L_{\alpha n}-1) \, ,
\label{qcan-range}
\end{eqnarray}
and the number $N^{SU(2)}$ is given in Eq. (\ref{F-beta}).

Finally, the 1D Hubbard model energy eigenvalues and related energy scales are provided.
The former have the following general functional form when expressed
in terms of the $\beta =c,\alpha n$ band momentum distribution functions $N_{\beta} (q_j)$
and number $M_{\alpha,-1/2}$ of unpaired spins $(\alpha =s)$ and unpaired $\eta$-spins $(\alpha =\eta)$ 
of projection $-1/2$,
\begin{equation}
E = \sum_{j=1}^{L}\left(N_{c} (q_j)\,E_c (q_j) + U/4 - \mu_{\eta}\right)
+ \sum_{\alpha=\eta,s}\sum_{n=1}^{\infty}\sum_{j=1}^{L_{\alpha n}}\,N_{\alpha n} (q_j)\,E_{\alpha n} (q_j) 
+ \sum_{\alpha=\eta,s}2\mu_{\alpha}\,M_{\alpha,-1/2} \, .
\label{E}
\end{equation}
Here, 
\begin{equation}
2\mu_{s} = 2\mu_B\,\vert H\vert 
\, ; \hspace{0.50cm} 2\mu_{\eta} = 2\vert\mu\vert \hspace{0.20cm}{\rm for}\hspace{0.20cm}  n_e \neq 1
\, ; \hspace{0.50cm} 2\mu_{\eta} = 2\mu_0 \hspace{0.20cm}{\rm for}\hspace{0.20cm}  n_e = 1 \, ,
\label{2mu-eta-s}
\end{equation}
and
\begin{eqnarray}
E_c (q_j) & = & - 2t\cos k^c (q_j) - U/2 + \mu_{\eta} - \mu_s \, ,
\nonumber \\
E_{\alpha n} (q_j) & = & 2n\mu_{\alpha} + 
\delta_{\alpha,\eta}\left(2t\sum_{\iota = \pm1}\sqrt{1-(\Lambda^{\eta n} (q_j) -i\,\iota\,nu)^2} - nU\right)
 \hspace{0.20cm}{\rm where}\hspace{0.20cm} \alpha = \eta,s  \hspace{0.20cm}{\rm and}\hspace{0.20cm} n =1,...,\infty \, . 
\label{spectra-E-an-c-0}
\end{eqnarray}
For each $u>0$ energy and momentum eigenstate, the momentum rapidity function $k^c (q_j)$,
related $c$-band rapidity $\Lambda^{c} (q_j) = [\sin k^c (q_j)]/u$, and the $n =1,...,\infty$ rapidity 
functions $\Lambda^{\eta n} (q_j)$ are the solutions of the TBA equations, Eqs. (\ref{Tapco1}) and (\ref{Tapco2}).

The energy scale $2\mu^0$ in Eq. (\ref{2mu-eta-s}) is the $n_e=1$ half-filling Mott-Hubbard 
gap \cite{Lieb,Lieb-03,Ovchi-70}. For $u>0$ it is an even function of the spin density $m$ that remains finite for 
the whole interval, $m\in [-1,1]$. For instance, at spin densities $m=0$ \cite{Lieb,Lieb-03} and $m=-1,1$ it reads,
\begin{eqnarray}
2\mu^0 & = & U -4t + 8t\int_0^{\infty}d\omega {J_1 (\omega)\over\omega\,(1+e^{2\omega u})} 
= {16\,t^2\over U}\int_1^{\infty}d\omega {\sqrt{\omega^2-1}\over\sinh\left({2\pi t\omega\over U}\right)} 
\hspace{0.20cm} {\rm for}\hspace{0.20cm}  m = 0 \, ,
\nonumber \\
& = & \sqrt{(4t)^2+U^2} - 4t 
\hspace{0.20cm}{\rm for}\hspace{0.20cm}  m = -1,1 \, ,
\label{2mu0}
\end{eqnarray}
respectively. Its $u\ll 1$ limiting behaviors \cite{Ovchi-70} 
are $2\mu^0 \approx (8/\pi)\,\sqrt{t\,U}\,e^{-2\pi \left({t\over U}\right)}$ at $m = 0$
and $2\mu^0 \approx U^2/8t$ for $m = \pm 1$. Its $u\gg 1$ behavior
is $2\mu^0 \approx (U - 4t)$ for the whole $m \in [-1,1]$ range.

For electronic densities $n_e\in ]0,1[$, spin density $m=0$, and the whole $u>0$ range 
the maximum $sn$ pseudoparticle pairing energy in Eq. (\ref{Wan-gen}) vanishes.
The maximum $\eta n$ pseudoparticle pairing energy in that equation has for 
such densities and limiting interaction values $u\rightarrow 0$ and $u\gg 1$ the following limiting behaviors,
\begin{eqnarray}
W_{\eta n}^{\rm pair}  & = & \vert\varepsilon_{\eta n}^0 (0)\vert =
4t\cos\left({\pi\over 2}n_e\right) = 2\vert\mu\vert 
\hspace{0.20cm}{\rm for}\hspace{0.20cm} u \rightarrow 0 \, ,
\nonumber \\
& = & {8(1-n_e)\,t^2\over n U}\left(1-{\sin (2\pi (1-n_e))\over 2\pi (1-n_e)}\right)
\hspace{0.20cm}{\rm for}\hspace{0.20cm} u \gg 1 \, , 
\label{Wetanu-UU}
\end{eqnarray}
respectively. 

For the electronic density interval $n_e\in ]0,1[$ and spin density $m\rightarrow n_e$, 
the maximum $sn$ and $\eta n$ pseudoparticle pairing energies in 
in Eq. (\ref{Wan-gen}) have the following analytical expressions
valid for the whole $u>0$ range,
\begin{eqnarray}
W_{sn}^{\rm pair} & = & \vert\varepsilon_{sn}^0 (0)\vert =
\sqrt{(4t)^2+(n U)^2}\,{1\over\pi}\arctan\left({\sqrt{(4t)^2+(n U)^2}\over (n U)}\tan (\pi n_e)\right)
\nonumber \\
& - & n U\,n_e - {4t\over\pi}\cos (\pi n_e) \arctan\left({4t\sin (\pi n_e)\over n U}\right) 
\hspace{0.20cm}{\rm for}\hspace{0.20cm}n_e \in ]0,1[\hspace{0.20cm}{\rm and}\hspace{0.20cm}m\rightarrow n_e \, ,
\label{Ws-nu}
\end{eqnarray}
and
\begin{eqnarray}
W_{\eta n}^{\rm pair} & = & \vert\varepsilon_{\eta n}^0 (0)\vert =
\sqrt{(4t)^2+(n U)^2}\,{1\over\pi}\arctan\left({\sqrt{(4t)^2+(n U)^2}\over (n U)}\tan (\pi (1-n_e))\right)
\nonumber \\
& - & n U\,(1-n_e) - {4t\over\pi}\cos (\pi (1-n_e)) \arctan\left({4t\sin (\pi (1-n_e))\over n U}\right) 
\hspace{0.20cm}{\rm for}\hspace{0.20cm}n_e \in ]0,1[\hspace{0.20cm}{\rm and}\hspace{0.20cm}m\rightarrow n_e \, ,
\label{Weta-nu}
\end{eqnarray}
respectively. For $u\rightarrow 0$ and $u\gg 1$ these expressions simplify to,
\begin{eqnarray}
W_{sn}^{\rm pair} & = & \vert\varepsilon_{sn}^0 (0)\vert = 4t\sin^2 \left({\pi n_e\over 2}\right) = 2\mu_B\,H_c 
\hspace{0.20cm}{\rm for}\hspace{0.20cm} u \rightarrow 0 \, ,
\nonumber \\
& = & {8n_e\,t^2\over n U}\left(1-{\sin (2\pi n_e)\over 2\pi n_e}\right) =  {1\over n}\,2\mu_B\,H_c
\hspace{0.20cm}{\rm for}\hspace{0.20cm} u \gg 1 \, , 
\label{Wsn-U0-Ul}
\end{eqnarray}
and
\begin{eqnarray}
W_{\eta n}^{\rm pair} & = & \vert\varepsilon_{\eta n}^0 (0)\vert = 4t\sin^2 \left({\pi (1-n_e)\over 2}\right) = 2\vert\mu\vert 
\hspace{0.20cm}{\rm for}\hspace{0.20cm} u \rightarrow 0 \, ,
\nonumber \\
& = & {8(1-n_e)\,t^2\over n U}\left(1-{\sin (2\pi (1-n_e))\over 2\pi (1-n_e)}\right)
\hspace{0.20cm}{\rm for}\hspace{0.20cm} u \gg 1 \, ,
\label{Wetanu-UUmne}
\end{eqnarray}
respectively. 

\section{Number of state representations of the spin-$1/2$ $XXX$ chain and 1D Hubbard model symmetries}
\label{HMSymmetry}

The spin-$1/2$ $XXX$ chain has a global spin $SU(2)$ symmetry whose number of independent state representations
is $2^{L}$. Out of such $\sum_{2S=0\,({\rm integers})}^{L}\,{\cal{N}}(S) = 2^{L}$ state representations, there is
for a given spin $S$ a number ${\cal{N}}(S) = (2S+1)\,{\cal{N}}_{\rm singlet} (S)$ of representations. Those 
correspond to $(2S+1)$ multiplet configurations and a number, 
\begin{equation}
{\cal{N}}_{\rm singlet} (S) = {L\choose L/2-S}-{L\choose L/2-S-1} \, ,
\label{NsingletS}
\end{equation}
of singlet configurations. In Appendix A of Ref. \cite{Takahashi-71} it is shown
that for LWSs the dimension, Eq. (\ref{NsingletS}), can alternatively be written for each $S$-fixed 
subspace as,
\begin{equation}
{\cal{N}}_{\rm singlet} (S) = \sum_{\{N_{n}\}}\,\prod_{n =1}^{\infty} {L_n\choose N_n} \, .
\label{Nsinglet-MM}
\end{equation}
As justified in Section \ref{pseudoRoots}, this expression also applies to the non-LWSs 
belonging to the same $SU(2)$ tower as the LWS it refers to.
Here $\sum_{\{N_{n}\}}$ is a summation over all sets of the numbers $\{N_{n}\}$ corresponding to the same number of
spin-singlet pairs, $\Pi=\sum_{n=1}^{\infty}n\,N_n=(L-2S)/2$, Eq. (\ref{Nsingletpairs}).
The equality of the dimensions in Eqs. (\ref{NsingletS}) and (\ref{Nsinglet-MM}) confirms
that the Hilbert space of the spin-$1/2$ $XXX$ chain, Eq. (\ref{Hchain}), is spanned by a 
number $2^L$ of energy eigenstates. It equals that of its symmetry independent state representations.

As reported in Section \ref{alphanpseudop}, the 1D Hubbard model Hilbert-space dimension $4^L$ equals 
the number of independent state representations of its global $[SU(2)\otimes SU(2)\otimes U(1)]/Z_2^2$ 
symmetry. This is the second issue addressed in this Appendix. Such a model $c$ pseudoparticles, spins $1/2$, and $\eta$-spins $1/2$ 
configurations that generate an energy eigenstate are a superposition of local original lattice occupancy 
configurations. The rotated-electron occupancies of a number $L_{\eta}=L-N_c$ of original lattice sites separate 
into two degrees of freedom: Those of the $c$ lattice $U(1)$ symmetry associated with $N_c^h=L_{\eta}$ $c$ band
holes and the $\eta$-spin $SU(2)$ symmetry degrees of freedom associated with
$L_{\eta}=L-N_c$ $\eta$-spins $1/2$, respectively. The degrees
of freedom of rotated-electron occupancies of the remaining $L_{s}=N_c$ original lattice sites also separate
into two degrees of freedom: Those of the $c$ lattice $U(1)$ symmetry associated with $N_c=L_{s}$ $c$ pseudoparticles
and the spin $SU(2)$ symmetry degrees of freedom associated with $L_{s} = N_c$ spins $1/2$, respectively. 

On the one hand, each $\alpha n$-pairs configuration occupies a number $2n$ of original lattice sites.
The set of such configurations of an energy eigenstate thus occupy a number 
$2\Pi_{\alpha}=\sum_{n=1}^{\infty} 2n\,N_{\alpha n}$ of original lattice sites. On the other hand, each of the 
$M_{\alpha}=2S_{\alpha}$ unpaired spins $1/2$ ($\alpha =s$) and unpaired $\eta$-spins $1/2$ ($\alpha =\eta$) 
singly occupies an original lattice site. Similarly, each of the $N_c$ $c$ pseudoparticles singly occupies an original lattice site.
The remaining $N_c^h=L-N_c$ sites remain unoccupied in what their $c$ lattice $U(1)$ symmetry degrees
of freedom is concerned. Therefore, for an energy eigenstate with fixed 
$N_c\in [0,L]$, spin $S_s$, and $\eta$-spin $S_{\eta}$ values, the following 
number of original lattice sites sum rules are fulfilled,
\begin{equation}
L = \sum_{\alpha=\eta,s}(M_{\alpha} + 2\Pi_{\alpha}) = \sum_{\alpha=\eta,s}(2S_{\alpha} + \sum_{n=1}^{\infty}2n\,N_{\alpha n}) 
\hspace{0.20cm}{\rm and}\hspace{0.20cm} L = N_c^h + N_c \, .
\label{LL}
\end{equation}
They refer to the two $SU(2)$ symmetries degrees of freedom of the $L$ original lattice sites occupancies and their
$c$ lattice $U(1)$ symmetry degrees of freedom, respectively.

As for the spin-$1/2$ $XXX$ chain, there is a strong requirement for each $\alpha n$-string referring to an 
$\alpha n$-pairs configuration. Such a configuration involves a number $2n$ of spins $1/2$ ($\alpha =s$) and 
$\eta$-spins $1/2$ ($\alpha =\eta$) within $l = 1,...,n$ $\alpha$-singlet pairs. The requirement
under consideration is that in the dimension of any $S_{\alpha}$-fixed subspace ${\cal{N}} (S_{\alpha}) = (2S_{\alpha}+1)\,{\cal{N}}_{\rm singlet} (S_{\alpha})$,
the number of independent $\alpha$-singlet configurations ${\cal{N}}_{\rm singlet} (S_{\alpha})$ is {\it exactly} the same when 
obtained from the counting of the following two types of apparently different
configurations: (i) Two $\alpha =\eta,s$ $SU(2)$ group state representations associated with
the spins $1/2$ ($\alpha =s$) and $\eta$-spins $1/2$ ($\alpha =\eta$) independent configurations with the same 
spin and $\eta$-spin, respectively, $S_{\alpha}$; (ii) The independent $n=1,...,\infty$ bands $\{q_j\}$ occupancy configurations
of the sets of $N_{\alpha n}$ TBA $\alpha n$-strings that obey the sum rule 
$\sum_{n=1}^{\infty}n\,N_{\alpha n} = (L_{\alpha}-2S_{\alpha})/2$, Eq. (\ref{sum-Nseta}).
(The factor $(2S_{\alpha}+1)$ in the dimension ${\cal{N}} (S_{\alpha}) = (2S_{\alpha}+1)\,{\cal{N}}_{\rm singlet} (S_{\alpha})$ refers to
the number of multiplet configurations of the $M_{\alpha} = 2S_{\alpha}$ unpaired spins $1/2$ ($\alpha =s$) and 
unpaired $\eta$-spins $1/2$ ($\alpha =\eta$) that are not paired
within $\alpha n$-pairs configurations $\alpha$-singlet pairs.)

On the one hand, it follows directly from the $SU(2)$ symmetry algebra that number of independent spin ($\alpha =s$) and 
$\eta$-spin ($\alpha =\eta$) $SU(2)$ symmetry state representations 
of the 1D Hubbard model in a fixed-$N_c$ and fixed-$S_{\alpha}$ subspace is
given by ${\cal N}(S_{\alpha},L_{\alpha})=(2S_{\alpha} +1)\,{\cal{N}}_{\rm singlet} (S_{\alpha},L_{\alpha})$. Here,
\begin{equation}
{\cal{N}}_{\rm singlet} (S_{\alpha},L_{\alpha}) =  {L_{\alpha} \choose \Pi_{\alpha}}-
{L_{\alpha}\choose \Pi_{\alpha}-1}\hspace{0.20cm}{\rm for}\hspace{0.20cm}\alpha=\eta, s \, ,
\label{N-singlet}
\end{equation} 
is the corresponding number of independent spin ($\alpha =s$) and $\eta$-spin ($\alpha =\eta$) 
$\alpha$-singlet state representations.

On the other hand, as for the spin-$1/2$ $XXX$ chain, the value of the number $N_{\alpha n}^h = L_{\alpha n} - N_{\alpha n}$ of 
$\alpha n$-band holes that naturally emerges 
from the TBA, Eq. (\ref{N-h-an}), ensures that for each $S_{\alpha}$-fixed subspace the $\alpha$-singlet
dimension ${\cal{N}}_{\rm singlet} (S_{\alpha})$ given in Eq. (\ref{N-singlet}) can indeed alternatively be written as, 
\begin{equation}
{\cal{N}}_{\rm singlet} (S_{\alpha},N_c) =
\sum_{\{N_{\alpha n}\}}\, \prod_{n =1}^{\infty}\,{L_{\alpha n}\choose N_{\alpha n}}\hspace{0.20cm}{\rm for}\hspace{0.20cm}\alpha=\eta, s \, .
\label{Ncs-cpb}
\end{equation} 
The summation $\sum_{\{N_{\alpha n}\}}$ runs here over all sets of $\alpha n$-strings numbers $\{N_{\alpha n}\}$ 
corresponding to the same fixed spin ($\alpha =s$) and $\eta$-spin ($\alpha =\eta$)
$S_{\alpha}=L_{\alpha}/2 - \sum_{n=1}^{\infty}n\,N_{\alpha n}$. This is imposed by the exact sum rule, Eq. (\ref{sum-Nseta}).

The demonstration in Appendix A of Ref. \cite{Takahashi-71} for spin LWSs of the spin-$1/2$ $XXX$ chain
of the equality of the dimensions given in Eqs. (\ref{N-singlet}) and (\ref{Ncs-cpb}), respectively,
also applies to the 1D Hubbard model. Specifically, it applies to that model spin LWSs ($\alpha =s$) and $\eta$-spins LWSs 
($\alpha =\eta$). This also holds for the multiplet towers of non-LWSs generated from $S_{\alpha}>0$ LWSs. 
All $2S_{\alpha}+1$ states of such a tower have indeed exactly the same $\alpha$-singlet configurations as the 
corresponding $S_{\alpha}>0$ LWS.

In each subspace with fixed values for $L_{s} = N_c$, $L_{\eta} = N_c^h = L-N_c$, $S_s$, and $S_{\eta}$, there are   
${\cal{N}}(S_{\eta},L_{\eta}) \times {\cal{N}}(S_{s},L_{s}) \times d_c (N_c)$ state representations of the
$SU(2)\otimes SU(2)\otimes U(1)$ symmetry in the model two $SU(2)\otimes SU(2)\otimes U(1)]/Z_2^2$ symmetry. Here,
\begin{equation}
d_c ={L\choose N_c} = {L\choose N_c^h} \, ,
\label{dc}
\end{equation}
gives the number of independent $c$ pseudoparticles occupancy configurations. It equals that of state representations
of the $c$ lattice $U(1)$ symmetry in the subspace under consideration. 

The following completeness sum rule has been obtained in Refs. \cite{Completeness,Complete2,Complete3} by 
use of the $\alpha =\eta,s$ dimensions ${\cal{N}}_{\rm singlet} (S_{\alpha},N_c)$ and $c$ dimension $d_c$, 
Eqs. (\ref{Ncs-cpb}) and (\ref{dc}), respectively,
\begin{eqnarray}
4^{L} & = & \sum_{\substack{N_{c}=0\\({\rm integers})}}^{L} 
\sum_{\substack{2S_{\eta}=0\\({\rm integers})}}^{L_{\eta}=L-N_c}\,
\sum_{\substack{2S_{s}=0\\({\rm integers})}}^{L_{s}=N_c} C (N_c,S_{\eta},S_{s})\,
{\cal{N}}(S_{\eta},L_{\eta}) \times {\cal{N}}(S_{s},L_{s}) \times d_c (N_c) \, ,
\nonumber \\
& & C (N_c,S_{\eta},S_{s}) = \vert\cos\left({\pi\over 2}(2S_{\eta}+N_c)\right)\cos\left({\pi\over 2}(2S_{s}+N_c)\right)\vert = 0,1\, .
\label{Ntot-dess}
\end{eqnarray}
The role of the phase factor, $C (N_c,S_{\eta},S_{s})=0,1$, is to select the allowed independent representations of the model
two $SU(2)$ symmetries. 

The main issue under consideration here is the equality of the dimensions given in Eqs. (\ref{N-singlet}) and (\ref{Ncs-cpb}),.
Beyond the results of Refs. \cite{Completeness,Complete2,Complete3}, it shows that the number of independent state representations 
of the 1D Hubbard model global $[SU(2)\otimes SU(2)\otimes U(1)]/Z_2^2$ symmetry exactly equals its Hilbert-space 
dimension $4^L$. 

\section{Spin-$1/2$ $XXX$ chain specific heat in the critical regime}
\label{XXXconfig2}

Here it is shown that in the $2\mu_B\vert H-H_c\vert\ll k_B T$ limit where the expression, Eq. (\ref{cV-hc}), is valid it 
is exactly the same as that obtained from the scaling function of the specific heat in the critical regime considered
in Ref. \cite{He-17}. The latter is denoted here by $c_V/L$. In the units used in that reference it reads,
\begin{equation}
{c_V\over L} = \sqrt{T\over\pi J}\left(-{3\over 8}\,f^s_{3/2} + {1\over 2}{\Delta\over T}\,f^s_{1/2}
- {1\over 2}\left({\Delta\over T}\right)^2\,f^s_{-1/2}\right) \, ,
\label{CvHe}
\end{equation}
where $f^s_n = {\rm Li}_n \left(-e^{\Delta\over T}\right)$, $\Delta = 4J - h$, and ${\rm Li}_n (x) = \sum_{l=1}^{\infty}x^l/l^n$.

Up to the first order in $\Delta/T\ll 1$ that the expression, Eq. (\ref{cV-hc}), refers to, only the first two terms in Eq. (\ref{CvHe})
contribute through the following expansions,
\begin{eqnarray}
f^s_{3/2} & \approx & {\rm Li}_{3/2} (-1-\Delta/T) \approx
{\rm Li}_{3/2} (-1) - {\Delta\over T}{\partial {\rm Li}_{3/2} (x)\over \partial x}\vert_{x=-1}
= - \sum_{l=1}^{\infty}{(-1)^{l-1}\over l^{3/2}} - {\Delta\over T}\sum_{l=1}^{\infty}{(-1)^{l-1}\over l^{1/2}} 
\nonumber \\
& = & - {1\over\sqrt{2}}(\sqrt{2}-1)\,\zeta (3/2) + {\Delta\over T}(\sqrt{2}-1)\,\zeta (1/2) \, ,
\nonumber \\
f^s_{1/2} & \approx & {\rm Li}_{1/2} (-1) = - \sum_{l=1}^{\infty}{(-1)^{l-1}\over l^{1/2}} 
= (\sqrt{2}-1)\,\zeta (1/2) \, .
\label{fsfs}
\end{eqnarray}

On the one hand, the use of these expansions in the expression, Eq. (\ref{CvHe}), leads to the following specific heat
expansion up to first order in $\Delta/T\ll 1$,
\begin{equation}
{c_V\over L} = \sqrt{T\over\pi J}\left({3\over 8\sqrt{2}}\,(\sqrt{2}-1)\,\zeta (3/2)
+ {1\over 8}\,{\Delta\over T}\,(\sqrt{2}-1)\,\zeta (1/2)\right) \, .
\label{CvHeExp}
\end{equation}

On the other hand, the use in Eq. (\ref{cV-hc}) of the coefficient expressions in 
Eq. (\ref{c012}) accounting for that $\Gamma (1/2) = \sqrt{\pi}$ and
$\Gamma (3/2) = \sqrt{\pi}/2$ leads to,
\begin{equation}
{c_V\over L} = k_B\,\sqrt{2k_BT\over\pi J}\left({3\over 8\sqrt{2}}\,(\sqrt{2}-1)\,\zeta (3/2)
+ {1\over 8}\,{2\mu_B(H_c-H)\over T}\,(\sqrt{2}-1)\,\zeta (1/2)\right) \, .
\label{cV-hcEXP}
\end{equation}

The studies of Ref. \cite{He-17} and those of this review use different units. From analysis of the corresponding
Hamiltonian expressions, one finds that the parameters $k_B$, $J$, and $2\mu_B$ used in this review 
correspond in the units of Ref. \cite{He-17} to $1$, $2J$, and $1$, respectively. Hence
$H_c =J/\mu_B$ becomes $4J$. Finally, under the corresponding units transformations
$k_B\rightarrow 1$, $J\rightarrow 2J$, $2\mu_B\rightarrow 1$, and $H_c\rightarrow 4J$, 
the expansion, Eq. (\ref{cV-hcEXP}), becomes exactly that given in Eq. (\ref{CvHeExp}).

\section{Spin-$1/2$ $XXX$ chain current expression, susceptibility, and zero-temperature stiffness}
\label{XXXconfig3}

The spin-$1/2$ $XXX$ chain spin susceptibility $\chi$ and zero-temperature spin stiffness $D$ expressions in Eq. (\ref{chiD})
are derived in this Appendix by means of procedures that resemble those of a Fermi liquid. The latter
stiffness is related to the current operator expectation values. We start by confirming that in the TL their
usual expression obtained from the BA for LWSs can be written in the $n$-bands holes representation,
as given in Eq. (\ref{J-part}).

Such LWSs current operator expectation values can be derived from the $\Phi/L$ dependence
of the energy eigenvalues $E(\Phi/L)$ of the spin-$1/2$ $XXX$ chain in a uniform vector potential $\Phi/L$,
Eq. (A2) of Ref. \cite{CTD-15}. The spin currents are then given by $\langle \hat{J}^z\rangle = d E(\Phi/L)/d(\Phi/L)\vert_{\Phi=0}$. 
This straightforwardly leads to \cite{CTD-15,CT-17}
$\langle\hat{J}^z_{LWS} (l_{\rm r},S)\rangle = \sum_{n=1}^{\infty}\sum_{j=1}^{L_n}\,N_n (q_j)\,\,j_n (q_j)$.
Here $j_n (q_j)$ is given by $j_n (q_j) = - j_n^h (q_j)$, Eq. (\ref{jn-fn}).

In order to confirm the equality within the TL of this $\langle\hat{J}^z_{LWS} (l_{\rm r},S)\rangle$ expression and
that in Eq. (\ref{J-part}), it is useful to replace the set of $n$-bands discrete momentum values $\{q_j\}$
by a continuum momentum variable $q\in [q_n^-,q_n^+]$. 
The elementary currents $ j_n^h (q_j)$ in Eq. (\ref{jn-fn}) can 
then be exactly written as $j_n^h (q_j) = - 2J\,d\cos k^n (q)/dq$ 
for $q \in [q_n^-,q_n^+]$. It was used here that $d k^n (q)/d q=1/[2\pi\sigma^n (k^n (q))]$.
The equality under consideration requires that
$\sum_{n=1}^{\infty}\sum_{j=1}^{L_n}\,j_n^h (q_j) = 0$.
This quantity can be written as,
\begin{equation}
- {L\over 2\pi}\sum_{n=1}^{\infty}\int_{q_n^-}^{q_n^+}\,2J{d\over dq} \cos k^n (q)
= - {L\over 2\pi}\sum_{n=1}^{\infty}\sum_{\iota = \pm}(\iota)\,2J \cos k^n (q_n^{\iota}) = 0 \, .
\label{J-part0TL}
\end{equation}
It indeed vanishes. Here the relation $q_n^- = -q_n^+$ was used.

The spin susceptibility is controlled by transitions between ground states referring to different
canonical ensembles. As given in Eq. (\ref{N0qHm}), those are not populated by $n$-pseudoparticles with
$n>1$ pairs. Hence here we consider the spin chain in the subspace spanned by energy eigenstates
that are not populated by such pseudoparticles. In that subspace the general energy functional, Eq. (\ref{DEnHm}), simplifies to,
\begin{eqnarray}
\delta E & = & \sum_{j=1}^{L_1}\varepsilon (q_j)\delta N (q_j) 
+ {1\over L}\sum_{j=1}^{L_1}\sum_{j'=1}^{L_{1}}{1\over 2}\,f (q_j,q_{j'})\,\delta N (q_j)\delta N (q_{j'}) 
\nonumber \\
& = & {L\over 2\pi}\int_{-k_{F\uparrow}}^{k_{F\uparrow}}dq \varepsilon (q)\delta N (q) 
+ {L\over 4\pi^2}\int_{-k_{F\uparrow}}^{k_{F\uparrow}}dq\int_{-k_{F\uparrow}}^{k_{F\uparrow}}dq'{1\over 2}\,
f (q,q')\,\delta N (q)\delta N (q') \, .
\label{DEnHmN1}
\end{eqnarray}
Here $\delta N (q_j)\equiv \delta N_1 (q_j)$, $\varepsilon (q_j)\equiv \varepsilon_1 (q_j)$, $f (q_j,q_{j'})\equiv f_{1\,1} (q_j,q_{j'})$.
Within the TL, we have again replaced in the second expression the discrete momentum values such that
$q_{j+1}-q_j=2\pi/L$ by continuum momentum variables.

As in a Fermi liquid, the energy contributions of second order in the $n=1$ band momentum distribution deviations in 
Eq. (\ref{DEnHmN1}) lead to corrections in the $n=1$ energy dispersion $\varepsilon^0 (q)\equiv \varepsilon_1^0 (q)$, Eq. (\ref{varepsilon-nHm}) for $n=1$.
Up to first order in these deviations one finds,
\begin{equation}
\breve{\varepsilon}^0 (q) = \varepsilon^0 (q) + {1\over 2\pi}\int_{-k_{F\uparrow}}^{k_{F\uparrow}} 
d q' \,f  (q,q')\,\delta N (q') \, .
\label{varepsilon-nHmN1}
\end{equation}

The derivation of the spin susceptibility involves small deviations $\delta q_F =\delta k_{F\downarrow}$
in the $n=1$ band Fermi momentum associated with transitions to ground states. For them
that band momentum distribution in Eq. (\ref{N0qHm}) reads,
\begin{equation}
N (q) = \theta (k_{F\downarrow} + \delta k_{F\downarrow} - \vert q\vert) \, .
\label{N0qHmDEV}
\end{equation}
Expanding this distribution around that of the initial ground state, leads
to $N (q) = \theta (k_{F\downarrow} - \vert q\vert) + \delta N (q)$
where the deviation $\delta N (q)$ is given by,
\begin{equation}
\delta N (q) = \delta (k_{F\downarrow} - \vert q\vert)\,\delta k_{F\downarrow} \, .
\label{deltaN1F}
\end{equation}
Here $\delta (x)$ is the usual delta function. 

Inserting the deviations, Eq. (\ref{deltaN1F}), in the energy dispersion,
Eq. (\ref{varepsilon-nHmN1}), and replacing the obtained expressions in 
Eq. (\ref{hm}), which defines the magnetization curve, leads to,
\begin{equation}
{\partial h(m)\over \partial m}  = - {1\over 2\mu_B}\left(v + {1\over 2\pi}\sum_{\iota = \pm}
f  (k_{F\downarrow},\iota k_{F\downarrow})\right){\delta k_{F\downarrow}\over\delta m}
= - {v\,(\xi^0)^2\over 2\mu_B}\,{\delta k_{F\downarrow}\over\delta m} 
= - {v^0\over 2\mu_B}\,{\delta k_{F\downarrow}\over\delta m} \, .
\label{derivhm}
\end{equation}
Here $v^0 = v +  {1\over 2\pi}\sum_{\iota = \pm}\,f (k_{F\downarrow},\iota k_{F\downarrow}) = v\,(\xi^0)^2$ 
and $\xi^0 = 1 + \Phi (k_{F\downarrow},k_{F\downarrow}) + \Phi (k_{F\downarrow},-k_{F\downarrow})$
where $i = 0,1$, $v \equiv v_1 (k_{F\downarrow})$ and $\Phi\equiv \Phi_{1\,1}$. 

From the use of the relation $\delta k_{F\downarrow} = -\pi \delta m$ one finds
$\delta k_{F\downarrow}/\delta m = -\pi$. The use of Eq. (\ref{derivhm}) with
$\delta k_{F\downarrow}/\delta m = -\pi$ in the spin susceptibility expression
$\chi = 2\mu_B/(\partial h(m)/\partial m)$ readily leads to $\chi = 4\mu_B^2/(\pi\,v_{0})$.
This is the expression given in Eq. (\ref{chiD}).

The derivation of the zero-temperature spin stiffness $D$ involves again the spin-$1/2$ $XXX$ chain in the subspace spanned 
by energy eigenstates that are not populated by pseudoparticles with $n>1$ spin-singlet pairs. For simplicity, the $n=1$ $n$-pseudoparticles 
are here called pseudoparticles. We consider low-frequency excitations involving a small density of 
pseudoparticles with momentum $q$ in the vicinity of the $n=1$ band Fermi points $\pm k_{F\downarrow}$.
As in a Fermi liquid, such excitations are described by deviations $\delta N (q;x,t)$. They depend explicitly
on both position $x$ and time $t$. The corresponding momentum distribution functions reads,
\begin{equation}
N (q) = \theta (k_{F\downarrow} - \vert q\vert) + \delta N (q;x,t) \, .
\label{Nqxt}
\end{equation}
These distribution functions describe true low-frequency excitations of the spin chain. The particular
form of the inhomogeneous time-dependent pseudoparticle deviations that describe these excitations
can be obtained by solving kinetic equations. Those are introduced below.

Now the energy dispersion $\varepsilon (q_j)\equiv \varepsilon_1 (q_j)$ in Eq. (\ref{varepsilon-nHm}) for $n=1$
becomes a local function of the pseudoparticle deviations $\delta N (q;x,t)$. To first order in these
deviations it is given by,
\begin{equation}
\breve{\varepsilon} (q;x,t) = \varepsilon (q) + {1\over 2\pi}\int_{-k_{F\uparrow}}^{k_{F\uparrow}} 
d q' \,f  (q,q')\,\delta N (q';x,t) \, .
\label{varepsilon-qxt}
\end{equation}
To compute the flow of pseudoparticles through each side of a small volume element (1D segment), one considers 
the balance of the flow inward and outward. This leads to the following kinetic equation,
\begin{equation}
{\partial N (q;x,t)\over\partial t} + {\partial N (q;x,t)\over\partial x} {\partial\breve{\varepsilon} (q;x,t)\over\partial q} 
- {\partial N (q;x,t)\over\partial q} {\partial\breve{\varepsilon} (q;x,t)\over\partial x} = 0 \, .
\label{KinEq}
\end{equation}

It is useful to introduce a weak inhomogeneous magnetic-field probe ${1\over 2}h(x,t)$. It is associated
with the system conserved spin projection $S^z$. This requires an additional term on the left-hand
side of Eq. (\ref{KinEq}). It is given by ${\partial N (q;x,t)\over\partial q}{\cal{F}}(q;x,t)$. Here
${\cal{F}}(q;x,t)=-{\partial {1\over 2}h(x,t)\over\partial x}$ is the force felt by the pseudoparticles due
to the applied magnetic-field probe, ${1\over 2}h(x,t)$. 

The expression in Eq. (\ref{varepsilon-qxt}) is only valid for excitations involving a small density 
of pseudoparticles. From the use of Eq. (\ref{Nqxt}) in
the kinetic equation, Eq. (\ref{KinEq}), keeping contributions up to first order in the deviations and
introducing the term associated with the coupling to the external probe, we obtain,
\begin{equation}
{\partial\,\delta N (q;x,t)\over\partial t} + v (q)\,{\partial\,\delta N (q;x,t)\over\partial x} 
- {\partial N^0 (q)\over\partial q}\left({1\over 2\pi}\int_{-k_{F\uparrow}}^{k_{F\uparrow}} 
d q' \,f  (q,q')\,{\partial\,\delta N (q';x,t)\over\partial x} + {\partial\,{1\over 2}h(x,t)\over\partial x}\right) = 0 \, .
\label{DeltaKinEq}
\end{equation}
Here $N^0 (q) = \theta (k_{F\downarrow} - \vert q\vert)$.

We now consider excitations that are periodic in space and time with wave vector $k$ and angular frequency
$\omega$. They are characterized by the distribution function,
\begin{equation}
\delta N (q;x,t) = \delta N (q;k,\omega)\,e^{i(kx - \omega t)} + {\rm c.c.} \, .
\label{dNkom}
\end{equation}
The magnetic-field probe $h(x,t)/2$ is also assumed to be periodic in space and time,
$h(x,t)/2 = [h(k,\omega)/2]\,e^{i(kx - \omega t)} + {\rm c.c.}$.

The real part of the spin conductivity can be written for $\omega\rightarrow 0$ as \cite{Shastry-90},
\begin{equation}
{\rm Re}\,\sigma (\omega) = {\rm Re}\,\left(\lim_{k\rightarrow 0}i\left({1\over 2}\right)^2{\omega\over k^2}\chi (k,\omega)\right) \, .
\label{sigma}
\end{equation}
$\chi (k,\omega)$ is here the response function to the magnetic-field probe,
$\chi (k,\omega) =\delta\langle S^z (k,\omega)\rangle/[h(k,\omega)/2]$,
for small $k$ and low $\omega$. To derive this response function one uses the distribution, Eq. (\ref{dNkom}), and 
the above applied magnetic-field probe $h(x,t)/2$ in the kinetic equation, Eq. (\ref{DeltaKinEq}).
The solution of the obtained equation leads after some manipulations to the following form for the 
response function $\chi (k,\omega)$ that is valid for small $k$ and low $\omega$,
\begin{equation}
\chi (k,\omega) = - {1\over\pi}{(4k)^2\,v^1\over v^2k^2 - (\omega + i\delta)} \, .
\label{chikomexp}
\end{equation}
Here $v^1 = v + {1\over 2\pi}\sum_{\iota = \pm} (\iota) f (k_{F\downarrow},\iota k_{F\downarrow}) 
= v\,(\xi^1)^2$. 

Finally, from the use of this expression on the right-hand side of Eq. (\ref{sigma}),
one arrives after some straightforward algebra to ${\rm Re}\,\sigma (\omega) = 4v^1\,\delta (\omega)$.
From the equality $2\pi\,D\,\delta (\omega)=4v^1\,\delta (\omega)$ where $D$ is the spin
stiffness, one finds that at zero temperature it reads $D = 2v_1/\pi$. This is the expression 
given in Eq. (\ref{chiD}).

%%%%%%%%%%%%%%%%%%%%%%%%%%%%%%%%%%%%%%%%%%%%%%%%%%%%%%%%%%%%%%%%
\section{The 1D Hubbard model three fractionalized effective lattices}
\label{FEL}

In this Appendix the 1D Hubbard model $c$ effective lattice and the squeezed spin and $\eta$-spin spin effective lattices 
considered in Section \ref{specificRE} for $u>0$ are shown to naturally emerge from the $u\rightarrow\infty$ 
model properties. Such properties refer to the
electron occupancy configurations that generate the states $\vert  l_{\rm r},l_{\eta s},\infty \rangle$.
Such properties apply for $u>0$ to the rotated electrons that populate the corresponding states
$\vert l_{\rm r},l_{\eta s},u\rangle = {\hat{V}}^{\dag}\vert  l_{\rm r},l_{\eta s},\infty \rangle$
of the same $V (u)$-set.

In the $u\rightarrow\infty$ limit the rotated electrons become electrons and the 
rotated-electron numbers apply as well to electron
occupancies. The corresponding electron spatial coordinates are used in Eq. (2.23) of Ref. \cite{Woy}
for the wave functions of the LWSs $\vert  l_{\rm r},l_{\eta s}^0,\infty \rangle$. Here
it is denoted by $\Psi_{l_{\rm r},l_{\eta s}^0,\infty} (x^s,...;x^d,...;x^{s\downarrow},...) = 
\langle  x^s,...;x^d,...;x^{s\downarrow},...\vert l_{\rm r},l_{\eta s}^0,\infty\rangle$. In their argument,
$x^s,...;x^d,...;x^{s\downarrow},...$ is a shorten notation for the spatial coordinates 
$x^s,...,x^s_{N_R^{s,0}};x^d,...,x_{N_{R,-1/2}^{\eta,0}}^d;x^{s\downarrow},...,x_{N_{R,-1/2}^{s,0}}^{s\downarrow}$
of the $N_R^{s,0}$ singly occupied sites, $N_{R,-1/2}^{\eta,0}$ doubly occupied sites, and
$N_{R,-1/2}^{s,0}$ spin-down singly occupied sites, respectively. 

One straightforwardly finds that, given two arbitrary operators ${\tilde{M}}={\hat{V}}^{\dag}\,{\hat{M}}\,{\hat{V}}$ and 
${\tilde{N}}={\hat{V}}^{\dag}\,{\hat{N}_e}\,{\hat{V}}$, the matrix-element relations
$\langle l_{\rm r},l_{\eta s},u \vert {\tilde{M}}{\tilde{N}}\vert  0_{\rm elec}\rangle =
\langle  l_{\rm r},l_{\eta s},\infty\vert {\hat{M}}{\hat{N}_e}\vert 0_{\rm elec}\rangle$ and
$\langle l_{\rm r},l_{\eta s},u\vert {\tilde{M}}{\tilde{N}}\vert  l_{\rm r}',l_{\eta s}',u \rangle =
\langle l_{\rm r},l_{\eta s},\infty\vert {\hat{M}}{\hat{N}_e}\vert  l_{\rm r}',l_{\eta s}',\infty\rangle$ hold.
Here the electron vacuum invariance ${\hat{V}}\vert 0_{\rm elec}\rangle=\vert 0_{\rm elec}\rangle$ and 
the $V (u)$-set of states transformation $\vert  l_{\rm r},l_{\eta s},u\rangle={\hat{V}}^{\dag}\vert  l_{\rm r},l_{\eta s},\infty \rangle$ were used.
These matrix-element relations reveal that finite-$u$ correlators of two operators, ${\tilde{M}}$ and ${\tilde{N}}$, exactly equal those of the
corresponding two unrotated operators, ${\hat{M}}$ and ${\hat{N}_e}$, in the $u\rightarrow\infty$ limit.

We denote by $\Psi_{l_{\rm r},l_{\eta s}^0,rt} (x^s,...;x^d,...;x^{s\downarrow},...) = 
\langle  x^s,...;x^d,...;x^{s\downarrow},...\vert l_{\rm r},l_{\eta s}^0,u\rangle$ the $N_e$-rotated-electron wave function
of a finite-$u$ LWS $\vert l_{\rm r},l_{\eta s}^0,u\rangle ={\hat{V}}^{\dag}\vert  l_{\rm r},l_{\eta s}^0,\infty \rangle$. 
It is a function of corresponding rotated-electron spatial coordinates.
From the use of the above matrix-element relations 
one finds that $\Psi_{l_{\rm r},l_{\eta s}^0,rt} (x^s,...;x^d,...;x^{s\downarrow},...) = 
\Psi_{l_{\rm r},l_{\eta s}^0,\infty} (x^s,...;x^d,...;x^{s\downarrow},...)$ for the whole finite-$u$ range.

Combining the equalities $\Psi_{l_{\rm r},l_{\eta s},rt} =\Psi_{l_{\rm r},l_{\eta s}^0,rt}$ for finite $u$ and
$\Psi_{l_{\rm r},l_{\eta s}^0,rt} =\Psi_{l_{\rm r},l_{\eta s}^0,\infty}$ with the expression given in
Eq. (2.23) of Ref. \cite{Woy} for $\Psi_{l_{\rm r},l_{\eta s}^0,\infty}$ leads to,
\begin{equation}
\Psi_{l_{\rm r},l_{\eta s},rt} (x^s,...;x^d,...;x^{s\downarrow}) = 
{1\over\sqrt{{\cal{C}}_{rt}}}\left(\phi_{U(1)}^{c} (x^s,...)\times \phi^{\eta}_{SU(2)} (x^d,...) \times
\phi^{s}_{SU(2)} (x^{s\downarrow},...)\right) \, .
\label{amplitude-1D}
\end{equation}
Here $\phi_{U(1)}^{c} (x^s,...) = (-1)^Q\,\det\left(e^{i k_{Pj}^{\infty}x^s_{Qj}}\right)$ is a $c$ pseudofermion Slater determinant.
The $c$ pseudofermions occupy exactly the same sites as the $c$ pseudoparticles. Moreover, 
$\phi^{\eta}_{SU(2)} (x^d,...) = (-1)^{(N^0-2S_c)/2}\,\phi_1 (x^d,...)$ refers to the $\eta$-spins $1/2$, 
$\phi^{s}_{SU(2)} (x^{s\downarrow},...) =\phi_2  (x^{s\downarrow},...)$ corresponds to the spins $1/2$,
and ${\cal{C}}_{rt}$ is a normalization constant. Since such wave functions refer to rotated electrons, the
BA rapidities on which they depend are $u$ independent. They are actually those of the electrons in the 
$u\rightarrow\infty$ limit.

For a LWS the spatial coordinates of the wave functions $\phi_{U(1)}^{c} (x^s,...)$, $\phi^{\eta}_{SU(2)} (x^d,...)$, and 
$\phi^{s}_{SU(2)} (x^{s\downarrow},...)$ refer in the TL to well-defined lattice sites subsets of the $c$ effective lattice, 
squeezed $\eta$-spin effective lattice, and squeezed spin effective, respectively. 
The $c$ effective lattice is identical to the original lattice. On the contrary, the numbers of sites $L_{\eta}$ 
and $L_s$ of the squeezed $\eta$-spin and spin effective lattices, respectively, are in general smaller
than $L$, as given in Eq. (\ref{Na-eta-s}).

The subset of the $c$ effective lattice $N_c$ sites of spatial coordinates $x^s,...$ in the argument
of $\phi_{U(1)}^{c} (x^s,...)$ refers to those occupied. The remaining $N_c^h = L-N_c$ sites whose spatial coordinates are
uniquely defined are unoccupied. The $L_{\eta}=N_c^h$ sites of the 
squeezed $\eta$-spin effective lattice correspond to those of the original lattice that have the same
spatial coordinates as the $c$ effective lattice unoccupied sites. The subset of $N_{R,-1/2}^{\eta,0}$ sites
of spatial coordinates $x^d,...$ in $\phi^{\eta}_{SU(2)} (x^d,...)$
corresponds to those occupied by $\eta$-spins of projection $-1/2$. The remaining
$N_{R,+1/2}^{\eta,0}=L_{\eta}-N_{R,-1/2}^{\eta,0}$ sites whose spatial coordinates are those left over
are occupied by $\eta$-spins of projection $+1/2$.

Similarly, the $L_{s}=N_c$ sites of the squeezed spin effective lattice refer to those of the original lattice that have the same
spatial coordinates as the $c$ effective lattice occupied sites. The subset of $N_{R,-1/2}^{s,0}$ sites
of spatial coordinates $x^{s\downarrow},...$ in $\phi^{s}_{SU(2)} (x^{s\downarrow},...)$
corresponds to those occupied by spins of projection $-1/2$. The remaining
$N_{R,+1/2}^{s,0}=L_{s}-N_{s,-1/2}^{\eta,0}$ sites whose spatial coordinates are
uniquely defined are occupied by spins of projection $+1/2$.
Within the present TL, the spin effective lattice (and $\eta$-spin effective lattice) has $j=1,...,L_s$ sites 
located at $x = a_s\,j$. (and $j=1,...,L_{\eta}$ sites located at $x = a_{\eta}\,j$ where $j=1,...,L_{\eta}$.)
The spacings $a_s$ and $a_{\eta}$ are given in Eq. (\ref{a-alpha}).

\section{The 1D Hubbard model $s1$ pseudoparticle operator representation}
\label{s1pseudoOR}

The $s1$ pseudoparticle operator representation considered in Section \ref{alphanpseudop}
for the 1D Hubbard model is valid in the {\it extended Takahashi subspaces}. A related
{\it Takahashi subspace} is spanned by all LWSs with the same fixed values for the 
set of numbers $N_c$ and $\{N_{\alpha n}\}$ for $\alpha =\eta,s$ and $n=1,...,\infty$. 
The corresponding extended Takahashi subspace is spanned by such LWSs and the
$(2S_{\eta}+1)\times (2S_{s}+1)-1$ non-LWSs generated from each of them as given in Eq.  (\ref{Gstate-BAstate}).
A general $\alpha n$ pseudoparticle operator representation can be introduced for the 1D Hubbard model in an 
{\it extended Takahashi subspace}. For the results reviewed in this paper only the
$s1$ pseudoparticle operator representation is though needed.

That in fixed-$L_{s1}$ extended Takahashi subspaces the local $s1$ pseudoparticle operators obey a fermionic 
algebra, can be confirmed in terms of their statistical interactions \cite{Haldane-91}. 
The local $s1$ pseudoparticle creation and annihilation operators may be written as,
\begin{equation}
f^{\dag}_{j,s1} = e^{i\phi_{j,s1}}\,g^{\dag}_{j,s1}  \hspace{0.20cm}{\rm and}\hspace{0.20cm}  
f_{j,s1} = (f^{\dag}_{j,s1})^{\dag}  \hspace{0.20cm}{\rm for}\hspace{0.20cm} j = 1,...,L_{s1} \, .
\label{ffs1j}
\end{equation}
Here $\phi_{j,s1} = \sum_{j'\neq j}f^{\dag}_{j',s1}$ and the operator $g^{\dag}_{j,s1}$ obeys a hard-core bosonic algebra. 
This algebra is justified by the corresponding statistical interaction vanishing for the model in fixed-$L_{s1}$ extended 
Takahashi subspaces. The $s1$ effective lattice has 
been constructed inherently to that algebra being of hard-core type for the operators $g^{\dag}_{j,s1}$ and $g_{j,s1}$. 
Therefore, through a Jordan-Wigner 
transformation, $f^{\dag}_{j,s1} = e^{i\phi_{j,s1}}\,g^{\dag}_{j,s1}$ (see for instance Ref. \cite{Wang-92}), the operators
$f^{\dag}_{j,s1}$ and $f_{j,s1} = (f^{\dag}_{j,s1})^{\dag}$ in Eq. (\ref{ffs1j}) obey indeed a Fermionic algebra, Eq. (\ref{ffs1}).

%%%%%%%%%%%%%%%%%%%%%%%%%%%%%%%%%%%%%%%%%%%%%%%%%%%%%%%%%%%%%%%%
\section{Integral equations that define the 1D Hubbard model pseudofermion rapidity phase shifts}
\label{PSIE}

The rapidity phase shifts, Eq. (\ref{Phi-barPhi}), are in units of $2\pi$ uniquely defined by the
following integral equations \cite{Carmelo-97-B},
\begin{equation}
\bar{\Phi }_{s1\,c}\left(r,r'\right) = -{1\over\pi}\arctan(r-r') + \int_{-r^0_s}^{r^0_s}
dr''\,G(r,r'')\,{\bar{\Phi }}_{s1\,c}\left(r'',r'\right) \, ,
\label{Phis1c-m}
\end{equation}
\begin{equation}
\bar{\Phi }_{s1\,\eta n}\left(r,r'\right) =  -
{1\over\pi^2}\int_{-r^0_c}^{r^0_c} dr''{\arctan\Bigl({r''-r'\over n}\Bigr)\over{1+(r-r'')^2}} +
\int_{-r^0_s}^{r^0_s} dr''\,G(r,r'')\,{\bar{\Phi}}_{s1\,\eta n}\left(r'',r'\right) \, , 
\label{Phis1cn-m}
\end{equation}
\begin{eqnarray}
\bar{\Phi }_{s1\,sn}\left(r,r'\right) & = & {\delta_{1 ,n}\over\pi}\arctan\Bigl({r-r'\over 2}\Bigl) + 
{(1-\delta_{1,n})\over\pi}\Bigl(\arctan\Bigl({r-r'\over n-1}\Bigl) + \arctan\Bigl({r-r'\over n+1}\Bigl)\Bigr)
\nonumber \\
& - &  {1\over\pi^2}\int_{-r^0_c}^{r^0_c} dr''{\arctan\Bigl({r''-r'\over n}\Bigr)\over{1+(r-r'')^2}} +
\int_{-r^0_s}^{r^0_s} dr''\,G(r,r'')\,{\bar{\Phi}}_{s1\,s1}\left(r'',r'\right) \, ,
\label{Phis1sn-m}
\end{eqnarray}
\begin{equation}
\bar{\Phi }_{c\,c}\left(r,r'\right) = {1\over\pi}\int_{-r^0_s}^{r^0_s} dr''{\bar{\Phi}_{s1\,c}\left(r'',r'\right) \over {1+(r-r'')^2}} \, ,
\label{Phicc-m}
\end{equation}
\begin{equation}
\bar{\Phi }_{c\,\alpha n}\left(r,r'\right) = -{1\over\pi}\arctan\Bigl({r-r'\over n}\Bigr) +
{1\over\pi}\int_{-r^0_s}^{r^0_s} dr''{\bar{\Phi}_{s1\,\alpha n}\left(r'',r'\right) \over {1+(r-r'')^2}} \, ,
\ \hspace{0.20cm}{\rm where}\hspace{0.20cm} \alpha = \eta, s \, ,
\label{Phiccn-m}
\end{equation}
\begin{equation}
{\bar{\Phi }}_{\eta n\,c}\left(r,r'\right) = {1\over\pi}\arctan\Bigl({r-r'\over {n}}\Bigr) 
- {1\over\pi}\int_{-r_c^0}^{+r_c^0} dr''{{\bar{\Phi}}_{c\,c}\left(r'',r'\right) \over {n[1+({r-r''\over {n}})^2]}} \, , 
\label{Phicnc-m}
\end{equation}
\begin{equation}
\bar{\Phi }_{\eta n\,\eta n'}\left(r,r'\right) = {1\over 2\pi}\,\Theta_{n,n'}(r-r') -
{1\over\pi}\int_{-r_c^0}^{+r_c^0} dr''{\bar{\Phi }_{c\,\eta n'}\left(r'',r'\right) \over {n[1+({r-r''\over n})^2]}} \, , 
\label{Phicncn-m}
\end{equation}
\begin{equation}
\bar{\Phi }_{\eta n\,sn'}\left(r,r'\right) = 
- {1\over\pi}\,\int_{-r_c^0}^{+r_c^0} dr''{\bar{\Phi }_{c\,sn'}\left(r'',r'\right) \over {n[1+({r-r''\over n})^2]}} \, , 
\label{Phicnsn-m}
\end{equation}
\begin{equation}
{\bar{\Phi }}_{sn\,c}\left(r,r'\right) = - {1\over\pi}\arctan\Bigl({r-r'\over {n}}\Bigr) +
{1\over\pi}\int_{-r^0_c}^{r^0_c} dr''{{\bar{\Phi }}_{c,c}\left(r'',r'\right) \over {n[1+({r-r''\over n})^2]}} 
- \int_{-r^0_s}^{r^0_s} dr''{\bar{\Phi }}_{s1\,c}\left(r'',r'\right) {\Theta^{[1]}_{n,1}(r-r'')\over{2\pi}} 
\hspace{0.20cm}{\rm for}\hspace{0.20cm} n > 1 \, , 
\label{Phisnc-m}
\end{equation}
\begin{equation}
{\bar{\Phi }}_{sn\,\eta n'}\left(r,r'\right) =
{1\over\pi}\int_{-r^0_c}^{r^0_c} dr''{{\bar{\Phi}}_{c,\eta n'}\left(r'',r'\right) \over {n[1+({r-r''\over n})^2]}} 
- \int_{-r^0_s}^{r^0_s} dr''{\bar{\Phi}}_{s1\,\eta n'}\left(r'',r'\right) {\Theta^{[1]}_{n,1}(r-r'')\over {2\pi}} 
\hspace{0.20cm}{\rm for}\hspace{0.20cm} n > 1 \, , 
\label{Phisncn-m}
\end{equation}
\begin{equation}
{\bar{\Phi }}_{sn\,sn'}\left(r,r'\right) = {1\over 2\pi}\,\Theta_{n,n'}(r-r') 
+ {1\over\pi}\int_{-r^0_c}^{r^0_c} dr''{{\bar{\Phi}}_{c\,sn'}\left(r'',r'\right) \over {n[1+({r-r''\over n})^2]}} 
- \int_{-r^0_s}^{r^0_s} dr''{\bar{\Phi}}_{s1\,sn'}\left(r'',r'\right) {\Theta^{[1]}_{n,1}(r-r'')\over{2\pi}} \, . 
\label{Phisnsn-m}
\end{equation}

In the above equations, the parameters $r_c^0$ and $r_s^0$ are given in Eq. (\ref{QB-r0rs}),
the functions $\Theta_{n,n'}(x)$ and $\Theta^{[1]}_{n,n'}(x)$ are defined in
Eqs. (\ref{Theta}) and (\ref{The1}) of Appendix \ref{TBAconfig}, respectively, and the kernel $G(r,r')$ 
reads \cite{Carmelo-92-B},
\begin{eqnarray}
G(r,r') & = & - {1\over{2\pi}}\left[{1\over{1+((r-r')/2)^2}}\right]
\nonumber \\
& \times & 
\left(1 - {1\over 2\pi}\sum_{\iota = \pm1}(\iota)\left(\arctan(r + \iota\,r^0_c)
+ \arctan(r' + \iota\,r^0_c)+{1\over (r-r')}
\ln {1+(r + \iota\,r^0_c)^2\over 1+(r' + \iota\,r^0_c)^2}\right)\right) \, .
\label{G}
\end{eqnarray}

The relation of the $m\rightarrow 0$ parameter $\xi_{0}$ in Eq. (\ref{ZZ-gen}) to phase shifts is obtained by
combining such equation with Eq.  (\ref{x-aa}). This gives,
\begin{eqnarray}
\lim_{m\rightarrow 0}\,Z^1 & = & 
\left[\begin{array}{cc}
\xi_{0} & \xi_{0}/2 \\
0 & 1/\sqrt{2} 
\end{array}\right] =
\left[\begin{array}{cc}
1 & 0 \\
0 & 1
\end{array}\right] +
\sum_{\iota = \pm}\left[\begin{array}{cc}
\Phi_{c\,c}(\iota 2k_F,2k_F) & \Phi_{c\,s1}(\iota 2k_F,k_F) \\
\Phi_{s1\,c}(\iota k_F,2k_F) & \Phi_{s1\,s1}(\iota k_F,k_F)  
\end{array}\right] \, ,
\nonumber \\
\lim_{m\rightarrow 0}\,Z^0 & = & \left[\begin{array}{cc}
1/\xi_{0} & 0 \\
-\xi_{0}/2 & \sqrt{2} 
\end{array}\right] =
\left[\begin{array}{cc}
1 & 0 \\
0 & 1
\end{array}\right] +
\sum_{\iota = \pm}(\iota)\left[\begin{array}{cc}
\Phi_{c,c}(\iota 2k_F\,2k_F) & \Phi_{c\,s1}(\iota 2k_F,k_F) \\
\Phi_{s1,c}(\iota k_F\,2k_F) & \Phi_{s1\,s1}(\iota k_F,k_F)  
\end{array}\right] \, .
\label{ZZPS}
\end{eqnarray}
Here $Z^0 = ((Z^1)^{-1})^T$. In the above equations we have accounted for that the $c$ and $s1$ band Fermi points in Eq. (\ref{q0Fcs}) 
read $q_{Fc} = 2k_F$ and $q_{Fs1} =k_F$, respectively, in the $m\rightarrow 0$ limit

Conversely, in the $m\rightarrow 0$ limit the $c$ and $s1$ pseudofermion phase shifts with both momenta 
at the Fermi points can be expressed in terms of only the parameter $\xi_{0}$ as follows,
\begin{eqnarray}
2\pi\Phi_{c\,c}(\iota\,2k_F,2k_F) = \iota\,2\pi\Phi_{c,c}(2k_F,\iota\,2k_F) 
& = & {\pi\,(\xi_{0} -1)^2\over \xi_{0}}\hspace{0.20cm}{\rm for}\hspace{0.25cm}\iota = + \, ,
\nonumber \\
& = & {\pi\,(\xi_{0}^2 -1)\over \xi_{0}}\hspace{0.20cm}{\rm for}\hspace{0.25cm}\iota = - \, ,
\nonumber \\
2\pi\Phi_{c\,s1}(\iota\,2k_F,k_F,) = \iota\,2\pi\Phi_{c\,s1}(2k_F,\iota\,k_F,)  
& = & {\pi\over 2}\,\xi_{0}\hspace{0.20cm}{\rm for}\hspace{0.25cm}\iota = \pm \, .
\label{PhiFP}
\end{eqnarray}

%%%%%%%%%%%%%%%%%%%%%%%%%%%%%%%%%%%%%%%%%%%%%%%%%%%%%%%%%%%%%%%%
\section{$\delta S_{\alpha} = \mp n$ $\alpha$-multiplet elementary processes associated with creation and
annihilation of one $\alpha n$ pseudoparticle in the spin density $m\rightarrow n_e$ limit}
\label{Wm*}

For electronic densities in the interval $n_e\in ]0,1[$, spin density $m\rightarrow n_e$, and the whole $u>0$ range
the energy dispersions $\varepsilon_{\alpha n}^0 (q_j)$ and $\varepsilon_{\alpha n} (q_j)$ are for small
momentum $q_j$ $(\alpha =s)$ and small momentum deviations $(q_j \mp q_{\eta n})$ $(\alpha =\eta)$ given by,
\begin{eqnarray}
\varepsilon_{sn}^0 (q_j) & \approx & {q_j^2\over 2m_{sn}^*} - W_{sn} = {q_j^2\over 2m_{sn}^*} - W_{sn}^{\rm pair}
\hspace{0.20cm}{\rm for}\hspace{0.20cm} q_j \approx 0
\hspace{0.20cm}{\rm and}\hspace{0.20cm}\varepsilon_{sn} (q_j) = \varepsilon_{sn}^0 (q_j) + n\,2\mu_B H \, ,
\nonumber \\
\varepsilon_{\eta n}^0 (q_j) & \approx & {(q_j \mp q_{\eta n})^2\over 2m_{\eta n}^*}  
\hspace{0.20cm}{\rm for}\hspace{0.20cm}(q_j \mp q_{\eta n}) \approx 0
\hspace{0.20cm}{\rm and}\hspace{0.20cm}\varepsilon_{\eta n} (q_j) = \varepsilon_{\eta n}^0 (q_j) + n\,2\mu \, .
\label{bandssnetansq}
\end{eqnarray}
Here $W_{sn}=W_{sn}^{\rm pair}$ is given in Eq. (\ref{Ws-nu}) of Appendix \ref{TBAconfig}.

The effective masses $m_{sn}^*$ and $m_{\eta n}^*$ in these expressions
have the following analytical expressions that are functions of $n_e$ and $U/t$,
\begin{equation}
m_{sn}^* = {n U\over 4t^2}{{\sqrt{(4t)^2+n U^2}\over n U}{1\over\pi}\arctan\left({\sqrt{(4t)^2+(n U)^2}\over n U}\tan (\pi n_e)\right)
\over 1 - {\sqrt{(4t)^2+ (n U)^2}\over n U}{1\over 1 + \left({4t\sin (\pi n_e)\over n U}\right)^2}
{\sin (2\pi n_e)\over 2\arctan\left({\sqrt{(4t)^2+ (n U)^2}\over n U}\tan (\pi n_e)\right)}} 
\hspace{0.20cm}{\rm for}\hspace{0.20cm}n_e \in ]0,1[\hspace{0.20cm}{\rm and}\hspace{0.20cm}m\rightarrow n_e \, ,
\label{msnstar}
\end{equation}
and 
\begin{equation}
m_{\eta n}^* = {n U\over 4t^2}{{\sqrt{(4t)^2+n U^2}\over n U}{1\over\pi}\arctan\left({\sqrt{(4t)^2+(n U)^2}\over n U}\tan (\pi (1-n_e))\right)
\over 1 - {\sqrt{(4t)^2+ (n U)^2}\over n U}{1\over 1 + \left({4t\sin (\pi (1-n_e))\over n U}\right)^2}
{\sin (2\pi (1-n_e))\over 2\arctan\left({\sqrt{(4t)^2+ (n U)^2}\over n U}\tan (\pi (1-n_e))\right)}} 
\hspace{0.20cm}{\rm for}\hspace{0.20cm}n_e \in ]0,1[\hspace{0.20cm}{\rm and}\hspace{0.20cm}m\rightarrow n_e \, ,
\label{metanstar}
\end{equation}
respectively. In the $u\rightarrow 0$ and $u\gg 1$ limits these expressions read,
\begin{eqnarray}
m_{sn}^* & = & {1\over 2t}\hspace{0.20cm}{\rm for}\hspace{0.20cm}u\rightarrow 0 
\nonumber \\
m_{sn}^* & = & {n U\over 4t^2}{n_e\over\left(1 - {\sin (2\pi n_e)\over 2\pi n_e}\right)}
\hspace{0.20cm}{\rm for}\hspace{0.20cm}u \gg 1 \, ,
\label{msnlimits}
\end{eqnarray}
and
\begin{eqnarray}
m_{\eta n}^* & = & {1\over 2t}\hspace{0.20cm}{\rm for}\hspace{0.20cm}u\rightarrow 0 
\nonumber \\
m_{\eta n}^* & = & {n U\over 4t^2}{(1-n_e)\over\left(1 - {\sin (2\pi (1-n_e))\over 2\pi (1-n_e)}\right)}
\hspace{0.20cm}{\rm for}\hspace{0.20cm}u \gg 1 \, ,
\label{metanlimits}
\end{eqnarray}
respectively. In the case of the $s1$ band, the energy dispersion, Eq. (\ref{bandssnetansq}) for $\alpha n=s1$,  
is that given in Eq. (\ref{vares1smallq}). The corresponding effective triplet mass is provided in Eq. (\ref{ms1star}). 

For $h\approx H_c$ and electronic densities $n_e\in ]0,1[$ not too close to $0$ and $1$, 
the elementary process associated with creation of one $s n$ pseudoparticle or one 
$\eta n$ pseudoparticle onto the ground state leads to a spin or $\eta$-spin deviation $\delta S_s = -n=-1,...,-L_s/2$ or
$\delta S_{\eta} = -n =-1,...,-L_{\eta}/2$, respectively. Under such an elementary process, a number 
$2n=2,4,...,L_s$ or $2n=2,4,...,L_{\eta}$ of initial-state unpaired spins $1/2$ or unpaired $\eta$-spins $1/2$, respectively, become paired within the 
final-state pseudoparticle spin-singlet $s n$-pairs configuration or $\eta$-spin-singlet $\eta n$-pairs configuration,
respectively. 

The opposite elementary process associated with the annihilation of one 
$s n$ pseudoparticle $(\alpha =s)$ or one $\eta n$ pseudoparticle $(\alpha =\eta)$ to return to the initial state, involves
the breaking of all its $n$ $\alpha$-singlet pairs. This gives rise to the emergence of
$2n$ unpaired spins $1/2$ or unpaired $\eta$-spins $1/2$, respectively, in the new final
state. Such an opposite elementary process leads to a spin or $\eta$-spin deviation $\delta S_s = n=1,...,L_s/2$ or
$\delta S_{\eta} = n=1,...,L_{\eta}/2$, respectively.

Each above mentioned $\delta S_{\alpha} = -n$ elementary process has a minimum excitation 
energy given by $\Delta_{sn}^{\rm min} = \varepsilon_{sn} (0)= n\,2\mu_B H - W_{sn}^{\rm pair}$ for $\alpha =s$
or $\Delta_{\eta n}^{\rm min} = \varepsilon_{\eta n} (\pm q_{\eta n})
= n\,2\mu$ for $\alpha =\eta$. For $n>1$ the gap $\Delta_{sn}^{\rm min}$ increases from 
$\Delta_{sn}^{\rm min}=(n-1)\, 2\mu_B H + 2\mu_B (H-H_c)
\approx (n-1)\, 2\mu_B H$ for $u\rightarrow 0$ to $\Delta_{sn}^{\rm min}=(n-1/n)\, 2\mu_B H + 2\mu_B (H-H_c)/n
\approx (n-1/n)\, 2\mu_B H$ for $u\gg 1$ and $\Delta_{\eta n}^{\rm min}$ reads
$\Delta_{\eta n}^{\rm min} = n\,(U + 4t\cos (\pi n_e)) + n\,2\mu_B (H-H_c) \approx n\,(U + 4t\cos (\pi n_e))$.

Hence within the low-temperature crossover critical regime considered in Section \ref{exc-spectraQL2}, for which
$2\mu_B\vert H-H_c\vert\ll k_B T$, only the spin-triplet channel associated with elementary
$\delta S_s = \pm 1$ spin-triplet processes with $\delta S_s = -1$ is available. Its excitation energy reads 
$\Delta_{s1}^{\rm min}=2\mu_B (H-H_c)$. The next two minimum gaps read $\Delta_{s2}^{\rm min} = \varepsilon_{s2} (0)
= 4\mu_B h - W_{s2}^{\rm pair}$ for a spin $\delta S_s = -2$ elementary process,
which is the minimum gap $\Delta_{s}^{\rm min}$, Eqs. (\ref{Deltasmin}) and (\ref{Deltasminlimyts}),
and $\Delta_{\eta 1}^{\rm min} = (U + 4t\cos (\pi n_e))$ for an $\eta$-spin-triplet $\delta S_{\eta} = -1$ elementary process.
Those refer though to high-energy processes such that $\Delta_{s2}^{\rm min}\gg 2\mu_B (H-H_c)$ and
$\Delta_{\eta 1}^{\rm min}\gg 2\mu_B (H-H_c)$.

Finally, one finds that $W_{\alpha n}^{\rm pair}$ and $m_{\alpha n}^*$ are related yet different quantities whose interplay 
partially controls the $\delta S_{\alpha} = \pm n$ elementary processes under consideration. This is achieved
from inspection for electronic densities $n_e\in ]0,1[$ and spin density $m\rightarrow n_e$
of the form of both the energy dispersions in Eq. (\ref{bandssnetansq}), maximum pairing energies 
$W_{\alpha n}^{\rm pair}$ in Eq. (\ref{Wan-gen}) and Eqs. (\ref{Ws-nu})-(\ref{Wetanu-UUmne}) of Appendix \ref{TBAconfig},
and effective masses $m_{\alpha n}^*$ in Eqs. (\ref{ms1star}) and (\ref{ms1muhculim}) and
Eqs. (\ref{msnstar})-(\ref{metanlimits}).

This applies to the spin-triplet $S_s = \pm 1$ elementary processes. Those contribute to the low-temperature
specific heat expressions, Eqs. (\ref{cV-HMhcGCR}) and (\ref{cV-HMhc}), in the crossover critical regime. It
refers to a small field window  $2\mu_B\vert H-H_c\vert\ll k_B T$ around $H_c$. In this case, the
spin-triplet effective mass $m_{s1}^*$, Eqs. (\ref{ms1star}) and (\ref{ms1muhculim}), and the maximum
pairing energy $W_{s1}^{\rm pair}=W_{s1}=2\mu_B H_c$, Eq. (\ref{muhc}), associated with the spin-singlet configuration binding of the two 
paired spins $1/2$ within the $s1$ pseudoparticle are different yet related quantities. They are given by
$m_{s1}^*= {\partial^2 \varepsilon_{s1}^0 (q)\over \partial q^2}\vert_{q=0}$
and $W_{s1}=(\varepsilon_{s1}^0 (q_{s1})-\varepsilon_{s1}^0 (0)) = -\varepsilon_{s1}^0 (0) = 2\mu_B H_c$, respectively.
Here $q_{s1} = 2k_F = \pi n_e$ for $m\rightarrow n_e$.

%%%%%%%%%%%%%%%%%%%%%%%%%%%%%%%%%%%%%%%%%%%%%%%%%%%%%%%%%%%%%%%%
\section{Derivation of the $\beta =c,s1$ pseudofermion spectral function within the TL and example of momentum-dependent exponents}
\label{SIPDT}

The 1D Hubbard model $\beta =c,s1$ pseudofermion spectral function general expression, Eq. (\ref{BQ-gen}), 
involves the $\beta =c,s1$ lowest peak weight $A^{(0,0)}_{\beta}$ and relative weight
$a_{\beta}=a_{\beta}(m_{\beta,\,+1},\,m_{\beta,\,-1})$. After a suitable algebra
similar to that reported in Ref. \cite{Karlo-97} for the $u\rightarrow\infty$ spin-less fermion spectral function,
one finds that the former weight refers to a Slater determinant of $\beta =c,s1$ pseudofermion operators. It involves
$\beta =c,s1$ pseudofermion anticommutators associated with two $\beta$ pseudofermions of canonical momentum 
${\bar{q}}_j$ and ${\bar{q}_{j'}}$, respectively. Here ${\bar{q}}_j$ and ${\bar{q}}_{j'}=q_{j'}$ 
correspond to a PS excited state and the corresponding ground-state $\beta$ band, respectively. 
One then finds the anticommutators \cite{V-1},
\begin{equation}
\{f^{\dag }_{{\bar{q}}_j,\beta},f_{{\bar{q}}_{j'},\beta}\} =
{1\over L_{\beta}}\,e^{-i({\bar{q}}_j-{\bar{q}}_{j'})/
2}\,e^{i\,2\pi\Phi_{\beta}^T(q_j)/2}\,{\sin\Bigl(2\pi\Phi_{\beta}^T (q_j)/
2\Bigr)\over\sin ([{\bar{q}}_j-{\bar{q}}_{j'}]/2)} \hspace{0.20cm}{\rm and}\hspace{0.20cm}
2\pi\Phi_{\beta}^T (q_j) = 2\pi\Phi_{\beta}^0 + 2\pi\Phi_{\beta} (q_j) \, , 
\label{pfacrGS}
\end{equation}
and $\{f^{\dag}_{{\bar{q}}_{j},\beta},f^{\dag}_{{\bar{q}}_{j'},\beta}\} = \{f_{{\bar{q}}_{j},\beta},f_{{\bar{q}}_{j'},\beta}\}=0$.
Here $2\pi\Phi_{\beta}^0$, Eq. (\ref{pican}), is the non-scattering part of the overall $\beta$ pseudofermion
phase shift denoted by $2\pi\Phi_{\beta}^T (q_j)$ in this equation.

The use of the Slater determinant of $\beta =c,s1$ pseudofermion operators that involves such anticommutators 
leads after some algebra to the following general expression for the $\beta =c,s1$ lowest peak weight 
$A^{(0,0)}_{\beta}$ \cite{V-1,CarCadez-16,CarCadez-17},
\begin{eqnarray}
A^{(0,0)}_{\beta} & = & \Big({1\over
L}\Bigr)^{2N^{\odot}_{\beta}}\, \prod_{j=1}^{L_{\beta}}\,
\sin^2\Bigl({\pi\over 2}\left(1- (1-2\Phi^T_{\beta}(q_j))N_{\beta}^{\odot}(q_j)\right)\Big)\, \prod_{j=1}^{L_{\beta}-1}\,
\Bigl(\sin\Bigl({\pi j\over L}\Bigr)\Bigr)^{2(L_{\beta} -j)} 
\nonumber \\
& \times &
\prod_{i=1}^{L_{\beta}}\prod_{j=1}^{L_{\beta}}\,\theta (j-i)\,
\sin^2\left({\pi\over 2}\left(1 - \left(1 - {(2(j-i) + 2\Phi^T_{\beta}({q}_j) - 2\Phi^T_{\beta}({q}_i))
\over L}\right)N_{\beta}^{\odot}({q}_j)N_{\beta}^{\odot}({q}_i)\right)\right) 
\nonumber \\
& \times &
\prod_{i=1}^{L_{\beta}}\prod_{j=1}^{L_{\beta}}\,{1\over
\sin^2\left({\pi\over 2}\left(1 - \left(1 - {2(j-i) + 2\Phi^T_{\beta}({q}_j)\over L}\right)
N_{\beta}^{\odot}({q}_i)N_{\beta}^{\odot}({q}_j)\right)\right)}  
\hspace{0.20cm}{\rm where}\hspace{0.20cm}  \beta = c, s1 \, .
\label{A00}
\end{eqnarray}
The numbers of $\beta =c,s1$ band discrete momentum values, $L_{\beta}$,
$\beta =c,s1$ pseudofermions, $N_{\beta}^{\odot}=\sum_{j=1}^{L_{\beta}}N_{\beta}^{\odot}({q}_j)$,
and the corresponding $\beta$ band momentum distribution function, $N_{\beta}^{\odot}({q}_j)$, are
in this expression those of the PS excited state generated by the processes (A) and (B) as defined
in Section \ref{PDT}. Moreover, $\Phi^T_{\beta}({q}_j)$ is the phase-shift functional in Eq. (\ref{pfacrGS}) in units of $2\pi$.

The general expression of the relative weights $a_{\beta}=a_{\beta}(m_{\beta,\,+1},\,m_{\beta,\,-1})$ in Eq. (\ref{BQ-gen}),  
reads \cite{V-1,CarCadez-16,CarCadez-17},
\begin{equation}
a_{\beta}(m_{\beta,\,+1},m_{\beta,\,-1})=\Bigl(\prod_{\iota = \pm}
a_{\beta,\iota}(m_{\beta,\iota})\Bigr)
\Bigl(1+{\cal{O}}\Bigl(\ln L/L\Bigr)\Bigr) 
\hspace{0.20cm}{\rm where}\hspace{0.20cm} \beta = c, s1 \, ,
\label{aNNDP}
\end{equation}
where,
\begin{equation}
a_{\beta,\iota}(m_{\beta,\iota}) = \prod_{j=1}^{m_{\beta,\iota}}
{(2\Delta_{\beta}^{\iota} + j -1)\over j} = \frac{\Gamma (m_{\beta,\iota} +
2\Delta_{\beta}^{\iota})}{\Gamma (m_{\beta,\iota}+1)\,
\Gamma (2\Delta_{\beta}^{\iota})} 
 \hspace{0.20cm}{\rm where}\hspace{0.20cm} 
\beta = c, s1 \hspace{0.20cm}{\rm and}\hspace{0.20cm} \iota = \pm \, ,
\label{aNDP}
\end{equation}
and $\Gamma (x)$ is the usual gamma function. The relative weights, Eq. (\ref{aNNDP}), are associated 
with the tower of excited energy eigenstates generated by the processes (C) as defined in Section \ref{PDT}.

For $m_{\beta,\iota}=1$, Eq. (\ref{aNDP}) leads to,
\begin{equation}
a_{\beta,\iota}(1) = 2\Delta_{\beta}^{\iota} =\left({\delta {\bar{q}}_{F\beta}^{\iota}\over (2\pi/L)}\right)^2 
\hspace{0.20cm}{\rm where}\hspace{0.20cm}\beta = c, s1  \hspace{0.20cm}{\rm and}\hspace{0.20cm}  \iota = \pm \, . 
\label{a10DP-iota}
\end{equation}
Here $a_{\beta,\iota}(1)$ is the relative weight of the $\alpha,\iota$ pseudofermion spectral function $m_{\beta,\iota}=1$ peaks.
It can be written as given in Eq. (\ref{functional}). Moreover, 
$\delta {\bar{q}}_{F\beta}^{\iota}/(2\pi/L) =\iota\,\delta N^F_{\beta,\iota}+\Phi_{\beta} (\iota q_{F\beta})$
is the excited-state canonical momentum $\beta =c,s1;\iota = \pm$ Fermi-point deviation.
The $\beta = c, s1$ weights $a_{\beta,\iota}(1)$ correspond to the particular cases $a_{\beta} (1,\,0)= 2\Delta_{\beta}^{+1}$
and $a_{\beta} (0,\,1) = 2\Delta_{\beta}^{-1} $ of the general relative weights, Eq. (\ref{aNNDP}), 

The $\delta$-functions in the pseudofermion spectral function expression, Eq. (\ref{BQ-gen}),
impose the important equality $((L/4\pi\,v_{\beta})(\omega' +\iota\,v_{\beta}\,k')-\Delta_{\beta}^{\iota})=m_{\beta,\iota}$.
Hence $((L/4\pi\,v_{\beta})(\omega' +\iota\,v_{\beta}\,k')-\Delta_{\beta}^{\iota})$ is proportional to $L$. This implies that 
for any arbitrarily small $k'$ and $\omega'$ values for which $0<(\omega' +\iota\,v\,k')/(4\pi v)\ll 1$ the corresponding 
values of the $\iota = \pm$ integer numbers, 
\begin{equation}
m_{\iota}={L\over 4\pi\,v_{\beta}}\left(\omega' +\iota\,v_{\beta}\,k')-\Delta_{\beta}^{\iota}\right) \, ,
\label{miota}
\end{equation}
are in the TL such that $m_{\beta,\iota}\gg 1$. Hence in the TL the $\beta,\iota$ relative weight, Eq. (\ref{aNDP}), has
the following asymptotic behavior \cite{V-1,CarCadez-16,CarCadez-17},
\begin{equation}
a_{\beta,\iota} (m_{\beta,\iota}) \approx \frac{1}{\Gamma (2\Delta_{\beta}^{\iota})}
\Bigl(m_{\beta,\iota}+\Delta_{\beta}^{\iota}\Bigr)^{2\Delta_{\beta}^{\iota}-1}
\hspace{0.20cm}{\rm for}\hspace{0.20cm}2\Delta_{\beta}^{\iota}\neq 0 
\hspace{0.20cm}{\rm where}\hspace{0.20cm} \beta = c, s1 \hspace{0.20cm}{\rm and}\hspace{0.20cm} \iota = \pm \, . 
\label{f}
\end{equation}
Furthermore, in the TL the $\beta =c,s1$ lowest peak weight $A^{(0,0)}_{\beta}$, Eq. (\ref{A00}), 
can be written as,
\begin{equation}
A^{(0,0)}_{\beta} = {F^{(0,0)}_{\beta}\over (L\,S_{\beta})^{-1+2\Delta_{\beta}^{+1} +2\Delta_{\beta}^{-1}}}
 \hspace{0.20cm}{\rm where}\hspace{0.20cm} \beta = c,\,s1 \, .
\label{F00}
\end{equation}
Here $F^{(0,0)}_{\beta}$ and $S_{\beta}$ are in the TL independent of $L$ and
$2\Delta_{c}^{+1}$, $2\Delta_{c}^{-1}$, $2\Delta_{s1}^{+1}$, and $2\Delta_{s1}^{-1}$
are the four functionals, Eq. (\ref{a10DP-iota}). 

In the general case in which the values of such four functionals are finite, 
one finds from the use of Eq. (\ref{f}) in the $\beta =c,s1$ pseudofermion spectral function 
expression, Eq. (\ref{BQ-gen}), that in the TL it can be written as,
\begin{equation}
B_{Q_{\beta}} (k',\omega') = {L\over 4\pi v_{\beta}}\,
A^{(0,0)}_{\beta}\,\prod_{\iota = \pm}\,a_{\beta,\iota}
\Bigl({\omega' +\iota\,v_{\beta}\,k'\over 4\pi v_{\beta}/L}\Bigr) 
\hspace{0.20cm}{\rm where}\hspace{0.20cm} \beta = c,s1 \, .
\label{B-J-i-sum-GG=}
\end{equation}
Further use in this expression of Eqs. (\ref{f}) and (\ref{F00}) leads finally to
the following expression for the $\beta =c,s1$ pseudofermion spectral function, Eq. (\ref{BQ-gen}), 
valid in the TL,
\begin{equation}
B_{Q_{\beta}} (k',\omega') = 
{F^{(0,0)}_{\beta}\over 4\pi\,v_{\beta}\,S_{\beta}}\,
\prod_{\iota = \pm}\,{\Theta (\omega' +\iota\,v_{\beta}\,k')\over 
\Gamma (2\Delta_{\beta}^{\iota})}\,
\Bigl({\omega' +\iota\,v_{\beta}\,k'\over 4\pi \,v_{\beta}\,S_{\beta}}\Bigr)^{-1 +2\Delta_{\beta}^{\iota}} 
\hspace{0.20cm}{\rm where}\hspace{0.20cm} \beta = c,s1 \, .
\label{B-J-i-sum-GG}
\end{equation}
Here $F^{(0,0)}_{\beta}$ and $S_{\beta}$ are the $L$ independent quantities in the $A^{(0,0)}_{\beta}$ expression,
Eq. (\ref{F00}). (The product $S_{c}\times S_{s1}$ reads $1$ both in the $u\rightarrow 0$ and $u\rightarrow\infty$ limits.) 

In applications of the 1D Hubbard model PDT to the description of the ARPES of 1D metallic states in physical systems, 
the one-electron removal spectral function at zero spin density is that of interest. For electronic densities in the range 
$n_e\in [0,1]$, that spectral function has three main branch lines 
called $c$, $c'$, and $s1$ branch line. Their energy spectra have the following simple expressions
in terms of the $\beta =c$ and $\beta =s1$ bands energy dispersions 
$\varepsilon_{\beta} (q_j)$ defined in Eq. (\ref{epsilon-q}),  
\begin{eqnarray}
\omega_c (k) & = & \varepsilon_c (\vert k\vert + k_F)\leq 0
\nonumber \\
k & = & -{\rm sgn}\{k\} k_F - q \in [-k_F,k_F] 
\nonumber \\
\omega_{c'} (k) & = & \varepsilon_c (\vert k\vert - k_F) \leq 0
\nonumber \\
k & = & {\rm sgn}\{k\} k_F - q \in [-3k_F,3k_F] \, .  
\nonumber \\
\omega_{s1} (k) & = & \varepsilon_{s1} (k) = \varepsilon_{s1} (k) \leq 0
\nonumber \\
k & = & -q \in [-k_F,k_F] \, .
\label{3spectraOESF}
\end{eqnarray}
Here $k$ is the one-electron excitation momentum.

The corresponding PDT momentum dependent exponents in Eq. (\ref{branch-l}) 
that control the line shape of the one-electron removal spectral function 
near the $c$, $c'$, and $s1$ branch lines are given by,
\begin{eqnarray}
\zeta_c (k) & = & -{1\over 2} + \sum_{\iota=\pm1}\left({\xi_0\over 4} - \Phi_{c\,c}(\iota 2k_F,q)\right)^2  
\nonumber \\
k & = & \in [-k_F,k_F] \, , 
\nonumber \\
q & = & -{\rm sgn}\{k\} k_F - k \in [-2k_F,-k_F] \hspace{0.2cm}{\rm and}\hspace{0.1cm}\in [k_F,2k_F] \, ,
\nonumber \\
\zeta_{c'} (k) & = & -{1\over 2} + \sum_{\iota=\pm1}\left({\xi_0\over 4} - \Phi_{c\,c}(\iota 2k_F,q)\right)^2  
\nonumber \\
k & = & \in [-3k_F,3k_F] \, , 
\nonumber \\
q & = & {\rm sgn}\{k\} k_F - k \in [-2k_F,k_F]\hspace{0.2cm}{\rm and}\hspace{0.1cm}\in[-k_F,2k_F] \, .
\nonumber \\
\zeta_{s1} (k) & = & -1 + \sum_{\iota=\pm1}\left({\iota\over 2\xi_0} + \Phi_{c\,s1}(\iota 2k_F,q')\right)^2  
\nonumber \\
k & \in & [-k_F,k_F]\hspace{0.25cm}{\rm and}\hspace{0.25cm}q' = -k \in  [-k_F,k_F] \, .
\label{3expoOESF}
\end{eqnarray}
The parameter $\xi_0$ appearing here is given by $\xi_{0} = \xi_{0} (r_c^0)$ where the function $\xi_{0} (r)$ is the unique
solution of the integral equation, Eq. (74) of Ref \cite{Carmelo-92-B} with $x=r$.
It has limiting values $\xi_0=\sqrt{2}$ for $u\rightarrow 0$ and $\xi_0=1$ for $u\rightarrow\infty$. 
The $c$ pseudofermion phase shifts $\Phi_{c\,c}(\pm 2k_F,q)$ and $\Phi_{c\,s1}(\pm 2k_F,q')$ also appearing
in the exponents expressions provided in Eq. (\ref{3expoOESF}) 
are through the general relation $\Phi_{\beta\,\beta'}(q_j,q_{j'}) = \bar{\Phi }_{\beta\,\beta'} \left(r,r'\right)$, 
Eq. (\ref{Phi-barPhi}), defined in terms of corresponding rapidity phase shifts $\bar{\Phi }_{c\,c} \left(\pm r^0_c,r\right)$
and $\bar{\Phi }_{c\,s1} \left(\pm r^0_c,r'\right)$, respectively. The latter are defined by the integral equations
given in Appendix \ref{PSIE}.

%%%%%%%%%%%%%%%%%%%%%%%%%%%%%%%%%%%%%%%%%%%%%%%%%%%%%%%%%%%%%%%%%%%%%%%%%%

\end{document}